\documentclass[longauth]{aa}
\usepackage{graphicx}
\usepackage{txfonts}
\usepackage[breaklinks=true]{hyperref}
\makeatletter
\renewcommand*\aa@pageof{, page \thepage{} of \pageref*{LastPage}}
\usepackage{amsmath}	
\usepackage{amssymb}	
\usepackage{natbib}
\usepackage{colortbl}
\usepackage{multirow}
\usepackage[normalem]{ulem}
\usepackage{longtable}

\graphicspath{{./}{Figures_subm/}}

\newcommand{\HI}{\textrm{H~{\textsc{i}}}}
\newcommand{\HII}{\textrm{H~{\textsc{ii}}}}
\newcommand{\HeI}{\textrm{He~{\textsc{i}}}}

\newcommand{\CI}{\textrm{C~{\textsc{i}}}}
\newcommand{\CII}{\textrm{C~{\textsc{ii}}}}

\newcommand{\FeII}{\textrm{Fe~{\textsc{ii}}}}
\newcommand{\FeIII}{\textrm{Fe~{\textsc{iii}}}}
\newcommand{\NI}{\textrm{N~{\textsc{i}}}}
\newcommand{\NiII}{\textrm{Ni~{\textsc{ii}}}}
\newcommand{\NiIII}{\textrm{Ni~{\textsc{iii}}}}

\newcommand{\OI}{\textrm{O~{\textsc{i}}}}

\newcommand{\SII}{\textrm{S~{\textsc{ii}}}}

\newcommand{\mum}{$\mu$m\xspace}
\newcommand{\molh}{H$_2$\xspace}
\newcommand{\oric}{\mbox{$\theta^1$ Ori C}\xspace}
\newcommand{\Ta}{\HII\, region\xspace}
\newcommand{\Tb}{atomic PDR\xspace}
\newcommand{\Tc}{DF~1\xspace}
\newcommand{\Td}{DF~2\xspace}
\newcommand{\Te}{DF~3\xspace}
\newcommand{\Tdf}{molecular PDR\xspace}

\newcommand{\uint}{${\rm erg\, cm^{-2}\,s^{-1}\,sr^{-1}}$}

\usepackage{color}
\definecolor{lightgrey}{rgb}{0.84, 0.84, 0.84}

\begin{document}

\title{PDRs4All III: JWST's NIR spectroscopic view of the Orion Bar}

       \author{Els Peeters \inst{1, 2, 3} \and
          {E}milie Habart\inst{4} \and
          Olivier Bern\'{e} \inst{5} \and
          Ameek Sidhu \inst{1,2} \and
          Ryan Chown \inst{1,2}\and
          Dries Van De Putte \inst{6} \and
          Boris Trahin \inst{4} \and
          Ilane	Schroetter \inst{5} \and
          Am\'elie Canin \inst{5} \and
          Felipe Alarc\'on \inst{7} \and
          Bethany Schefter \inst{1,2}
% ...
          }
        
\institute{Department of Physics \& Astronomy, The University of Western Ontario, London ON N6A 3K7, Canada; \email{pis@pdrs4all.org}      \and 
Institute for Earth and Space Exploration, The University of Western Ontario, London ON N6A 3K7, Canada          \and 
Carl Sagan Center, SETI Institute, 339 Bernardo Avenue, Suite 200, Mountain View, CA 94043, USA         \and
Institut d'Astrophysique Spatiale, Universit\'e Paris-Saclay, CNRS,  B$\hat{a}$timent 121, 91405 Orsay Cedex, France        \and 
Institut de Recherche en Astrophysique et Plan\\'etologie, Universit\'e Toulouse III - Paul Sabatier, CNRS, CNES, 9 Av. du colonel Roche, 31028 Toulouse Cedex 04, France         \and 
Space Telescope Science Institute, 3700 San Martin Drive, Baltimore, MD 21218, USA        \and
Department of Astronomy, University of Michigan, 1085 South University Avenue, Ann Arbor, MI 48109, USA
% ...
}

   \date{Received xxx, 2023; accepted xxx}

 
  \abstract
  {JWST has taken the sharpest and most sensitive infrared (IR) spectral imaging observations ever of the Orion Bar photodissociation region (PDR), which is part of the nearest massive star-forming region the Orion Nebula, and often considered to be the ``prototypical'' strongly illuminated PDR.}
  {We investigate the impact of radiative feedback from massive stars on their natal cloud and focus on the transition from the \HII\ region to the atomic PDR (crossing the ionisation front (IF)), and the subsequent transition to the molecular PDR (crossing the dissociation front (DF)). Given the prevalence of PDRs in the interstellar medium and their dominant contribution to IR radiation, understanding the response of the PDR gas to far-ultraviolet (FUV) photons and the associated physical and chemical processes is fundamental to our understanding of star- and planet formation and for the interpretation of any unresolved PDR as seen by JWST.}
   {We use high-resolution near-IR integral field spectroscopic data from NIRSpec on JWST to observe the Orion Bar PDR as part of the PDRs4All JWST Early Release Science Program. We construct a $3\arcsec\times25\arcsec$ spatio-spectral mosaic covering $0.97-5.27$~\mum at a spectral resolution R of $\sim$2700 and an angular resolution of $0.075\arcsec-0.173\arcsec$. To study the properties of key regions captured in this mosaic, we extract five template spectra in apertures centered on the three \molh dissociation fronts, the atomic PDR, and the \HII\ region. This wealth of detailed spatial-spectral information is analysed in terms of variations in the physical conditions--incident UV field, density, and temperature---of the PDR gas.}
   {The NIRSpec data reveal a forest of lines including, but not limited to, \HeI, \HI, and \CI\ recombination lines, ionic lines (e.g., \FeIII, \FeII), \ion{O}{i} and \ion{N}{i} fluorescence lines, Aromatic Infrared Bands (AIBs including aromatic CH, aliphatic CH, and their CD counterparts), CO$_2$ ice, pure rotational and ro-vibrational lines from \molh, and ro-vibrational lines HD, CO, and CH$^+$, most of them detected for the first time towards a PDR. Their spatial distribution resolves the H and He ionisation structure in the Huygens region, gives insight into the geometry of the Bar, and confirms the large-scale stratification of PDRs. In addition, we observe numerous smaller scale structures whose typical size decreases with distance from \oric and IR lines from \CI, if solely arising from radiative recombination and cascade,  reveal very high gas temperatures (a few 1000~K) consistent with the hot irradiated surface of small-scale dense clumps deep inside the PDR. The morphology of the Bar, in particular, the H$_2$ lines reveals multiple, prominent filaments which exhibit different characteristics. This leaves the impression of a ``terraced'' transition from the predominantly atomic surface region to the CO-rich molecular zone deeper in. We attribute the different characteristics of the \molh filaments to their varying depth into the PDR and, in some cases, not reaching the C$^+$/C/CO transition. 
   These observations thus reveal what local conditions are required to drive the physical and chemical processes needed to explain the different characteristics of the DFs and the photochemical evolution of the AIB carriers.}
   {This study showcases the discovery space created by JWST to further our understanding of the impact radiation from young stars has on their natal molecular cloud and proto-planetary disk, which touches on star- and planet formation as well as galaxy evolution.}


   \maketitle
%

\section{Introduction}
\label{sec:intro}

Massive stars output enormous amounts of radiative and mechanical energy into the interstellar medium (ISM) during their main sequence lifetimes. This energy injection shapes the global properties of the ISM, such as its structure, thermal balance, chemistry and ionisation state. 
Negative stellar feedback plays a critical role in secular galaxy evolution as it suppresses star formation \citep{williams1997, hopkins2012, kim2013}, while positive stellar feedback results in swept-up gas and dust from which future stars can form on timescales  $\lesssim 0.15$ Myr \citep[e.g.][]{Elmegreen:77, Preibisch:99, Koenig:08, Kirsanova:08, Ojha:11, Egorov:14, Egorov:17}. 

Most of this interaction between massive stars and their surroundings occurs in photodissociation regions\footnote{Also sometimes called ``photon-dominated regions'' \citep[e.g.][]{Sternberg95}.} (PDRs), where stellar FUV radiation ($6-13.6$~eV) drives the physical and chemical processes \citep{tielens:85, tielens:85b}. While PDRs were initially associated with young massive stars \citep{tielens:85}, PDRs are also found in the diffuse ISM \citep{wolfire_neutral_2003}, reflection nebulae \citep[e.g.][]{Burton90, Sheffer11}, planetary nebulae \citep{bernard2005physical}, surfaces of proto-planetary disks \citep{vicente2013polycyclic}, pillars \citep{McLeod}, globules \citep{Reiter:19}, and molecular clouds. PDRs produce a significant fraction of the ISM radiative emission of galaxies, in particular in star forming galaxies, (Ultra-)Luminous IR galaxies (ULIRGs) and galactic nuclei. Indeed, the neutral ISM and most of the molecular ISM, where most of the ISM mass is found, resides in PDRs \citep{wolfire2022}. Consequently, understanding PDRs is a key prerequisite for understanding star- and planet-formation and the large scale ecology of the ISM of galaxies and its relationship to galaxy evolution. 

The large-scale PDR structure is stratified with temperatures decreasing from 10$^4$ at the PDR front to a few 100s of K in the atomic PDR, and, crossing the \molh dissociation front, to a few tens of K deep into the molecular PDR. While models have been very successful in explaining the observed large-scale structure of PDRs \citep{tielens:85b, Sternberg89, abgrall, lebourlot, Rollig2007, wolfire2022}, recent high-angular resolution ALMA and Keck observations have revealed a  varying PDR front and highly structured PDR \citep{goicoechea, Habart2022}. The highly structured nature of the molecular PDR layers betrays the dynamic action of the evaporation flow that advects material from the molecular cloud, through the PDR and the ionisation front, into the ionised gas. Hence, observations at high angular resolution are required to resolve the small-scale structure to fully understand the processes responsible for shaping PDRs.  

The IR is key in understanding PDRs. Indeed, IR spectra of PDRs are extremely rich--they feature a plethora of strong \HI\ recombination lines, fine-structure lines from atomic and ionised gas, rotational and ro-vibrational emission from \molh and other small molecules, as well as broad emission bands commonly referred to as Aromatic Infrared Bands (AIBs), all superimposed on undulating continuum emission. This spectral diversity provides ample diagnostics to characterise the physical and chemical anatomy of PDRs and to characterise the photochemical evolution of molecules and dust \citep[e.g.][]{Marconi98, Walmsley00, Sheffer11, Pilleri12, Habart2022}. However, past IR observations had insufficient angular resolution to resolve the small-scale structure of PDRs or were limited by the spectral resolution and/or wavelength coverage or both. The unprecedented capabilities of JWST allow, for the first time, to combine high spatial resolution (0.075\arcsec\ to 0.173\arcsec) with medium spectral resolution and large IR wavelength coverage for PDR studies. Such observations thus provide the critical PDR diagnostics at an angular resolution that enables probing the highly structured PDR anatomy and investigate the intricate combination of physical, chemical, and dynamical processes at play in shaping the PDR anatomy. 

The PDRs4All Early Release Science (ERS) program (ID1288)\footnote{https://pdrs4all.org} fully exploits JWST's angular resolution by observing the nearest massive star-forming region, the Orion Nebula, in the NIRCAM and MIRI imaging mode and NIRSpec and MIRI spectral mapping mode \citep{pdrs4all}. This unique data set will serve as the reference data set for PDRs in the next decades and will facilitate the interpretation of numerous JWST observations. Indeed, given the prevalence of PDRs in the Universe and the strong IR emission of PDRs, much of the emission (to be) observed by JWST is from (unresolved) PDRs. 

This work presents the first analysis of the PDRs4All NIRSpec data set and accompanies the PDRs4All NIRCam and MIRI imaging paper \citep{Habart:im}, the NIRSpec proplyd paper \citep{Berne:proplyd}, the MIRI MRS PAH paper \citep{Chown:23}, and the MIRI MRS gas lines paper \citep{vandeputte23}. This paper is organised as follows. First, we describe the characteristics of the PDR, the Bar, as deduced from earlier studies in Sect.~\ref{sec:OB}. This is followed by a description of the observations, data reduction and flux measurements in Sect.~\ref{sec:obs-reduction}. A spectral inventory is given in Sect.~\ref{sec:spectral-characteristics}. Then, we discuss the spatial variation of gas and dust tracers and thus the PDR structure and anatomy in Sect.~\ref{sec:spatial-variation}. We analyze the \HI\ and \HeI\ recombination lines, the fluorescence lines, and the \molh and \CI\ emission to determine the physical conditions in the Bar in Sect.~\ref{sec:physparams}. Last, we discuss the Bar structure in Sect.~\ref{sec:discussion} and give a summary and conclusions in Sect.~\ref{sec:conclusions}.

\section{The Bar}
\label{sec:OB}
The Bar is a rim of the Orion molecular cloud core \mbox{(OMC-1)}, the closest site of ongoing massive star-formation\footnote{The most commonly adopted distance to the Bar is 414\,pc \citep{Menten:07} although recent GAIA observations suggest slightly lower values \citep{Kounkel2018,Gross:18}. We refer to \citet{Habart:im} for a discussion. In this paper we adopt a distance of 414\,pc. Hence, 1\arcsec\, roughly corresponds to 0.002~pc.} \citep[e.g.,][]{Genzel89,Bally:08}. The Bar is often referred as the ``Bright Bar" or ``Orion Bar" \citep[e.g.][]{Elliott1974,Tielens93,ODell:20}. In the following, we name it the ``Bar".
The outskirts of \mbox{OMC-1} are primarily illuminated by strong UV radiation from the \mbox{O7V-type} star \mbox{$\theta^1$ Ori C} \citep{Sota:11}, the most massive star of the Trapezium cluster, at the center of the Orion Nebula and $\sim$2$'$ north east of the Bar \citep[e.g.,][]{Stacey93,Luhman94,Odell01,Goicoechea15}. Intense ionizing radiation and strong winds from \mbox{$\theta^1$ Ori C} \citep[two main forms of stellar feedback in the region;][]{Gudel08,Pabst19} power and shape the Orion Nebula, which is a blister \HII~region that is eating its way into the natal molecular cloud (located behind the cluster in our line of sight). The strong stellar UV radiation has carved out a large cavity in the background molecular cloud, where the inner concave regions tilt to form the Bar \cite[e.g.,][]{Odell01}. 

 The Bright Bar historically refers to the elongated rim near the ionisation front (IF) that separates the edge of the molecular cloud from the surrounding \HII~region, with \mbox{$n_e$\,$\simeq$\,5$\times$10$^3$\,cm$^{-3}$} and \mbox{$T_e$\,$\simeq$\,9$\times$10$^3$\,K} at the IF \citep[e.g.,][]{Weilbacher15}. The UV radiation impinging on the IF is \mbox{(1-4)$\times$10$^4$} times the mean interstellar field \citep[e.g.,][]{Marconi98}. Some areas of the Bar may also be illuminated by the \mbox{O5V-type} star \mbox{$\theta^2$ Ori A}, on the near side of the cluster \citep{ODell17}. Beyond the IF, only far-UV (FUV) photons with energies below 13.6\,eV pervade the Bar. This marks the beginning of the PDR. Because of its high temperatures and nearly edge-on orientation on the sky (with a tilt angle of about 4$^{\circ}$, which leads to limb-brightening effects), the Bar shines at all wavelengths from optical to radio. Indeed, this PDR is the prototypical source to study the physical and chemical stratification caused by strong FUV radiation \citep[e.g.,][]{Tielens93,Hoger95, jansen1995millimeter,Wiel09}. 

The first layers of the Bar PDR are predominantly neutral and atomic, meaning \mbox{[H]\,$>$\,[H$_2$]\,$\gg$\,[H$^+$]} \citep{vanderWerf13,Henney2021}. The so-called ``atomic PDR'' zone presents a plethora of IR atomic emission lines from low ionisation potential elements \citep[recombination lines, forbidden lines, etc.; e.g.,][]{Walmsley00}. This warm (several hundred~K) and moderately dense ($n_{\rm H}$ of a few 10$^4$\,cm$^{-3}$) gas is mainly heated by photoelectrons ejected from Polycyclic Aromatic Hydrocarbons (PAHs), and mainly cooled by far-IR (FIR) [\CII]\,158\,$\mu$m and [\OI]\,63\,$\mu$m fine-structure lines \citep[e.g.,][]{Tielens93,Herrmann97,Bernard-Salas12,Ossenkopf13}. In addition, this extended atomic PDR zone coincides with the peak of very bright AIB emission \citep[e.g.,][]{Bregman89, Sellgren90, Tielens93, Giard94, Knight21}.  

At about 15$''$ ($\sim$0.03\,pc) from the IF, the flux of FUV photons is sufficiently attenuated that most of the hydrogen becomes molecular. This position marks the critical H/H$_2$ transition zone, the dissociation front (DF). The DF displays a forest of IR rotational H$_2$ and HI~21\,cm emission lines \citep[e.g.,][]{Parmar91, Luhman94, vanderWerf96, Allers05, Shaw09, vanderWerf13}. In addition, H$_2$ lines from \mbox{FUV-pumped} vibrationally excited levels up to $v$\,=\,10 are  detected \citep{Kaplan17, Kaplan21}. Reactive molecular ions such as CH$^+$, SH$^+$, CO$^+$ or OH$^+$ start to form close to the DF \citep[e.g.,][]{Stoerzer95, Fuente03, Nagy13, Tak13, Goicoechea17}. The first steps of  PDR chemistry are triggered by the presence of vibrationally excited H$_2$, whose internal energy overcomes the endoergicities and energy barriers of key gas-phase reactions \citep[e.g.,][]{Goicoechea22} and, thus, initiates the formation of molecular hydrides \citep[e.g.,][]{tielens:85,Sternberg95,Agundez10}.

The transition from C$^+$ to C to CO is expected to take place  beyond the DF, where the PDR becomes mostly molecular. That is, \mbox{[H$_2$]\,$\gg$\,[H]}. However, observations have not accurately settled the exact position of the \mbox{C$^+$/C/CO} transition zone \citep[e.g.,][]{Tauber95,Wyrowski97,Cuadrado19,Salas19}. The CO gas temperature just beyond the DF is \mbox{$T_{\rm k}$\,$\simeq$\,200-300\,K} \citep[][]{Habart10,Joblin18} and decreases further into the molecular cloud. This confirms the presence of a sharp (gas and dust) temperature gradient from the \HII~region interface to the molecular cloud interior \citep{Arab12,Salgado16}. Despite the strong irradiation conditions, the so-called ``molecular PDR'' (\mbox{$n_{\rm H}$\,$\simeq$\,10$^5$-10$^6$\,cm$^{-3}$}) shows a rich chemical composition, including a large variety of small hydrocarbons and complex organic species \citep[e.g.,][]{Hoger95,Simon97,Peeters:pads:04,Leurini06,Cuadrado15,Cuadrado17}. Dense clumps (\mbox{$n_{\rm H}$\,$\simeq$\,10$^7$\,cm$^{-3}$}) with angular sizes of $\sim$5$''$ ($\sim$2000\,au) are known to exist deeper inside the Bar \citep[e.g.,][]{Tauber94,vanderWerf96,YoungOwl00,Lis03}. However, it is not clear whether these clumps will ultimately form stars and whether smaller (sub-arcsecond) clumps can exist closer to the DF \citep[e.g.,][]{Gorti02,Andree17}. If they exist, additional heating by collisional de-excitation of vibrationally excited H$_2$ will keep their irradiated surfaces very hot, at several thousands~K \mbox{\citep[e.g.,][]{Burton90}}.

Unfortunately, most of our knowledge of the Bar comes from modest angular resolution observations ($\sim$5$''$-40$''$), especially at FIR to radio wavelengths, that do not spatially resolve the main transition zones of the PDR. Consequently, their fundamental structures: homogeneous versus clumpy, physical conditions, chemical composition, and role of dynamical effects are not fully known. ALMA provided the first \mbox{$\sim$\,1\,$''$} resolution images of the CO and HCO$^+$ emission \citep{goicoechea}. Instead of an homogeneous PDR with well-defined and spatially separated \mbox{H/H$_2$} and \mbox{C$^+$/C/CO} transition zones, ALMA revealed rich small-scale structures (akin to filaments and globulettes), sharp-edges, and uncovered the molecular emission from a protoplanetary disk  \citep[\mbox{203-506};][]{Bally00,Champion17}. Even spatially sharper IR photometric images with the Keck telescope (using adaptative optics) uncovered the presence of not a single, but several small-scale photodissociation fronts \citep{Habart2022}. Our JWST/NIRSpec integral field observations across the Bar, from the \HII~region to the main molecular dissociation fronts (and including the protoplanetary disks \mbox{203-504} and \mbox{203-506}), represent the first sub-arcsecond spectroscopic study of this prototypical PDR. This study complements our first JWST photometric images of the Bar \citep{Habart:im}, which show an unprecedented view of the region, revealing very complex small-scale PDR structures, and ridges, and a 3D terraced distribution of multiple dissociation fronts that contrasts with the classical 1D view of the H/H$_2$ and \mbox{C$^+$/C/CO} transition zones of a PDR.

\begin{figure*}
\begin{center}
\includegraphics[width=\textwidth]{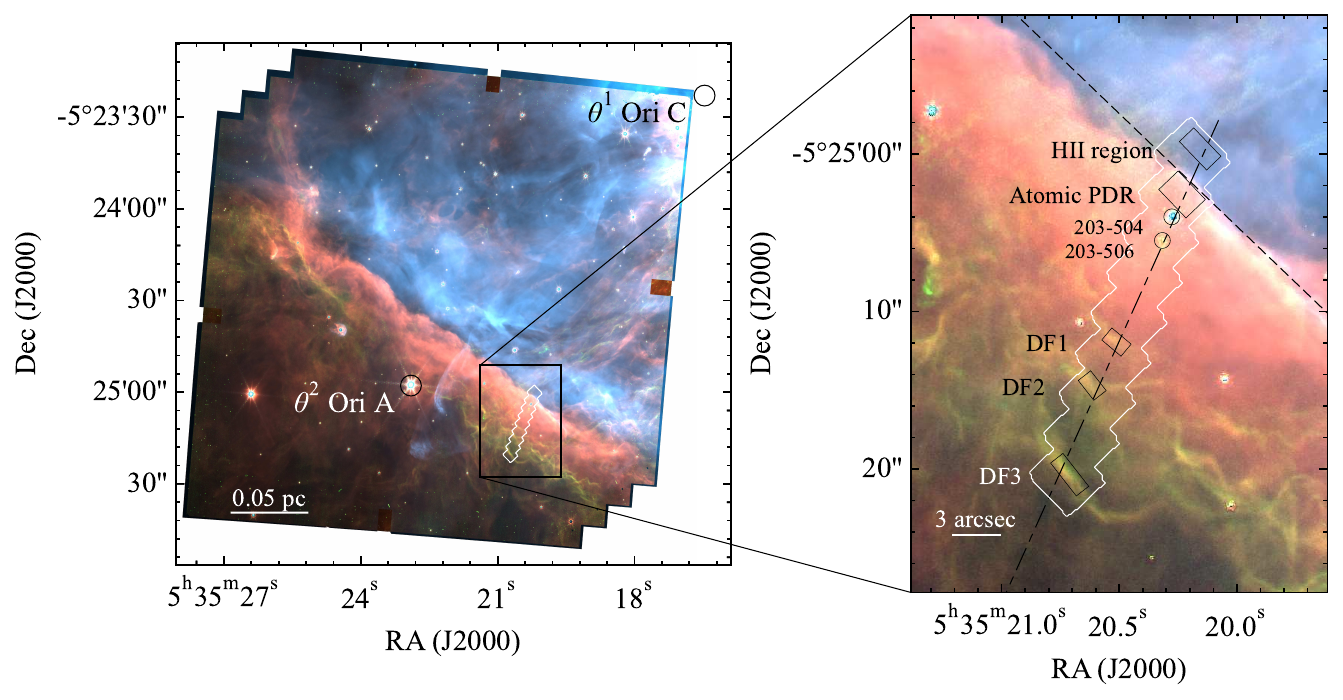}
\caption{Composite NIRCam image of the Bar showing the NIRSpec mosaic footprint (white boundary). The composite image is composed of F335M (AIB emission) in red, F470N-F480M (H$_2$ emission) in green, and F187N (Paschen $\alpha$ emission) in blue \citep{Habart:im}. Bright stars \oric and $\theta^2$~Ori~A are indicated with black circles in the left panel. In the right panel the five black boxes indicate the apertures used to extract our five template spectra. The dot-dashed line indicates the cut perpendicular to the Bar (position angle, PA, of 155.79\textdegree), while the dashed line indicates the position of the ionisation front in the NIRSpec FOV (PA=46.21\textdegree). The protoplanetary disks 203-504 and 203-506 are indicated with black circles.} 
\label{fig:FOV}
\end{center}
\end{figure*}

\section{Observations, data reduction and analysis}
\label{sec:obs-reduction}

\subsection{Observations}
\label{subsec:observations}

The observations are part of the Early Release Science program PDRs4All: Radiative feedback from massive stars \citep[ID: 1288, PIs: Bern\'{e}, Habart, Peeters;][]{pdrs4all}. We obtained our observations using the NIRSpec instrument \citep{nirspec} onboard the JWST \citep{gardner_JWST2006} in the integral field unit (IFU) mode \citep{IFUnirspec}, which provides spatially resolved imaging spectroscopy. This resulted in a $9\times1$ mosaic covering $3\arcsec\times25\arcsec$ and centered on position $\alpha\, (J2000)=$ 05 35 20.4749, $\delta\, (J2000)=$ -05 25 10.45 with a position angle (PA) of 43.74$^{\circ}$ in the 0.97--5.27 $\mu$m range at a angular resolution of $0.075\arcsec$ to $0.173\arcsec$ and with a pixel size of 0.1\arcsec$\times$0.1\arcsec. The field of view (FOV) of the NIRSpec mosaic is shown in Fig.~\ref{fig:FOV}. We also obtained  background observations using a single pointing centered on position $\alpha\, (J2000)=$ 05 27 19.400, $\delta\, (J2000)=$ -05 32 04.40. For our science and background observations, we used the three high spectral resolution, R $\sim2700$, gratings (G140H, G235H, and G395H) covering the wavelength range from 0.97 to 5.27~\mum, the NRSRAPID readout mode (as this mode is appropriate for bright sources), and a 4-point dither pattern. To quantify the leakage of the Micro-Shutter Array (MSA), we used the most accurate strategy for taking imprint exposures to date, that is we obtained imprint exposures with the same exposure time as the science (and background) exposures at all dither positions. Five groups per integration with one integration per exposure are used, for a total on-source integration time of 257.7 s.

\subsection{Data Reduction}
\label{subsec:reduction}
We download the uncalibrated Level 1 science and background NIRSpec IFU data from the MAST portal. We reduce the data using the JWST Science Calibration Pipeline (version 1.10.2.dev26+g8f690fdc) and context jwst\_1084.pmap of the Calibration References Data System (CRDS). First, we perform detector-level corrections on science, background, and imprint exposures in the Detector 1 step. Then we correct the resulting rate files for 1/f correction using the algorithm provided by the helpdesk. The algorithm measures the pattern in the unilluminated pixels in a given rate file using a column-by-column rolling median basis and subtracts it from the data in the illuminated pixels of that file. These cleaned rate files are then used as input for calibrations of individual exposures in the Spec 2 step. Finally, we combine all exposures to build cubes in Spec 3. We note that we disabled the outlier detection step in Spec 3 because it introduces artefacts and removed bright lines from our data. Lastly, we point out that we reduce the data of each pointing in a given spectral segment separately, yielding 27 spectral cubes from 9 pointings in three spectral segments at the end of the Spec3 step of the pipeline. 

While the pipeline is able to produce a mosaic that combines all pointings over the full wavelength range, this results in some undesirable artefacts such as stripes near pointing edges. By building the mosaic outside of the pipeline, we can apply calibration factors to single-segment cubes (to improve the overall flux calibration), and we can specify precisely how we want to deal with overlapping data (spatially and spectrally). 

Starting with spectral cubes generated by the \textit{JWST} pipeline, we create the final mosaic using the following approach. 
\begin{enumerate}
    \item We use the Astropy-affiliated package for image reprojection \texttt{reproject}, and the \texttt{reproject.find\_optimal\_celestial\_wcs} routine on all of the 27 input cubes (9 pointings, 3 segments each) to determine the World Coordinate System (WCS) information and array shape of the final cube.
    \item Next, we use \texttt{reproject.reproject\_exact} to reproject every wavelength-plane of these 27 cubes to the final cube shape and WCS. 
    \item A physical gap between NIRSpec's detectors leads to a gap of missing wavelengths in each IFU cube \citep[for details see][]{boker2022}. The wavelength gap spans bluer wavelengths in the Northern part of each pointing and smoothly shifts to redder wavelengths toward the Southern part of each pointing. As best as possible, we use data from adjacent pointings to fill in these gaps. For spaxels that are covered by two partially-overlapping pointings, we coadd the overlapping spectra unless one of the spaxels is either missing flux, or is within a 9-pixel distance from the respective pointing edge (a 9 pixel distance was chosen to avoid edge effects). If a spaxel fills in flux that is missing from an overlapping spaxel due to a wavelength gap, we use \texttt{specutils.manipulation.FluxConservingResampler} to resample the fluxes onto the wavelength grid of the cube where flux is missing. We resample onto the wavelength grid of the pointing with missing flux because the number of spaxels that require interpolation is a small minority of the total number of spaxels, and because all spaxels in the mosaic need to be on the same wavelength grid -- only spaxels from the cube that are affected by the wavelength gap need to be dealt with separately. As a consequence, not all pixels in the extraction aperture contribute at a given wavelength in the wavelength gap region. 
    For cases where one of the two spatially overlapping spaxels is within 9 pixels of the edge of its respective pointing while that of the other pointing is not, the spaxel that is closer to the edge of its pointing is ignored -- at that location, the mosaic contains the spaxel that is further from its pointing edge. 
    \item We then ensure that the NIRSpec flux calibration is accurate by computing synthetic NIRCam images from the NIRSpec cubes, then reproject background-subtracted NIRCam images onto the NIRSpec pixel grid, and then compare the synthetic NIRCam flux against the true NIRCam flux in each pixel. Similar to the imaging/IFU cross-calibration method of \citep{kraemer2022} using \textit{Spitzer/IRAC} and \textit{Spitzer/IRS} data, we perform a linear regression on the pixel-by-pixel synthetic NIRCam flux vs. true NIRCam flux. The best-fit slope of this relationship (for each NIRCam filter) is our estimate of the cross-calibration factor between NIRCam and NIRSpec. The best-fit parameters are tabulated in Table~\ref{tab:xcal} \citep[see][for details]{Chown:23}. 
    \item We then multiply the G235H and G395H NIRSpec mosaics and their uncertainties by their respective calibration factors. Since we do not have background-subtracted NIRCam data in any filters that overlap with the G140H wavelength range, we are unable to assess the flux calibration of that segment in the same way. We multiply the G140H mosaic by the calibration factor for the G235H segment. We tested an approach where each G140H spaxel was scaled to match the flux in the overlapping G235H spaxel, but we found this approach to be unreliable due to the presence of data reduction-related artefacts.

\end{enumerate}

The applied reduction process produces very high quality data. However, after processing the raw data, some artefacts remain in the data. The remaining artefacts that are present include:
\begin{enumerate}
    \item  Bright circular artefacts that are localised in wavelength (a few spectral bins) and in position (roughly circular, a few pixels wide, and in the same positions on the detector).
    \item Vertical stripes at the edge of each pointing (N-NW to S-SE direction) with lower flux, likely due to the fact that a path loss correction using flight data cannot be performed using the available reference files.
    \item A sinusoidal wave pattern in the uncertainty data of the three segments.
    \item Fluxes within a few wavelength bins of a gap are generally unreliable.
    \item A roughly sinusoidal wave pattern in the surface brightness and/or broad absorption/emission features in gratings G140H and G235H in the NRS2 detector \footnote{The NRS1 (NRS2) detector covers wavelengths beyond (below) the wavelength gap in each segment.}. As these are not present in the grating covering the subsequent wavelength range in the NRS1 detector, this is likely residual $1/f$ noise (the effects of $1/f$ noise are more pronounced on the NRS2 detector compared to the NRS1 detector).
\end{enumerate}
We mask out the bright circular artefacts, and replace bad data from vertical stripes with better data when co-adding adjacent pointings as mentioned above. 

We extract spectra in five apertures (Fig.~\ref{fig:FOV} and Table~\ref{tab:aper}) by applying a 3-$\sigma$ cut to remove bad data and calculating the inverse-variance weighted average of spaxels within each aperture. We use five large extraction apertures positioned in front of the ionisation front (IF), at the peak of the PAH emission and at the three \HI/\molh dissociation fronts (\Tc, \Td, \Te) in the mosaic. The resulting spectra thus serve as templates for the \Ta, the \Tb and the \Tdf. In addition, to measure quantities from weaker lines as a function of distance from \oric, we spatially rebin the spectral mosaic to a 2x2 pixel scale prior to fitting the lines.

\begin{figure*}
\begin{center}
\resizebox{.98\hsize}{!}{
\includegraphics{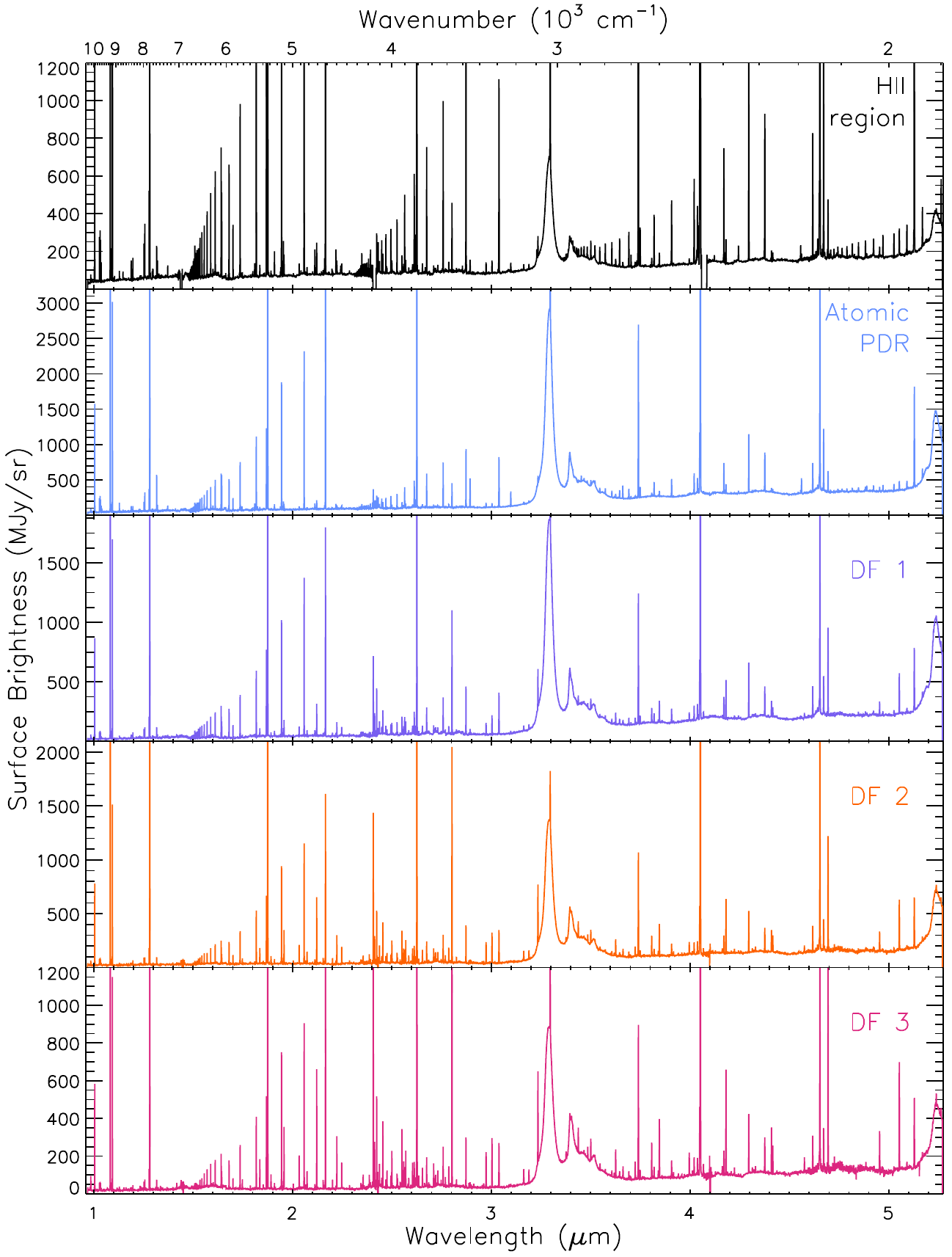}}
\caption{Template spectra representing from top to bottom the \Ta, the \Tb, and the dissociation fronts \Tc, \Td, and \Te. }
\label{fig:template}
\end{center}
\end{figure*}

\subsection{Flux measurements}
\label{subsec:analysis}

We measure the flux of selected emission lines by fitting a Gaussian line profile of a fixed full width at half maximum (FWHM) set by the spectral resolution at that wavelength. We determine the spectral resolution by using the resolution curves for the G140H, G235H, and G395H gratings given in the Jdox\footnote{https://jwst-docs.stsci.edu/jwst-near-infrared-spectrograph/nirspec-instrumentation/nirspec-dispersers-and-filters}. Before performing a fit, we subtract a linear continuum. We visually select a wavelength range for continuum determination that is devoid of emission lines. Given the presence of artefacts in the data set and the fact that the wavelength range for fitting is very small, we find that subtracting an offset for the continuum instead of a linear continuum works better when measuring fluxes across the entire spectral map. Lines located on top of the strong 3.3~\mum AIB are fit simultaneously with the AIB emission (see Sect.~\ref{subsubsect:AIBint} for details).

For the uncertainties on the measured fluxes, we use the flux uncertainties from the Gaussian fit of the line which takes into account the uncertainties on the surface brightness provided by the data reduction pipeline. We note that the uncertainties resulting from the pipeline are too low and therefore, the quoted flux uncertainties likely underestimate the true uncertainties. 
To assess the influence of the artefacts on the line fluxes, we compare selected line fluxes with their corresponding NIRCam filter combination \citep[see][their Fig.~11]{Habart:im} and conclude that the agreement with NIRCam is excellent. Indeed, despite remaining systematic artefacts in the data set, the change in intensities of Br~$\alpha$, Pa~$\alpha$, [\FeII] at 2.1644~\mum, \molh 0-0 S(9), \molh 1-0 S(1) and total AIB emission across the PDR matches their spatial behaviour as observed by NIRCam very well. In addition, the Br~$\alpha$, Pa~$\alpha$, \molh 0-0 S(9), and total AIB emission agree within 3\% on the absolute scale with their corresponding (continuum subtracted) NIRCam filter (note the AIB filter does not require continuum subtraction). Deviations are larger for the [\FeII] at 2.1644~\mum, and \molh 1-0 S(1) emission as these transitions do not dominate the emission captured by the corresponding (continuum subtracted) NIRCam filter.

\begin{figure*}
\begin{center}
\resizebox{\hsize}{!}{
\includegraphics[clip,trim =1cm 1cm 1cm 1cm]{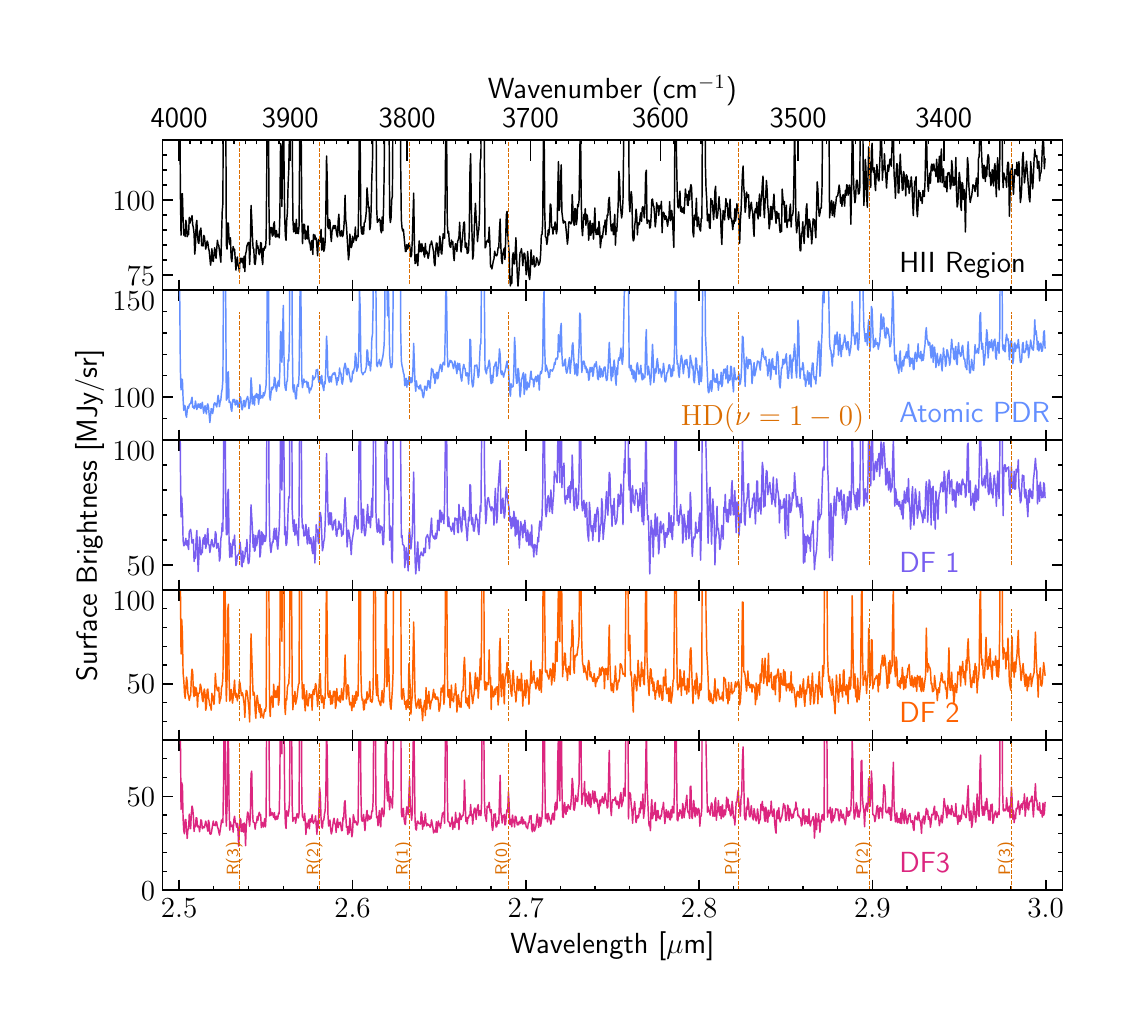}
\includegraphics[clip,trim =1.cm 1cm 1cm 1cm]{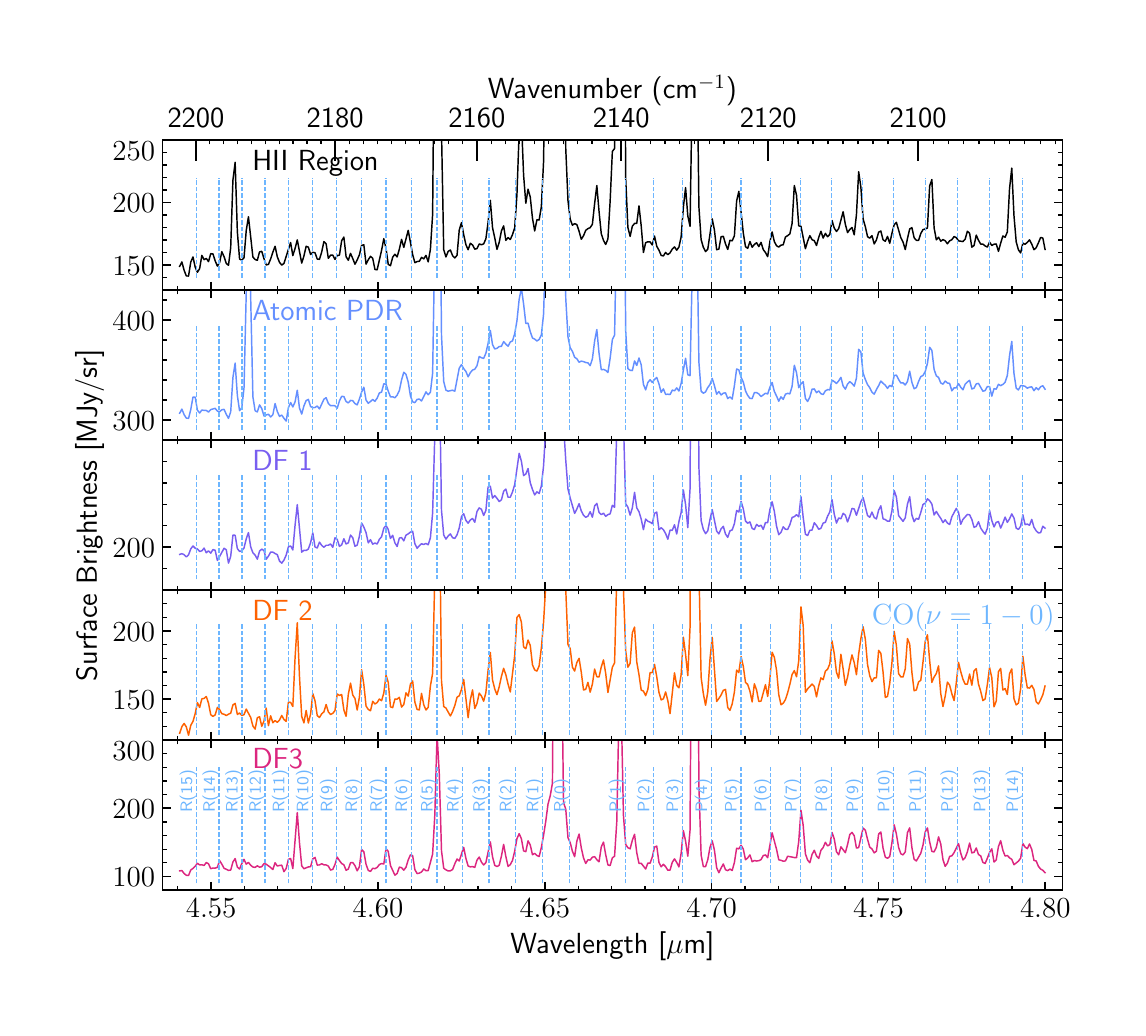}}
\caption{\textbf{Left:} Detection of the HD $v=1-0$ ro-vibrational lines at $\sim$2.6~\mum in the \Tb and molecular PDR. \textbf{Right:} Detection of the CO $v=1-0$ band centered at 4.7\,$\mu$m in the molecular PDR. For the CO $v=2-1$ band detection, see Figs.~\ref{fig:app:template7} and~\ref{fig:app:template8}.}
\label{fig:HDCOdetect}
\end{center}
\end{figure*}

\begin{figure*}
\begin{center}
\resizebox{.8\hsize}{!}{
\includegraphics[clip,trim =1.cm 1cm 1cm .5cm]{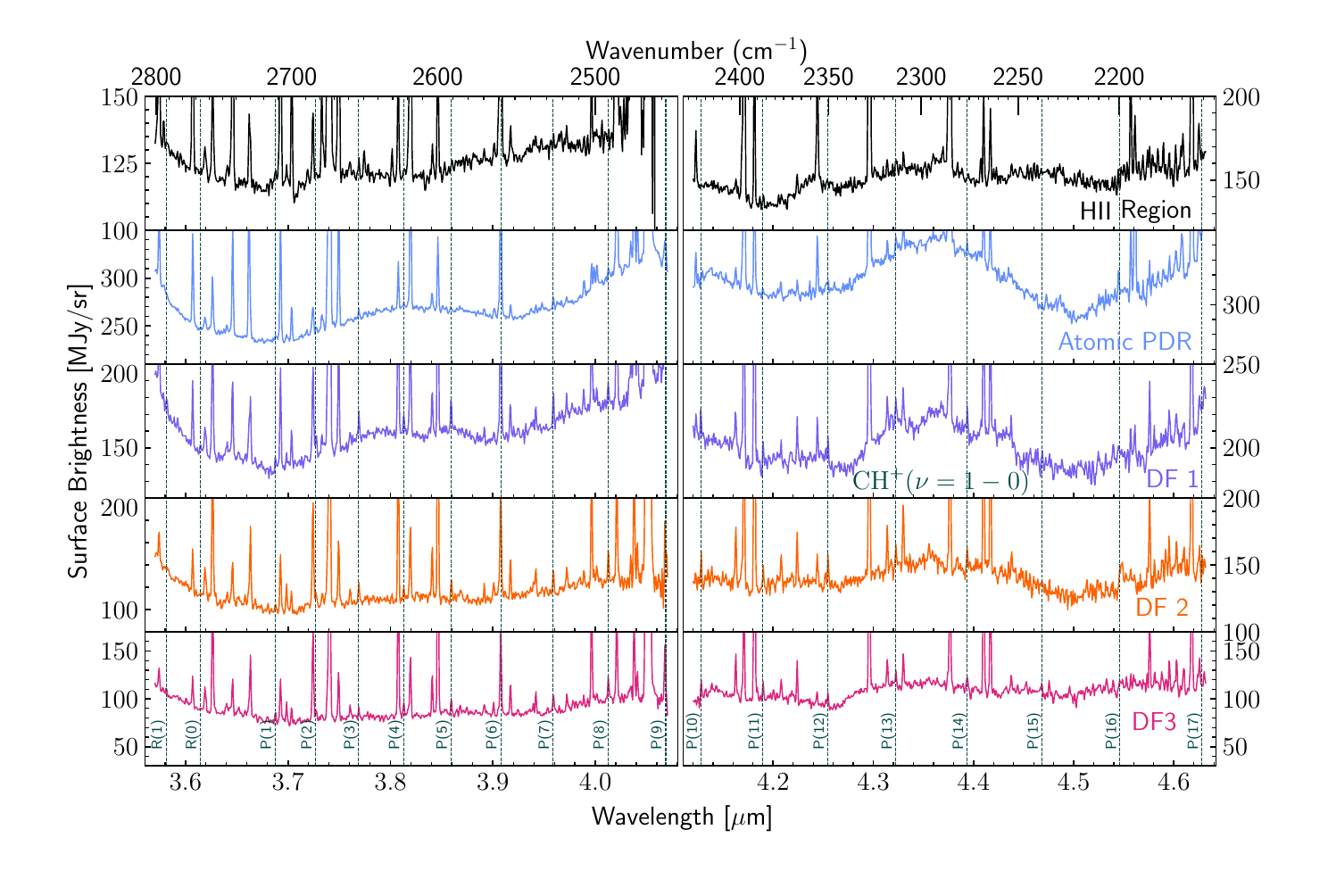}}
\caption{ Detection of the CH$^+$ $v=1-0$ ro-vibrational lines in the \Tb, \Tc, \Td and \Te.  }
\label{fig:CH+detect}
\end{center}
\end{figure*}

\section{Spectral inventory}
\label{sec:spectral-characteristics}

The five template spectra probing the \Ta, the \Tb, and the \Tdf (see Sect.~\ref{subsec:reduction}) are shown in Fig.~\ref{fig:template} and in more detail with line labels in Figs.~\ref{fig:app:template1}--~\ref{fig:app:template8}. These $0.97-5.27$~\mum near-IR (NIR) spectra reveal a spectacular richness of spectral lines and bands on top of weak continuum emission. In particular at the shortest wavelengths, numerous (blended) emission lines are present to the point of the line confusion limit. 

Across the mosaic, \HI\, recombination lines are detected from the Paschen series (up to principal quantum number n$_{u}=7$ in all template spectra) and the Bracket, Pfund, Humphreys, and n$_l=7$ series (up to n$_{u}=25-30$ in the \Tdf; n$_{u}=45-50$ in the \Ta and \Tb). We also detect numerous \HeI\ recombination lines as well as emission lines from \FeII\, and \FeIII, \CI\, recombination lines, \OI\, and \NI\, fluorescent emission and the [KrIII] 2.1986~\mum transition. 
In addition to atomic and ionic lines, the NIR spectra of the Bar show many high-energy ro-vibrational lines from simple molecules (H$_2$, HD, CO, and CH$^+$). These lines are generally faint and become apparent deeper inside the molecular layers of the PDR (\mbox{\Tc, \Td, and \Te templates}), where most of the hydrogen is locked up in H$_2$.

\begin{figure}
\begin{center}
\resizebox{\hsize}{!}{
\includegraphics{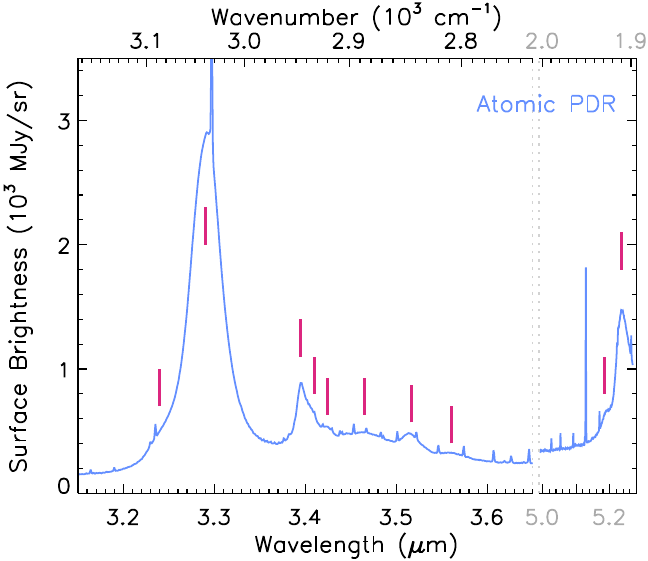}}
\caption{Components of the AIB emission detected in the \Tb.}
\label{fig:pahtemplate}
\end{center}
\end{figure}

\begin{figure*}
\begin{center}
\resizebox{\hsize}{!}{
\includegraphics{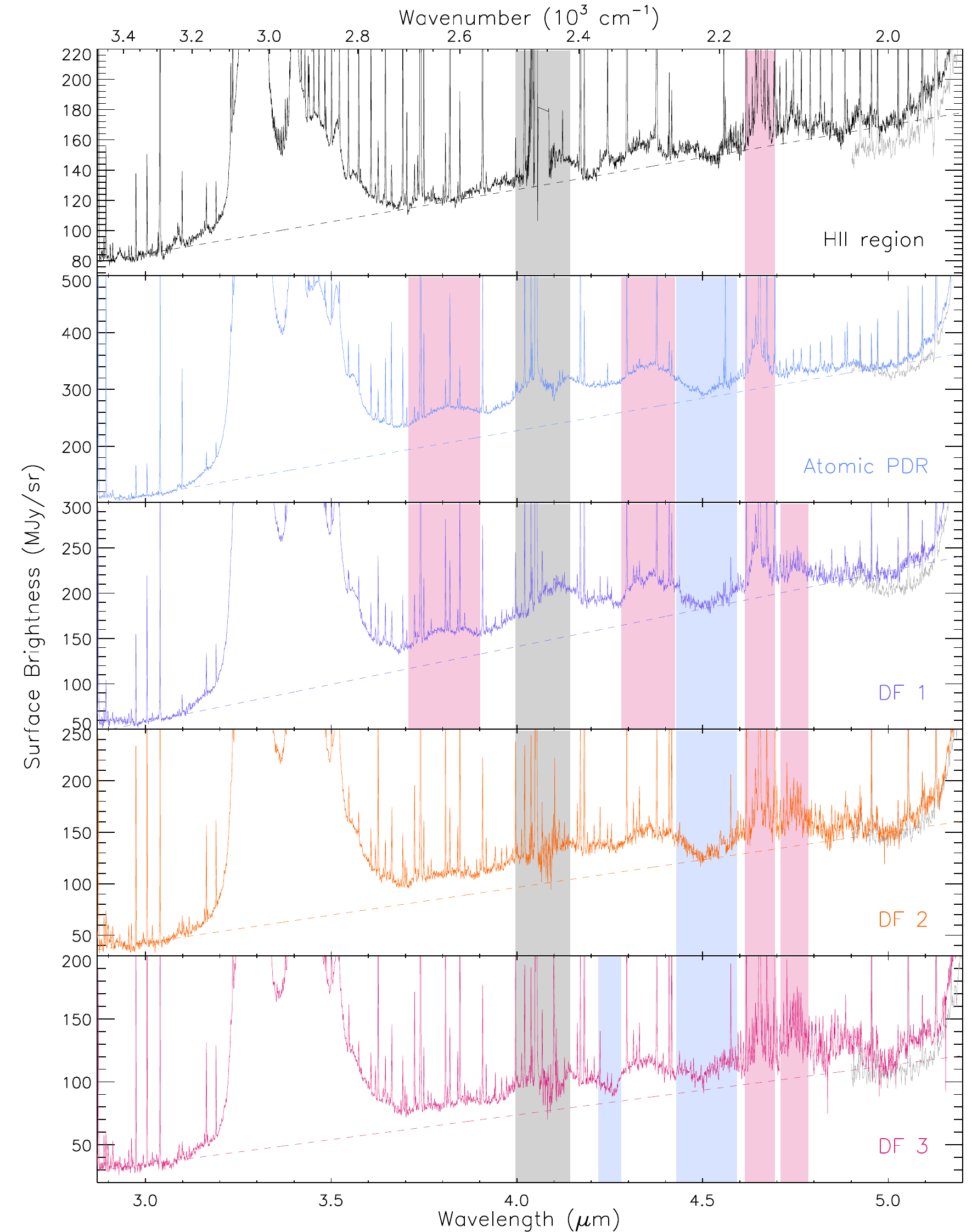}}
\caption{Illustration of excess broad band emission between $3-5$~\mum. Close to the wavelength gap, fluxes are often unreliable (Sect.~\ref{subsec:reduction}; indicated by the grey shaded area). The MIRI spectra are shown in light grey. The dashed lines show a linear continuum matched with the data near 2.98 and 4.99~\mum. Red (blue) shaded boxes indicate tentative emission (absorption) bands.
See Sect.~\ref{sec:spectral-characteristics} for a discussion.}
\label{fig:tentativebands}
\end{center}
\end{figure*}

The molecular emission is dominated by a forest of H$_2$ lines from vibrationally excited bands (\mbox{$v=1-0$}, \mbox{$v=2-1$}, etc.), with detections up to \mbox{$v$\,= 6}. \cite{Kaplan21}  previously reported on the detection of some of these lines \mbox{(in the $1.45-2.45\,\mu$m range)} from ground-based observations at higher spectral resolution than NIRSpec (R $\sim$45000), but at significantly lower angular resolution (0.3\arcsec\ pixel scale).
These vibrationally excited levels are populated by far-UV (FUV) pumping in the Lyman and Werner bands of H$_2$, followed by radiative and collisional de-excitation \citep[][]{Black76,Sternberg89,Burton90}. Interestingly, we also detect ro-vibrational lines of the HD isotopologue in the $v=1-0$ band at $\sim$2.6\,$\mu$m (Fig.~\ref{fig:HDCOdetect}).

In addition, we detect high-$J$ H$_2$ pure rotational lines in the ground vibrational state, up to \mbox{$v=0-0$ $S$(19)}, involving very high-energy rotational levels; \mbox{$E_u/k$\,$\simeq$\,30,000\,K}. Moreover, we report on the first detection of H$_2$ pure rotational lines within the vibrationally excited states \mbox{$v$\,=\,1} (up to \mbox{$v=1-1$ $S$(19)}) and \mbox{$v$\,=\,2} ($S$(9)). These highly excited rotational levels are populated by the radiative and collisional de-excitation of FUV-pumped levels. 
 
Quite unexpectedly we report on the first detection, toward an interstellar PDR, of the CO $v=1-0$ and $v=2-1$ bands centred at 4.7\,$\mu$m (Fig.~\ref{fig:HDCOdetect}). Detected ro-vibrational lines are faint but seen up to high $J$ values.
 This implies that rotational levels within the vibrational state $v=2$,  with energies of about \mbox{$E/k$\,$\simeq$7000\,K}, are populated in the PDR. These are substantially higher energies than those of the highest-$J$ pure rotational line ($v=0-0$, $J=23-22$) detected in the far-IR \citep[\mbox{$E_u/k$\,$\simeq$1500\,K;}][]{Joblin18}.

Concerning other hydride molecules previously detected in the Bar through rotational spectroscopy of the $v=0$ state \citep[e.g.,][]{Gerin16}, we detect CH$^+$ $v=1-0$ ro-vibrational lines at \mbox{$\sim$3\,$\mu$m} \citep[see spectroscopic analysis in][]{Changala21}. 
Far-IR pure rotational lines of CH$^+$ up to \mbox{$v=0-0$, $J=5-6$} were first
detected by \cite{Nagy13} at the much lower ($\sim$10$''$) angular resolution with the Herschel space telescope.
Here we detect CH$^+$ $v=1-0$ ro-vibrational lines toward both the \Tb and the molecular PDR (\Tc, \Td, and \Te; Fig.~\ref{fig:CH+detect}). This likely indicates that small molecular fractions of H/H$_2$ are enough to form sufficient CH$^+$ and to excite the $v=1-0$ band through chemical formation pumping \citep[CH$^+$ is a very reactive molecular ion; e.g., ][]{Nagy13,Godard13}. The NIR CH$^+$ $v=1-0$ band has stronger $P$-branch lines than $R$-branch lines (nearly undetected), a spectroscopic behaviour previously reported and explained by \cite{Neufeld21} toward the planetary nebula NGC\,7027. For a detailed analysis of the CH$^+$ $v=1-0$, we refer the reader to Zannese et al. (in prep.). 

We observe weak continuum emission with increasing surface brightness towards longer wavelengths. The continuum emission does not increase in surface brightness towards the shortest wavelengths, indicating there is not a strong contribution from scattered light. 

The observations also display strong aromatic infrared bands (AIBs; Fig.~\ref{fig:pahtemplate}). The strong 3.29~\mum AIB along with weaker bands at 3.25, 3.40, 3.46, 3.52, 3.56~\mum are perched on top of a broad plateau \citep[see also][]{geballe, Sloan:97}. We report that the 3.40~\mum band is composed of three sub-components centred at 3.395, 3.403, and 3.424~\mum. An additional AIB band is (partially) detected at 5.236~\mum. This band has a blue shoulder peaking near 5.18~\mum. While the bands at $\sim$3.4~\mum are aliphatic in nature (see Sect.~\ref{subsec:AIB}), we will refer to these bands as the AIBs in the remainder of the paper. 

While the detection of weak broad features is very challenging given the current calibration  (Sect.~\ref{subsec:reduction}), Fig.~\ref{fig:tentativebands} reveals broad structures in the 3.5-5.2~\mum range\footnote{We note that the NRS1 detector covers wavelengths up to $3.983-4.099$~\mum, whereas the less reliable NRS2 detector covers wavelengths larger than $4.086-4.203$~\mum depending on the IFU virtual slit. However, no artefacts are known at the positions of these structures (per the JWST helpdesk).}. All templates, except the \Ta template, show enhanced emission (with respect to a linear continuum) from $\sim$3.1~\mum to $\sim$4.9~\mum. We rule out an artificial decrease in flux near 4.5~\mum, which is much less pronounced in the \Ta template, because the template spectra match the MIRI-MRS continuum (starting at 4.9~\mum) very well indicating the flux levels near 5~\mum are accurate. Consequently, the templates tentatively show a lack of emission or an absorption feature near 4.5~\mum. We are unaware of any known absorption feature near 4.5~\mum of similar width and thus favour the interpretation of a lack in emission. If born out, such an extended emission (band) has not been seen before, likely due to the lower angular resolution and incomplete wavelength coverage, and, in some cases, low data quality. A detailed investigation of its characteristics will be presented in a forthcoming paper. The templates show an asymmetric band (with a red wing) centred at 4.644~\mum and potentially an asymmetric band (with a red wing) centred at 4.746~\mum (see also Appendix~\ref{app:PADs}). The CD stretching mode in deuterated PAHs, occurs between 4.54-4.75~\mum \citep{Hudgins:04, Buragohain:15, Yang:20, Yang:21, Allamandola:21}. In particular, the CD stretch in PAHs to which D atoms are added (DPAHs) occurs near 4.6~\mum (2170~cm$^{-1}$), whereas the CD stretch in deuterated methyl groups near 4.7~\mum (2130~cm$^{-1}$). This band has been observed in Orion and a few other \HII\ regions \citep{Peeters:pads:04, Onaka:14, Doney:16, Onaka:22}, although the band profile could not be resolved due to the low angular and spectral resolution. 
The \Tb and, perhaps, \Tc also show two broad emission bands centred near 3.8~\mum (with a width of $\sim$0.5~\mum) and near 4.35~\mum. These bands coincide with a nitrile (-CN) stretch at 4.38~\mum (2280~cm$^{-1}$) and at 4.52~\mum (2220~cm$^{-1}$), and the CD stretch for PAHs in which a peripheral H atom is replaced by a D atom (PADs), near 4.4~\mum \citep[$4.3-4.5$~\mum range;][]{Hudgins:04, Allamandola:21}\footnote{Note that this also coincides with an aldehyde (-CHO) stretch at 3.8~\mum (2600~cm$^{-1}$), however, this would also give a C=O stretch near 5.9~\mum that is not detected \citep{Champion17}.}. Lastly, \Te shows a potential band in absorption near 4.27~\mum that could arise from the C=O antisymmetric stretching mode in CO$_2$ ice. However, the presence of CO$_2$ ice in the PDR is unlikely given the physical conditions in the dissociation front (e.g. hot gas temperature, warm grains, low A$_V$) and the apparent lack of $H_2O$ ice.  Improved data reduction and/or further observations may have to confirm the reality of this feature.

\subsection{Reference line list}

To facilitate the identification of detected lines in the present observations as well as in future JWST observations, we prepared a line list based on model calculations using the Cloudy \citep{cloudy} and Meudon PDR codes \citep{lepetit}. A detailed description of these model calculations can be found in \citet[][]{pdrs4all}. These model calculations include lines with  intensities greater than $5 \times 10^{-10}$\,W\,m$^{-2}$\,sr$^{-1}$. After the data arrived, we expanded the line list to include all  detected lines. The resulting line list includes hydrogen recombination lines with the upper principal quantum number up to 50 and all the lines of molecular hydrogen listed in \citet{Roueff19}. We also include [\FeIII], [\NiII], and [\NiIII] lines, which were not included in the Cloudy simulation but are detected in the ionised region \citep[see also][]{vandeputte23}. \OI~ and \NI~ fluorescence lines are added with the criteria that the Einstein-A coefficients are larger than $5 \times 10^5$\,s$^{-1}$ and $0.9 \times 10^5$\,s$^{-1}$ for \NI~ and \OI, respectively. The criteria are chosen to include all detected lines. For molecular lines, we include the rovibrational lines of HD up to $v$=3  and those of CO with $v=1-0$ and $2-1$. Some of these transitions are also detected in the molecular PDR. In addition, the lines of CH$^+$ and OH are included since they are detected in the proplyd 203-506, which is located within the NIRSpec mosaic \citep{Berne:proplyd2, zannese2023}. 
Finally, the list also contains several dust bands, including strong absorption bands of ice species, major AIBs, and C$_{60}$.

The line list contains the transition wavelength in vacuum, assignment of the transition, upper level energy, and Einstein A-coefficient. Atomic transition data are taken from the atomic line database at University of Kentucky \citep{vanhoof2018}\footnote{\url{https://www.pa.uky.edu/~peter/newpage/}}. For molecular hydrogen, we refer to \citet{Roueff19}. We adopt the CO and HD data from the HITRAN \citep{Hitran:22}\footnote{\url{https://hitran.org}}. 
The data of \citet{Changala21} for CH$^+$ and those of \citet{Tabone2021} and Tabone et al. (in preparation) are used with \citet{yousefi2018} and \citet{brooke2016} for the OH lines. Values in \citet{brieva2016} are used for C$_{60}$, while the data in Sect.~\ref{subsec:AIB} and \citet{Chown:23} are taken for the AIBs. For the ice species, we refer to \citet{gibb2004} and \citet{boogert2015}.

The present line list contains nearly 7000 lines and dust bands, which are potentially detectable by NIRSpec and MIRI/MRS observations. The list is available in the science enabling products at the PDRs4All website \url{https://pdrs4all.org}.  Note that there still remain several unidentified lines in the spectra \citep[see also][]{vandeputte23}.

\begin{figure*}
\begin{center}
\resizebox{.99\hsize}{!}{
\includegraphics{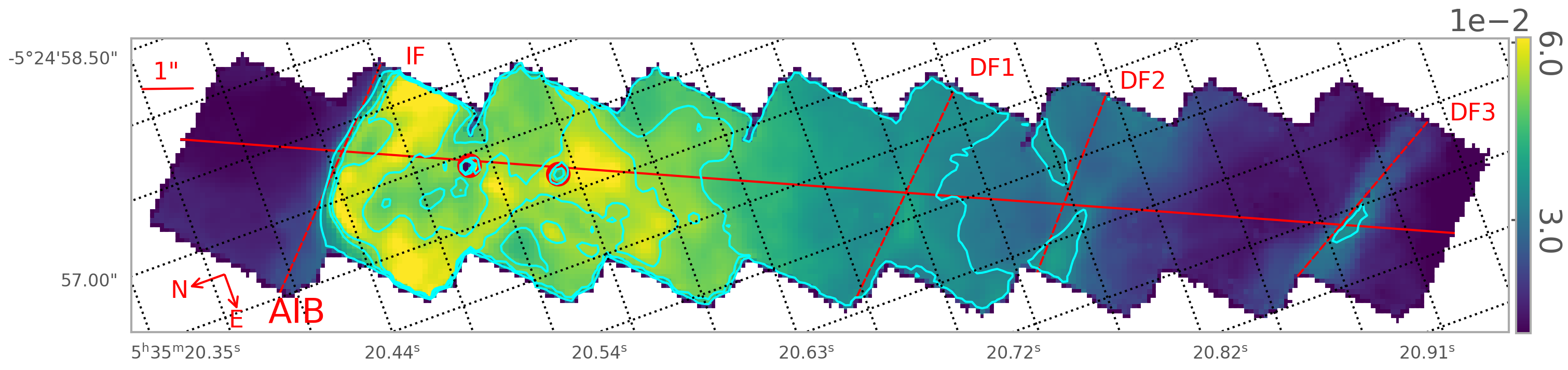}}
\resizebox{.99\hsize}{!}{
\includegraphics{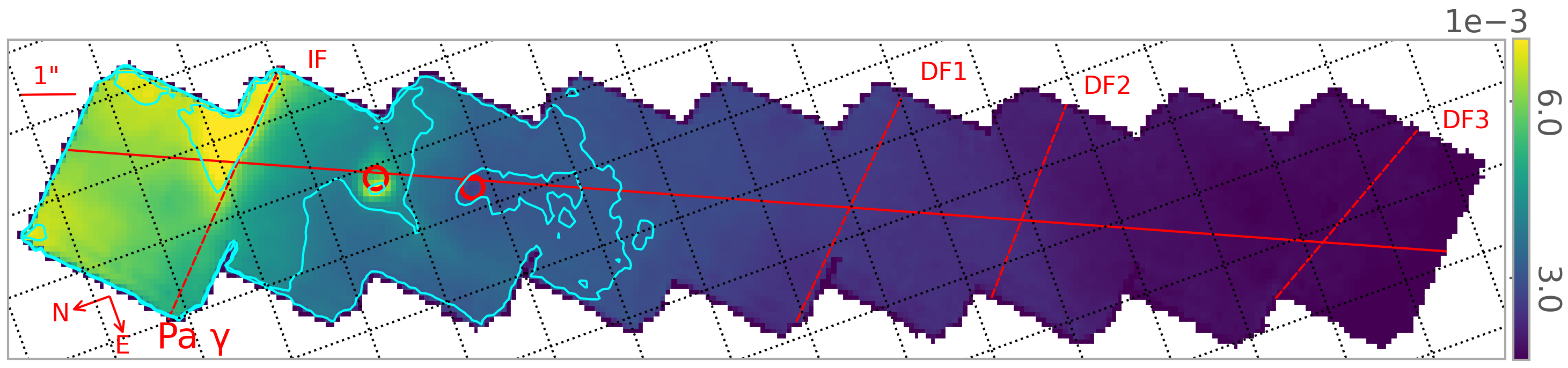}
\includegraphics{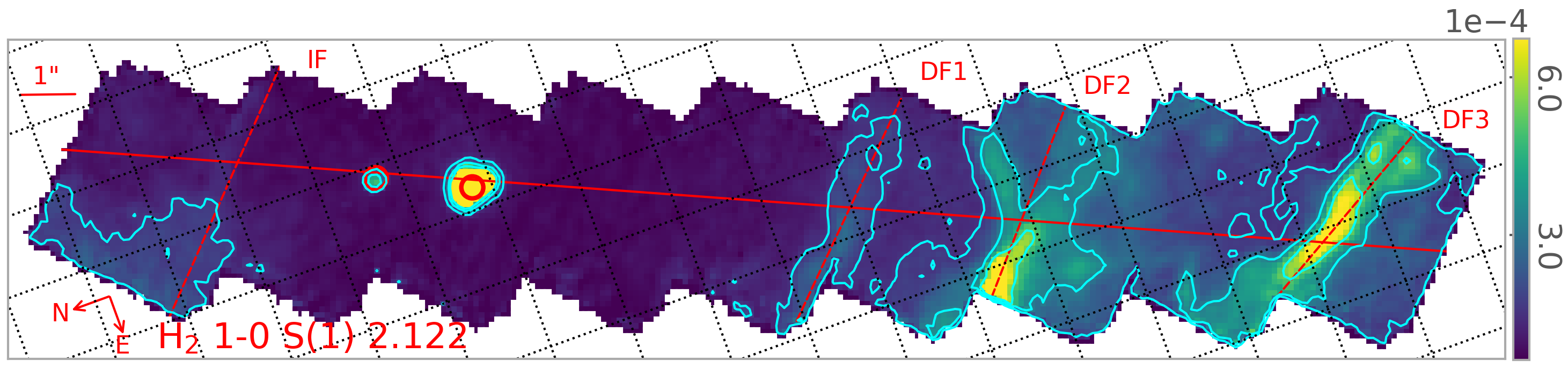}}
\resizebox{.99\hsize}{!}{
\includegraphics{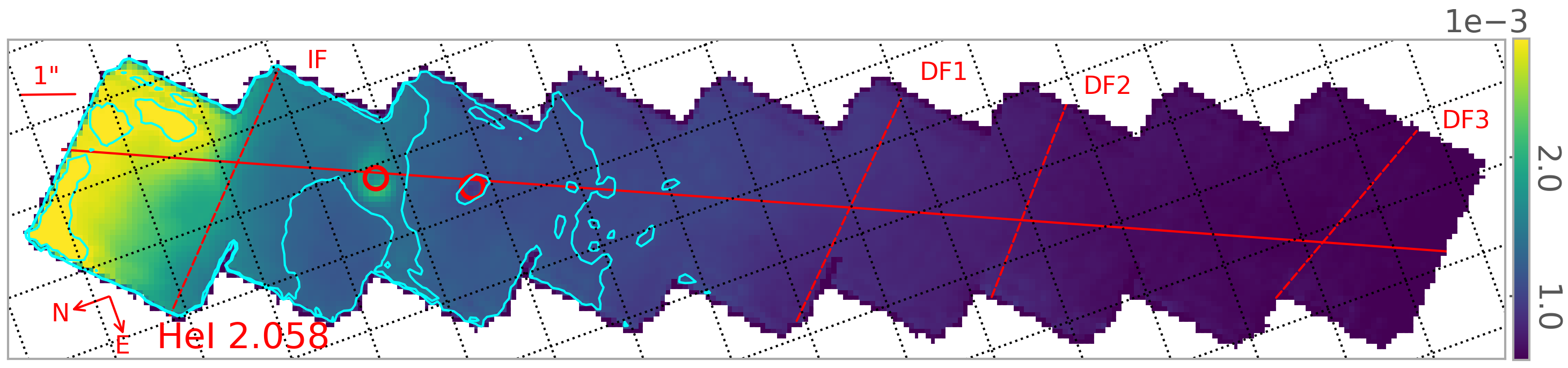}
\includegraphics{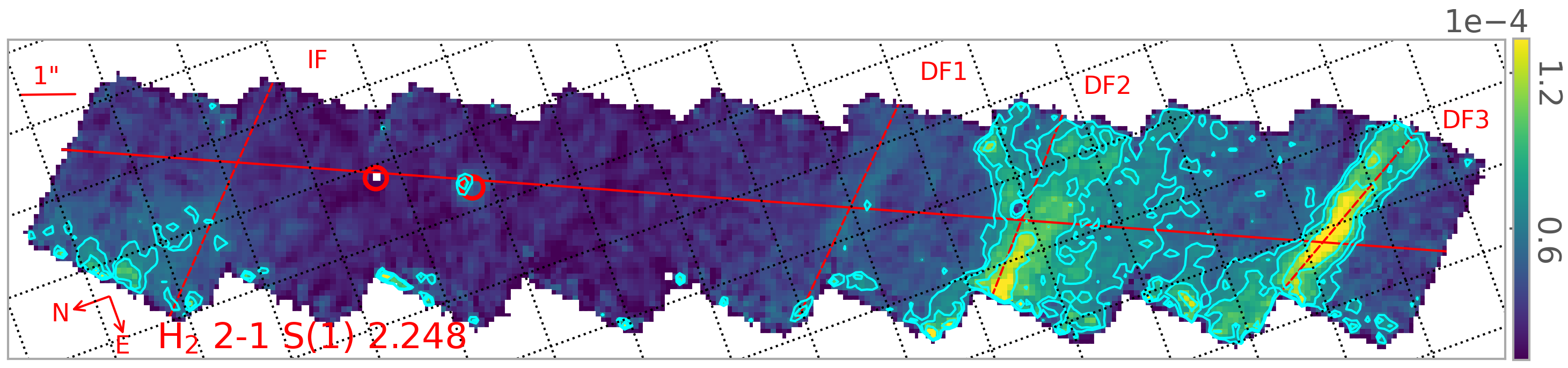}}
\resizebox{.99\hsize}{!}{
\includegraphics{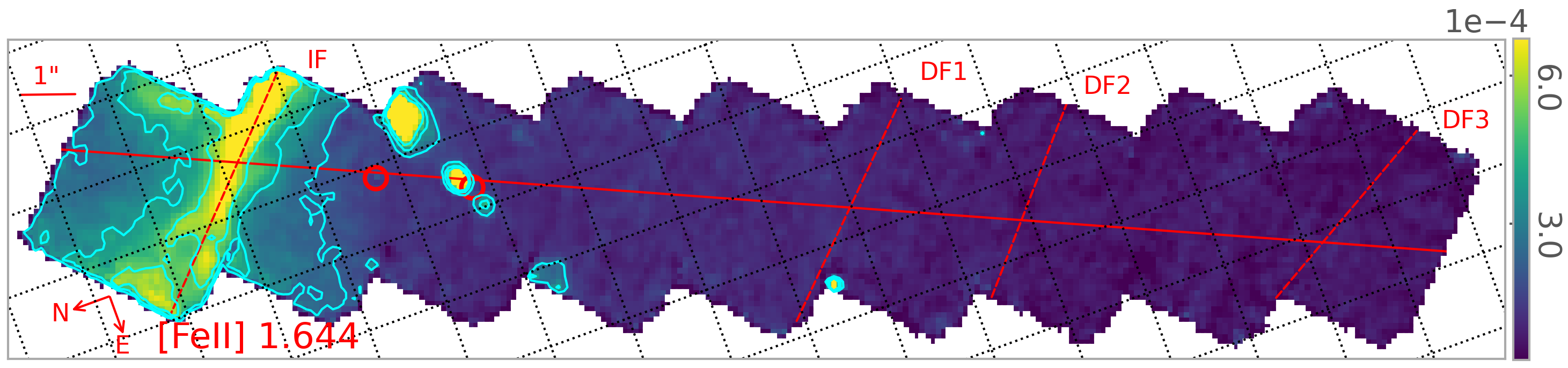}
\includegraphics{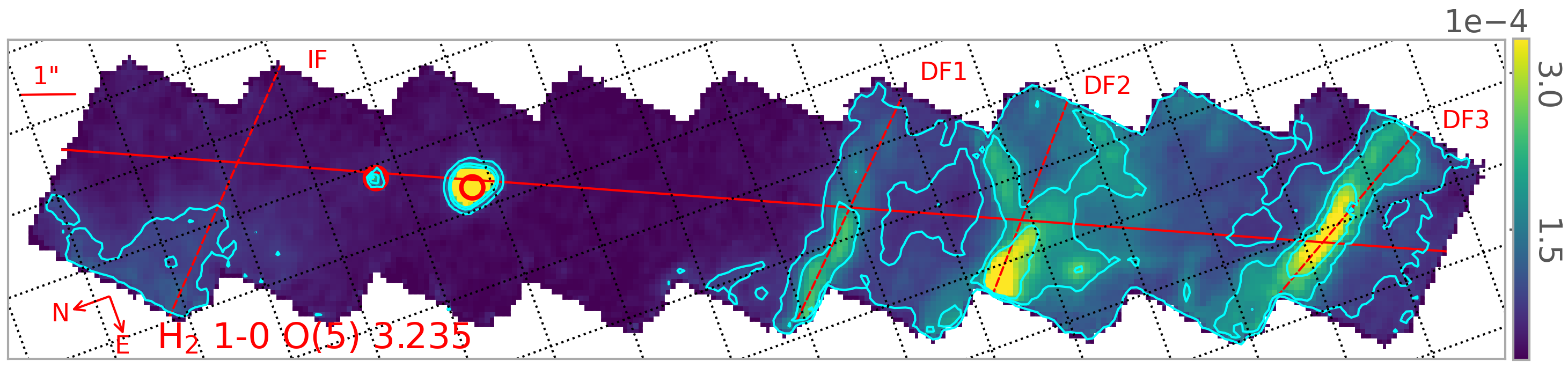}}
\resizebox{.99\hsize}{!}{
\includegraphics{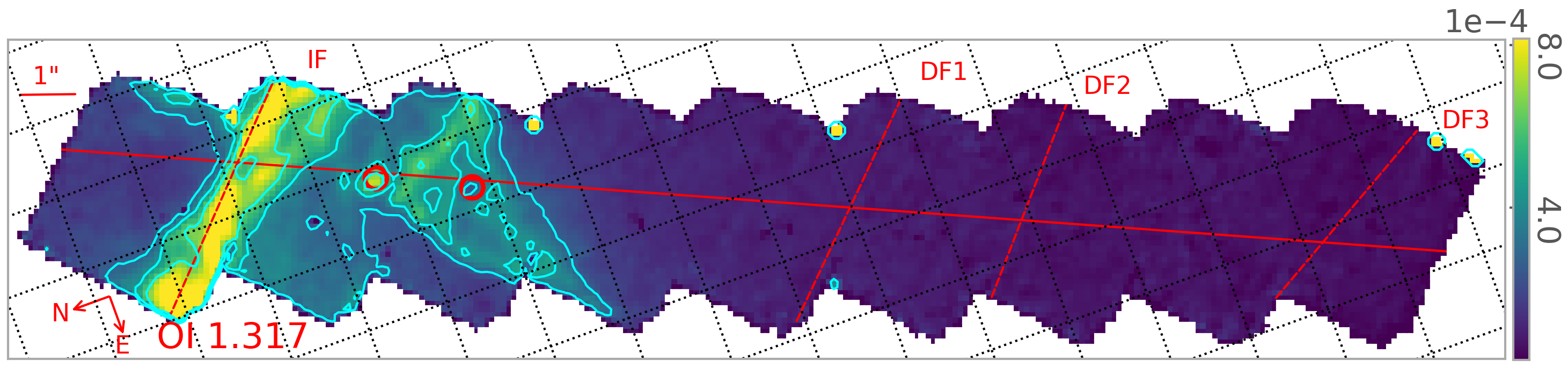}
\includegraphics{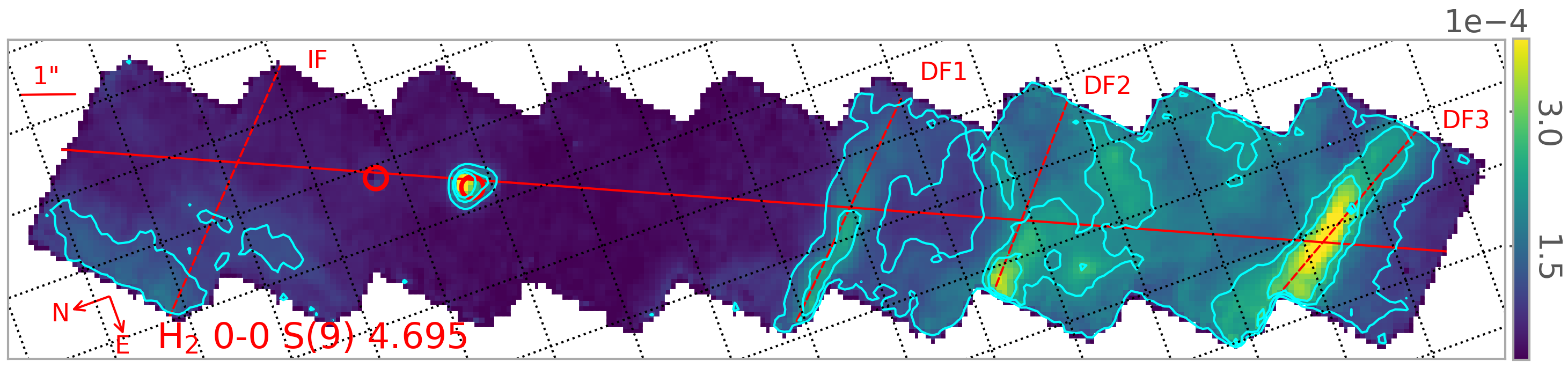}}
\resizebox{.99\hsize}{!}{
\includegraphics[scale=0.225]{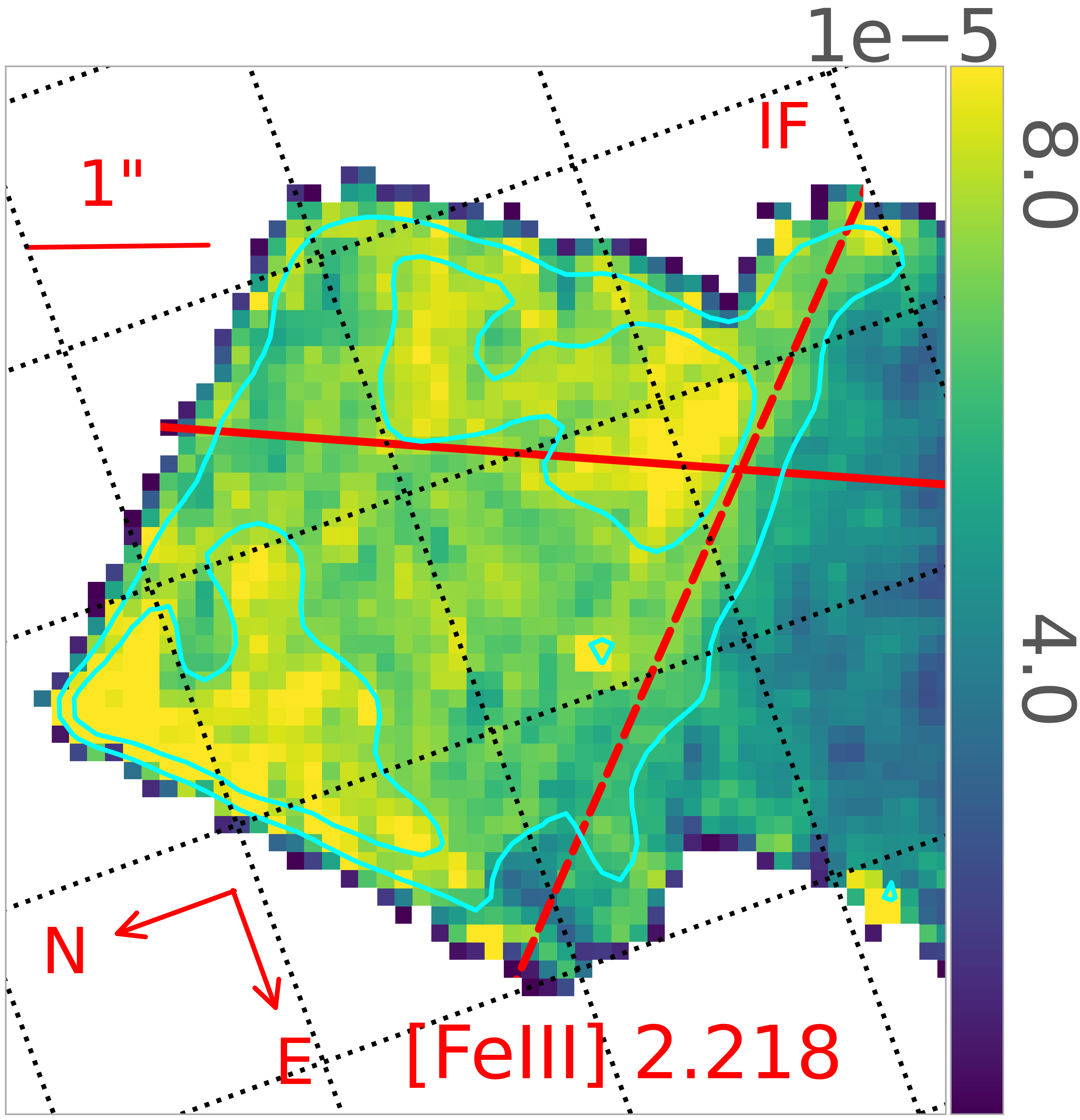}
\includegraphics{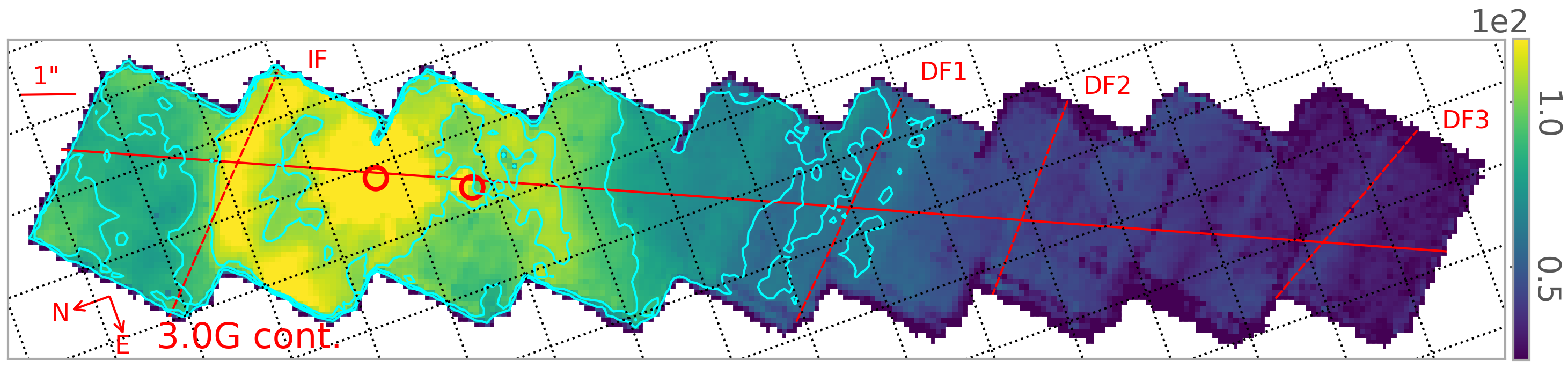}}
\caption{The Bar as seen in selected transitions in units of ${\rm erg\, cm^{-2}\,s^{-1}\,sr^{-1}}$. No extinction correction is applied. Starting at the top, we show an image of the total AIB emission (i.e. sum of all AIB components in the $3.2-3.7$~\mum range). Following that, (from top to bottom) we show Pa~$\gamma$, \HeI\ 2.058~\mum, \mbox{[\FeII] 1.644~\mum}, \OI\ 1.317~\mum, and \mbox{[\FeIII] 3.22~\mum} in the left column and \molh 1-0 S(1), \molh 2-1 S(1), \molh 1-0 O(5), \molh 0-0 S(9), and the continuum from the Gaussian decomposition at 3~\mum (${\rm MJy\,sr^{-1}}$) in the right column. We set the colour range from the bottom 0.5~\% to the top 99.5~\% intensity levels of the data for each map (across the entire NIRSpec mosaic), excluding values of zero, edge pixels, and the two proplyds (as well as the surrounding region of the proplyds for the continuum). White pixels inside the mosaic indicate values of zero reflecting issues with the data. The nearly horizontal red line indicates the NIRSpec cut and the nearly vertical red lines indicate from left to right the DFs (\Te, \Td, \Tc) and the IF. The two proplyds are indicated by the circles. Across the entire NIRSpec mosaic, contours show the 52, 75, and 90~\% intensity levels of the data for AIB, 65, 80, 98~\% for Pa~$\gamma$ and \HeI\, 77.9, 90.2, 96~\% intensity levels for \OI\ 1.317~\mum and \mbox{[\FeII] 1.644~\mum}, 87.7, 94.9~\% intensity levels for \mbox{[\FeIII] 2.218~\mum}, 60, 90, 98~\% intensity levels for \molh (excluding \molh 2-1 S(1), which uses 80, 90, 98~\% intensity levels), and the 50, 68, 85~\% intensity levels for the continuum at 3~\mum. We note that a smoothing is applied to the \mbox{[\FeIII] 2.218~\mum} contour levels.}
\label{fig:maps}
\end{center}
\end{figure*}

\begin{figure*}
\begin{center}
\resizebox{.99\hsize}{!}{
\includegraphics{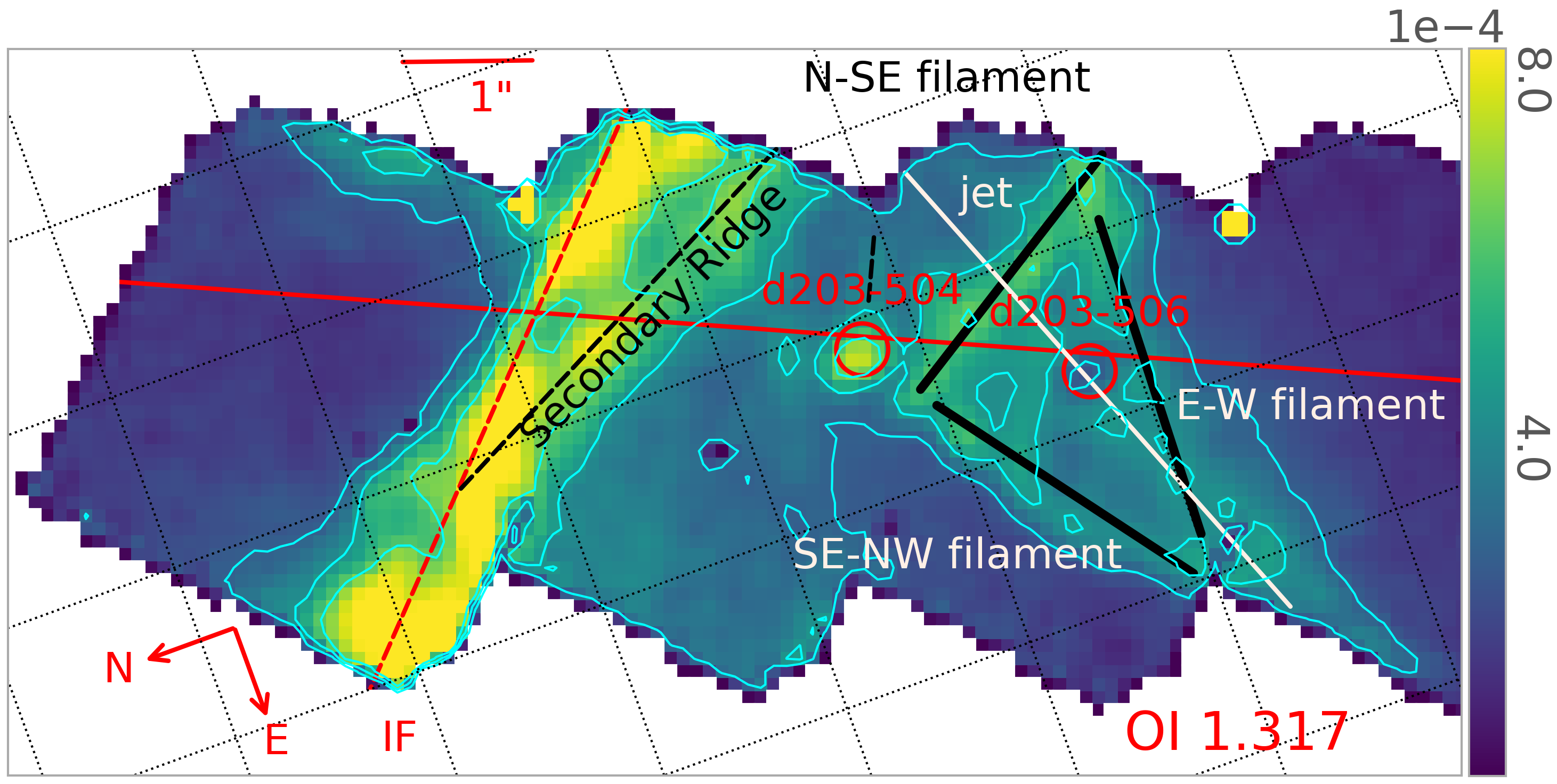}
\includegraphics{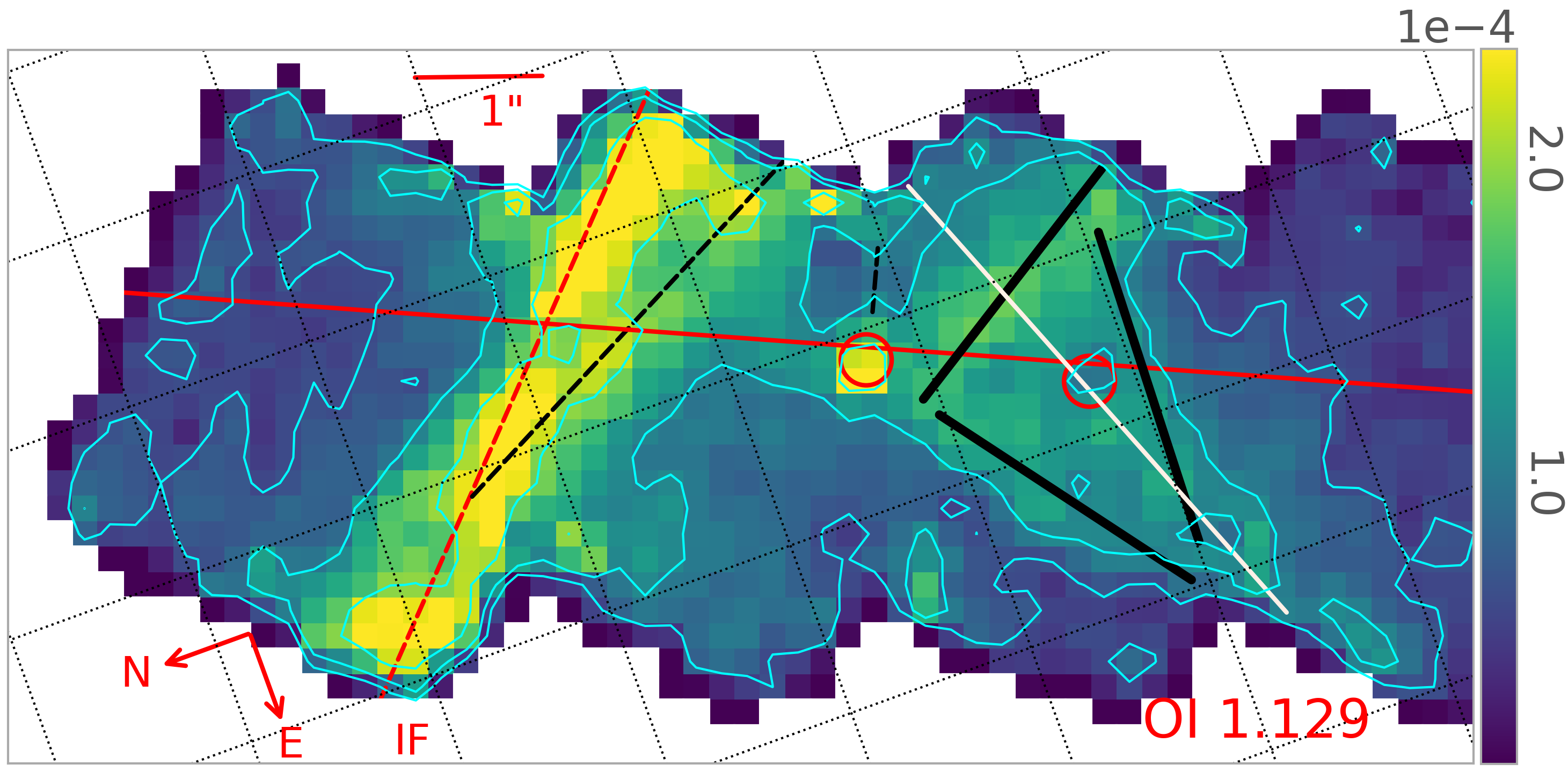}}
\resizebox{.99\hsize}{!}{
\includegraphics{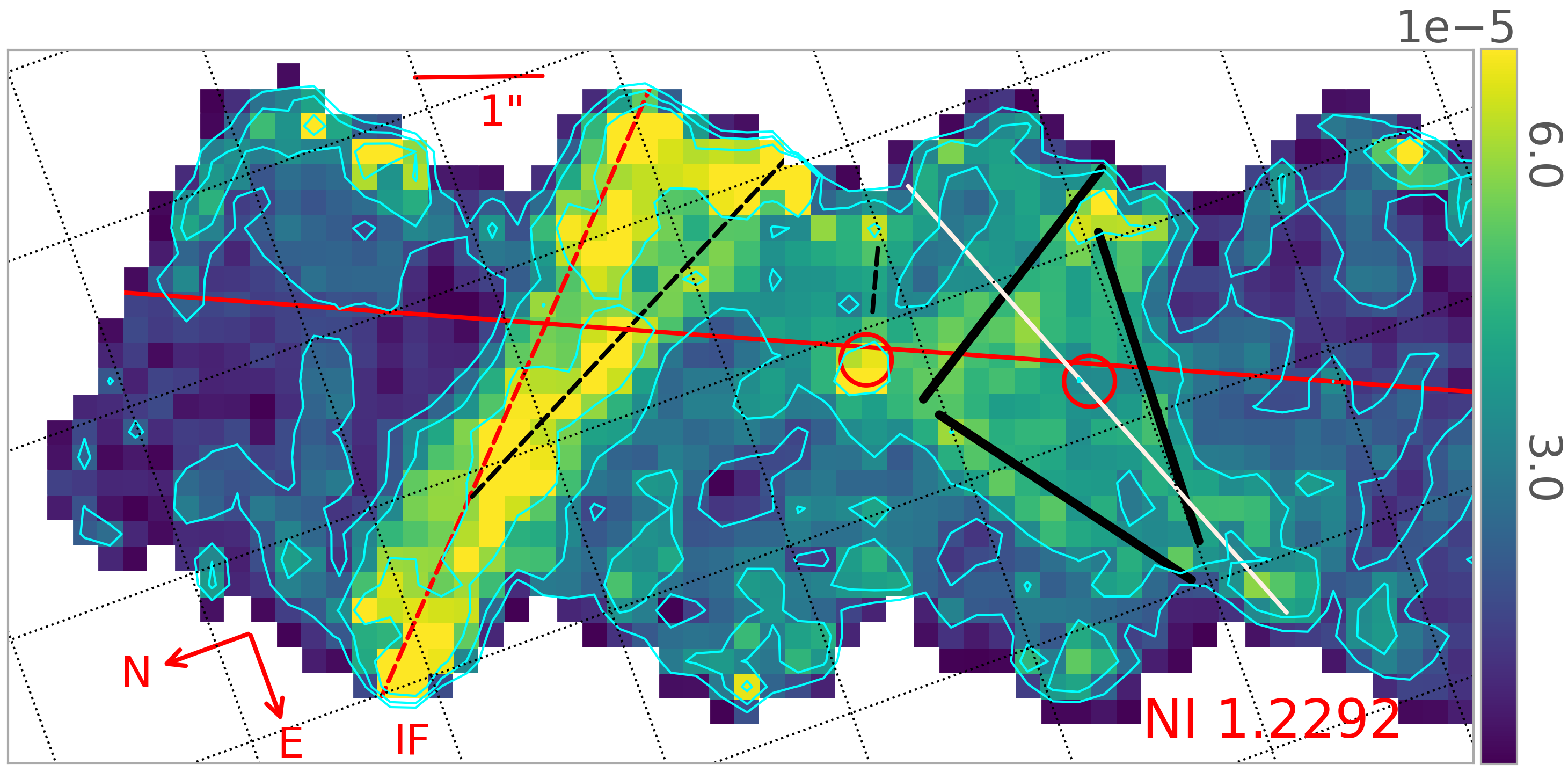}
\includegraphics{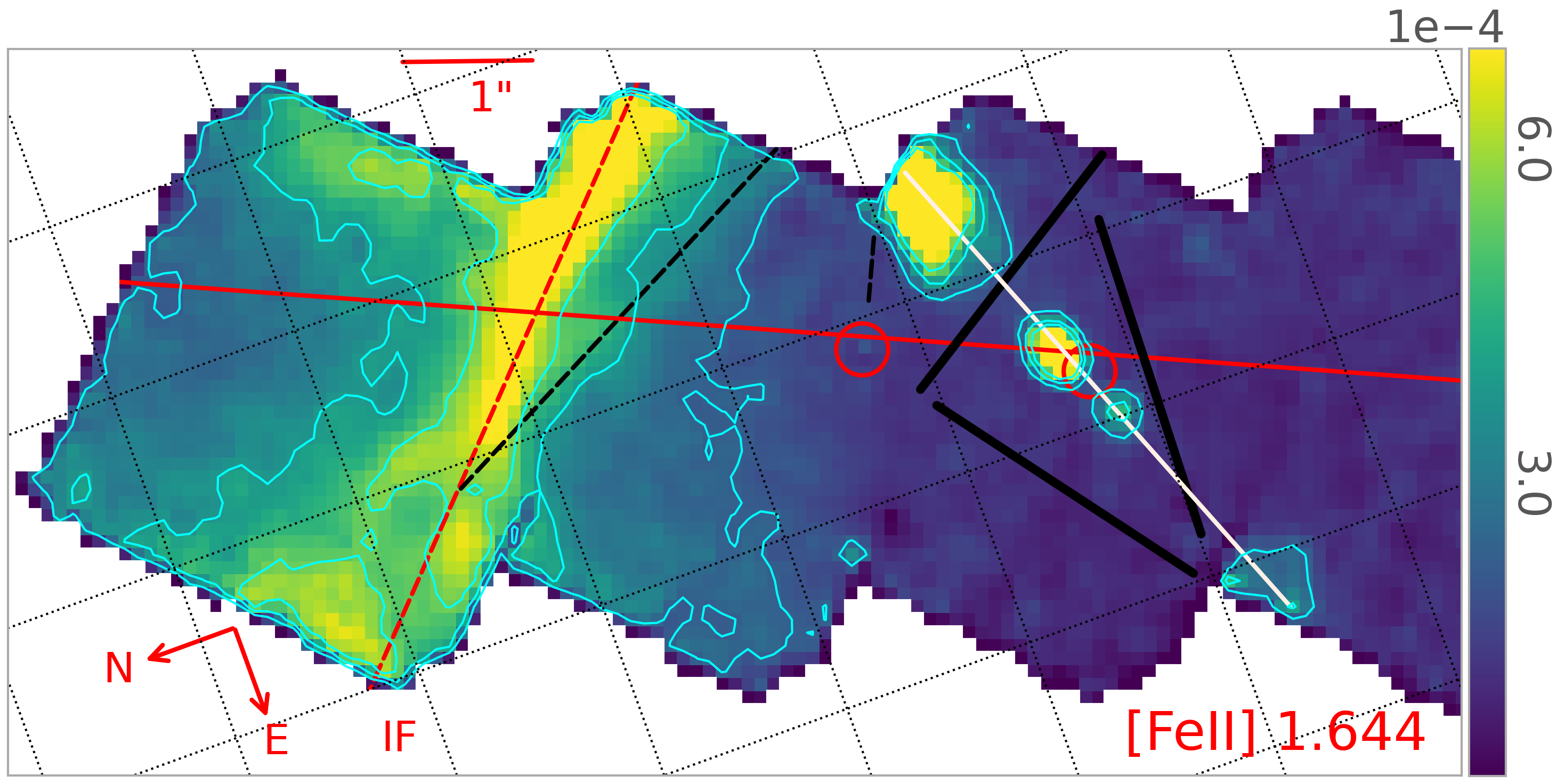}}
\caption{Maps of \OI\, 1.317~\mum, 1.129~\mum (top) as well as \NI \, 1.2292~\mum and \mbox{[\FeII] 1.644~\mum} (bottom) across part of the NIRSpec mosaic (see the \OI\, 1.317~\mum and \mbox{[\FeII] 1.644~\mum} maps in Fig.~\ref{fig:maps} for the full mosaic).  Similar filamentary structure beyond the IF is seen in the three fluorescent lines but not in the \mbox{[\FeII] 1.644~\mum} line. We set the colour range from the bottom 0.5~\% to the top 99.5~\% intensity levels of the data for each map (across the entire NIRSpec mosaic), excluding values of zero, edge pixels, and the two proplyds. White pixels inside the mosaic indicate values of zero reflecting issues with the data. The nearly horizontal red line indicates the NIRSpec cut and the nearly vertical red line indicates the IF. The two proplyds are indicated by the circles (left panel), the jet associated with proplyd 203-504 by a gold line and the filaments by red lines. Contours show the 77.9, 90.2, and 96~\% intensity levels of the data for \OI\, 1.317~\mum, 70, 85, and 97~\% intensity levels for \OI\, 1.129~\mum and 55, 81, and 96~\% intensity levels for \NI \, 1.2292 \mum, and 80, 90.2, and 96~\% intensity levels of the data for \mbox{[\FeII] 1.644~\mum} (across the entire NIRSpec mosaic). }
\label{fig:fluorescence}
\end{center}
\end{figure*}

\begin{figure}
\begin{center}
\resizebox{\hsize}{!}{
\includegraphics{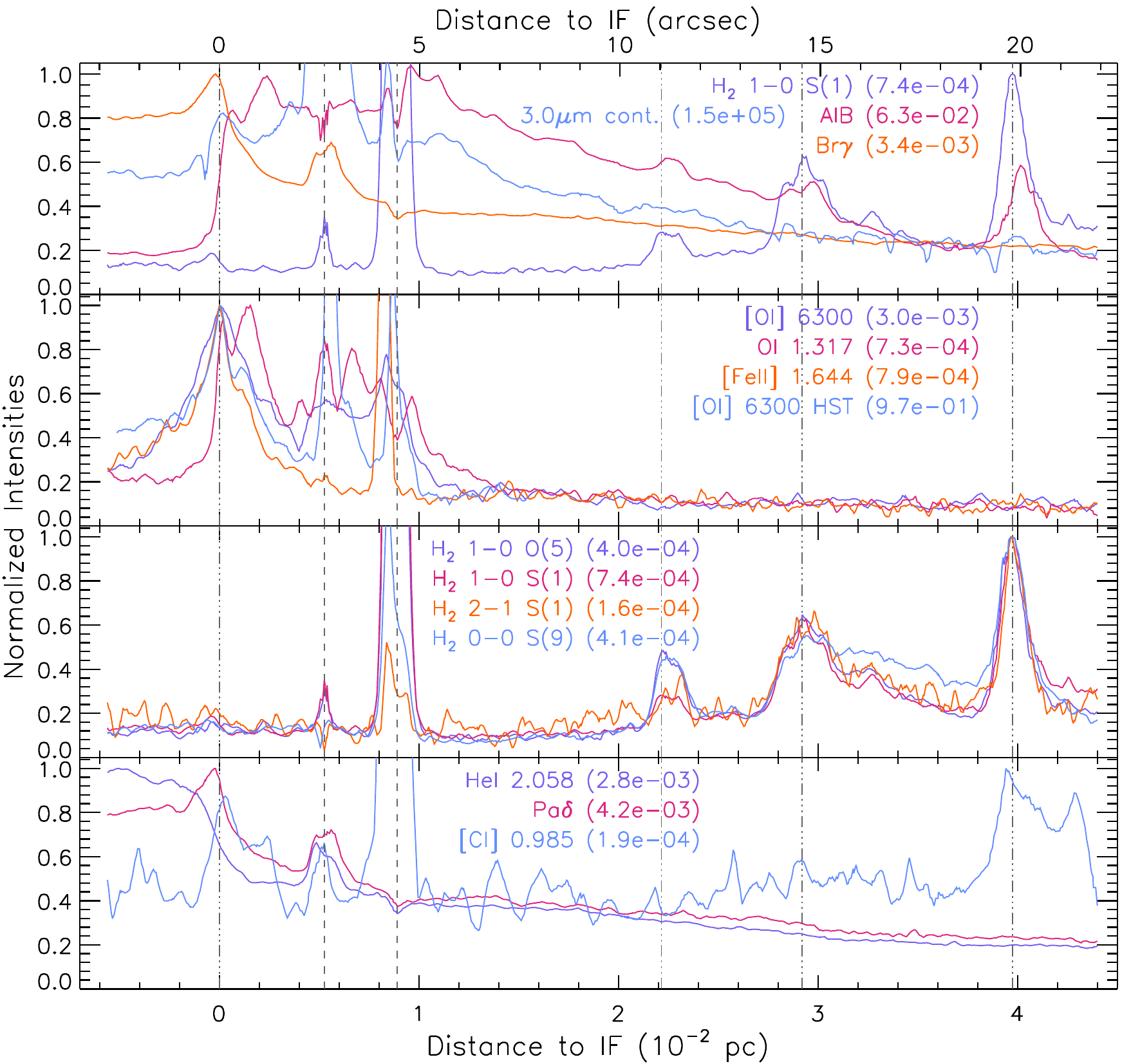}}
\caption{Normalised line intensities as a function of distance to the IF (0.228~pc or 113.4\arcsec\ from \oric) along a cut crossing the NIRSpec mosaic (see also Figs.~\ref{fig:FOV} and~\ref{fig:maps} for the location of the cut). As the cut is not perpendicular to the IF and distances are given along the cut, a correction factor of cos(19.58\textdegree)=0.942 needs to be applied to obtain a perpendicular distance from the IF. No extinction correction is applied. For reference, we show the extinction corrected \mbox{[\OI] 6300~${\mathring{\mathrm{A}}}$} emission observed by \citet{Weilbacher15} and the \mbox{[\OI] 6300~${\mathring{\mathrm{A}}}$} emission observed by \citet{Bally00} along the same cut. The dash-dot-dot-dot vertical lines indicate the position of the IF, \Tc, \Td, and \Te respectively from left to right. The dashed vertical lines indicated the location of the proplyds 203-504 (left) and 203-506 (right). Units are \uint\ except for the 3.0~\mum continuum which is in MJy/sr. }
\label{fig:cut}
\end{center}
\end{figure}

\section{Spatial variation of gas and dust tracers}
\label{sec:spatial-variation}

Fig.~\ref{fig:maps} shows maps of the intensity variation of selected gas and dust tracers. Surface brightness profiles along a cut across the NIRSpec mosaic (depicted in Fig.~\ref{fig:FOV}, PA=155.79\textdegree) are shown in Fig.~\ref{fig:cut}. We discuss the Bar PDR in Sect.~\ref{subsec:OB} and the emission associated with the two proto-planetary disks in Sect.~\ref{subsec:proplyds}. 

\subsection{Variations in the Bar}
\label{subsec:OB}

A layered structure is observed as we move away from \oric. The ionisation front (IF), the atomic PDR as traced by AIB emission, and the \molh emission peak at increasing distances, consistent with earlier studies \citep[e.g.][]{Tielens93, Marconi98, Walmsley00, Goicoechea15, Habart:im}. However, given the angular resolution of NIRSpec, we can now observe and resolve this anatomy at sub-arcsec scales across the 0.97--5.27~\mum wavelength range, which reveals filaments and ridges not seen before. For the following discussion, we define the IF by the peak intensity of the \mbox{[\OI] 6300~${\mathring{\mathrm{A}}}$} and \mbox{[\FeII] 1.644~\mum} emission lines at 0.228~pc (113.4\arcsec; PA=46.21\textdegree) from \oric and the dissociation fronts (DFs) by the maximum intensities of the \molh emission at 0.250, 0.257, and 0.267~pc (124.4\arcsec, 127.9\arcsec, 133.2\arcsec) from \oric (see below). We use the physical parameters given in Table~\ref{tab:param_OB} and Fig.~\ref{fig:geometry} for the different regions in the Bar. 

The \HI\, recombination lines trace the \Ta. Their emission is detected throughout the mosaic, consistent with the presence of a foreground \HII\ region in front of the atomic and molecular PDR (Fig.~\ref{fig:geometry}). 
  Overall, the \HI\, recombination lines show the same morphology. The \HI\ emission is strongest in the \Ta, peaking slightly before the IF (by 0.1\arcsec\, or \mbox{$0.02\, \times \,10^{-2}\,{\rm pc}$}) and then decreasing steeply up to \mbox{$\sim0.9\times\,10^{-2}\,{\rm pc}$} ($\sim$4.5\arcsec) from the IF, before levelling off at longer distances.
In addition to this overall morphology with distance from \oric, the \HI\, emission shows structure on smaller scales in the \HII\, region (in front of the IF) and its peak intensity near the IF is enhanced in the south-western half compared to that in the north-eastern half. Further structure is observed near the proplyds (see Sect.~\ref{subsec:proplyds}).   

The AIB emission traces the atomic PDR. The transition from the \HII\, to the \HI\ region, as traced by the AIB emission, is very sharp, with a change in surface brightness of up to $\sim$65\% over a distance of $\sim$1\arcsec. The AIB emission remains roughly constant up to \mbox{$\sim1.3\times\,10^{-2}\,{\rm pc}$} ($\sim$6.5\arcsec) from the IF, after which it gradually decreases. It exhibits local maxima near the proplyds (Sect.~\ref{subsec:proplyds}). Additional local maxima are detected near the three dissociation fronts, but these maxima are slightly displaced with respect to the dissociation fronts by \mbox{0.02, 0.06, and 0.04$\times10^{-2}$~pc} (0.1\arcsec, 0.3\arcsec, and 0.2\arcsec) towards the south for \Tc, \Td, and \Te, respectively. 
The AIB emission is highly structured across the atomic PDR. Specifically, the emission just past the IF is highly variable and shows additional local maxima at a distance of $\sim$2\arcsec\, from the northern part of the IF covered by the mosaic, in the S-SE direction of proplyd 203-504, and south of prolyd 203-506 (by about 0.2 to 1.2\arcsec). Lastly, in front of the IF, the AIB emission originating from the background face-on PDR, OMC-1, is enhanced in the eastern half of the FOV. This enhanced emission is part of a larger structure seen in the NIRCam AIB image (Fig.~\ref{fig:FOV}, red colour) and is not correlated with the foreground extinction (see Sect.~\ref{subsec:HI}). This suggests that the background PDR, OMC-1, displays an irregular surface that is affecting the amount of UV-excitation of the AIB carriers, as previously noted by, for example, \citet{Salgado16}.   

The \molh emission traces the dissociation fronts. The morphology of the H$_{2}$ lines (H$_{2}$ 0-0 S(9), 1-0 S(1), 2-1S(1), and 1-0 O(5) at 4.695, 2.122, 2.248, and 3.235~\mum, respectively) are consistent with one another. H$_{2}$ emission is observed throughout the mosaic, with higher intensities found in the lower half of the mosaic that trace the molecular PDR. Enhanced emission is also observed in the molecular PDR at \mbox{$2.21$, $2.92$, and $3.97 \times\,10^{-2}\,{\rm pc}$} (11.03\arcsec, 14.55\arcsec, 19.80\arcsec) from the IF, which subsequently defines three dissociation fronts in the edge-on PDR. The intensity of these four H$_{2}$ lines peak at \Te, followed by \Td and \Tc. Towards the northern part of the atomic PDR, we detect enhanced H$_{2}$ emission in the eastern half of the mosaic. Additionally, as for the AIB emission, in the region in front of the IF, we detect enhanced \molh emission in the eastern half of the mosaic. We note that this emission is originating from the background face-on PDR in the molecular cloud, OMC-1. 

The peak [\FeII]~1.644~\mum emission is co-spatial with the peak \mbox{[\OI] 6300~${\mathring{\mathrm{A}}}$} emission. This is consistent with  observations with MUSE by \citet[][extinction-corrected]{Weilbacher15} and by HST from \citet[not extinction-corrected]{Bally00}, indicating [\FeII] is an excellent tracer of the IF. The drop in [\FeII] intensity away from the IF is sharper than the drop in the MUSE \mbox{[\OI] 6300~${\mathring{\mathrm{A}}}$} intensity due to the lower angular resolution of the latter. The IF towards the West of the mosaic is more pronounced (i.e., larger intensity variation over a smaller area) than towards the East, consistent with the sharper transitions seen along a cut perpendicular to the bar, but West of the NIRSpec mosaic \citep{Habart:im}. On smaller spatial scales, both tracers exhibit almost identical profiles along the IF. 

The [\FeIII] 3.229~\mum emission is detected in front of the IF and thus inside the \Ta. The sharp drop in [\FeII] intensity towards \oric is likely caused by the transition from \FeII\ to predominantly \FeIII\ in front of the IF.

The overall morphology of the \HeI\ emission lines is the same, although small differences are observed between the measured transitions (see Sect.~\ref{subsec:He}). \HeI\ emission is observed throughout the mosaic, peaking closest to \oric, and displays a different spatial profile compared to that of the \HI\ emission. Closest to \oric, its intensity is roughly constant or decreases slightly with distance from \oric. We note that the NIRSpec mosaic covers only a small part of the Huygens \HII\, Region and, thus, likely misses the real \HeI\ emission peak. Subsequently, it drops rapidly. This sudden decrease in intensity starts before the IF, where the \HI\, intensity starts to increase. In contrast to [\FeII] and \HI, this rapid drop is less sharp and already transitions to a slow steady decline at $0.15 \times 10^{-2}\,{\rm pc}$ (0.74\arcsec) after the IF. 

The \OI\, 1.317~\mum emission peaks just beyond the IF and, overall, drops off sharply with increasing distance from the IF. This transition arises from \OI\ in the neutral gas  that is UV-pumped to its upper level, resulting in peak emission just beyond the IF. We do not observe enhanced emission in the direction of $\theta^2\, {\rm Ori\, A}$ (located to the East of the mosaic), suggesting that \oric is the main source of UV radiation. The structure seen in the [\FeII] emission along the IF is mimicked (but offset) by the \OI\, 1.317~\mum emission. The latter also exhibit a secondary, slightly weaker, ridge south of the primary ridge, at a small angle with the primary ridge (about 14 degrees), as well as enhanced emission towards the Eastern edge of the IF, neither of which are prominently seen in [\FeII] or \mbox{[\OI] 6300~${\mathring{\mathrm{A}}}$}, albeit both show slightly enhanced emission near the peak of the secondary ridge (also offset). The \OI\, 1.317~\mum emission displays filamentary structure between $3-6$\arcsec\, south of the IF, south of the proplyd 203-504, and surrounding the proplyd 203-506 (Fig.~\ref{fig:fluorescence}). These filaments' projections resemble a triangle with the strongest filament (N-SE filament) being parallel to the secondary ridge. These filaments are not seen in [\FeII] emission, which traces the IF. However, the SE-NW and the N-SE filaments 
do show weak emission in \mbox{[\OI] 6300~${\mathring{\mathrm{A}}}$} \citep{Bally00} and enhanced AIB emission is seen in part of the SE-NW filament. Two other fluorescent transitions within our wavelength coverage, the \OI\, 1.129~\mum and \NI\, 1.2292~\mum transitions, exhibit the same spatial distribution as the \OI\, 1.317~\mum emission (Fig.~\ref{fig:fluorescence}). The filamentary structure in the SE-NW direction towards proplyd 203-504
is associated with this proplyd (Sect.~\ref{subsec:proplyds}). While the lower half of the filamentary structure directed in the E-W direction 
may be aligned with the southern jet from proplyd 203-506 as seen in [\FeII], the top half is clearly offset from this jet by $\sim$16 degrees. We note this filament is slightly bent. 

The spatial distribution of the [\CI] 0.985~\mum line is unique (Fig.~\ref{fig:cut}): the observed variations in its intensity are only half of those seen in other tracers (excluding the proplyds) and it exhibits a ``flat'' profile with local maxima just beyond the IF, and at the proplyds, \Td, and \Te (though not at \Tc). We thus confirm that the \CI\, originates in the neutral gas beyond the IF. The [\CI] line shows enhancements at 2 out of the 3 \molh dissociation fronts, similar to the results of \citet{Walmsley00}, but over a much larger distance scale. In contrast to \citet[][due to their lower angular resolution]{Walmsley00}, [\CI]~0.985~\mum exhibits a local maximum just beyond the IF (co-spatial with the double ridge seen in \mbox{\OI\ 1.317~\mum}), which is absent in \molh and reflects the much smaller scale size near the IF.

Lastly, the continuum emission at 3~\mum is strong in the \Ta due to free-free emission, free-bound emission and emission from dust inside the \HII\, region as this line-of-sight crosses the Huygen's region \citep[][]{Salgado16} (see Figs.~\ref{fig:maps} and ~\ref{fig:cut}). An increase in intensity occurs near the IF, resulting in a local peak, co-spatial with the \mbox{\OI 1.317~\mum} peak emission. The continuum emission remains strong throughout the atomic PDR and displays small scale variations that mimic those seen in the AIB emission. In addition, very strong continuum emission is detected towards the proplyd 204-504. The continuum emission slowly decreases deeper into the molecular PDR 
due to the geometrical dilution of the radiation field

\begin{figure*}
\begin{center}
\resizebox{.99\hsize}{!}{
\includegraphics{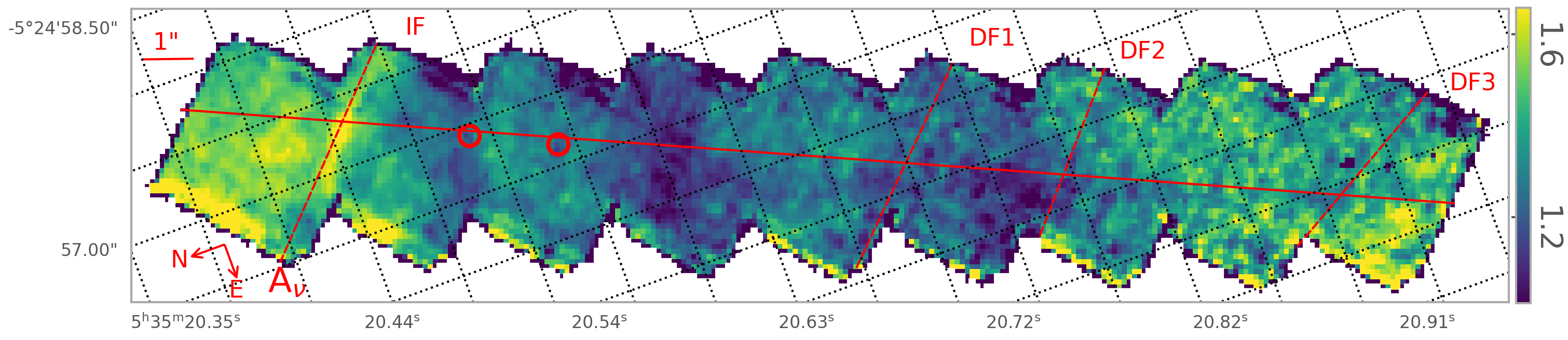}}
\resizebox{.99\hsize}{!}{
\includegraphics{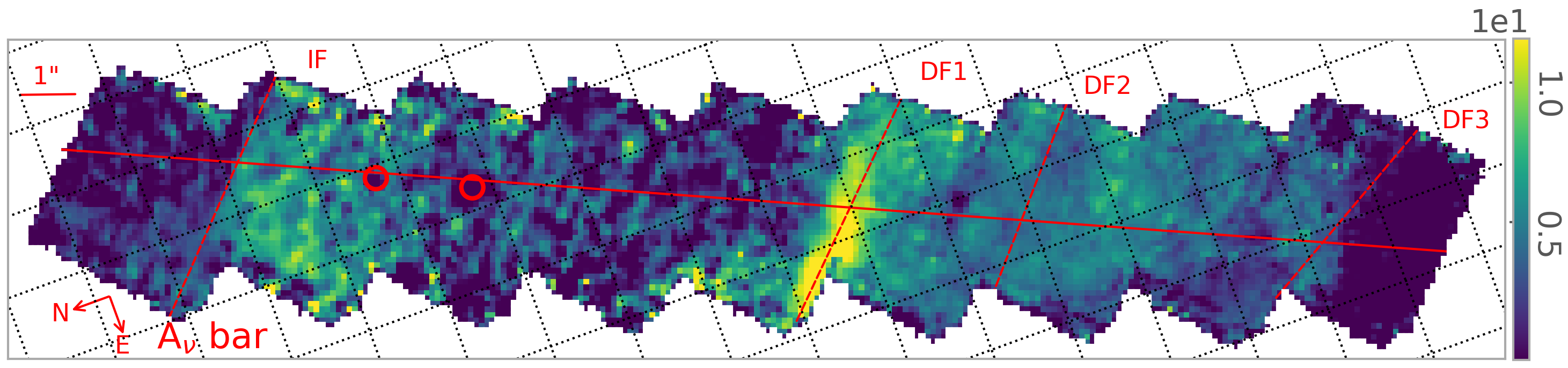}
\includegraphics{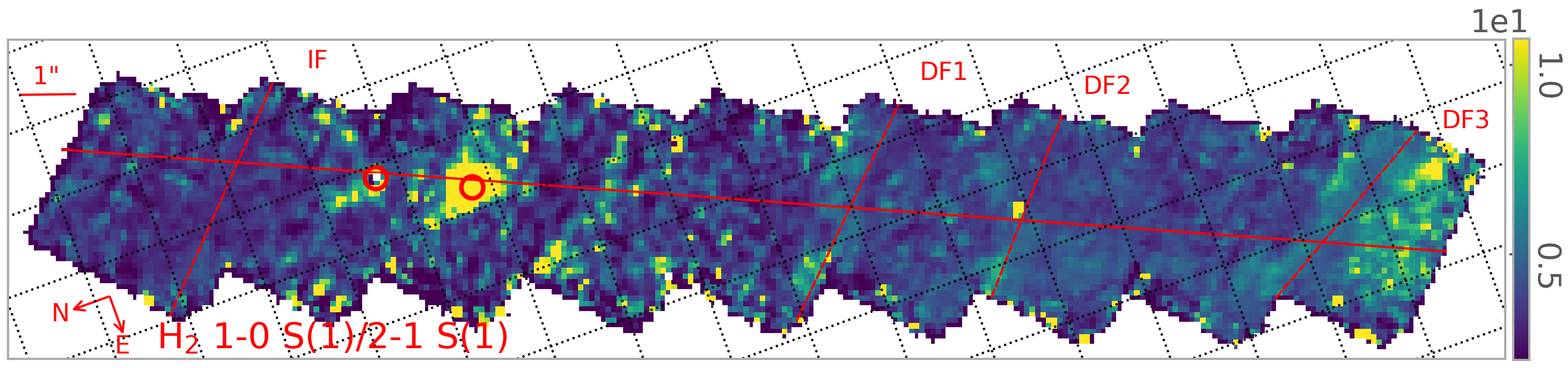}}
\resizebox{.99\hsize}{!}{
\includegraphics{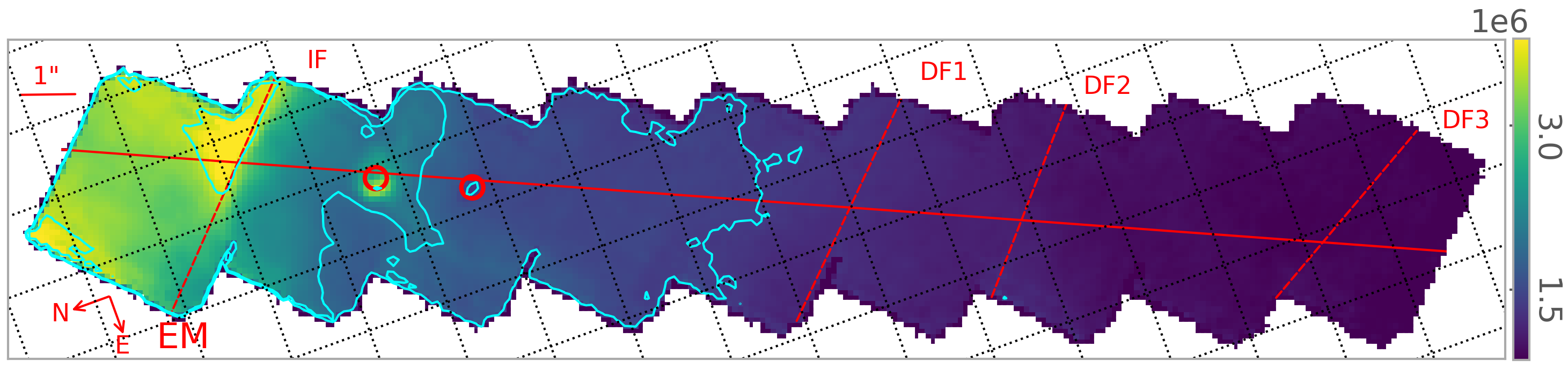}
\includegraphics{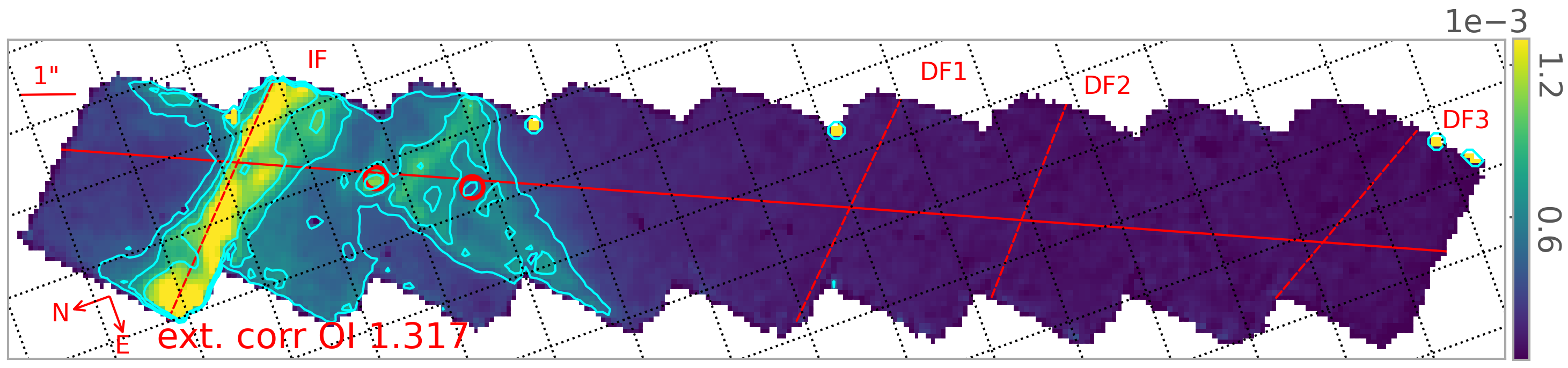}}
\caption{Starting at the top, we show an image of the visual line-of-sight fore-ground extinction A$_V$ (Sect.~\ref{subsec:HI}). Next, from left to right, we show  the visual line-of-sight internal PDR extinction A$_{V_{bar}}$ and the \molh 1-0 S(1)/2-1 S(1) ratio.
Lastly, we show the Emission Measure (EM; in units of pc cm$^{-6}$, Sect.~\ref{subsec:HI}) and the extinction corrected \mbox{\OI\ 1.317~\mum} (ext. corr. \mbox{\OI\ 1.317~\mum}; in units of ${\rm erg\, cm^{-2}\,s^{-1}\,sr^{-1}}$), respectively. We set the colour range from the bottom 0.5~\% to the top 99.5~\% of data for each map, excluding values of zero, edge pixels, and the two proplyds. White pixels inside the mosaic indicate values of zero reflecting issues with the data. The nearly horizontal red line indicates the NIRSpec cut and the nearly vertical red lines indicate from left to right the DFs (\Te, \Td, \Tc) and the IF. The two proplyds are indicated by the circles. Contours show the 55, 78, and 98~\% of the data for EM, and 77.9, 90.2, and 96 for \mbox{\OI\ 1.317~\mum}.}
\label{fig:maps_parameters}
   \end{center}
\end{figure*}

\begin{figure}
\begin{center}
\resizebox{\hsize}{!}{
\includegraphics{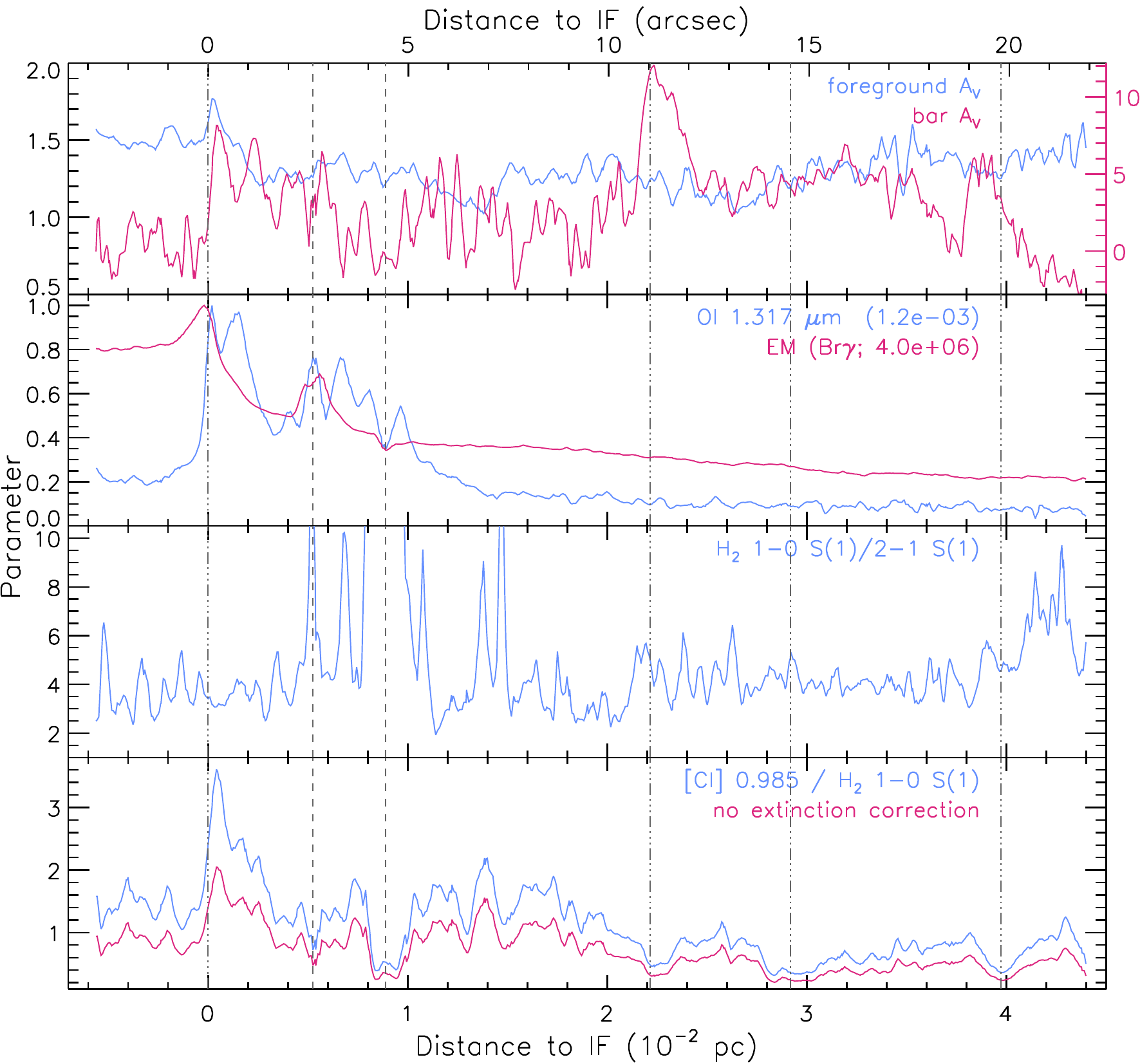}}
\caption{Foreground and internal PDR extinction (top panel), \mbox{\OI \, 1.317~\mum} line flux and emission measure (second panel) and line flux ratios (lower two panels) as a function of distance to the IF (0.228~pc or 113.4\arcsec\ from \oric) along a cut crossing the NIRSpec mosaic (see Fig.~\ref{fig:FOV}). As the cut is not perpendicular to the IF and distances are given along the cut, a correction factor of cos(19.58\textdegree)=0.942 needs to be applied to obtain a perpendicular distance from the IF. Foreground extinction correction is applied to determine bar $A_V$ ($A_{V_{bar}}$) and for the line intensities and ratios depicted in blue colour. The dash-dot-dot-dot vertical lines indicate the position of the IF, \Tc, \Td, and \Te respectively from left to right. The dashed vertical lines indicated the location of the proplyds 203-504 (left) and 203-506 (right). }
\label{fig:cut_parameters}
   \end{center}
\end{figure}

\subsection{Variations in protoplanetary disks}
\label{subsec:proplyds}

Two proto-planetary disks are present within the NIRSpec mosaic: the bright proplyd 203-504 and the silhouette disk 203-506 \citep[Fig.~\ref{fig:FOV},][]{Bally00}. 

The 203-504 proplyd shows strong emission from \HI\, recombination lines, He~I, the fluorescent \OI\, and \NI\, lines, [\CI] 0.985~\mum, and \molh, and enhanced AIB emission. A filamentary structure starting at this proplyd and extending towards the SE direction is seen in the \HI\, recombination lines. This SE-NW filament has been reported by \citet{Bally00} in H$\alpha$ and \mbox{[\OI] 6300~${\mathring{\mathrm{A}}}$}, being identified as a monopolar jet associated with the proplyd. The fluorescent lines also show a filamentary structure associated with this jet, but while this emission is parallel to the filament seen in \HI, it is offset to the west by $\sim0.2$\arcsec. No enhanced [\FeII] emission is observed in this filament. Enhanced emission in the \HI\, recombination lines is also seen towards the SSW of the proplyd. 

The proplyd 203-506 is discussed in detail by \citet{Berne:proplyd} and \citet{Berne:proplyd2}. In addition to these authors' results, we report that the proplyd exhibits strong emission in \mbox{[\OI] 6300~${\mathring{\mathrm{A}}}$} and [\CI]0.985, and the disk is seen in absorption in He~I and the fluorescent \OI\, and \NI\, lines. The [\FeII] emission is bright perpendicularly to the major axis of the silhouette disk 203-506, and on both sides, tracing the launching zone of a faint collimated jet observed before in \mbox{[\OI] 6300~${\mathring{\mathrm{A}}}$} HST images \citep{Bally00}. Extending in the jet direction, toward the North-West and South-East, we observe enhanced [\FeII] emission, with the North-West component being much brighter than the South-East component. This is suggestive of a Herbig-Haro object being associated with the proplyd jet.

\section{Deriving physical parameters}\label{sec:physparams}

\subsection{\HI\ recombination lines}
\label{subsec:HI}

The \HI\, recombination lines provide an estimate of the foreground extinction by comparing the observed ratios with those from case B recombination theory assuming an electron temperature of 10000~K\footnote{Here we have adopted an electron temperature for our analysis of 10000~K. Adopting a temperature instead of 9000~K (8500~K) would increase the theoretical Br~$\gamma$/Pa~$\delta$ line ratio by $\sim$1.4\% (2.2\%) resulting in an average decrease in $A_V$ of 3.5\% (5.2\%).}, an electron density of $n_e = 1000\, {\rm cm^{-3}}$ and no radiation field \citep{Prozesky:18}.  We use the ratio of Paschen~$\delta$ and Brackett~$\gamma$. Both lines are among the strongest \HI\ lines observed with detector NRS1 and have a large wavelength difference. We adopt the NIR extinction curve from \citet{gordon2023} for an $R_V=5.5$ \citep{Cardelli89}.  The resulting foreground visual extinction, $A_V$, varies between roughly 0.9 and 1.9 magnitude, corresponding to c(H$\beta$)\footnote{c(H$\beta) = log(\frac{I_{0, H\beta}}{I_{obs, H\beta}})$ with $I_{0, H\beta}$ the surface brightness in the absence of extinction and $I_{obs, H\beta}$ the observed surface brightness.} of 0.4-0.8 when using the extinction curve of \citet[][Figs.~\ref{fig:maps_parameters} and~\ref{fig:cut_parameters}]{Blagrave:07}.
To first order, the derived foreground extinction decreases slightly with distance from \oric up to about a distance of \mbox{$\sim 2.6 \times\,10^{-2}\,{\rm pc}$} ($\sim$13\arcsec) from the IF, after which it slowly increases. It is also slightly structured, with weak local minima near, roughly, \mbox{$0.4$, $1.4$, and $2.6 \times\,10^{-2}\,{\rm pc}$}  ($\sim$2\arcsec, $\sim$7\arcsec, and $\sim$13\arcsec) from the IF. 
The derived extinction values are overall consistent with \citet[][]{Weilbacher15}, but the morphology differs slightly. Specifically, neither the increase in the extinction past roughly \mbox{$\sim 2.6 \times\,10^{-2}\,{\rm pc}$} ($\sim$13\arcsec) from the IF nor the local minima are observed by \citet[][]{Weilbacher15}. In addition, \citet[][]{Weilbacher15} derived a slightly increased extinction at the western side of the NIRSpec mosaic in the atomic PDR. These discrepancies likely results from the combination of 1) the lower angular resolution observations of \citet[][seeing of 0.67\arcsec\, to 1.25\arcsec]{Weilbacher15} resulting in a spatial averaging weighted by the emissivities, and 2) remaining systematic artefacts in the data set that are not captured by the comparison of these line fluxes in the individual dither observations (which agree within $0.5\%$).

A caveat of the presented extinction analysis is the following. The extinction is a summation of absorption and scattering out of the line-of-sight. For the Bar, we have an extended background light source
and the extinction is due to foreground extended dust. If the background light source and dust layer are both uniformly distributed, the light scattered outside the line-of-sight will be compensated by the light scattered in the line-of-sight for a spherically symmetric, unresolved source. Therefore, the net attenuation will only be due to absorption. However, the actual situation is more complex. Both the background light source and the dust are distributed heterogeneously and it is impossible to estimate the actual geometry. But in general, the attenuation will be reduced by the light scattered into the line-of-sight \citep[c.f.][]{Code:73}. While the general trend of the determined extinction, $A_V$, should not be affected very much, it does affect the quantitative value of $A_V$ obtained from the standard extinction curve, which is estimated from observations of background stars. As the albedo is still around $\sim$0.5 in the NIR, the absorption will only be a half of extinction. The spectral dependence is also different between the absorption and scattering. However, this should not make a large difference because of the short spectral range considered and small attenuation in question. 
Despite this caveat, we will use these derived values for the foreground extinction to the Bar and employ the NIR extinction curve for an $R_V=5.5$ from \citet{gordon2023}.

We obtain an estimate of the emission measure, EM, from the Brackett~$\gamma$ surface brightness using Eq.~\ref{eq:EM}. The emission measure ranges within $0.84-4.15 \times 10^6\,{\rm pc\, cm^{-6}}$ (Figs.~\ref{fig:maps_parameters} and~\ref{fig:cut_parameters}; see Table~\ref{tab:physical_conditions} for the five templates). The obtained EM in front of, and at, the IF are consistent with those reported by \citet{Walmsley00} for similar regions north-east of the NIRSpec mosaic (their positions A and B). Assuming a depth of the ionised bar of 20\arcsec\  \citep[0.05~pc;][]{Walmsley00}, we obtain a rms electron density of the order of $9000\ {\rm cm^{-3}}$.

\begin{table*}[]
     \caption{Physical conditions derived for the five templates. \molh excitation temperatures and column densities are given in Table~\ref{tab:ext_param}.}
    \label{tab:physical_conditions}
   \begin{center}
    \begin{tabular}{llclllll}
    \hline \hline \\[-5pt]
     & & Sect. & \multicolumn{1}{c}{\HII} & \multicolumn{1}{c}{Atomic} & \multicolumn{1}{c}{\Tc}& \multicolumn{1}{c}{\Td}& \multicolumn{1}{c}{\Te}\\
     & & & \multicolumn{1}{c}{region} & \multicolumn{1}{c}{PDR} & \\
    \hline \\[-10pt]
    A$_V$\,\tablefootmark{1} & & \ref{subsec:HI}&  1.5 & 1.3 & 1.2 & 1.2 & 1.4\\
    A$_{V_{bar}}$\,\tablefootmark{2}  && \ref{subsec:h2} & -- & 4.33 & 9.34 & 4.67 & 2.00 \\
    A$_{V_{bar},\, I}$\,\tablefootmark{3}  && \ref{subsec:h2} & 0.00 & 7.86 & 37.24 & 8.67 & 3.22 \\
    EM (Br$\gamma$)&[$10^6\,{\rm pc\, cm^{-6}}$]  & \ref{subsec:HI} & 3.3 & 2.3 & 1.2 & 1.1 & 0.9\\ 
    T (\CI)&[K] & \ref{subsec:Clines}, \ref{app:CI} &  $\sim$2500 &  $\sim$2300 & $\sim$2900  & $\sim$6800  & $\sim$5600  \\
    EM$\times$T$^{-0.6}$ (\CI) &[cm$^{-6}$\,pc\, K$^{-0.6}$] & \ref{subsec:Clines}, \ref{app:CI} & 1034$\pm$18& 2635$\pm$40& 725$\pm$11& 938$\pm$15& 1589$\pm$17 \\
    EM (\CI)&[$10^5\,{\rm pc\, cm^{-6}}$]  & \ref{subsec:Clines}, \ref{app:CI} & 1.12$_{0.25}^{0.26}$ & 2.73$_{0.56}^{0.63}$ & 0.86$^{0.47}_{0.42}$& 1.86$^{0.83}_{0.60}$ & 2.82$^{0.49}_{0.37}$ \\
    3.4/3.29 AIB\,\tablefootmark{4} & & \ref{subsubsect:AIBint} & 0.11 & 0.09 & 0.10&  0.15 & 0.18   \\
    4.64/$\Sigma$AIB\,\tablefootmark{5} & [$10^{-3}$]& \ref{subsubsect:AIBint} & 8.2 & 6.3 & 6.1 & 2.8 & 2.4 \\
    4.64/3.29 AIB & [$10^{-3}$]& \ref{subsubsect:AIBint} & 16.0 & 12.1 & 11.9 & 6.0 & 5.6 \\  
    4.64/3.40 AIB & [$10^{-2}$]& \ref{subsubsect:AIBint} & 20.5 & 22.5 & 15.8 & 5.4 & 4.1 \\
   FWHM 3.29 AIB & [${\rm cm^{-1}}$] & \ref{subsubsect:AIBprof} & 39.3 & 37.4 & 38.1 & 41.2 & 42.4\\[5pt]
   \hline
    \end{tabular}
   \end{center}
   \tablefoot{
   \tablefoottext{1}{foreground extinction;}
   \tablefoottext{2}{internal PDR extinction calculated using the foreground formalism (see Fig.~\ref{fig:geometry});}
   \tablefoottext{3}{internal PDR extinction calculated using the intermingled formalism;}
   \tablefoottext{4}{ratio of the integrated intensities of the (3.39+3.40+3.42)/(3.29) AIBs using the Gaussian decomposition method.
   \tablefoottext{5}{$\Sigma$AIB refers to the sum of all Gaussian AIB components in the $3.2-3.7$~\mum range.}}
    }
   
\end{table*}

\begin{figure}
\begin{center}
\resizebox{\hsize}{!}{
\includegraphics{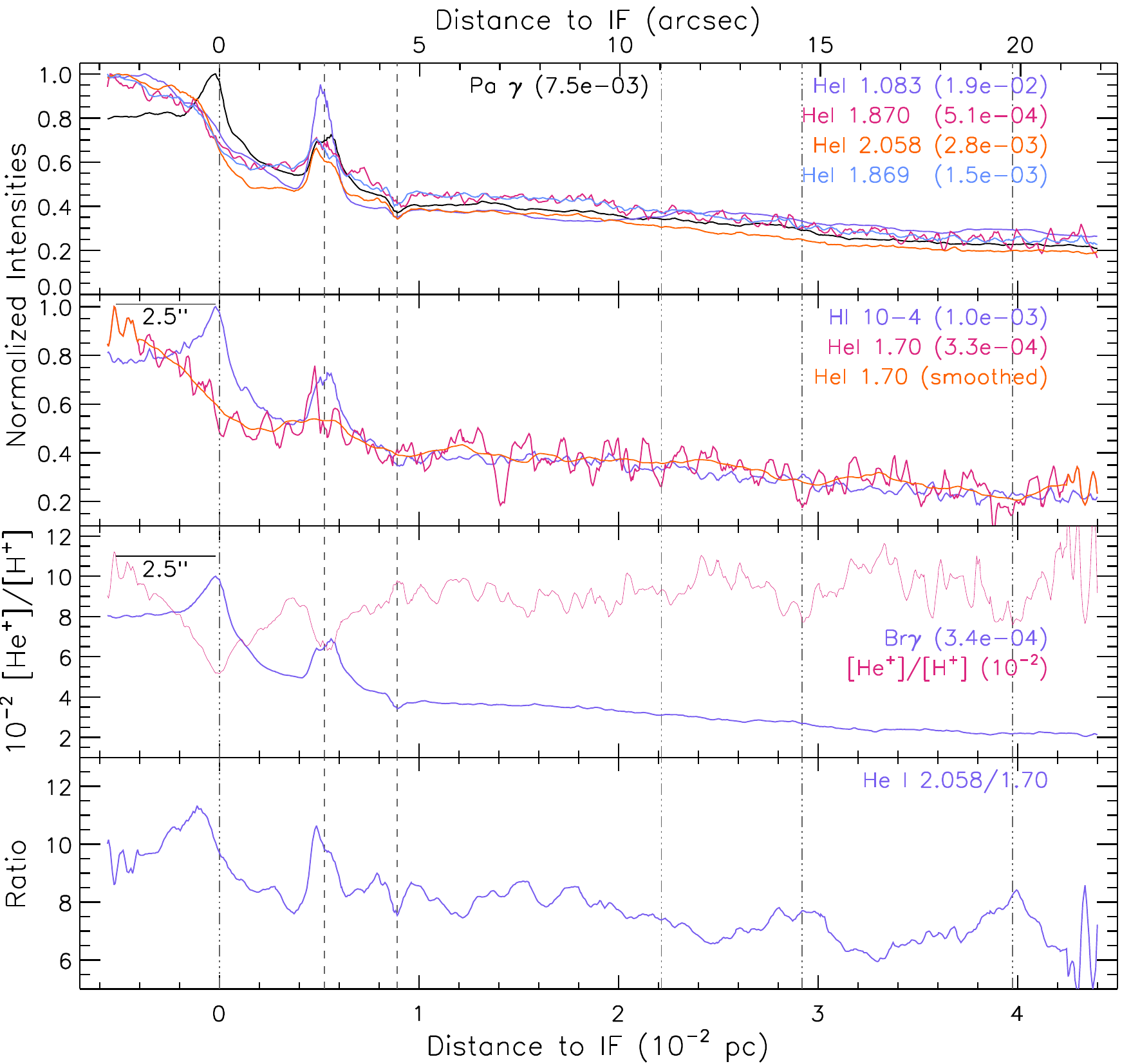}}
\caption{Analysis of the \HeI\ and \HI\ radial profiles as a function of distance to the IF (0.228~pc or 113.4\arcsec\ from \oric) along a cut crossing the NIRSpec mosaic (see Fig.~\ref{fig:FOV}). As the cut is not perpendicular to the IF and distances are given along the cut, a correction factor of cos(19.58\textdegree)=0.942 needs to be applied to obtain a perpendicular distance from the IF.  
Top two panels: Observed normalised line intensities of selected transitions. No extinction correction is applied. Units are in \uint.
Third panel: The [He$^+$]/[H$^+$] abundance. The normalised Br~$\gamma$ radial profile is shown for reference.
Bottom panel: The \HeI\ 2.058/1.70 radial profile. 
The dash-dot-dot-dot vertical lines indicate the position of the IF, \Tc, \Td, and \Te, respectively, from left to right. The dashed vertical lines indicated the location of the proplyds 203-504 (left) and 203-506 (right). }
\label{fig:he}
\end{center}
\end{figure}

\begin{figure}
\begin{center}
\resizebox{\hsize}{!}{%
\includegraphics{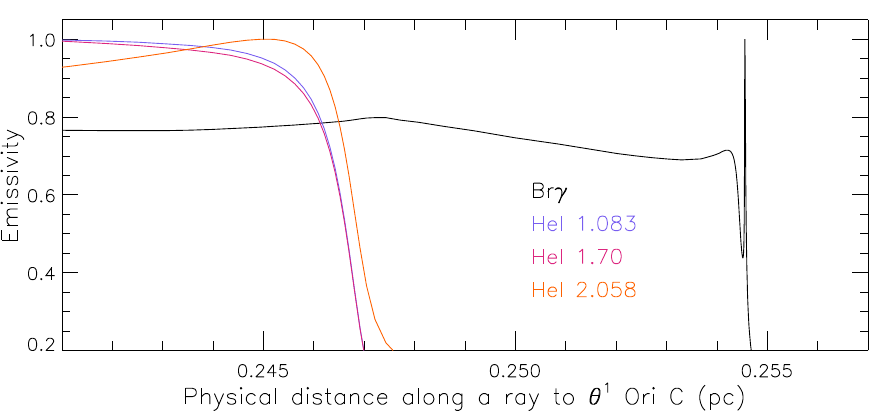}}
\caption{Emissivity profile of selected transitions as a function of the physical distance along a ray from \oric and perpendicular to the Bar given by a Cloudy model employing the physical parameters derived by \citet{Shaw09} and \citet{Pellegrini09}. See Sect.~\ref{subsec:He} for details.}
\label{fig:modelHe}
\end{center}
\end{figure}

\subsection{\texorpdfstring{HeI}{\HeI} recombination lines}
\label{subsec:He}

The distinct radial profiles for the \HeI\, and \HI\, recombination lines (Figs.~\ref{fig:cut} and~\ref{fig:he}) indicate, for the first time, that the He and H ionisation fronts are clearly separated in the Huygens region. While the location of the H-IF is well defined (i.e., peak \HI\ emission at 0.2274~pc or 113.276\arcsec\ from \oric), the location of the He-IF is somewhat uncertain as the NIRSpec radial profiles do not extend very deep into the Huygens Region. However, the \HeI\ 1.70~\mum intensity remains fairly constant up to about 15\arcsec\, away from the IF \citep{Marconi98}. Hence, we quantify the displacement between the H-IF and He-IF by the distance between the peak emission of the \HI\ and \HeI\ recombination lines and find a displacement of approximately \mbox{0.5$\times$10$^{-2}$\,pc} or \mbox{2.5\arcsec\ }(Fig.~\ref{fig:he}). Only the \HeI\ 1.083~\mum emission gives a slightly smaller displacement of \mbox{0.36$\times$10$^{-2}$\,pc} (1.8\arcsec). 

These results are consistent with Cloudy modeling \citep{cloudy} of the Orion Nebula. We adopt the model parameters derived from the detailed fits of the optical lines originating from the ionised gas by \citet{Shaw09} and \citet{Pellegrini09}\footnote{It should be noted that the analysis from \citet{Shaw09} and \citet{Pellegrini09} required the models to reproduce the projected distance of the IF, which was defined based on the peak emission of [\SII] at a projected distance from \oric of 111\arcsec. These authors adopt a distance of 437~pc for the Bar and thus the IF is located at a projected distance of 0.235~pc. In this paper, we instead define the IF by the peak intensity of the \mbox{[\OI] 6300~${\mathring{\mathrm{A}}}$} and \mbox{[\FeII] 1.644~\mum} emission lines at 113.4\arcsec\, or 0.228~pc from \oric (given the adopted distance of 414~pc). This small difference should, however, not considerably affect the spatial offset between the He and H ionisation front.}. 
In a radial direction from \oric, this model predicts the H IF at a distance of $\sim$0.254~pc and a displacement between the emissivity of selected H and He transitions of $\sim$0.007~pc (Fig.~\ref{fig:modelHe}). We note that the latter distances are physical distances along a ray into, and perpendicular to, the Bar from \oric as we did not model the corresponding radial surface brightness profiles of selected transitions as a function of the projected distance from \oric. These differences result in an offset between the model IF and our observations. We conclude therefore that the model calculations are in good agreement with the \HeI\, and \HI\, observations reported here.

Based on the radial profiles of \HeI\,1.70~\mum and \HI\,10-4, we can estimate the [He$^+$]/[H$^+$] abundance following \citet[][see Appendix~\ref{app:he}]{Marconi98}. Fig.~\ref{fig:he} shows the radial profile along the NIRSpec cut. Over most of the radial profile, the [He$^+$]/[H$^+$] abundance is fairly constant. However, due to NIRSpec's angular resolution, we also detect two strong dips. The dip at the IF reflects the displacement of the H-IF and He-IF, while the second dip is co-spatial with the proplyd 203-504. We note that the \HeI\ emission near this proplyd depends on the transition considered. Away from the IF and this proplyd, we derive a [He$^+$]/[H$^+$] abundance of $0.094\pm0.009$.  While previous observations did not resolve the difference in the H and He ionisation structure in the Huygens region, these studies obtain a similar [He$^+$]/[H$^+$] abundance because of spatial averaging \citep[e.g.,][]{Osterbrock:92, Esteban:98, Marconi98, Baldwin:00, Walmsley00, Esteban:94, Blagrave:07}. Recent refinements in the atomic data \citep{DelZanna:22} may necessitate a new analysis of the [He$^+$]/[H$^+$] abundance ratio to take full advantage of the high quality of the NIRSpec data.

We furthermore confirm the observed \HeI\ 2.058/1.70 ratio as detected by \citet{Marconi98}. This ratio ranges between about $7-10$ (Fig.~\ref{fig:he}). These authors argue that one can compare this ratio with the model predictions of \citet{Smits:96}, as the \HeI\ 1.70~\mum transition arises from the same upper level as the 4471~${\mathring{\mathrm{A}}}$\, transition used for normalisation in the model calculations. The observed value is significantly enhanced with respect to the model prediction, which \citet{Marconi98} potentially attributes to the neglect of collisional effects and line trapping. 

Lastly, we can obtain an estimate of the electron temperature, $T_e$, from the \HeI{} line ratios 2.1649/2.1137 and 2.1649/2.118 based on the diagnostic diagram of \citet[][their Fig. 7b]{MartinHernandez:03}. We find an electron temperature, $T_e$, of about $9000-9500\,{\rm K}$ and an optical depth of the 2$^3$S metatstable level, $\tau_{3890}$, of $\sim$0 for each of the template spectra except for \Tc for which we find an electron temperature, $T_e$, of about $8000\,{\rm K}$.

\begin{figure*}
\begin{center}
\resizebox{.9\hsize}{!}{%
\includegraphics{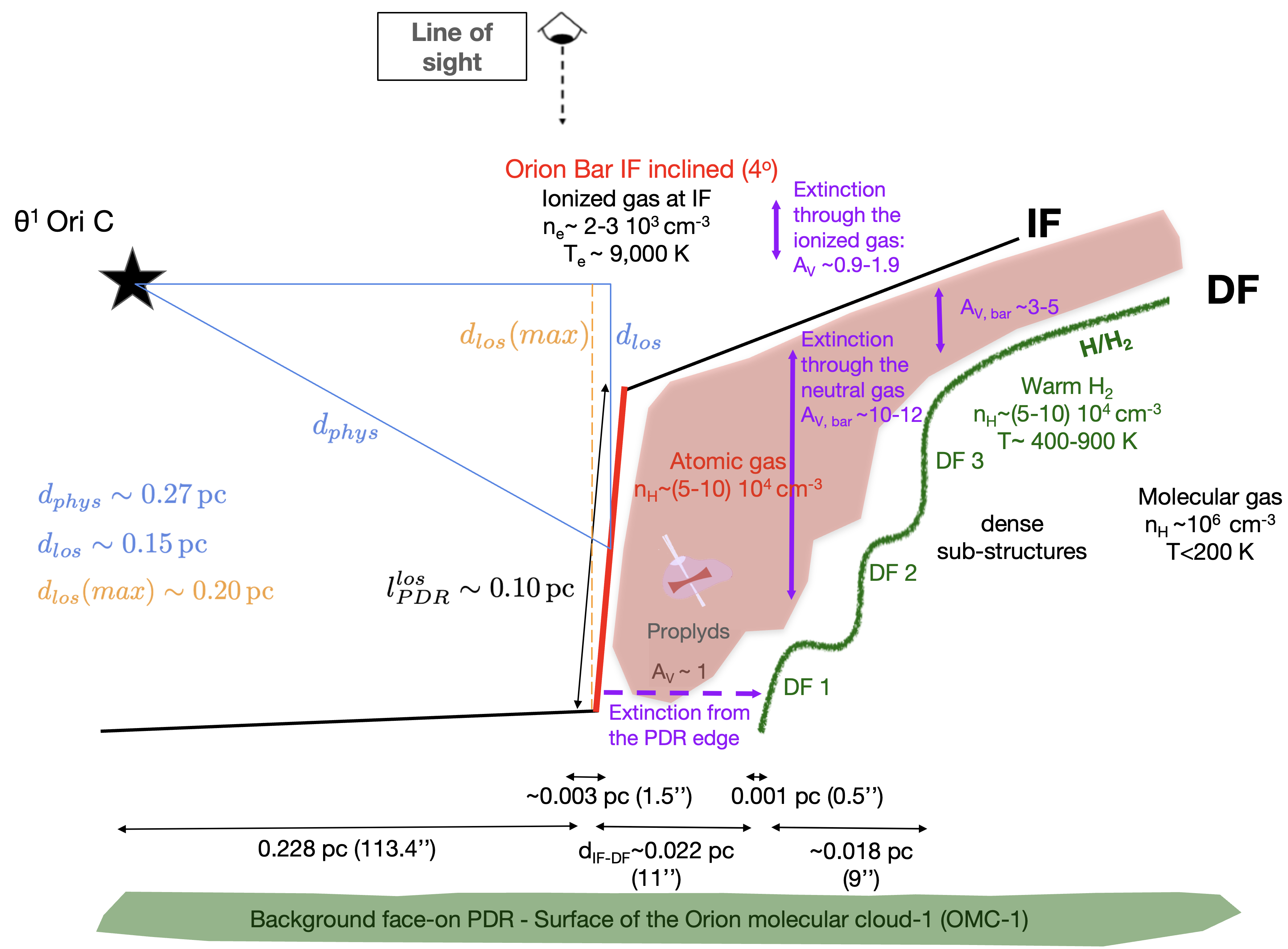}}
\caption{Schematic overview of the Bar as inferred from JWST and other observations \cite[e.g.][]{jansen1995millimeter,wen1995three,Odell01,Pellegrini09}. It shows the main features discussed in this paper and \citet[][based on NIRCam and MIRI imaging]{Habart:im}. Given the complexity of the PDR surface, parameters derived in this paper  are specific for the NIRSpec mosaic. Note that for clarity, the dimensions perpendicular to the bar are not to scale; the true spatial scales are explicitly given in the annotations. In addition, the sketch does not include foreground material, that includes a layer of ionised gas \citet{ODell:20} and, closer to the observer, layers that are grouped together under the designation as the Veil \citep[e.g.,][]{Rubin2011, Boersma2012, vanderWerf13, Pabst19, Pabst20}.}
\label{fig:geometry}
\end{center}
\end{figure*}

\begin{table*}
     \caption{Parameters employed or derived for the Orion Bar.  }
    \label{tab:param_OB}
   \begin{center}
    \begin{tabular}{lll}
    \\[-30pt]
    \hline \\[-10pt]
    \hline \\[-10pt]
    Parameter & Value & Reference\\
    \hline \\[-10pt]
    distance & 414$\pm$7~pc\tablefootmark{a} & \citet{Menten07}\\
     & 1\arcsec = 0.002~pc & \\
    projected distance, $d_{proj}$, between $\theta ^1$ Ori C and the IF\tablefootmark{b}  & 0.228~pc& Sect.~\ref{subsec:OB}\\
    physical distance, $d_{phys}$, between $\theta ^1$ Ori C and the IF & $\sim0.27$~pc & Sect.~\ref{subsec:O+N}\\
    transverse size $l_\mathrm{PDR}^\mathrm{los}$ & $\sim0.10$~pc & Sect.~\ref{subsec:O+N}\\
    line-of-sight distance, $d_{los}$, between $\theta ^1$ Ori C and the IF &  & Sect.~\ref{subsec:O+N}\\
        \hspace{2cm} Average & $\sim0.15$~pc & \\
         \hspace{2cm} Minimum & $\sim0.05$~pc & \\
        \hspace{2cm} Maximum & $\sim0.20$~pc & \\
      projected distance between He-IF and H-IF \tablefootmark{c}& \mbox{0.5 $\times$ 10$^{-2}$\,pc}; \mbox{2.5\arcsec} & Sect.~\ref{subsec:He}\\ 
    projected distance between IF and DF d$_{IF-DF}$\,\tablefootmark{c, d}  & \mbox{(2.2, 2.9, 4.0) $\times$ 10$^{-2}$\,pc} & Sect.~\ref{subsec:OB}\\
    & \mbox{11.0\arcsec, 14.5\arcsec, 19.8\arcsec} & \\
    $G_0$ at IF & $\sim(2.2-7.1) \times 10^4$ & Sect.~\ref{subsec:O+N}\\
    inclination angle $i$ & 4 ($1-8$) degrees & Sect.~\ref{subsec:O+N}, \citet{Salgado16}\\
     width IF at peak G$_0$ & \mbox{$(0.22-0.34) \, \times$ 10$^{-2}$\,pc} & Sect.~\ref{subsec:O+N}\\
     & $1.1\arcsec-1.7\arcsec$ & \\
    condition for face-on PDR to dominate G$_0$ & $d_{{\rm proj}}\ge$~0.24~pc & Sect.~\ref{subsec:O+N} \\
    FUV dust cross-section $\sigma _H$ & $6.5 \times 10^{-22}$  cm$^{2}$ / H & \citet{Cardelli89, Blagrave:07}\\
    &  & \citet{Schirmer2022}\\
    $R_V =A_V /E(B-V)$ & 5.5 & \citet{Cardelli89,Blagrave:07}\\
    $A_V /N_H$ & $3.5 \times 10^{-22}$~mag/cm$^{-2}$ & \citet{Cardelli89,Blagrave:07} \\
    total IR emission $L_{IR}$ & $9.5 \times 10^{3} \,{\rm L}_{\odot}$ & \citet{Salgado16}\\
    foreground extinction $A_V$ & $0.9-1.9$ mag & Sect.~\ref{subsec:HI}\\
    emission measure EM (\HI\ recombination lines) & $(0.84-4.15) \times 10^6\,{\rm pc\, cm^{-6}}$ &  Sect.~\ref{subsec:HI}\\
    rms electron density n$_{e, \mathrm{rms}}$ near IF (Br$\gamma$) & $\sim 9000\, {\rm cm^{-3}}$ &  Sect.~\ref{subsec:HI} \\
    density at the IF n$_e$\,\tablefootmark{d} & $2-3 \times 10^{3}$  cm$^{-3}$ & \citet{Weilbacher15}\\
    temperature at the IF T$_e$\,\tablefootmark{d} & $\sim 9 \times 10^3$ K & \citet{Weilbacher15}\\
    density in atomic PDR n$_H$ (AIB emission)\,\tablefootmark{d}& $(5-10) \times 10^{4}$  cm$^{-3}$ & \citet[][Sect.~5.1]{Habart:im} \\
    density from NIR H$_2$ n$_H$ & $(3.5) \times 10^{4}$ to 10$^{5}$ cm$^{-3}$& Sect.~\ref{subsec:h2} \\
    temperature at the DF $T$ & $\sim$ 400-700~K &  \citet{vandeputte23}\\ 
    & & \citet{Allers05}\\
    emission measure EM ([\CI]~0.984~\mum) & $(1.1 - 2.8) \times 10^5\,{\rm pc\, cm^{-6}}$ &  Sect.~\ref{subsec:Clines} \\
\hline \\[-25pt]    
\end{tabular}
   \end{center}
    \tablefoot{
    \tablefoottext{a}{see \citet{Habart:im} for a discussion on the adopted distance.}
   \tablefoottext{b}{IF is defined by the peak emission of the [\OI] 6300~${\mathring{\mathrm{A}}}$ and [\FeII] 1.644~\mum emission.}
   \tablefoottext{c}{Projected distance along the NIRSpec cut. As the cut is not perpendicular to the IF, a correction factor of cos(19.58\textdegree)=0.942 needs to be applied to obtain a perpendicular distance from the IF.}
   \tablefoottext{d}{Given the complexity of the PDR surface, parameters given are specific for the NIRSpec FOV. For average values across the entire Bar, see \citet{Habart:im}.}
   } 
\end{table*}

\subsection{OI and NI fluorescent emission}
\label{subsec:O+N}

The fluorescent lines of \OI\, at 1.129 and 1.317~\mum arise from the partially neutral gas in the ionisation front from UV-pumping by 1027 and 1040~${\mathring{\mathrm{A}}}$\, photons, respectively, followed by radiative decay. Hence, similar intensities are expected for both IR transitions. Yet, the \OI\, 1.317~\mum intensity is, on average, about 3 times the \OI\, 1.129~\mum intensity in the IF and the filamentary structure in the atomic PDRs (i.e., where the \OI\, 1.317~\mum intensity $> 15\sigma$). This is in contrast with previous results. Indeed, \citet{Marconi98} reported that the 1.129~\mum line is equal in strength to the 1.317~\mum line at two positions -- from which they concluded that pumping of the upper level of the 1.129~\mum transition by resonantly scattered Lyman~$\beta$ photons (1026~\AA) is negligible -- and stronger than the 1.317~\mum line at one position, which we do not see anywhere in the NIRSpec mosaic. Likewise, \citet{Walmsley00} observed similar line intensities at two positions, while a third position exhibited a stronger 1.317~\mum intensity, albeit with a lower factor than  observed in the NIRSpec data (1317/1129 $\sim$1.4). As these observations probe different positions of a highly structured PDR, seeing variations across the Bar is perhaps not that surprising. However, the origin of the enhanced \OI\, 1317/1129 ratio  observed here and in one position of \citet{Walmsley00} is unclear and requires further investigation.

The observed \OI\, 1.317~\mum intensity decreases with distance from the ionisation front (Figs.~\ref{fig:maps} and~\ref{fig:cut}). The location of the cut, crossing both the primary and secondary ridge, results in a very sharp, double-peaked profile just beyond the ionisation front. In addition, its radial profile shows multiple peaks between \mbox{0.34 and 1.07$\times$10$^{-2}$\,pc} from the IF (between 1.7\arcsec\ and 5.3\arcsec), followed by a slow drop at larger distances. We note that the observed \OI\, 1.317~\mum intensity follows the [\OI] 6300~\AA\, intensity to some degree (Fig.~\ref{fig:cut}). The ionisation front has a typical \HI\, column density of \mbox{$6\, \times\, 10^{18}\, {\rm cm^{-2}}$} \citep[][eq. 7.25]{Tielens_book05}, corresponding to an optical depth in the \OI\, UV pumping lines of $\sim$30 \citep{Marconi98}. Hence, the UV transitions that pump the \OI\, near-IR lines are on the logarithmic part of the curve of growth in the PDR itself. We attribute the multiple peaks to the presence of undulations in the Bar surface, which is tilted along the line of sight by about 4\textdegree\, \citep[Sect.~\ref{subsubsec:geometry};][]{Salgado16}. The subtle differences in the intensity distribution of the \OI\, 1.317~\mum line and the [\OI] 6300~\AA\, line may reflect local acceleration zones of the gas in the ionisation front that Doppler shift unattenuated stellar photons into the UV pumping wavelengths of \OI\, 1.317~\mum ([\OI] 6300~\AA\, emission is due to collisional excitation). We note that for an intrinsic line width of $3\, {\rm km/s}$, an acceleration by $1\, {\rm km/s}$ will increase the UV pumping photon flux by 30\%. We attribute the gradual decrease in both the \OI\, 1.317~\mum and the [\OI] 6300~\AA\, intensity beyond \mbox{1.07 $\times$ 10$^{-2}$\,pc} from the IF (5.3\arcsec; Fig.~\ref{fig:cut}) to the change over from an edge-on to a face-on geometry of the ionisation front, coupled with the geometric dilution of the incident UV field intensity. This transition from edge-on to face-on geometry occurs over a very small distance ($\sim$1\arcsec, \mbox{$6\,\times\, 10^{15}\,{\rm cm}$}). 
  Lastly, we ascribe part of the prominent triangular region of enhanced \OI\, 1.317~\mum at \mbox{$0.6-1.2$ $\times$ 10$^{-2}$\,pc} ($3-6$\arcsec) behind the ionisation front (Fig.~\ref{fig:fluorescence}) to a local gas acceleration zone. As there is no counterpart in the [\OI] 6300~\AA\, line for the E-W filament, this acceleration must occur inside the PDR itself.

\subsubsection{Strength of the FUV radiation at the PDR surface}
\label{subsubsec:G0}

The extinction corrected line intensity of fluorescent lines provide an estimate of the strength of the FUV radiation field, G$_0$. 
Applying the method outlined in App.~\ref{app:UVint} to the strongest fluorescent line (\OI\, 1.317~\mum line; Figs.~\ref{fig:maps_parameters} and~\ref{fig:cut_parameters}), we derive the strength of the FUV radiation field. 
The maximum strength of the FUV radiation field, G$_0$, ranges between $2.2-7.1\times 10^4$ 
across the IF seen in \mbox{\OI\ 1.317~\mum} emission (with a median value of $5.9\times 10^4$). This value is consistent with \citet{tielens:85b}, who derived G$_0$ from an analysis of the [\OI] and [\CII] atomic fine-structure lines observed toward the Bar PDR. \citet{Marconi98} derived a slightly smaller value for G$_0$ of 2.6 10$^4$ from their observations of the near-IR \OI\, fluorescent lines, reflecting a slightly lower 1.317~\mum intensity ($\sim1.4\times 10^{-4}\,$ versus a median value of $\sim7.7\times 10^{-4}\,$\uint\, across the IF in our mosaic) measured at a slightly different position. In addition, these authors also adopted a slightly higher extinction value ($A_V=2$ versus a median value of $A_V\sim1.64$ here, see Sect.~\ref{subsec:HI}) and a larger inclination angle $i$, that is the angle between the surface and the line-of-sight, ($\sin(i) =0.20$ versus 0.07, see Appendix~\ref{app:UVint}). 

\subsubsection{Bar geometry}
\label{subsubsec:geometry}
We can use the obtained values for $G_0$ to derive the geometry of the Bar following \citet[][Sect. 4.1]{Salgado16}. As the total IR emission is a measure for the amount of stellar radiation absorbed by the Bar, these authors equate the observed total IR emission to the product of the strength of the FUV radiation field, $G_0$, and the area at the surface of the PDR absorbing stellar radiation, $A_{abs}$. Assuming a rectangular absorbing area and using the observed size of the Bar (0.32~pc), the obtained absorbing area then provides an estimate of the transverse size of $l^{los}_{PDR}$. While \citet{Salgado16} adopts the $G_0$ reported by \citet{Marconi98}, we instead adopt the maximum value obtained for $G_0$ across the NIRSpec IF, as it provides the strongest constraints for the geometry. This results in a smaller absorbing area at the surface of the PDR, $A_{abs}$, of $0.033\,{\rm pc^2}$. This in turn results in a smaller transverse size of $l^{los}_{PDR}=0.10\,{\rm pc}$, which is consistent with model predictions of the Bar geometry \citep[e.g.][]{Tielens93, Pellegrini09}.

Based on the derived $l^{los}_{PDR}$, we can then derive an estimate of the inclination of the IF. The width of the IF, $l$, can be estimated from eq. 7.25 from \citet{Tielens_book05}. For a density of \mbox{$n=3 \, \times \, 10^{3}\,{\rm cm^{-3}}$}, we obtain a width of \mbox{$l=7\, \times \,10^{-4}\,{\rm pc}$}. Combined with the derived $l^{los}_{PDR}$ and the observed width in the \mbox{[\OI] 6300~\AA} emission from \citet{Weilbacher15} at the location where we observe the maximum $G_0$ value ($0.22-0.34\times10^{-2}$~pc; $1.1-1.7$\arcsec), we then obtain an inclination $i$ of $1\pm\,0.3^\circ$, indicating that the IF is almost completely seen edge-on. We note that the width of the IF varies considerably, even within the NIRSpec mosaic, thus further emphasizing the complexity of the PDR surface. 
Alternatively, we can get a rough measurement of the inclination $i$ from the decrease in the extinction corrected near-IR \OI\, emission. As the intensity of the near-IR \OI\, lines scales with $1/\sin(i)$ with respect to the intensity for a PDR seen face-on \citep{Marconi98} and ignoring geometrical dilution effects, the observed intensity contrast, a factor of $\sim7$, implies that the Bar is viewed under an angle of 8\textdegree. 
Thus, we have derived an inclination angle $i$ of 1\textdegree\, and 8\textdegree, consistent with previous estimates from \citet{jansen1995millimeter} and \citet[][; $<$3\textdegree]{Hoger95}, \citet[][, 7\textdegree]{Pellegrini09}, and \citep[][4\textdegree]{Salgado16} but slightly lower than the estimate from \citet[][10-15\textdegree]{Marconi98}. This indicates that the adopted inclination angle $i$ of 4\textdegree\, for the $G_0$ calculation is appropriate. An inclination of 1\textdegree\, will increase the $G_0$ values presented in this paper by a factor of 4 while an inclination of 8\textdegree\, will decrease the $G_0$ values presented in this paper by a factor of two\footnote{An inclination angle of 4\textdegree, as adopted in this paper, results in $sin(i)=0.07$, while an inclination angle of 1\textdegree\, results in $sin(i)=0.017$ and an inclination angle of 8\textdegree\, results in $sin(i)=0.14$)}.

The obtained absorbing area also provides an estimate of the physical distance between \oric and the IF, $d_{phys}$, for a given stellar luminosity from $A_{abs} = L_{IR} / (L_\star / (4\, \pi\, d_{phys}^2))$. We adopt the higher stellar FUV luminosity of $2.7\times 10^5\,{\rm L}_\odot$ employed by \citet{Salgado16}, as the lower stellar FUV luminosity does not provide physical results\footnote{The physical distance is smaller than the projected distance when adopting the lower stellar FUV luminosity.}. We obtain a physical distance between the \oric and the IF, $d_{phys}$, of $0.27\, {\rm pc}$. Given the physical size of the IF, we assume this distance reflects the physical distance from the mid-point of the IF to \oric (Fig.~\ref{fig:geometry}). Given a projected distance between \oric and the IF, $d_{proj}$, of $0.228\,{\rm pc}$, this constraints the distance between \oric and the mid-point of the IF projected along the line-of-sight, $d_{los}$, to $0.15\,{\rm pc}$ (Fig.~\ref{fig:geometry}). Consequently, \oric is located at a distance of $0.10\,{\rm pc}$ in front of the start of the IF along the line-of-sight and the depth of the Bar (end-point) from \oric along the line-of-sight is $0.20\,{\rm pc}$ (Fig.~\ref{fig:geometry}).

\subsection{Rotationally and vibrationally excited \texorpdfstring{\molh}{H2} emission}
\label{subsec:h2}

\begin{figure*}
    \centering
    \begin{tabular}{cc}

    \includegraphics[scale=0.40]{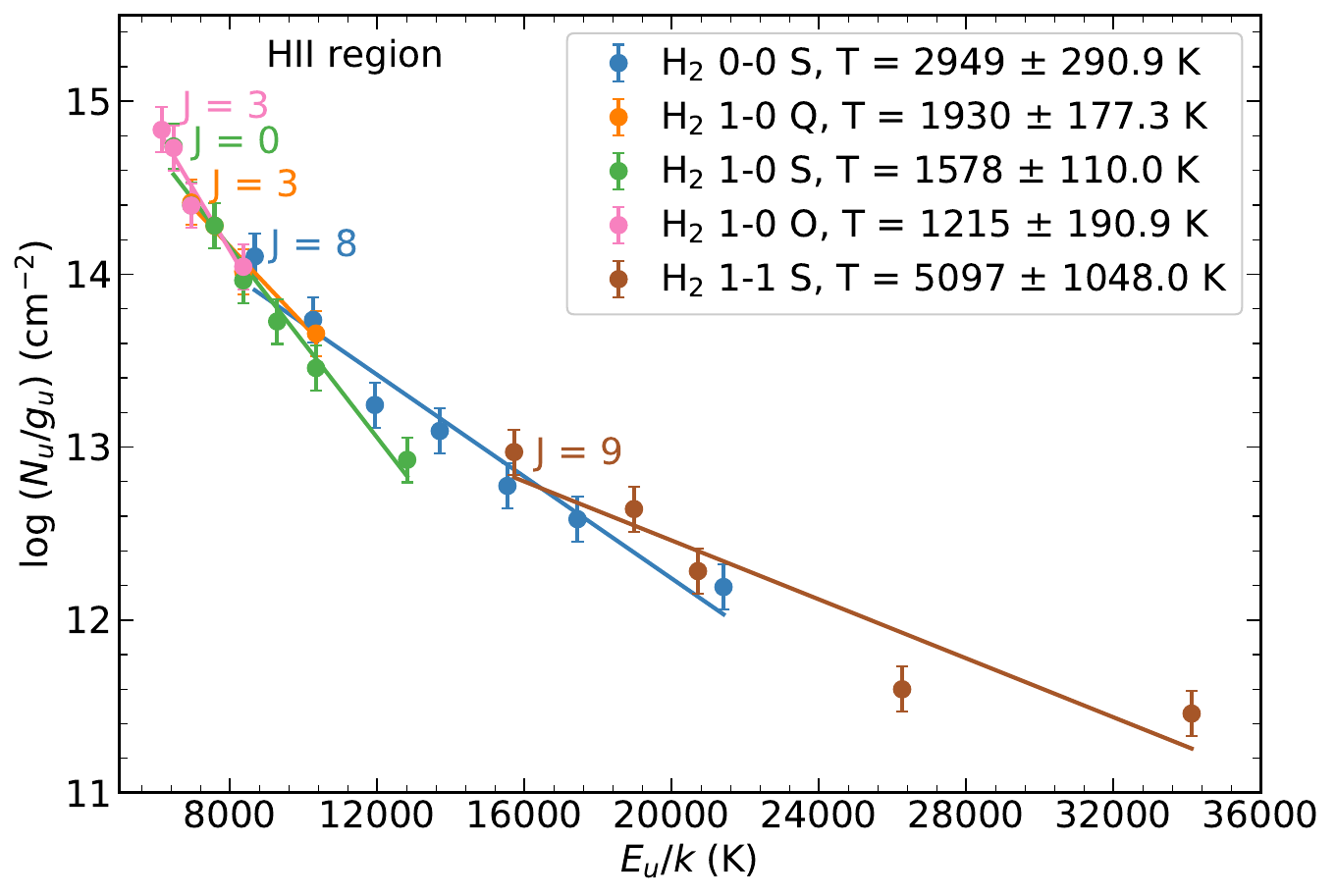} &  \includegraphics[scale=0.40]{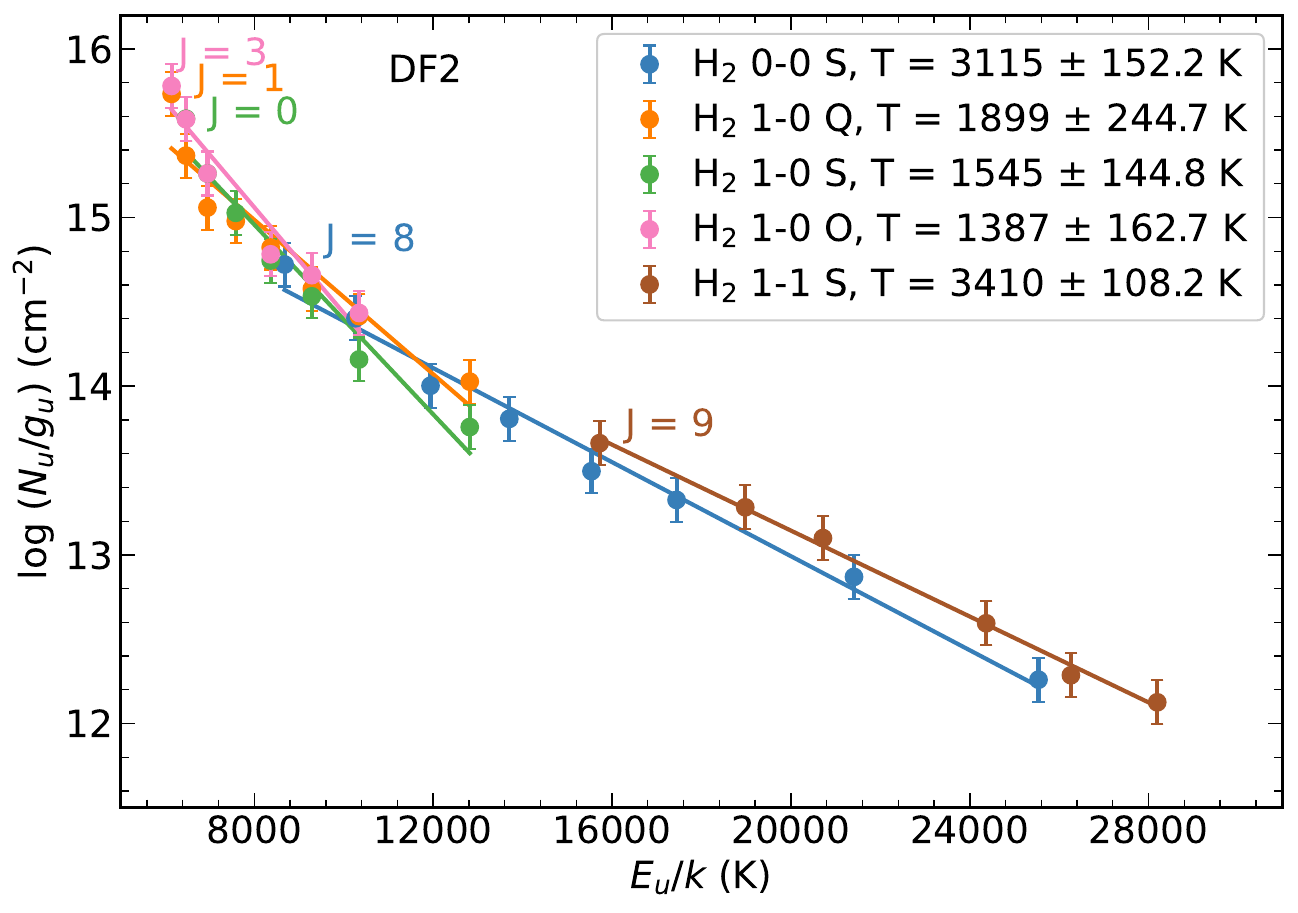}\\
    \includegraphics[scale=0.40]{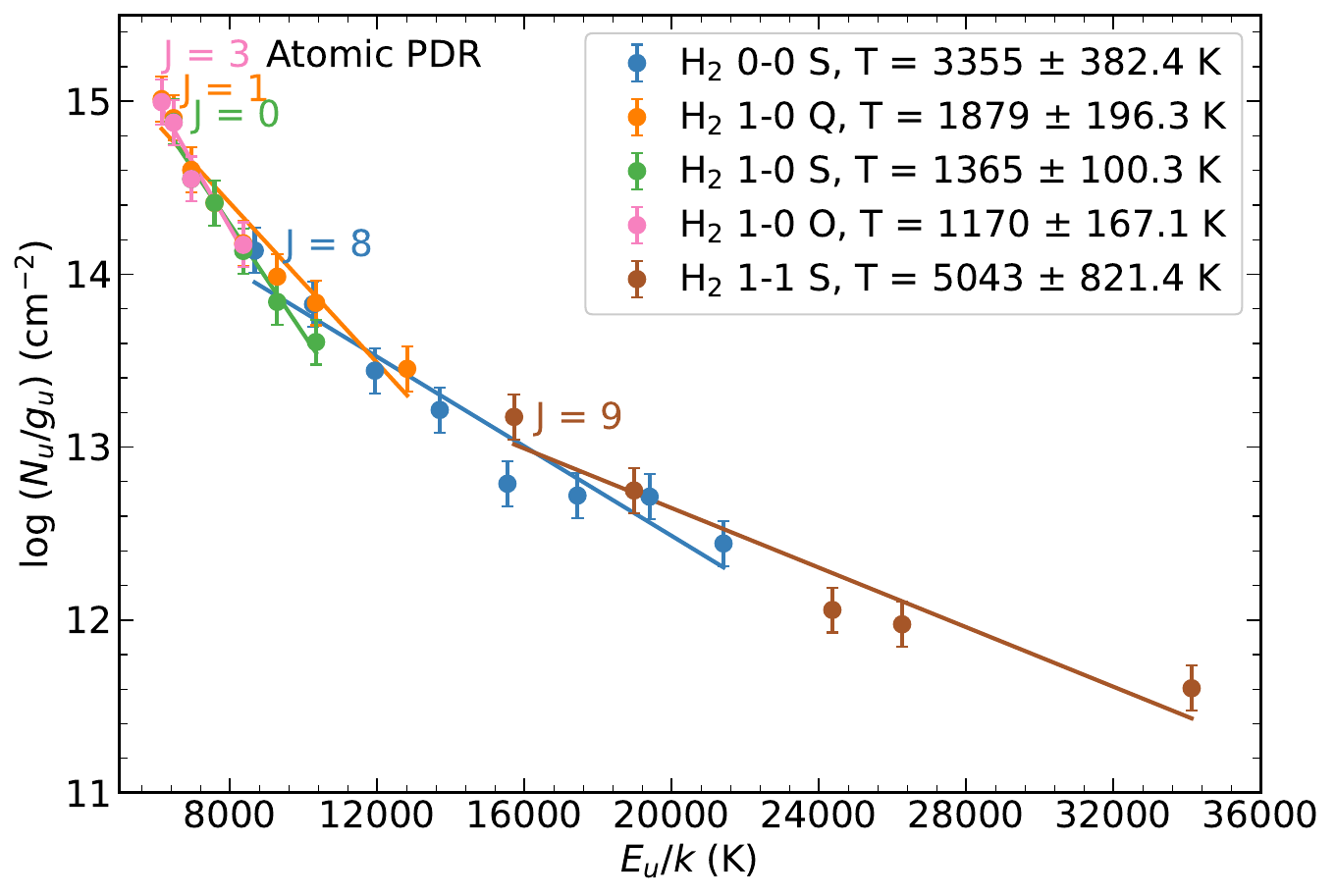} &
    \includegraphics[scale=0.40]{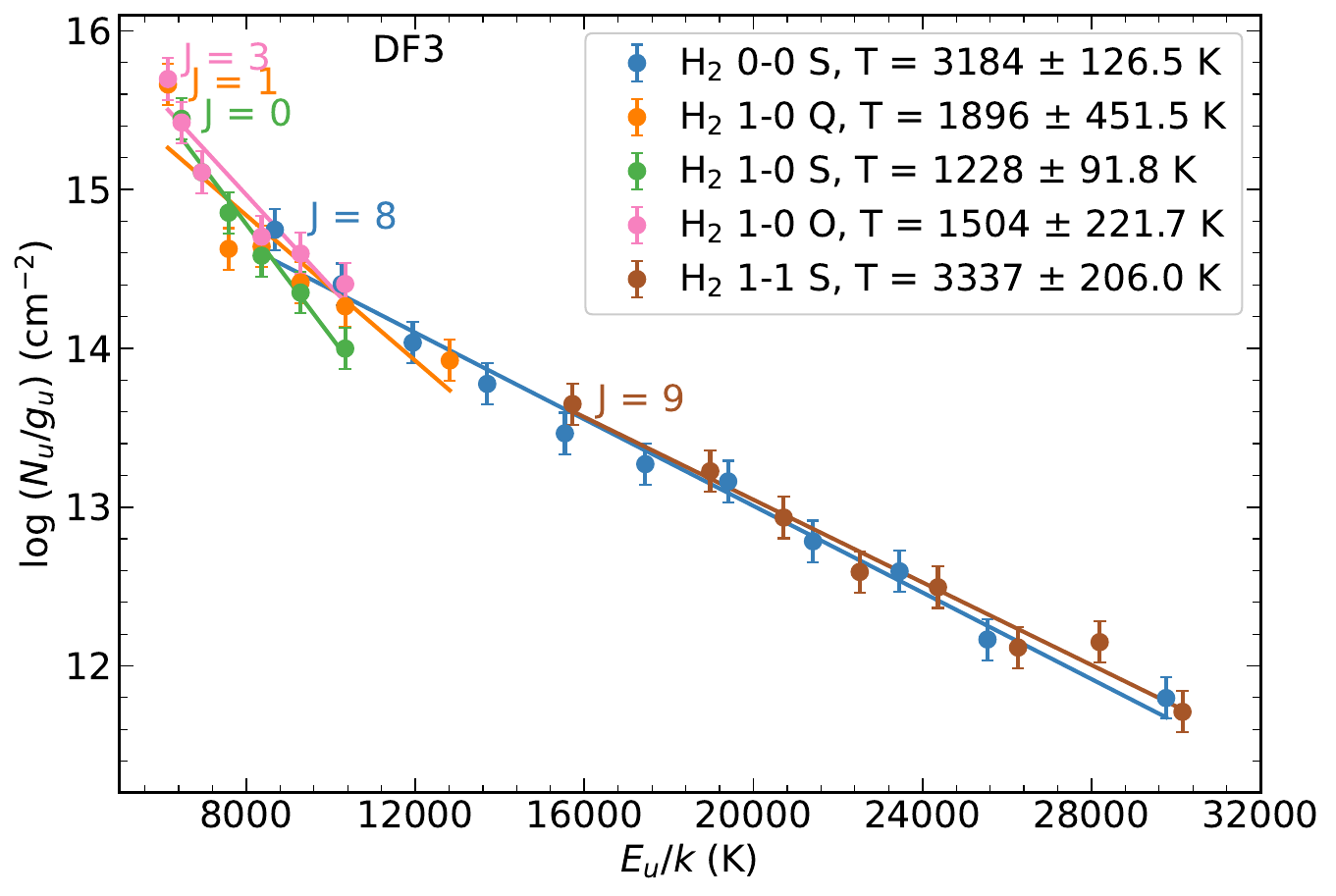} \\
    \includegraphics[scale=0.40]{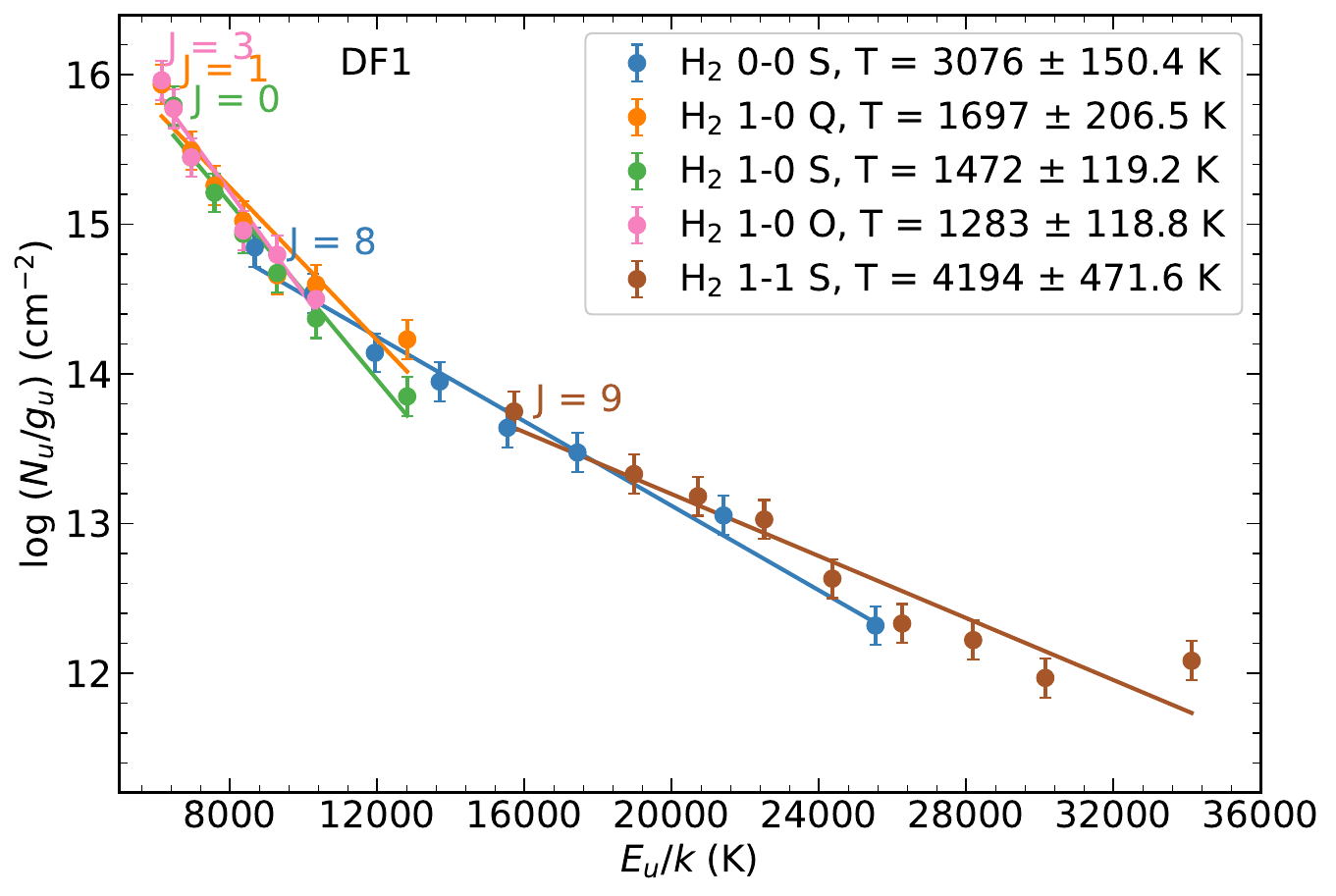}& \\
    
    \end{tabular}
    \caption{Excitation diagrams of \molh lines observed towards the \Ta, \Tb, \Tc, \Td, and \Te. With the exception of \Ta, the \molh lines in the excitation diagrams are first corrected for foreground visual extinction and subsequently for extinction within the PDR using the intermingled formalism. In \Ta the lines are only corrected for the foreground visual extinction. The excitation diagrams for the 0-0 S, 1-0 Q, 1-0 S, 1-0 0, and 1-1 S levels of \molh are fitted using a single temperature component of the gas, and the individual fits are represented by solid lines. The resulting temperature and column densities obtained from the fits are listed in Table~\ref{tab:h2_excitation_fits}. }
    \label{fig:ext_diag}
    \end{figure*}

\begin{table*}
    \centering
    \caption{The column densities and excitation temperatures obtained from fitting the excitation diagrams of the \molh lines observed towards \Ta, \Tb, \Tc, \Td, and \Te (Fig.~\ref{fig:ext_diag}). Except \Ta, all the \molh lines in the excitation diagrams are first corrected for foreground visual extinction and subsequently for extinction within the PDR using the foreground and intermingled formalisms. In \Ta the lines are only corrected for the foreground visual extinction.}
    \label{tab:ext_param}
    
    \begin{tabular}{c c c c c}
    \hline
    \hline\\[-5pt]
    & \multicolumn{2}{c}{Foreground formalism} & \multicolumn{2}{c}{Intermingled formalism}\\[5pt]
    \multicolumn{1}{c}{H$_{2}$ series} & N$_{\rm {H}_{2}}$ ($\times$10$^{18}$ cm$^{-2}$) & T (K) & N$_{\rm {H}_{2}}$ ($\times$10$^{18}$ cm$^{-2}$) & T (K) \\[5pt]
    \hline \\[-5pt]
    & \multicolumn{4}{c}{\textbf{\Ta}} \\
    0-0 S & 0.13 $\pm$ 0.06 & 2959 $\pm$ 293 & -- & -- \\
    1-0 Q & 0.50 $\pm$ 0.01  & 1926 $\pm$ 177 & -- & -- \\
    1-0 O & 3.38 $\pm$ 3.10   & 1200 $\pm$ 188 & -- & --\\
    1-0 S & 0.98 $\pm$ 0.39  & 1594 $\pm$ 112 & -- & -- \\
    1-1 S & 0.03 $\pm$ 0.02 & 5117 $\pm$ 1054 & -- & -- \\
    \hline \\[-5pt]
     & \multicolumn{4}{c}{\textbf{\Tb}} \\
    0-0 S & 0.12 $\pm$ 0.06 & 3381 $\pm$ 388 & 0.12 $\pm$ 0.06 & 3355 $\pm$ 382  \\
    1-0 Q & 1.06 $\pm$ 0.52 & 1869 $\pm$ 194 & 0.88 $\pm$ 0.43 & 1879 $\pm$ 196  \\
    1-0 O & 6.38 $\pm$ 5.61 & 1140 $\pm$ 162 & 5.05 $\pm$ 4.33 & 1170 $\pm$ 167 \\
    1-0 S & 2.55 $\pm$ 1.10 & 1414 $\pm$ 104 & 2.34 $\pm$ 1.04 & 1366 $\pm$ 100 \\
    1-1 S & 0.04 $\pm$ 0.03 & 5083 $\pm$ 829 & 0.04 $\pm$ 0.03 & 5043 $\pm$ 821 \\
    \hline\\[-5pt]
     & \multicolumn{4}{c}{\textbf{\Tc}} \\
    0-0 S & 0.44 $\pm$ 0.12 & 2974 $\pm$ 145 & 0.78 $\pm$ 0.20 & 3076 $\pm$ 150 \\
    1-0 Q & 3.17 $\pm$ 1.99 & 1706 $\pm$ 209 & 8.53 $\pm$ 5.36 & 1697 $\pm$ 207 \\
    1-0 O & 8.09 $\pm$ 4.48 & 1397 $\pm$ 134 & 11.8 $\pm$ 5.87 & 1283 $\pm$ 119 \\
    1-0 S & 4.24 $\pm$ 2.15 & 1481 $\pm$ 123 & 26.8 $\pm$ 15.6 & 1472 $\pm$ 119 \\
    1-1 S & 0.15 $\pm$ 0.10 & 4050 $\pm$ 450 & 0.25 $\pm$ 0.17 & 4194 $\pm$ 472 \\
    \hline\\[-5pt]
     & \multicolumn{4}{c}{\textbf{\Td}} \\
    0-0 S & 0.69 $\pm$ 0.18 & 3209 $\pm$ 158 & 0.54 $\pm$ 0.14 & 3116 $\pm$ 152 \\
    1-0 Q & 7.25 $\pm$ 4.30 & 1860 $\pm$ 235 & 3.20 $\pm$ 1.90 & 1899 $\pm$ 245 \\
    1-0 O & 31.2 $\pm$ 23.4 & 1266 $\pm$ 149 & 12.8 $\pm$ 8.73 & 1387 $\pm$ 163 \\
    1-0 S & 11.7 $\pm$ 5.74 & 1719 $\pm$ 160 & 6.28 $\pm$ 3.45 & 1545 $\pm$ 145 \\
    1-1 S & 0.61 $\pm$ 0.13 & 3513 $\pm$ 117 & 0.50 $\pm$ 0.11 & 3410 $\pm$ 108 \\
    \hline\\[-5pt]
     & \multicolumn{4}{c}{\textbf{\Te}} \\
    0-0 S & 0.53 $\pm$ 0.13 & 3202 $\pm$ 127 & 0.51 $\pm$ 0.12 & 3184 $\pm$ 127 \\
    1-0 Q & 2.71 $\pm$ 3.17 & 1888 $\pm$ 448 & 2.29 $\pm$ 2.68 & 1896 $\pm$ 452 \\
    1-0 O & 8.60 $\pm$ 6.92 & 1473 $\pm$ 216 & 7.18 $\pm$ 5.68 & 1504 $\pm$ 222 \\
    1-0 S & 13.9 $\pm$ 6.84 & 1253 $\pm$ 94  & 12.5 $\pm$ 6.31 & 1228 $\pm$ 92 \\
    1-1 S & 0.46 $\pm$ 0.20 & 3356 $\pm$ 208 & 0.44 $\pm$ 0.19 & 3337 $\pm$ 206 \\
    \hline
    \end{tabular}
    \label{tab:h2_excitation_fits}
\end{table*}
    
\molh emission is observed throughout the mosaic (see Fig.~\ref{fig:maps}). We detect a large number of ro-vibrational lines in our data. For a detailed inventory of the observed H$_{2}$ lines and the spatial distribution of their emission, we refer the reader to Sect.~\ref{sec:spectral-characteristics} and \ref{subsec:OB}, respectively. We report the measured fluxes for selected \molh lines detected in the five templates in Table~\ref{tab:fluxes}. We note that the \molh line intensity is strongest in \Te, followed by \Td, \Tc, \Tb, and \Ta. This variation in intensity is also observed in the surface brightness line cut across the NIRSpec mosaic presented in Fig.~\ref{fig:cut}. Moreover, there is an increase in \molh emission at the position of the two proplyds. The in-depth analysis of the \molh emission in proplyd 203-506 is discussed in \citet{Berne:proplyd}. Here, we focus on the \molh emission observed in the five templates. We highlight that the \molh emission observed in the \Ta originates from the background face-on PDR in OMC-1 (see Fig.~\ref{fig:geometry}). The emission from this background face-on PDR was previously observed with Herschel in other PDR tracers, particularly in high-J CO, CH$^+$ lines \citep{Parikka:2018} and [\OI] 63 and 145~\mum and [\CII] 158~\mum \citep{Bernard-Salas12}, as well as, with Spitzer in AIB emission \citep{Knight21:Orion}. In contrast, the \molh emission observed in the \Tb and the three dissociation fronts originates from the edge-on PDR in the Bar itself. 

The H$_{2}$ lines are powerful tools to probe the physical conditions of the emitting region. We use the observed lines to probe the extinction within the PDR and density throughout the mosaic. While we estimate the foreground visual extinction, A$_V$, using the \HI\, recombination lines (Sect.~\ref{subsec:HI}), we use the H$_{2}$ lines to measure the extinction in the neutral \HI\, region of the bar using the foreground and intermingled formalisms, referred to as A$_{V_{bar}}$ and A$_{V_{bar},\, I}$ respectively. The foreground formalism assumes that the dust is in front of the region emitting \molh, whereas the intermingled formalism assumes that the dust is mixed with the gas emitting \molh. Comparison of the ratio of lines that arise from the same upper v and J state with the corresponding intrinsic flux ratio gives a measure of amount of extinction within the PDR (A$_{V_{bar}}$ and A$_{V_{bar},\, I}$). We use the ratio of the \molh lines $1-0$ S(1) and $1-0$ O(5), which originates from the same upper level, corrected for foreground extinction, to estimate the internal PDR extinction. Table~\ref{tab:physical_conditions} shows the A$_{V_{bar}}$ and A$_{V_{bar},\, I}$ values obtained for the five templates using both the foreground and intermingled formalisms. Moreover, in Fig.~\ref{fig:cut_parameters}, we present the line cut across the mosaic of $A_{V_{bar}}$. We note that the internal PDR extinction is highest in the \Tc, followed by \Td, the \Tb, and \Te. In the \Ta, the internal PDR extinction is $\sim$ 0. 
The high value of the internal PDR extinction in \Tc implies that \Tc is further along the line of sight than \Td and \Te. Therefore, the column density along the line of sight increases for \Tc, which is more distant from the observer (but closer in projected distance from the ionisation front). The measured flux for the \molh lines corrected for the foreground and internal PDR extinction in the five templates is presented in Table~\ref{tab:fluxes}.

The H$_{2}$ line ratio of $1-0$ S(1)/$2-1$ S(1) is sensitive to the density. We present a line cut across the mosaic of this ratio in Fig.~\ref{fig:cut_parameters}. This ratio exhibits values ranging from $\sim$ $3-5$ across the mosaic and begins to increase beyond \Te. Furthermore, this ratio has large values in the vicinity of proplyds. To get a quantitative measure of the density, we fit the \molh lines corrected for the foreground and internal PDR extinction as well as the H$_{2}$ line ratio of $1-0$ S(1)/$2-1$ S(1) observed in the five templates employing the Meudon PDR Code \footnote{We note that the Meudon PDR code is part of the interstellar medium database, ISMDB, a web-based fitting tool to fit observations to PDR models. The ISMDB is one of the Science Enabling Products of the PDRS4All ERS program \citep{pdrs4all}}. The Meudon PDR code \citep{lepetit} can fit the observed line intensities to grids of model PDRs. In this paper, we employ the isochoric model to estimate the gas density. For the isochoric model, which assumes a constant gas density, we first fit all the H$_{2}$ line intensities corrected using both the foreground and intermingled formalisms, and then the ratio of $1-0$ S(1)/$2-1$ S(1) obtained with both formalisms, keeping the radiation field and gas density as free parameters. We fix the cloud size expressed as visual extinction A$_{V}$ to 10 for all the fits. Furthermore, since the H$_{2}$ emission in the atomic region and the dissociation fronts belongs to the edge-on PDR, which has an inclination ranging from $1-8^{\circ}$, we correct the observed fluxes associated with these regions for this geometrical effect. Adopting an inclination angle of $4^{\circ}$, we divide the observed fluxes by a factor of 14 (a geometrical factor in line intensities of $1/\sin(\theta)$ where $\theta$ is the inclination angle of the PDR. For $\theta=4^{\circ}$, the factor is equal to 14). We highlight that this is a pure geometrical factor that underestimates the real flux. We find that the extinction correction formalism (i.e. foreground or intermingled) does not influence the fit results. The best-fit model, considering all lines, results in a gas density of 3.5$\times 10^{4}$ cm$^{-3}$ in the \Ta, \mbox{$10^{3}-3.5 \times 10^{3}$ cm$^{-3}$} in the \Tb, $10^{4}$ cm$^{-3}$ in \Tc, \mbox{$10^{4}-3.5 \times 10^{4}$ cm$^{-3}$} in \Td, and \mbox{$3.5 \times 10^{3}- 10^{4}$ cm$^{-3}$} in \Te. When considering the line ratio, the best-fit model results in a gas density of \mbox{$3.5 \times 10^{4}- 10^{5}$ cm$^{-3}$} in the \Ta, \mbox{$10^{3}-3.5 \times 10^{3}$ cm$^{-3}$} in the \Tb, $3.5 \times 10^{4}$ cm$^{-3}$ in \Tc and \Td, and  \mbox{$10^{4}-3.5 \times 10^{4}$ cm$^{-3}$} in \Te. Finally, it is worth pointing out that adopting an inclination angle of $8^{\circ}$ leads to gas densities from the model fits that are consistently higher by a factor of 2.85 compared to an inclination angle of $4^{\circ}$. 

We furthermore analyse the H$_{2}$ excitation diagrams resulting. Fig.~\ref{fig:ext_diag} presents the excitation diagrams for each template spectrum, where we plot the upper state energy of the transition ($E_{u}$/k) versus the normalised column density ($N_{u}$/$g_{u}$), where $N_{u}$ is the upper state column density and $g_{u}$ is the statistical weight of the upper state energy level. We create these excitation diagrams using the H$_{2}$ fitting tool in the Photodissociation Region Toolbox \citep[PDRT;][]{2023AJ....165...25P}\footnote{The H$_{2}$ fitting tool is a Science Enabling Product of the PDRs4All program}. The tool allows for fitting a one- or two-temperature model and the ortho-to-para ratio (OPR). 
Here, we analyse the excitation diagrams of $v=0$ and $v=1$ vibrational levels for which lines are strong. To get an estimate of the excitation temperature and column density in the five templates, we fit the excitation diagram of the $0-0$ S, $1-0$ Q, $1-0$ S, $1-0$ O, and $1-1$ S series independently. We find that using a single temperature and the LTE OPR value of 3 gives the best fit results.  
We further note that for lines in the $v$ = 0 and 1 vibrational series analysed here, the dominant excitation mechanism is radiation -- IR radiative cascade of FUV-pumped \molh -- rather than collisions. Therefore, the temperatures resulting from these diagrams represent the excitation temperatures rather than the gas temperature. The excitation temperatures and the column densities obtained from fitting the excitation diagrams are presented in Table~\ref{tab:h2_excitation_fits}.
Lastly, we note that the excitation temperatures obtained from the excitation diagrams in \Td and \Te are similar to each other but distinct from that in \Tc implying that the physical properties of \Td and \Te differ from those of \Tc. A forthcoming paper will analyse the \molh excitation diagrams based on both the NIRSpec and MIRI IFU PDRs4All observations and thus will include both the collisionally excited levels and FUV pumped levels (Sidhu et al., in prep.). 

\subsection{CI emission lines}
\label{subsec:Clines}

Given the difference between the [\CI] 0.985~\mum line and the \mbox{[\FeII] 1.644~\mum} line delineating the IF and the resemblance between the [\CI] 0.985~\mum line and the \OI\ 1.317~\mum line just beyond the IF, we confirm that the \CI\, originates in the neutral gas beyond the IF. While there is a good resemblance between the [\CI] 0.985~\mum line and the \molh emission, enhanced [\CI] emission is observed just beyond the IF, similar as for the \OI\ 1.317~\mum emission. We find that the ratio of [\CI] 0.985/\molh 1-0 S(1) (not corrected for foreground extinction) varies between 0.21 and 2.05 and is $<$0.9 in the molecular PDR (Fig.~\ref{fig:cut_parameters}). This is a much smaller range than observed by \citet{Walmsley00}, who reported ratios between 0.2 and 6, and is likely due to their larger FOV. We also detect \CI\, recombination lines at 1.069 and 1.175~\mum (Fig.~\ref{fig:app:template1}), which, together with the [\CI] 0.983 and 0.985~\mum lines, provides an estimate for the electron temperature, $T_e$, and the gas density, $n_H$ (see Appendix~\ref{app:CI} for details).

\begin{figure*}
\begin{center}
\resizebox{\hsize}{!}{
\includegraphics{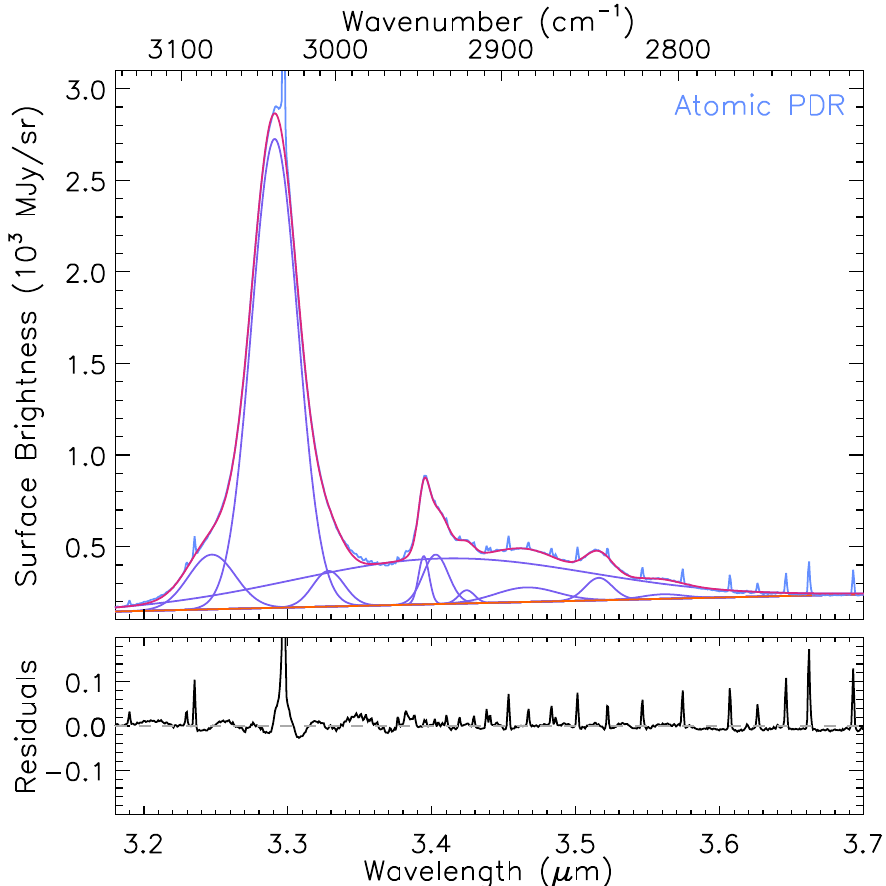}
\includegraphics{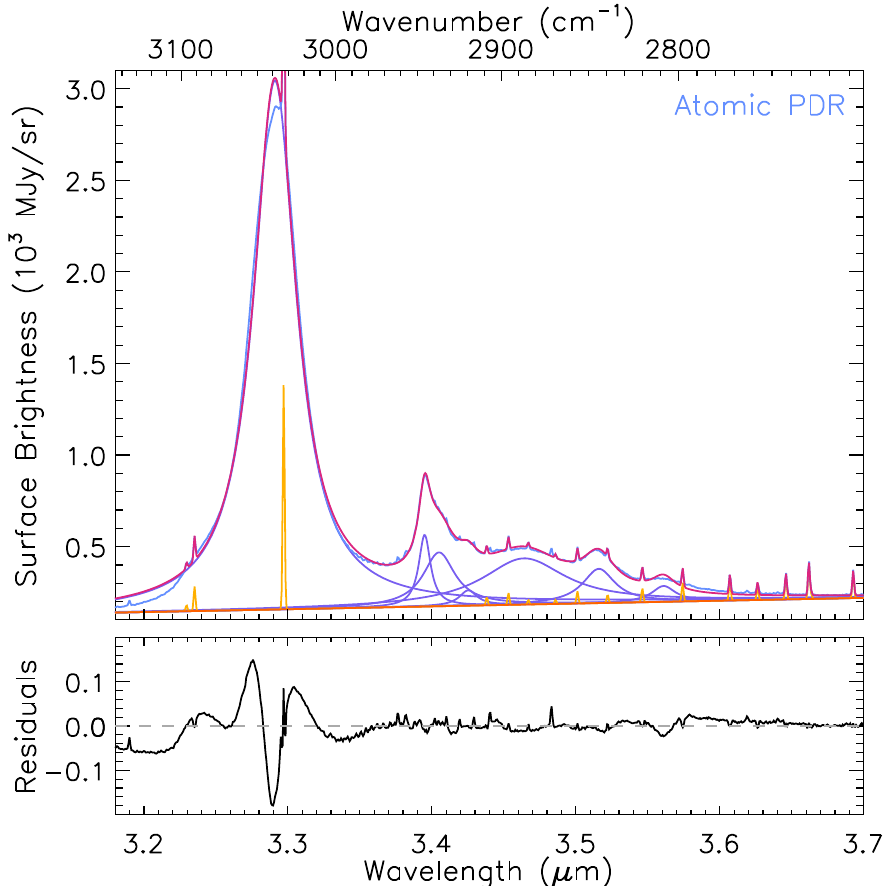}
}
\caption{Spectral decomposition of the AIB emission for the atomic PDR by a Gaussian decomposition (left) and by PAHFIT (right). The dust continuum emission is represented by the orange line, the AIB components by the purple profiles, and emission lines by the yellow Gaussians (for the PAHFIT decomposition; the Gaussian decomposition method removes the lines prior to fitting).  }
\label{fig:AIBfit}
\end{center}
\end{figure*}

First, we can determine whether case A or case B recombination theory applies based on the 1.0696/1.1759\footnote{The 1.0696 intensity is the sum of the 1.0687 and 1.0696~\mum line intensities and the 1.1759 intensity is the sum of the 1.1752 and 1.1758~\mum line intensities (Appendix~\ref{app:CI}).} line ratio as it is significantly distinct between both cases \citep{Escalante:90, Walmsley00}. However, the observed (extinction corrected) 1.0696/1.1759 line ratios give mixed results (Appendix~\ref{app:CI}). As the observed (extinction corrected) 0.984/1.0969\footnote{The 0.984 intensity is the sum of the 0.983 and 0.985~\mum line intensities (Appendix~\ref{app:CI}).} line ratios are consistent with either case A or case B, we estimate the optical depth of a UV resonance line for the high density conditions relevant for the Bar and find it to be optically thick (Appendix~\ref{app:CI}). We thus apply case B recombination theory.  
The observed extinction corrected 0.984/1.0969 line ratios then results in electron temperatures, $T_e$,  of  approximately 2500,
2300, 2900, 6800, and 5600 for, respectively, the \Ta, \Tb, \Tc, \Td, and \Te templates (Appendix~\ref{app:CI}). 
Given the uncertainties on the line ratio, the electron temperatures derived for the \Ta, the \Tb, and \Tc templates are similar to each other but clearly distinct from those derived for the \Td and \Te templates (which are also similar to each other within the uncertainties). We note that the emission lines we detect from \Tc may come from the face-on PDR of the background OMC-1 (Sect.~\ref{sec:discussion}). In agreement with the study by \citet{Walmsley00}, the derived electron temperatures are surprisingly high, much higher than temperatures derived from the pure rotational lines of \molh \citep[$\sim$400~K;][]{Allers05, vandeputte23} and millimeter-wave carbon recombination lines \citep[mmCRLs; $\sim$$500-600$~K;][]{Cuadrado19}. 

If all the \CI\ emission results from recombination (see below), these results can be interpret as the carbon NIR recombination lines arising from the hot irradiated surface of small clumps. Unfortunately, as the \CI\ 1.0696~\mum line is very weak, we cannot produce a line intensity map of this transition nor a map of the \CI\ electron temperature in order to visually detect clumps. 
Based on PDR modelling for a typical radiation field, $G_0$, of $5\times10^4$ and a wide range of densities\citep{Wolfire:10, wolfire2022}, we conclude that high gas densities ($n\approx 2-5\, \times 10^6\,{\rm cm^{-3}}$) are required to reproduce a mean electron temperature, $<T_e> = \int T n_e n_{C^+}T^{-0.6} dz / \int n_e n_{C^+}T^{-0.6} dz$, of $2000-3000$~K, similar to the derived electron temperatures for the \Ta, the \Tb, and \Tc templates.  However, these model calculations as well as model calculations for a radiation field, $G_0$, of $1\times10^5$ and a wide range of densities produce mean electron temperatures, $<T_e>$, that are below 4000~K which is significantly lower than those derived for \Td and \Te.  
The high temperature of the clump reflects the importance of heating by collisional de-excitation of UV pumped \molh vibrational levels at high densities \citep{Burton90}. As the dominant cooling transition ([\OI] 63~\mum) has a critical density of \mbox{$\sim3\times 10^{5}\, {\rm cm^{-3}}$}, cooling cannot keep up with the increased heating and the temperature has to rise to $\sim$5000~K to allow other cooling mechanisms to take over, including through dust radiative cooling and cooling through other (optical) gas lines. Given that the \molh emission comes from a deeper layer in the PDR with respect to the \CI\ emission, it naturally traces lower temperatures. Moreover, we note that the \CI\ recombination emission is weighted by the carbon emission measure (i.e. proportional to $n^2\, L$ with $n$ the gas density and $L$ the depth) and is thus more sensitive to higher density gas whereas the \molh pure rotational emission is sensitive to the column density, $N = n\,L$, rather than the density $n$. 
The required gas densities to reproduce the \CI\ electron temperature are significantly higher than the gas densities from the NIR \molh analysis (Sect.~\ref{subsec:h2}). This suggests then that the \molh pure rotational lines measure the temperature in the interclump gas and thus do not trace the high density clumps reflecting its sensitivity to the column density, $N$, rather than the density, $n$.
As the \CI\ emission arises from a very thin layer of a few thousand degree gas, we adopt A$_V$ of 0.5 for this layer (i.e. $N = 1\times 10^{21} {\rm cm^{-2}}$) to derive the gas density, n$_H$, from the 0.984~\mum line intensity (Eq.~\ref{eq:C:nH}). We obtain densities of $2.1$, $5.2$, $1.6$, $3.5$, and $5.3$ $\times\, 10^{8}\, {\rm cm^{-3}}$ for respectively, the \Ta, the \Tb, \Tc, \Td, and \Te templates. This is three to four orders of magnitude higher than the gas densities derived from the NIR \molh analysis (Sect.~\ref{subsec:h2}).
We note that the derived electron temperatures and gas densities for the clumps results in clumps' pressure that much exceeds the pressure of the inter clump medium. Hence, the clumps must be gravitational bound in order to survive or they are transient.

It is clear that the derived \CI\ electron temperatures and densities poses several issues. However, it relies on the assumption that the upper state 2p $^1$D$_2$ of the 0.982 and 0.985~\mum lines is populated solely by radiative recombination and cascade while additional excitation mechanisms would reduce both the derived temperatures and densities. Three other excitation mechanisms are possible. First, ultraviolet absorption and fluorescence via transitions at 1277.245~\AA\, (fluorescence fraction = 0.0106), 1280.135~\AA\, (fluorescence fraction = 0.00695), and 1656.928~\AA\, (fluorescence fraction = 0.000236). 
Second, direct electron-impact excitation \citep[\mbox{e$^-$ + 2p $^3$P $\rightarrow$ 2p $^1$D + e$^-$};][]{Zatsarinny:05, Zatsarinny:13}. 
Third, photodissociation of CO, which occurs via predissociation of far-UV lines \citep{vanDishoeck:88, Visser:09, Guan:21}. 
Combined with recombination, these three excitation mechanisms ensure that the near-IR [\CI] lines can be excited not only in the nebula and the ionised carbon zone of the atomic PDR, but also through the neutral carbon layer, and on into the molecular PDR where CO is photodissociated. The relative contribution of these excitation mechanisms and their influence on the \CI\ analysis as described above will be explored in a future paper.

\begin{figure*}
    \centering
    \resizebox{.80\hsize}{!}{
    \includegraphics[height=1.87cm]{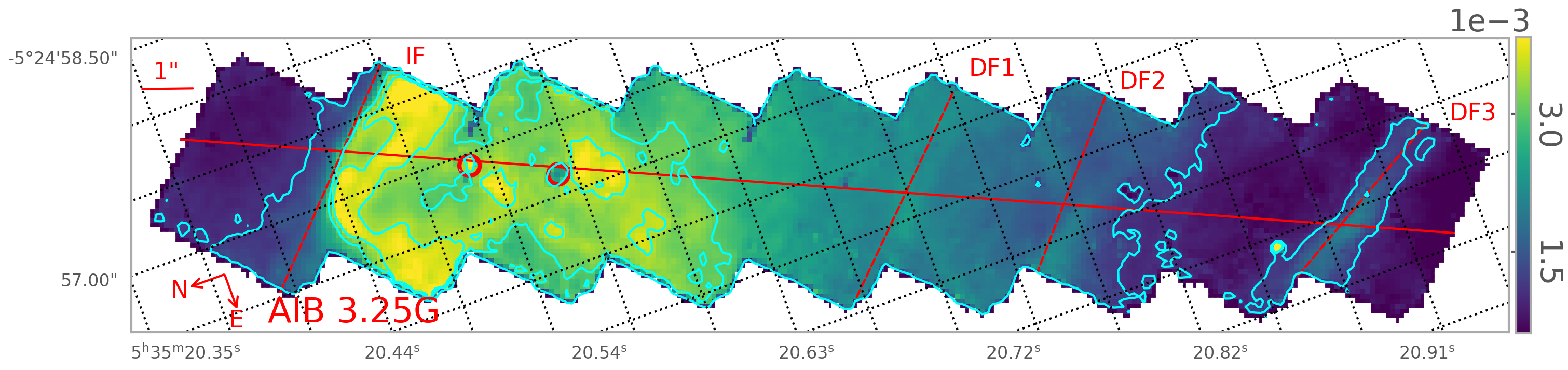}}\\
    \begin{tabular}{ccccccccccccccccc}
    \includegraphics[height=1.87cm]{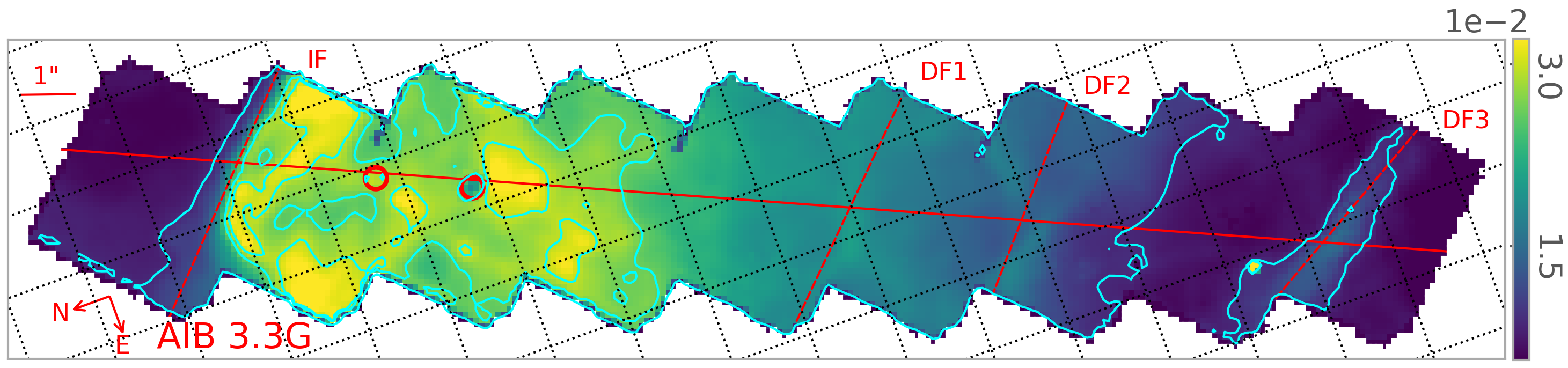} &
    \includegraphics[height=1.87cm]{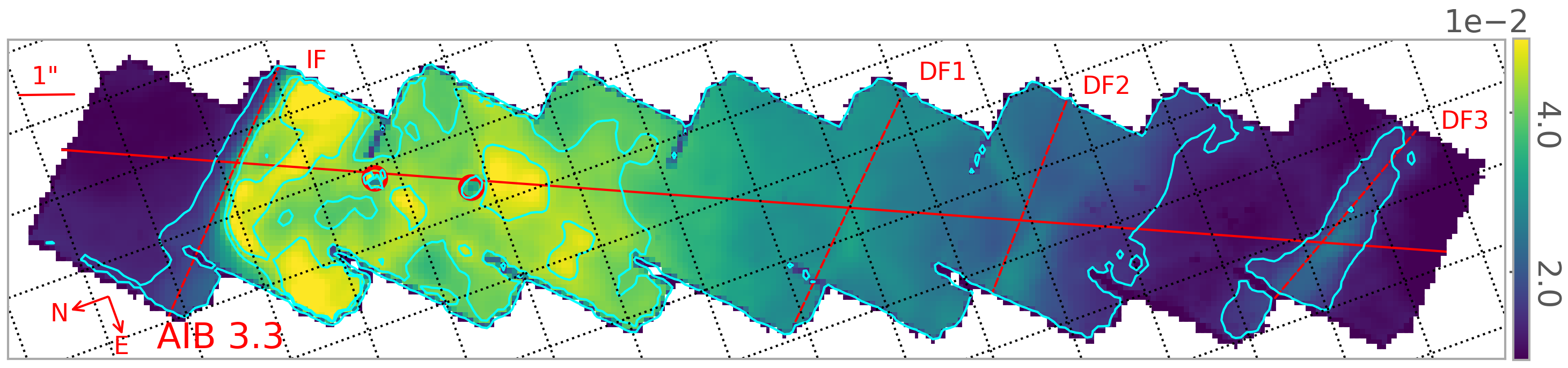}\\
    \includegraphics[height=1.87cm]{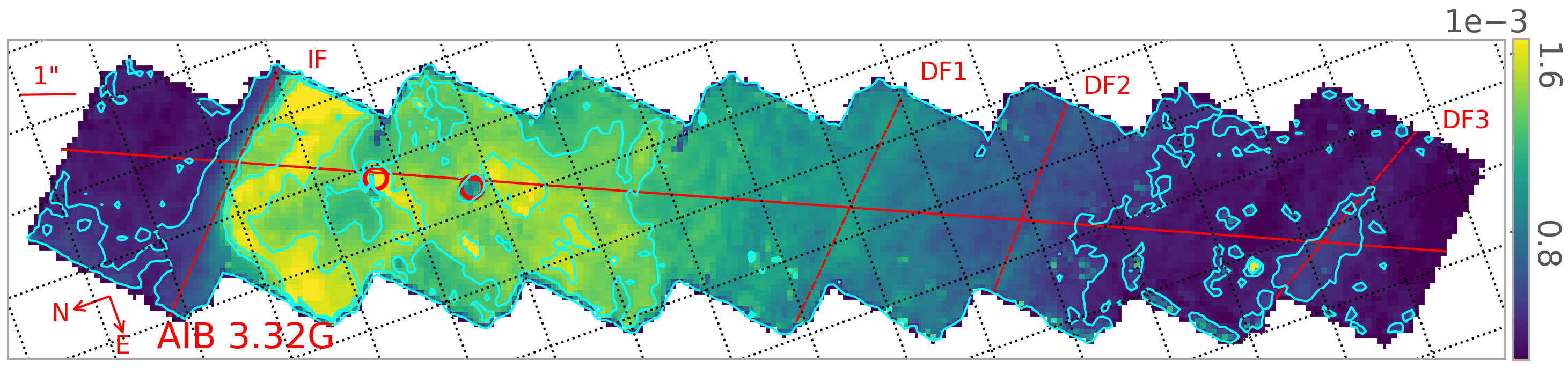} \\
    \includegraphics[height=1.87cm]{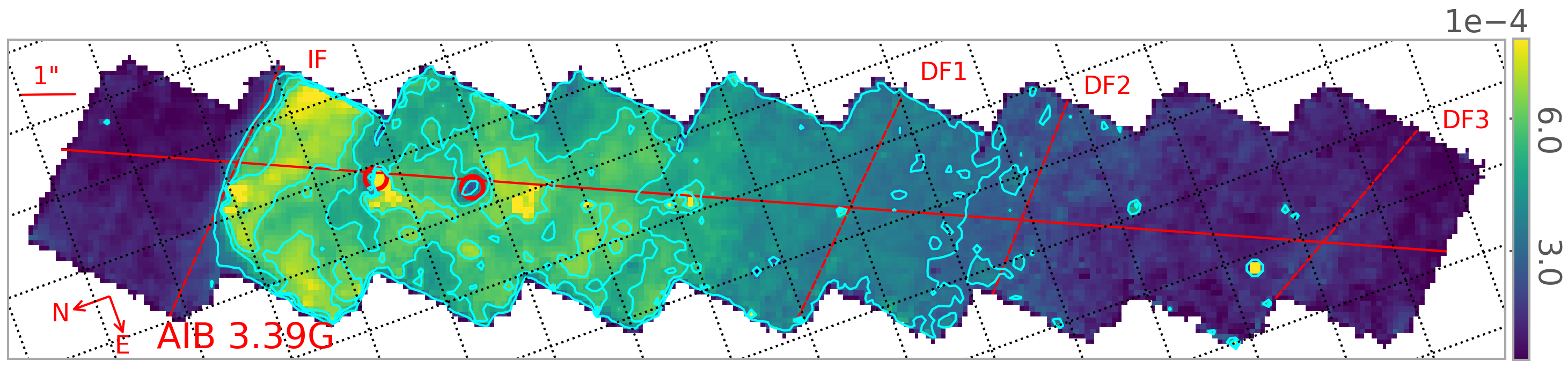}  &
    \includegraphics[height=1.87cm]{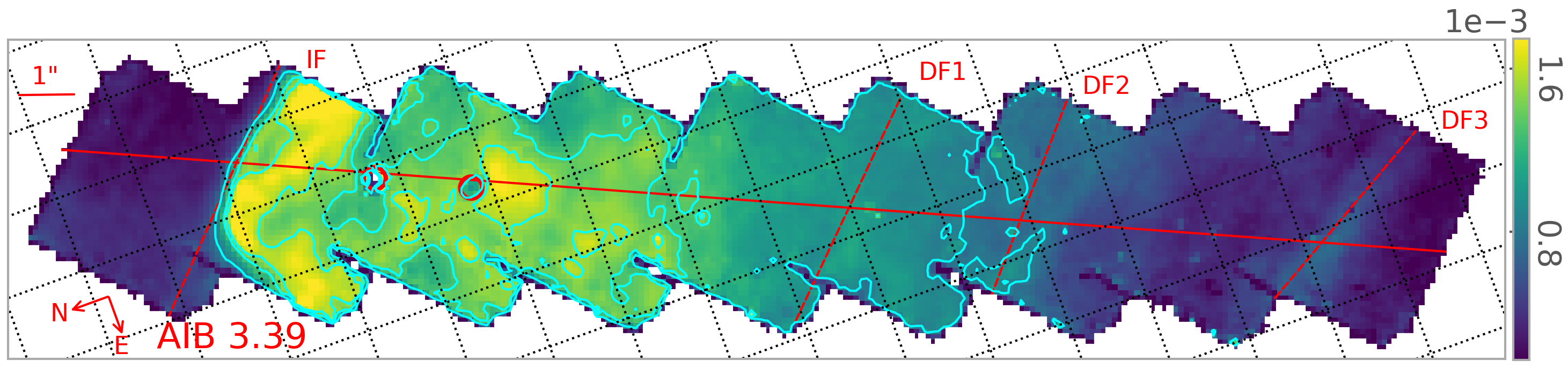}\\
    \includegraphics[height=1.87cm]{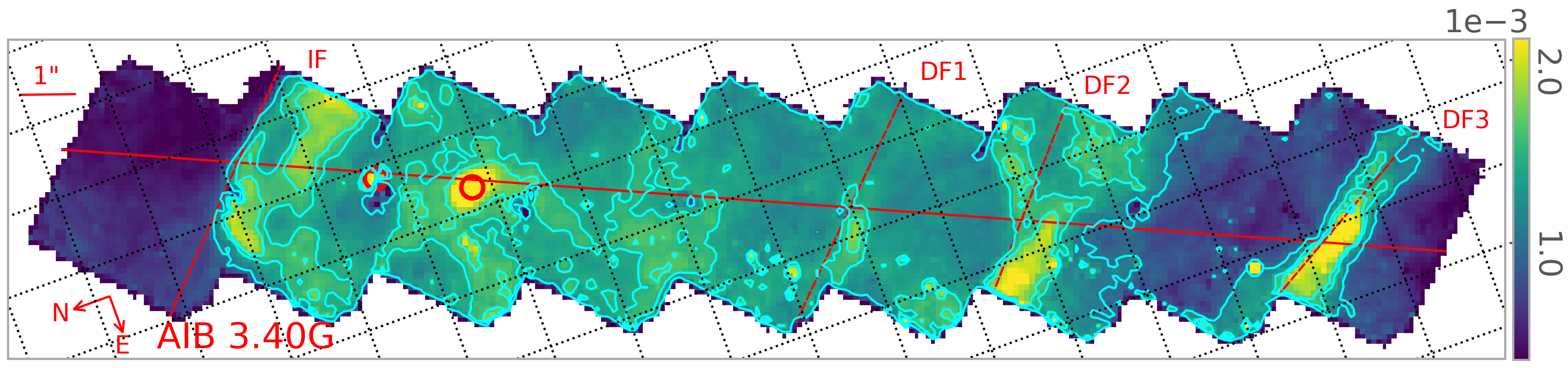} &
    \includegraphics[height=1.87cm]{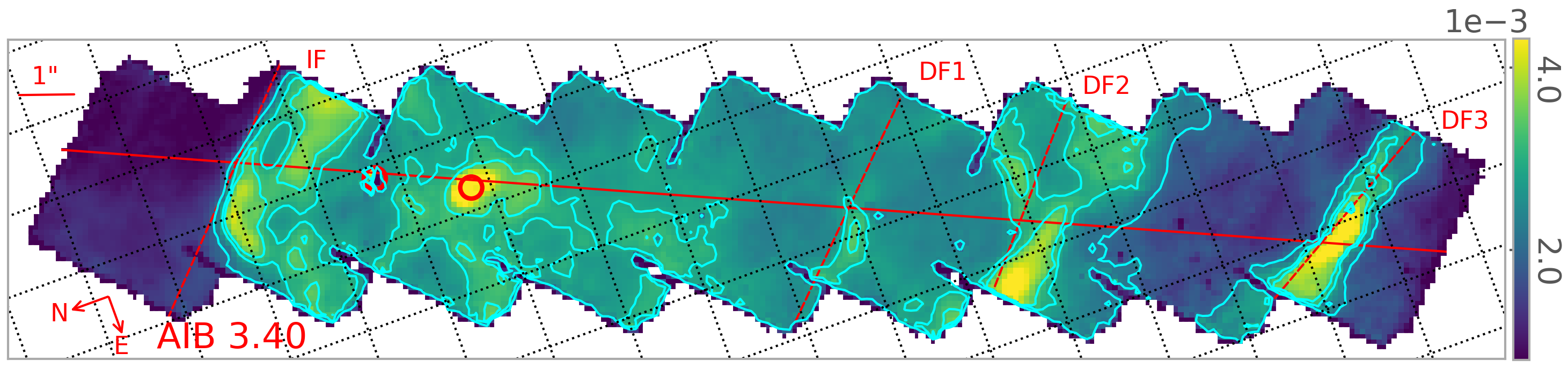}\\
    \includegraphics[height=1.87cm]{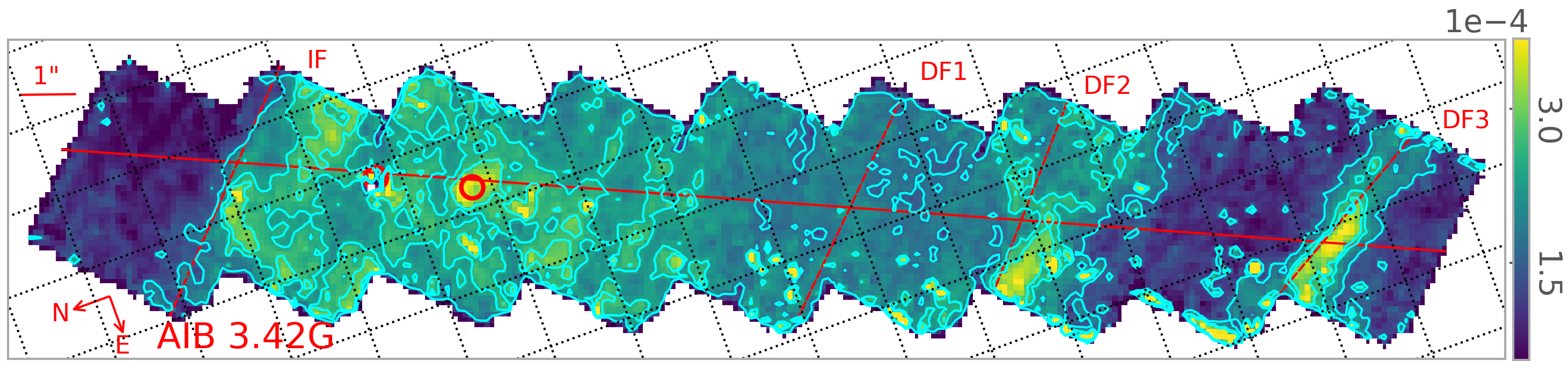} &
    \includegraphics[height=1.87cm]{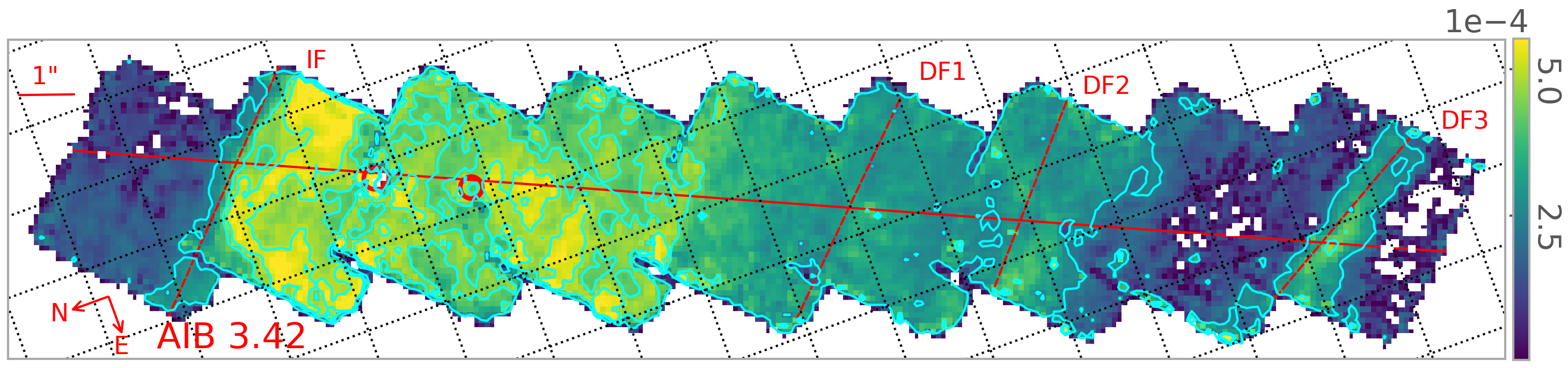}\\
    \label{fig:pahfitmap}
    \includegraphics[height=1.87cm]{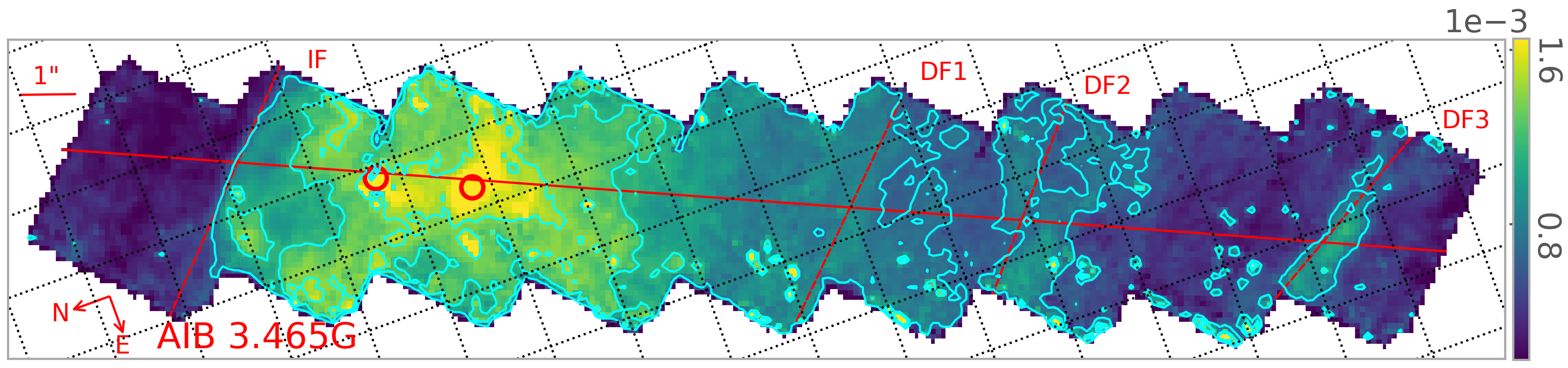} &
    \includegraphics[height=1.87cm]{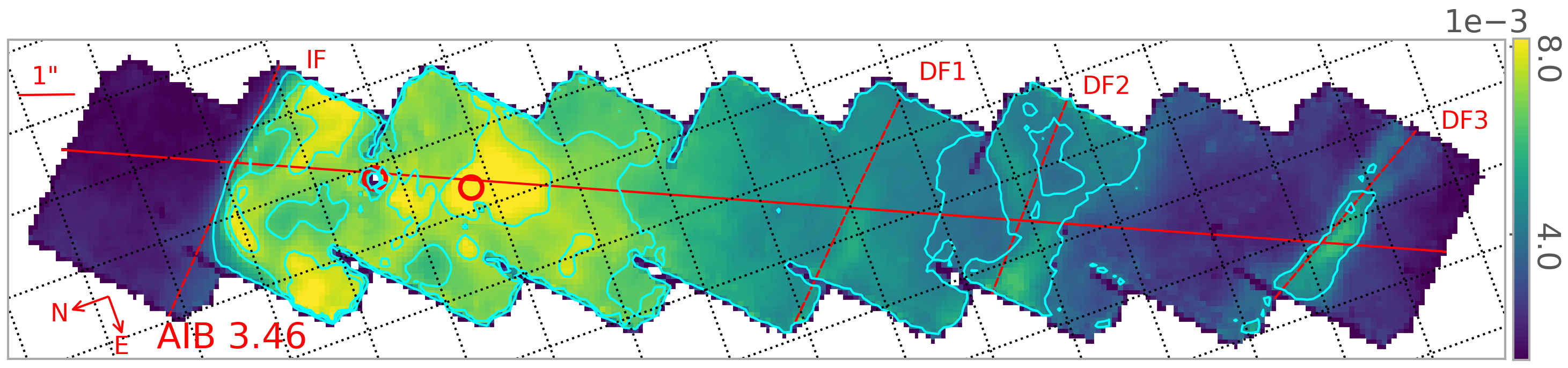}\\
    \includegraphics[height=1.87cm]{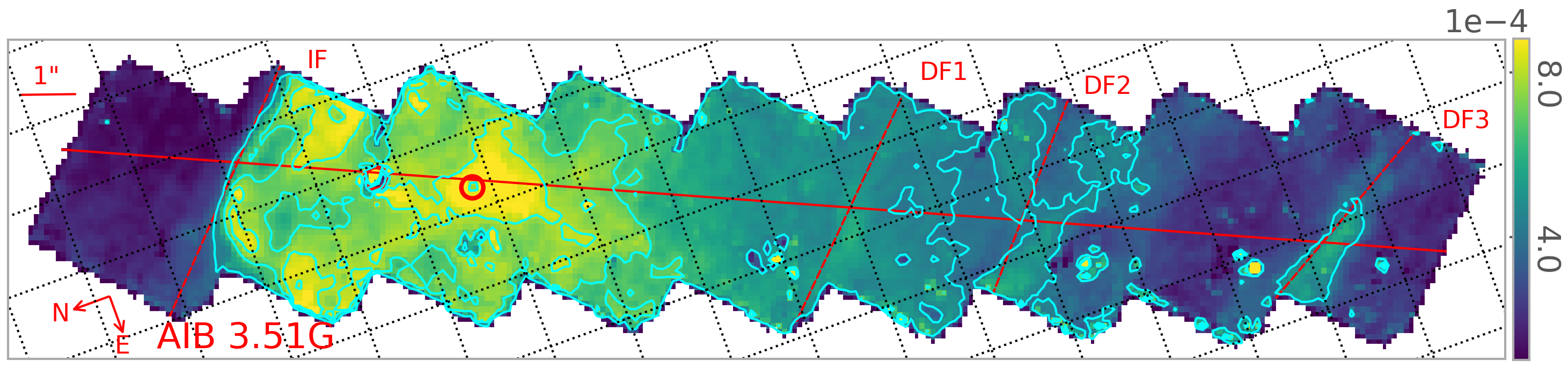} &
    \includegraphics[height=1.87cm]{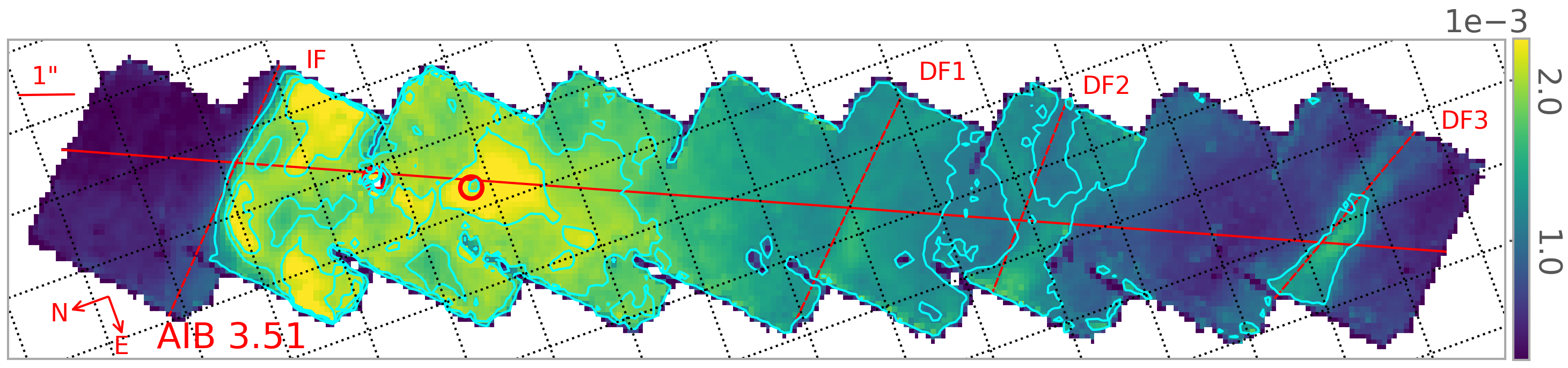}\\
    \includegraphics[height=1.87cm]{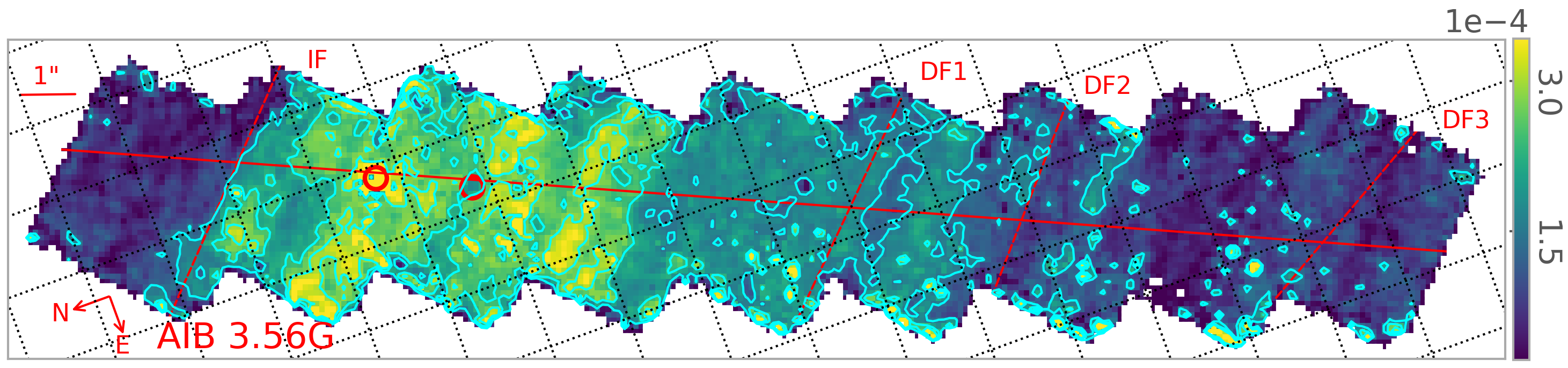} &
    \includegraphics[height=1.87cm]{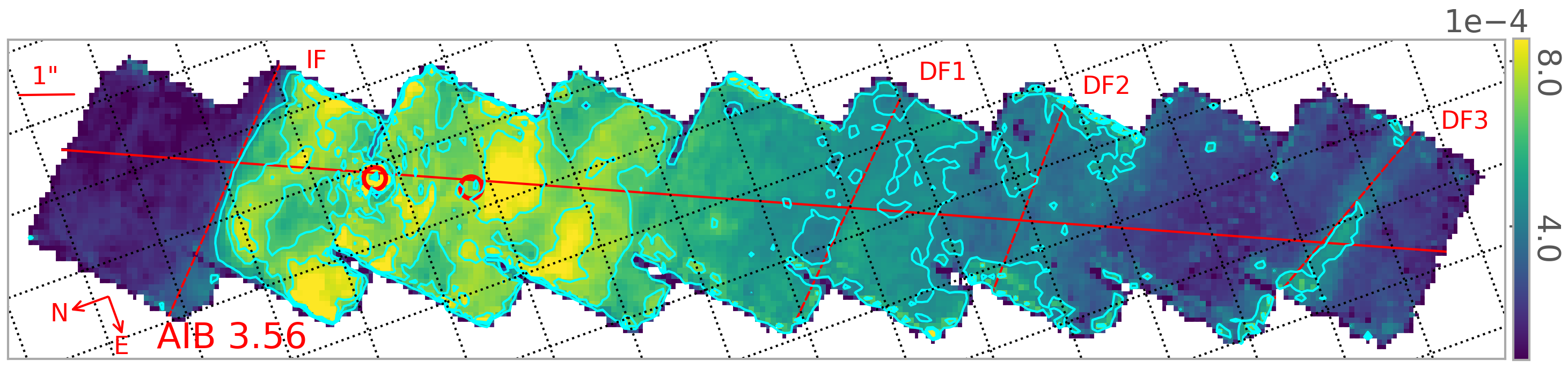}\\
    \includegraphics[height=1.87cm]{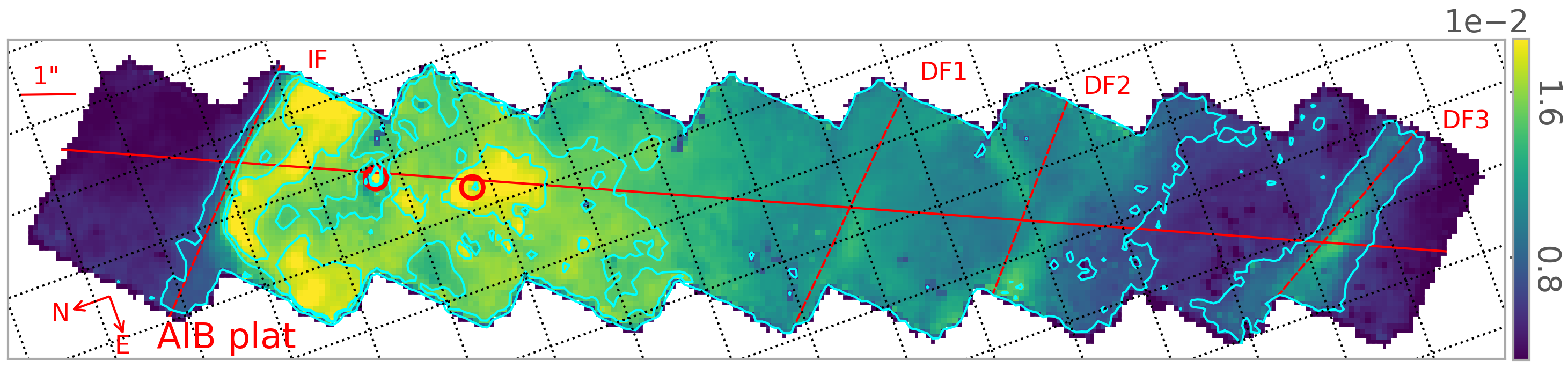} &
    \includegraphics[height=1.87cm]{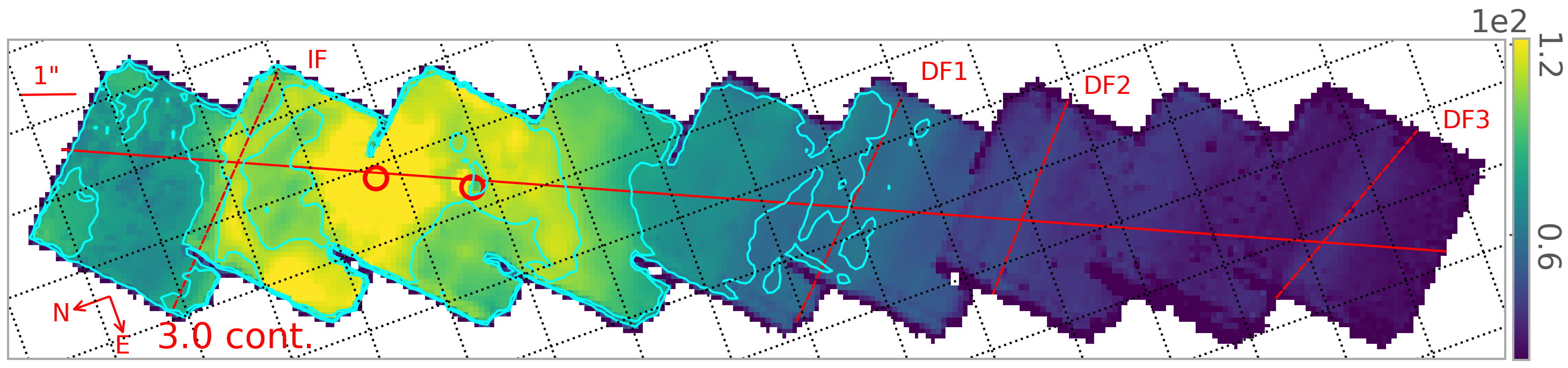}
    \end{tabular}
    \caption{Spatial distribution of the AIB components in the 3.2-3.7~\mum range from the Gaussian decomposition (top and left panels) and PAHFIT decomposition (right panels). ``AIB plat" refers to the plateau emission (bottom panel, left) and ``3.0 cont" to the continuum emission at 3.0~\mum (bottom panel, right). Units are ${\rm erg\, cm^{-2}\,s^{-1}\,sr^{-1}}$, except for the 3.0~\mum continuum map, which is in units of ${\rm MJy\,sr^{-1}}$. We set the colour range from the bottom 0.5~\% to the top 99.5~\% of data for each map, excluding values of zero, edge pixels, and the two proplyds (as well as the surrounding region of the proplyds for the continuum). White pixels inside the mosaic indicate values of zero reflecting either the component was not used in the fit or issues with the data. The nearly horizontal red line indicates the NIRSpec cut and the nearly vertical red lines indicate from left to right the DFs (\Te, \Td, \Tc) and the IF. The two proplyds are indicated by the circles. Contours show the 30, 78, 94~\%
    of the data for the 3.25G, 3.3/3.3G, and 3.32G components as well as the AIB plat, 50, 78, 93~\% for the 3.39/3.39G and 3.56/3.56G components, 35, 78, 93~\% for the 3.40/3.40G and 3.42/3.42G components, 45, 78, 94~\% for the 3.46/3.465G and 3.51/3.51G components, and 50, 68, 85~\% for the continuum emission at 3.0~\mum.}
    \label{fig:aibmaps}
\end{figure*}

\begin{figure*}
\begin{center}
\resizebox{\hsize}{!}{%
\includegraphics{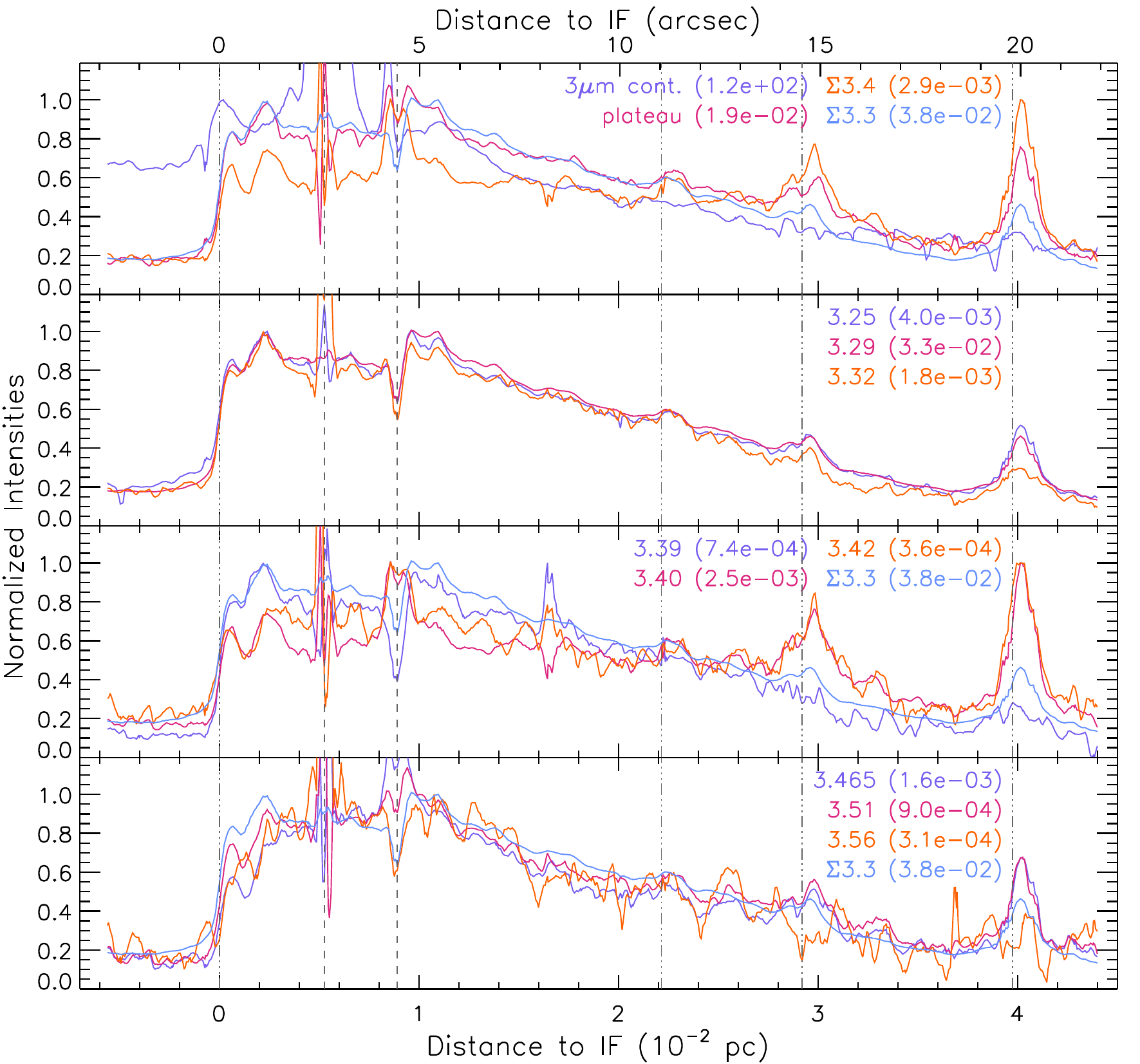}
\includegraphics{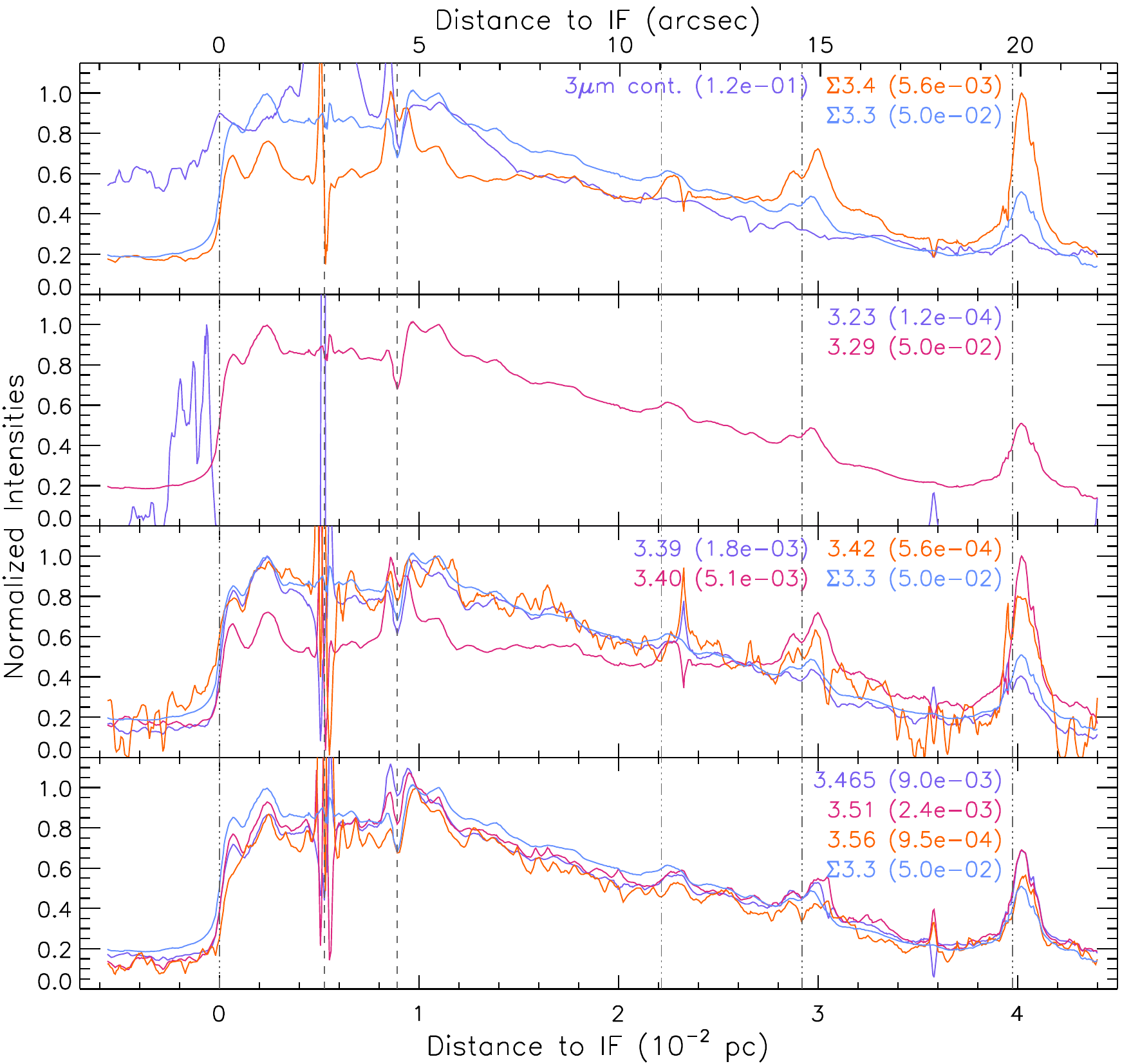}}
\caption{Normalised AIB and continuum intensities from the Gaussian decomposition (left) and the PAHFIT decomposition (right) as a function of distance to the IF (0.228~pc or 113.4\arcsec\ from \oric) along a cut crossing the NIRSpec mosaic (see Fig.~\ref{fig:FOV}). As the cut is not perpendicular to the IF and distances are given along the cut, a correction factor of cos(19.58\textdegree)=0.942 needs to be applied to obtain a perpendicular distance from the IF. $\Sigma3.3$ reflects the brightness of the 3.25, 3.29, 3.32~\mum or 3.23 and 3.29~\mum AIBs combined for, respectively, the Gaussian and PAHFIT decomposition. $\Sigma3.4$ reflects the brightness of the 3.40, and 3.42~\mum AIBs combined. No extinction correction is applied. The dash-dot-dot-dot vertical lines indicate the position of the IF, \Tc, \Td, and \Te, respectively, from left to right. The dashed vertical lines indicated the location of the proplyds 203-504 (left) and 203-506 (right). Units are \uint except for the 3~\mum continuum which is in units of ${\rm MJy\,sr^{-1}}$. }
\label{fig:aibcuts}
\end{center}
\end{figure*}

\subsection{AIB emission}
\label{subsec:AIB}

The components of the 3-4~\mum AIB emission in the Bar exhibit variations in intensity, which we discuss in Sect.~\ref{subsubsect:AIBint}, and variations in profiles, which we discuss in Sect.~\ref{subsubsect:AIBprof}. The comparison of the 3-4~\mum AIB emission to the mid-IR AIB bands is reported in \citet{Chown:23}. 

\subsubsection{AIB Intensity variations}
\label{subsubsect:AIBint}

We have performed two spectral decompositions of the AIB emission, which are applied to every pixel of the NIRSpec mosaic. First, we employ an updated version of PAHFIT \citep{PAHFIT2007, Lai:20}\footnote{available at \url{https://github.com/PAHFIT}}. In PAHFIT-based models, the AIBs are represented by Drude profiles. Second, we have employed a Gaussian decomposition of the AIB emission in the $3.2-3.7$~\mum region. The resulting fit for both methods reproduces the observations very well (Fig~\ref{fig:AIBfit}). In contrast to the PAHFIT method, which uses $7$ Drude profiles, the Gaussian decomposition also includes one Gaussian representing the underlying plateau emission and one Gaussian representing the extended red wing of the 3.29~\mum AIB. The remaining components are remarkably similar for both decomposition methods. Details on both spectral decompositions are given in Appendix~\ref{app:AIBdecomposition}. 
While neither decomposition method provides a physical decomposition, it allows for a systematic analysis of the AIB emission characteristics. The intensity maps of the AIB components are shown in Fig.~\ref{fig:aibmaps} and their emission along the cut in Fig.~\ref{fig:aibcuts}. 

Most AIBs (all but the 3.40 and 3.42~\mum AIBs) mimic the global morphology as seen by the total AIB emission (Fig.~\ref{fig:maps}): a sharp rise at the PDR front, a (broad) plateau in the atomic PDR with a width of about 6.5\arcsec (\mbox{$\sim 1.3\,\times\,10^{-2}\,{\rm pc}$}), after which a steady decline sets in. This decline levels off towards the south end of the mosaic/cut, reaching brightness levels as observed in the \Ta, where the AIB emission originates from the background face-on PDR OMC-1. The latter is in contrast with the 3~\mum continuum emission which is much higher in the \Ta than in the molecular PDR (see Sect.~\ref{subsec:OB}). The exceptions to the global AIB morphology are the 3.40~\mum AIB and the 3.42~\mum AIB (to a slightly lesser extent), which peak in \Te, instead of the \Tb, and show enhanced emission in \Td but not in \Tc compared to the 3.3~\mum AIB. The 3.4/3.29 integrated intensity ratio\footnote{calculated by the ratio of the (3.39+3.40+3.42)/(3.29) AIBs.} is 0.09-0.11 in the \Ta, the \Tb and \Tc templates, whereas values of 0.15 and 0.18 are observed in the \Td and \Te templates, respectively (for the Gaussian decomposition, Table~\ref{tab:physical_conditions}). In addition, while the 3.465 and 3.51~\mum AIBs peak in the \Tb, these AIBs also show enhanced emission in \Td and \Te, relative to that in the \Tb, with respect to the 3.29~\mum AIB. Hence, the 3.40, 3.42, 3.56, 3.465, and 3.51~\mum AIBs show a less steep increase at the PDR front than the 3.29~\mum AIB (of about 4 compared to an increase of about 5 seen in the 3.29~\mum AIB emission). The plateau emission (in the Gaussian decomposition) has a spatial distribution similar to the 3.29~\mum AIB, but shows slightly enhanced emission in \Td and \Te (albeit significantly less than the enhanced emission seen in the 3.40 and 3.42~\mum AIBs). Hence, based on its spatial distribution we cannot conclude whether or not the plateau is independent of the superposed features as is observed for the plateaus between 5-10, 10-15, and 15-20~\mum \citep{Bregman89, Peeters:12, peeters17, Stock2017}. 
Proplyd 203-506 stands out in the AIB maps. Namely, it shows enhanced emission in the 3.40 and 3.42~\mum AIBs with respect to the other AIBs, 
and a 3.3~\mum AIB intensity comparable to the lowest seen 3.3~\mum AIB intensity in the \Tb. 

Lastly, we tentatively detect the aromatic CD vibrational mode at 4.35~\mum in the \Tb and \Tc, and the aliphatic CD vibrational mode at 4.644~\mum in potentially all templates (see Sect.~\ref{sec:spectral-characteristics} and Appendix~\ref{app:PADs} for details). The aliphatic CD vibrational mode is strongest in the \Tb, followed by \Tc, the \Ta, \Td, and is weakest in \Te (Table~\ref{tab:AIBflux}; see Appendix~\ref{app:PADs} for details). As the band at 4.35~\mum is not well defined (see Sect.~\ref{sec:spectral-characteristics}), we refrain from estimating its intensity though we note that it is most easily discerned in the \Tb and \Tc that also exhibit the strongest 4.644~\mum band. The ratio of the 4.644~\mum band to the total AIB emission in the $3.2-3.7$~\mum range is 8.2, 6.3, 6.1, 2.8, and 2.4 $\times \, 10^{-3}$ for the \Ta, the \Tb, \Tc, \Td, and \Te templates (Table~\ref{tab:physical_conditions}) and thus is significantly lower in \Td and \Te compared to the other templates. A similar pattern is seen in the 4.644/3.29 AIB ratio whereas the  4.644/3.40 AIB ratio is highest in the \Tb, closely followed by the \Ta, subsequently \Tc and significantly lower in \Td and \Te (Table~\ref{tab:physical_conditions}). 

\subsubsection{Profile variations}
\label{subsubsect:AIBprof}

\begin{figure}[!h]
\begin{center}
\resizebox{\hsize}{!}{%
\includegraphics{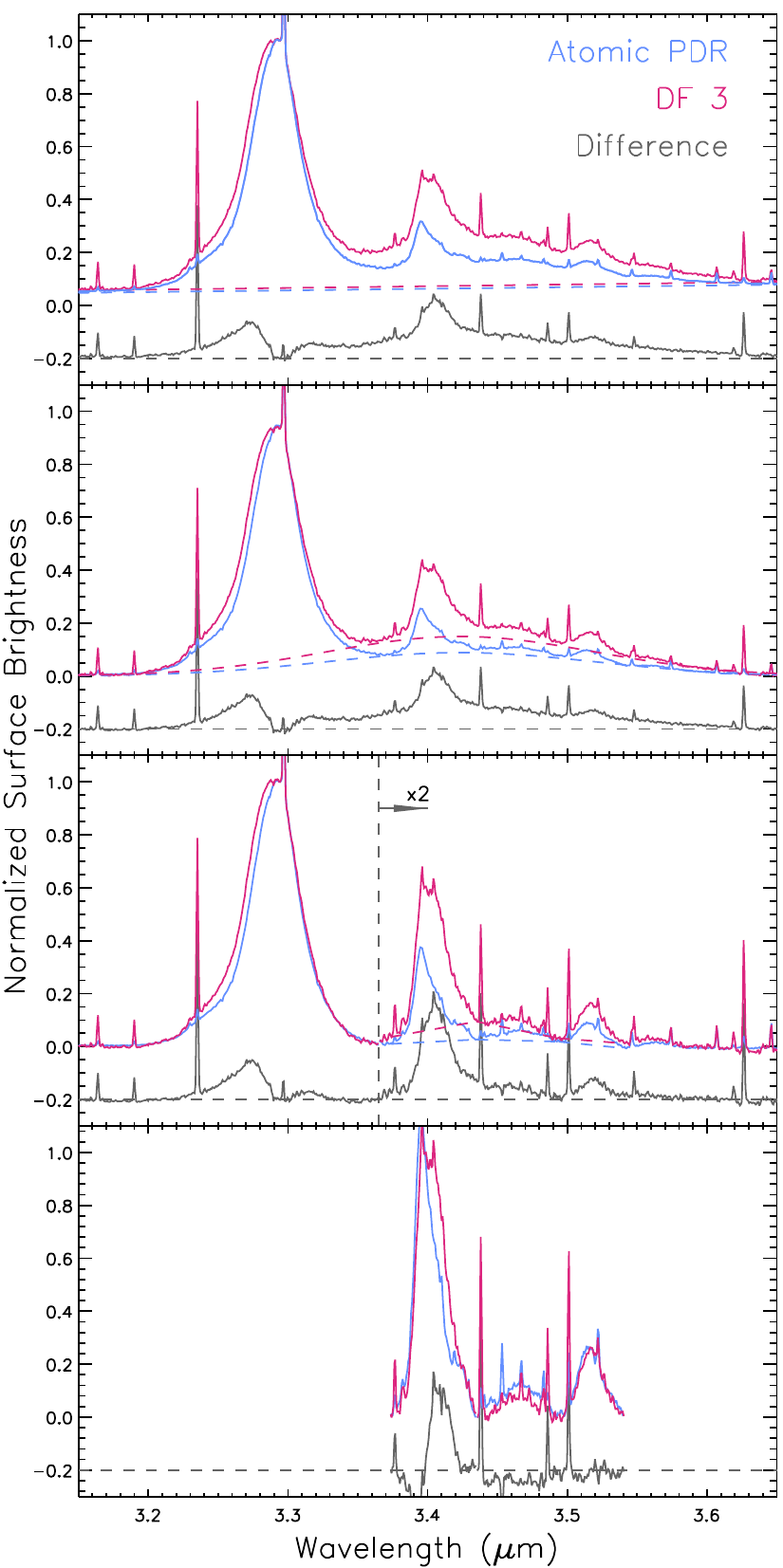}}
\caption{Comparison of the AIBs in the \Tb (blue) and \Te (red). The difference in their normalised emission is given in grey and the grey dashed line indicates the zero level for the difference spectrum.  Top to bottom: template spectra (solid lines) and their respective linear dust continuum (coloured dashed lines); continuum-subtracted template spectra (solid lines) and their respective plateau emission (coloured dashed lines); (continuum+plateau)-subtracted template spectra (solid lines) and their respective local continua for the weaker AIBs (coloured dashed lines);(continuum+plateau+local continuum)-subtracted template spectra. The template spectra are normalised on the peak intensity of the 3.29~\mum AIB in the top three panels, and of the 3.40~\mum AIB in the bottom panel.}
\label{fig:PAHvariation1}
\end{center}
\end{figure}

Not only do the relative AIB intensities change across the mosaic, so do the AIB band profiles (Fig.~\ref{fig:PAHvariation1}). The 3.29~\mum AIB has a variable width with the full-width-half-maximum (FWHM) ranging from \mbox{$37.4\,{\rm cm^{-1}}$} in the \Tb to \mbox{$42.4\,{\rm cm^{-1}}$} in \Te (see Table~\ref{tab:physical_conditions} for FWHM values for all templates). \Tc shows a band profile similar to that observed in the \Ta and the \Tb. The extra broadening in \Td and \Te relative to the other three templates is largely carried by the blue wing. As a consequence, the peak position seems to shift to slightly bluer wavelengths in the \Td and \Te templates albeit quantifying this shift is hampered by the atomic emission lines (Pf$\delta$, \HeI\ recombination lines) superposed on the peak of the AIB. Despite the observed profile variations, all templates exhibit a class A band profile in the classification scheme proposed by \citet{vandiedenhoven2004}, and thus showcase profile variability within class A. We note that also the mid-IR AIB band profiles exhibit variations \citep{Chown:23}. Similarly, despite the profile variations, the mid-IR AIBs exhibit a class A band profile except for the 11.2~\mum AIB which displays a class A profile in the \Ta, the \Tb, and \Tc and a class B profile in \Td and \Te \citep{Chown:23}.

Profile variations are also detected for the 3.4~\mum AIB. This band shows an asymmetric profile with a red wing and consists of three components (at 3.39, 3.40, 3.42~\mum, see Table~\ref{tab:AIBparam}). Comparison of the 3.4~\mum AIB profile in the template spectra indicate enhanced emission in the red wing of the 3.40~\mum component and thus enhanced broadening of the band in \Td and \Te. Similar to the 3.29~\mum AIB profile, \Tc displays a similar profile to that observed in the \Ta and the \Tb. Given that numerous \HI\  recombination lines (from the Humphreys series) and \molh lines are superposed on this AIB, a detailed analysis of the 3.4~\mum AIB will be performed in a forthcoming paper (Dartois et al., in prep.). 

\subsubsection{Comparison to previous observations}
\label{subsubsec:AIBliterature}

The spectral inventory of the AIB emission in the $3.2-3.7$~\mum range is consistent with prior high quality observations of the Orion Bar \citep{Sloan:97}. Likewise, the observed 3.4/3.29 intensity ratios (Table~\ref{tab:physical_conditions}) are consistent with earlier reports for the Orion Bar by \citet{geballe} and \citet{Sloan:97}. The latter authors also reported a widening (towards the red) of the 3.4~\mum AIB into the molecular PDR. To first order, this is consistent with our results but the strength of their ``excess'' emission (with respect to the 3.4~\mum AIB in the \Ta) peaks near 10\arcsec\ from the IF after which it steadily declines whereas we detect the broadest profile in \Te at a distance of 19.8\arcsec. In addition, as our observations have unparalleled angular resolution, the radial profiles of the AIB intensities with distance from \oric (Fig.~\ref{fig:aibcuts}) exhibit significant more detail compared to prior observations \citep{geballe, Sloan:97}. 

To our knowledge, the $3.1-4.9$~\mum broad emission or the 3.8~\mum band has not been observed before (Sect.~\ref{sec:spectral-characteristics}). In contrast, the bands near 4.4 and 4.6~\mum have been detected in the Orion Bar and were attributed to deuterated PAHs \citep{Peeters:pads:04, Onaka:14}. Following \citet[][i.e. assuming that the integrated absorbance values of the corresponding C-H/C-D modes are similar]{Peeters:pads:04}, we obtain an aliphatic D/H ratio (as probed by 4.64/3.40 bands) of 20.5, 22.5, 15.8, 5.4, and 4.1 $\times \, 10^{-2}$ in the \Ta, the \Tb, \Tc, \Td, and \Te templates, in contrast to the ratio of $\sim$ 1 reported by these authors. Instead, this derived aliphatic D/H ratio is more in line with their combined aliphatic and aromatic D/H ratio of 0.17\footnote{The combined aliphatic and aromatic ratio was probed by the (4.4+4.6)/(3.3+3.4+3.5) using the Gaussian decomposition. Using the nomenclature of this paper, this denominator reflects the sum of all Gaussian components in the $3.2-3.7$~\mum range except the plateau emission.}. A quantitative comparison with the results of \citet{Onaka:14} is not possible as these authors applied a different decomposition method\footnote{These authors used Lorentzian profiles for the 3.29, 3.41 and 3.48~\mum bands and did not assume a plateau component.}. However, these authors reported D/H ratios an order of magnitude lower than those of \citet{Peeters:pads:04}. Specifically, they reported an aliphatic D/H ratio (as probed by 4.64/(3.42+3.48) bands) of 0.04 which is of the same order of magnitude as our 4.64/(3.39+3.40+3.42+3.46) ratios using the PAHFIT decomposition (ranging from 0.03 to 0.007, though remember different decomposition methods are used).

\subsubsection{Photochemical evolution}
\label{subsubsec:AIBevolution}

The 3~\mum region is characteristic for the aromatic and aliphatic C-H stretching mode \citep[e.g.][]{ATB}. The aromatic CH stretching modes are very susceptible to resonances with combination bands \citep{Maltseva:2015,Maltseva:2016,Mackie:15,Mackie:16} and this interaction dominates their profiles \citep{mackie2022}. Smaller PAHs have more asymmetric profiles, i.e a less steep blue wing and enhanced red wing, reflecting their higher internal energies upon photon absorption \citep{Tielens:book2, mackie2022}. 
The observed widening of the asymmetric 3.4~\mum band in \Td and \Te thus may arise from an enhanced population of smaller PAHs. The corresponding broadening of the roughly symmetric 3.29~\mum band (largely driven by the blue wing) may then arise from the same enhanced population of smaller PAHs. Regions with broader AIB profiles (with respect to the \Tb) also show enhanced 3.4/3.29 intensity ratios which traces the aliphatic-to-aromatic ratio (or the degree of aromaticity).  The aliphatic-to-aromatic ratio, as traced by the 3.4/3.29 intensity ratio, is known to decrease with increasing intensity of the FUV radiation field \citep{geballe, Joblin:3umvsmethyl:96, Sloan:97, Mori:14,  pilleri2015} reflecting that aliphatic bonds are less stable than aromatic ones. 
Combined with the fact that smaller PAHs are less stable than larger ones, the observations point towards a more fragile population of complex hydrocarbons in \Td and \Te compared to the complex hydrocarbon population in the \Ta (i.e. the background PDR), the \Tb, and \Tc. This suggests a UV-driven photochemical evolution of the complex hydrocarbon population that eliminates the more fragile hydrocarbon species near the surface of the PDR which is subjected to a more intense FUV radiation field.  

In contrast, the tentative detection of high aliphatic D/H ratios near the surface of the PDR and low ratios deep in the molecular PDR argue against a slow loss of deuterated PAHs as material reaches the surface of the PDR; that is, the presence of PADs in the surface layer of the PDR is not inherited from its past (i.e. from the molecular cloud). Instead, the PAH D enhancement is a local effect that seems to be driven by UV radiation and/or density (as it is strongest in the \Tb  template).  In addition, we note that the aliphatic CD stretch is significantly stronger than the aromatic CD stretch, consistent with prior observations \citep{Peeters:pads:04, Onaka:14, Doney:16} suggesting that the D enhancement is more favourable on aliphatic than aromatic sites.  The expected difference in binding energy between hydrogen and deuterium is $\sim$440cm$^{-1}$ \citep{ATB, Wiersma:20} and, thus, too small to explain the observations. In addition, this energy difference leads to enhanced H scrambling and H loss compared to D while the molecule is exposed to a stronger UV field \citep{Wiersma:20}. Moreover, these authors found that D scrambling favours the migration to a strongly bound aromatic site (instead of an aliphatic site) which could lead to increased aromatic deuteration with respect to aliphatic deuteration. This is in contrast with the observations presented here. It should be noted that the interpretation or possible mechanism to increase aliphatic CD is still speculative and further investigations are warranted. Hence, the PAH D enhancement will be further explored in a forthcoming paper.

\section{Discussion}
\label{sec:discussion}

The Bar has served as the template PDR to develop the first PDR models \citep{tielens:85, tielens:85b}. These theoretical studies predicted a global stratification as a function of depth in the PDR. \citet{Tielens93} provided the first observational evidence of this stratification and reported an offset between the 3~\mum AIB emission, the \molh 1-0 S(1) emission, and the J=1-0 CO emission. The PDRs4All NIRSpec observations provide a more diverse and detailed picture of the Bar anatomy on spatial scales of $0.075\arcsec-0.173\arcsec$. The global stratification (and geometry) as seen with NIRSpec is summarised below, illustrated in Fig.~\ref{fig:geometry}, and quantified in Table~\ref{tab:param_OB}. Note that all distances are quoted along the NIRSpec cut (see Sect.~\ref{subsec:OB}). 

\begin{itemize}
\item Closest to \oric, we observe the He-IF at an approximate projected distance of \mbox{0.222~pc}. The He-IF is resolved and displaced from the H-IF by about \mbox{0.4-0.5$\times$10$^{-2}$\,pc} \mbox{(1.8\arcsec-2.5\arcsec)} in the Huygens region. The NIRSpec mosaic does not cover the peak emission from the \HeI\ recombination lines.
\item The H-IF is traced by the peak emission of the \HI\ recombination lines (at 113.276\arcsec\, or $d_{proj}=0.2274\,{\rm pc}$ from \oric). The width of the \HI\ peak is set by the fact that we see the slab under a small inclination (Sect.~\ref{subsubsec:geometry}). The sharp decrease towards \oric within the NIRSpec mosaic is likely caused by the gas being accelerated away once it is ionised. The \HI\ recombination line emission decreases away from \oric but does not go to zero due to the foreground \Ta (i.e., the \HII\ region in front of the atomic and molecular PDR along the line-of-sight). As for the \HeI\ emission, the NIRSpec mosaic does not cover the peak \HI\ emission within the \HII\ region which, is located about 24\arcsec\ in front of the H-IF \citep[e.g.,][]{Pellegrini09}.
\item The IF is well traced by the \mbox{[\OI] 6300~${\mathring{\mathrm{A}}}$} and \mbox{[\FeII] 1.644~\mum} emission. We note that this is displaced from the H-IF by 0.1\arcsec\, or \mbox{$0.02\, \times \,10^{-2}\,{\rm pc}$}. Enhanced emission of \mbox{[\OI] 6300~${\mathring{\mathrm{A}}}$} is also observed in the atomic PDR, which is mostly confined to the region surrounding the proplyds and the filaments. While part of this emission is associated with the proplyds and their jets, the strongest emission is seen in the N-SE filament, which is, to our knowledge, not associated with the proplyds and is parallel to the secondary ridge in the IF. In contrast, the \mbox{[\FeII] 1.644~\mum} only shows enhanced emission in the direction of the jet associated with proplyd 203-506. We further note that the N-SE filament is very strong in the \mbox{\OI\ 1.317~\mum} emission, but given the lack of enhanced emission in the \mbox{[\FeII] 1.644~\mum} line, undulation effects in the surface of the Bar as the prime reason of the observed enhancement in the \OI\ emission can be excluded. Instead, we attribute the enhanced emission to local acceleration zones (Sect.~\ref{subsec:O+N}). 
\item The AIB emission is an excellent tracer of the atomic PDR. The strength of the AIB emission is set by the strength of the FUV radiation field required for the excitation process and the column density of the carriers. The steep increase in its emission (up to 65\%) over a very small distance ($\sim$1\arcsec) centered at the IF thus indicates the (very sharp) onset of the atomic PDR. Along the NIRSpec cut, the AIB emission remains flat for about 6.5\arcsec\, or \mbox{$\sim 1.3\, \times \,10^{-2}\,{\rm pc}$}, after which it slowly decreases. As the strength of the FUV radiation decreases with depth into the PDR due to dust opacity, the lack of a decrease in AIB emission beyond the onset of the atomic PDR, as well as the small scale structure observed in the AIB emission, reveals a complexity of the atomic PDR in terms of geometry and small scale structure that is not captured by 1D PDR models. The AIB emission (primary ridge) peaks at 113.7\arcsec in contrast with earlier studies that reported the AIB emission peaking at 118 and 117\arcsec\ from \oric \citep{Salgado16, Knight21:Orion}. This is attributed to their use of lower angular resolution observations, as well as probing a different location on the Bar.  \citet{Habart:im} derived an atomic gas density of \mbox{$5-10\, \times\, 10^4\,{\rm cm^{-3}}$}.  
The (maximum) strength of the FUV radiation field impinging on the PDR front can be traced by the fluorescent lines (\OI\ 1.129, 1.317~\mum and \NI\ 1.2292~\mum) and varies across the PDR front between G$_0$ = 2.2 to $7.1\times 10^4$.
\item The transition from the atomic to the molecular PDR is highly structured and displays three dissociation fronts that are parallel to the IF \citep[see also][]{Habart:im}. The rise in the H$_2$ emission is very sharp (with factors of $\sim$ 3, 6, and 10 over a very small distance (0.5\arcsec). These ridges (at a distance of \mbox{(2.21, 2.92, 3.97) $\times$ 10$^{-2}$\,pc} or \mbox{11.03\arcsec, 14.55\arcsec, 19.80\arcsec}) represent edge-on portions of the DF, with \Td nearly coinciding with the average \molh emission in the Bar \citep{Habart:im}. Enhanced AIB emission is seen at the three DFs which is lightly displaced from the DFs by \mbox{(0.02, 0.06, 0.04) $\times$ 10$^{-2}$\,pc} (\mbox{0.1\arcsec, 0.3\arcsec, 0.2\arcsec}), in the direction away from the atomic PDR.
\end{itemize}

\begin{figure}
\begin{center}
\resizebox{\hsize}{!}{%
\includegraphics{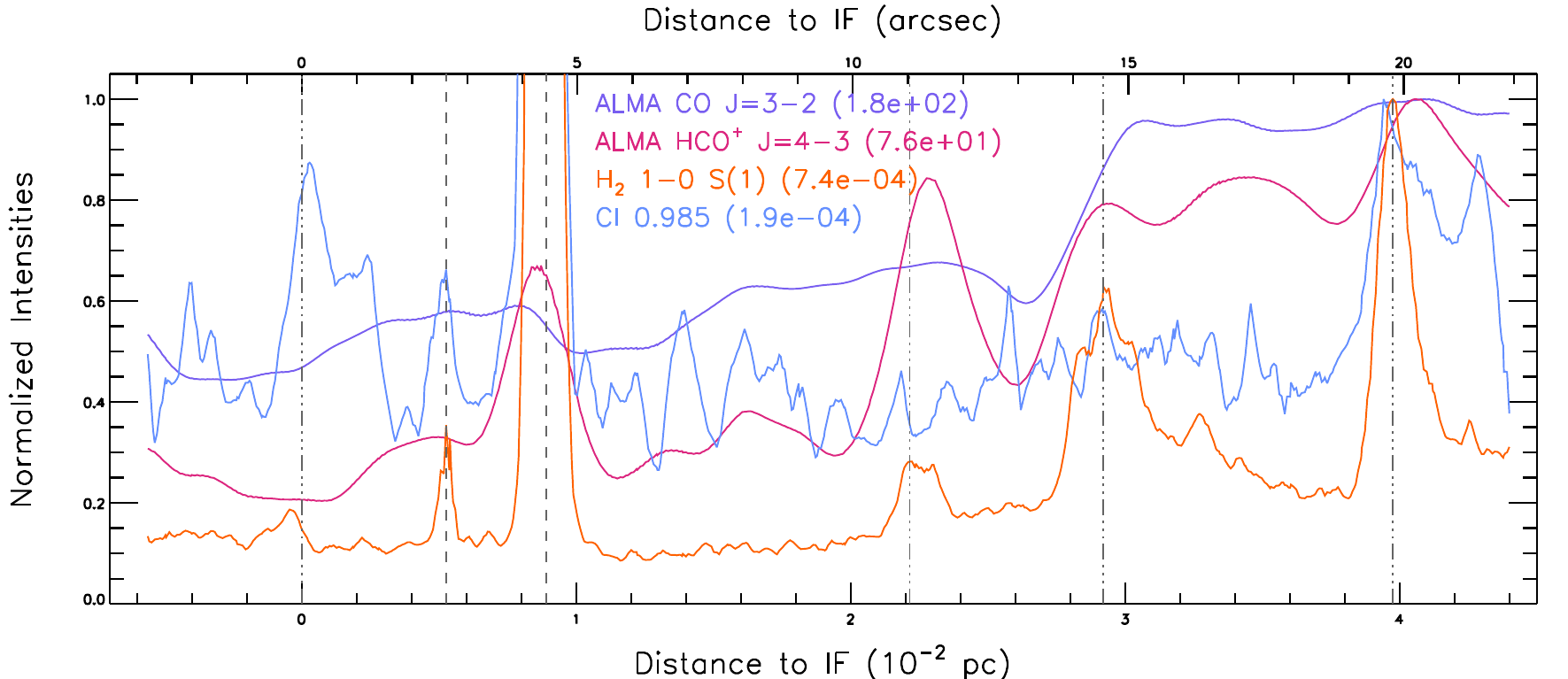}}
\caption{Comparison of the NIRSpec \molh 1-0 S(1) and \CI\ 0.985~\mum observations (in \uint)
 with ALMA observation
at 1\arcsec\, resolution of the HCO$^+$ $J$=4–3 line integrated intensity (in K km s$^{-1}$) and the CO $J$=3-2 line peak temperatures (T$_{peak}$) \citet{goicoechea}. Shown are normalised line intensities as a function of distance to the IF (0.228~pc or 113.4\arcsec\ from \oric) along a cut crossing the NIRSpec mosaic (see Fig.~\ref{fig:FOV}). As the cut is not perpendicular to the IF and distances are given along the cut, a correction factor of cos(19.58\textdegree)=0.942 needs to be applied to obtain a perpendicular distance from the IF. No extinction correction is applied. The dash-dot-dot-dot vertical lines indicate the position of the IF, \Tc, \Td, and \Te, respectively, from left to right. The dashed vertical lines indicated the location of the proplyds 203-504 (left) and 203-506 (right). }
\label{fig:HCO}
\end{center}
\end{figure}

On top of this large-scale morphology/stratification, we observe numerous smaller-scale structures. The typical size of these structures seem to be largest in the ionised gas tracers (a few arcsecs), whereas in the IF, the atomic and molecular PDR structures of sizes of a few 0.1\arcsec\ are observed. The IF and PDR front is thus highly irregular, non-uniform, and complex. This is also clearly demonstrated in the PDRs4All JWST images \citep{Habart:im}, as well as optical images \citep[e.g.][]{Weilbacher15, Henney2021}. As a consequence, all physical parameters, for example, derived in this paper are very precise, but, at the same time, inaccurate due to their dependence on the exact position (on 0.1\arcsec\ scale) of the intensity of the tracer used to obtain the physical parameter and the incredible small-scale variation observed in these tracers (as well as the assumptions used for the derivation). 
In addition, assuming the \CI\ emission arises solely from radiative recombination and cascade, the analysis of the \CI\ emission in the template spectra indicates the presence of very high density clumps embedded in a lower-density gas. Based on HCO$^+$ $J$=4-3 observations with an angular resolution of $\sim$1\arcsec, \citet{goicoechea} also reported that the gas density near the DF is very inhomogeneous and clumpy with small scale structure surrounding, and parallel with, the dissociation front (the DF set by these authors corresponds to \Td). Along the NIRSpec cut, this HCO$^+$ emission is strongest at \Te and displays slightly weaker emission near \Tc and \Td, whereas it is considerably weaker towards the \Ta and the \Tb (Fig.~\ref{fig:HCO}).

We further note that \Tc behaves uniquely  compared to \Td and \Te.  We summarise:  
\begin{enumerate} 
\item The extinction from the atomic PDR, $A_{V, bar}$, is about 2-3 times higher towards \Tc than towards \Td and \Te (Sect.~\ref{subsec:h2}). 
\item While \Tc is still visible in the AIB emission map, it is indiscernible in the NIRCAM F210M filter \citep[Sect.~\ref{subsec:OB},][]{Habart:im} nor does it exhibit signs of enhanced dust scattered light (Sect.~\ref{sec:spectral-characteristics}).
\item \Td and \Te display a slightly richer molecular inventory, for example, HD $v=1-0$ (Sect.~\ref{sec:spectral-characteristics}) and the presence of 6.850 and 6.943~\mum AIBs \citep{Chown:23}. 
\item The characteristics of the AIB emission towards \Tc are similar to those for the \Ta and the \Tb, but quite different from those observed in \Td and \Te. Specifically, 
\begin{enumerate}
    \item the FWHM of the 3.3 AIB is similar in the \Ta, the \Tb, and \Tc (about 38~cm$^{-1}$), but significantly lower than that of \Td and \Te (41-43~cm$^{-1}$, Sect.~\ref{subsec:AIB}). Similar broadening is observed in \Td and \Te for MIR AIBs \citep[5.25, 6.2, 7.7, 11.2, and 12.7~\mum; ][]{Chown:23}.
    \item the aliphatic-to-aromatic ratio, as probed by the 3.4/3.29 AIB, at \Tc is similar to that of the \Ta and the \Tb, but 1.5-1.75 times lower than in \Td and \Te (Sect.~\ref{subsec:AIB}).
    \item the 11.2~\mum AIB displays a class A profile in the \Ta, the \Tb, and \Tc and a class B profile in \Td and \Te \citep{Chown:23}. 
    \item the aliphatic D/H ratio in \Tc is similar to that in the \Tb and about twice that observed in \Td and \Te (Sect.~\ref{subsec:AIB}).
\end{enumerate}
\item Within the assumptions, the \CI\ emission indicates the presence of clumps. The electron temperature probed in \Tc is similar to that in the \Ta and the \Tb whereas it is significantly lower compared to the electron temperature probed in \Td and \Te. 
\item The CO $J$=3-2 emission intensity is considerably lower in \Tc than in \Td and \Te \citep[and similar to the \Tb and \Ta; Fig.~\ref{fig:HCO};][]{goicoechea}.  
\item The HCO$^+$ $J$=4-3 and CO $J$=3-2 emission velocities at \Tc have two components, one of which is more consistent with emission from the background molecular cloud, OMC-1, than from the Bar \citep{goicoechea}.
\end{enumerate}

Following \citet{goicoechea}, we attribute the small scale variations to preexisting, turbulently-driven density variations in OMC-1 that were amplified by the passage of the shock driven into this core by the stellar feedback of \oric,  perhaps guided by the magnetic field structure that will be enhanced parallel to the shock front. Such small-scale structures can be developed during the passage of the shock due to different types of instabilities \citep{Krasnobaev:16, Krasnobaev:17, Riashchikov:22}. It is tempting to speculate that the high density clumps are sites for future star formation, however \citet{goicoechea} concluded that their current mass is not sufficient to make them gravitationally unstable. 

Furthermore, based on \#1, \#2, and \#7, \citet{Habart:im} concluded that the DF 
displays a terraced-field-like structure where the three DFs are portion of the DF observed edge-on and with \Tc located at a larger distance (from us) compared to \Td and \Te (see Fig.~\ref{fig:geometry}). In addition, based on \#6 and \#7, \citet{goicoechea} suggested that the CO plumes present between the IF and their DF (i.e., \Td) could be CO gas flows that are photo-ablated from the molecular PDR into the atomic PDR. The location of \Tc between the IF and the steep increase seen in the CO $J$=3-2 emission intensity (\#6) thus indicates that the depth in the molecular cloud at \Tc is smaller than at \Td and \Te and does not reach the C$^+$/C/CO transition. At the same time, the HCO$^+$ observations indicate that \Tc (as well as \Td and \Te) is part of the compressed layers \citep[\#5,][]{goicoechea}. 
The distinct AIB properties in \Tc (similar to those in the \Ta and the \Tb templates) with respect to those observed in \Td and \Te then suggests that i) they are characteristic for depths in the PDR shortwards of the C$^+$/C/CO transition, or ii) they are due to an increased (and perhaps dominant) contribution of the atomic PDR to the line-of-sight emission towards \Tc, or iii) both. 

Finally, we note that the Bar is primarily illuminated by \oric. Beyond the Bar, \citet{ODell17} reported that the primary illuminating source is $\theta^2$ Ori A instead. The PDRs4All NIRCAM images \citep[][their Fig. 3]{Habart:im} suggest that the influence from $\theta^2$ Ori A is limited to its nearby environment in the direction of the Bar because of the high density in that direction (see the enhanced emission in a `box' surrounding $\theta^2$ Ori A in the F335M (red) and F470N (green) filters). In addition, from a qualitative perspective, our observations do not show any indication of additional or primarily ionisation due to the radiation field of $\theta^2$ Ori A. In fact, the decrease of [\OI]~1.317~\mum beyond \Tc is consistent with a geometrical dilution model centered on \oric (Sect~\ref{subsubsec:geometry}). A similar NIRSpec study of the immediate surrounding of $\theta^2$ Ori A would further clarify the response of the gas to the $\theta^2$ Ori A radiation field. 
We derived several geometrical distances relevant to obtain a comprehensive 3D picture of the Bar based on the [\OI]~1.317~\mum emission (Sect.~\ref{subsubsec:geometry}). As discussed above, these are specific for the NIRSpec mosaic and may change slightly for other positions on the Bar.

\section{Conclusions}
\label{sec:conclusions}

We present JWST NIRSpec IFU spectral imaging data of the proto-typical PDR in the Orion Nebula, the Bar. Our observations probe the $0.97-5.27$~\mum at a spectral resolution R of $\sim$2700 and approximately cover 3\arcsec\, by 25\arcsec\, at an angular resolution of $0.075\arcsec-0.173\arcsec$. At the distance of the Bar, this is equivalent to \mbox{$1.5-3.46\times\,10^{-4}\,{\rm pc}$}. As such, this unprecedented data set showcases both the large-scale and small-scale structure of the interstellar medium subjected to strong FUV radiation of nearby massive stars. In addition, our mosaic encompasses two proplyds, 203-504 and 203-506, and their associated jets and Herbig-Haro object.\\

These observations reveal a spectacular richness of spectral lines (over 800) and aromatic IR bands on top of weak continuum emission. We detect a forest of atomic and ionic lines as well as numerous \molh ro-vibrational lines. 
We furthermore report the detection of:
\begin{itemize}
    \item \molh pure rotational lines in the vibrational states $\nu=$0, 1, and 2
    \item ro-vibrational lines of HD $v=1-0$
    \item ro-vibrational lines of CO $v=1-0$, $v=2-1$
    \item ro-vibrational lines of CH$^+$ $v=1-0$
    \item vibrational emission of deuterated aromatic hydrocarbons
\end{itemize}
Most of these molecular lines are detected for the first time towards a PDR. We provide a line list to facilitate identification of observed lines in future JWST observations. 
We illustrate the immense diagnostic power provided by the combination of this treasure trove of emission lines and the unprecedented angular resolution through the analysis of the spatial distribution of selected line/band intensities and determine the variations in the physical conditions of the PDR gas and the evolution of complex hydrocarbons.\\

The observations furthermore reveal the anatomy of the Bar: a large-scale morphology or stratification with distance from \oric\ {\it and} numerous smaller-scale structure, some of which were inaccessible with earlier IR observations. The typical size of these structures is largest in the ionised gas tracers and smallest for the molecular gas tracers. We highlight in particular:
\begin{itemize}
    \item the spatially resolved He-IF and the H-IF in the Huygens region for the first time.
    \item the presence of three dissociation fronts (DFs) which show different characteristics. The increasing internal PDR extinction suggests each of the DFs is located increasingly further from us. \citet{Habart:im} posited that the DF surface is a terrace-field-like structure seen from above in which the 3 DFs are seen as edge-on portions of the DF surface. 
    \item the presence of hot (T$\ge$2000~K) irradiated surfaces of dense clumps as indicated by the \CI\ emission assuming it solely arises from radiative recombination and cascade.
    \item the constant density in the atomic PDR.
    \item the varying aromatic-to-aliphatic ratio and width of the AIBs showcasing the photochemical evolution of the AIB carriers, which is driven by the FUV radiation field.
    \item the presence of deuterated aromatic hydrocarbons with considerably stronger intensity in the surface layer of the PDR compared to the molecular PDR indicating the D-enhancement is not inherited but rather a local effect.
    \item  enhanced filamentary \mbox{\OI\ 1.317~\mum} emission in the jets associated with the two proplyds and in the atomic PDR, which may reflect a local gas acceleration zone.
\end{itemize}

Our results showcase the complexity of PDRs, and provide very strong constraints on the evolution of the physico-chemical conditions at the critical H$^+$/H$^0$/\molh transition and the external boundary conditions of dense molecular condensations. As such, the PDRs4All data set serves as the benchmark to extend PDR models in to the JWST era. The analysis of this data set and the numerous tools developed by the PDRs4All team\footnote{https://pdrs4all.org} will assist observers in the analysis of future observations of (unresolved) PDRs, in particular extragalactic objects, while at the same time highlighting the issues encountered when only limited spatial resolution is available.

\begin{acknowledgements}

This work is based on observations made with the NASA/ESA/CSA James Webb Space Telescope. The data were obtained from the Mikulski Archive for Space Telescopes at the Space Telescope Science Institute, which is operated by the Association of Universities for Research in Astronomy, Inc., under NASA contract NAS 5-03127 for JWST. These observations are associated with program \#1288.
Support for program \#1288 was provided by NASA through a grant from the Space Telescope Science Institute, which is operated by the Association of Universities for Research in Astronomy, Inc., under NASA contract NAS 5-03127.

EP and JC acknowledge support from the University of Western Ontario, the Institute for Earth and Space Exploration, the Canadian Space Agency (CSA, 22JWGO1-16), and the Natural Sciences and Engineering Research Council of Canada. JRG and SC thank the Spanish MCINN for funding support under grant 
\mbox{PID2019-106110GB-I00}.
CB is grateful for an appointment at NASA Ames Research Center through the San Jos\'e State University Research Foundation (80NSSC22M0107).
TO acknowledges support from JSPS Bilateral Program, Grant Number 120219939. MGW acknowledges partial support from JWST Theory grant 
JWST-AR-01557.001-A.
Work by YO and MR is carried out within the Collaborative Research Centre 956, sub-project C1, funded by the Deutsche Forschungsgemeinschaft (DFG) – project ID 184018867.

I.A. is funded by a fellowship in the Program of Academic Research Projects Management, PRPI-USP.
MSK is funded by "Russian Science Foundation, grant 21-12-00373.
P.M. acknowledges grants EUR2021-122006, TED2021-129416A-I00 and PID2021-125309OA-I00 funded by MCIN/AEI/ 10.13039/501100011033 and European Union NextGenerationEU/PRTR.
NN is funded by the United Arab Emirates University (UAEU) through UAEU Program for Advanced Research (UPAR) grant G00003479. 
AP acknowledges financial assistance from the Banaras Hindu University’s IoE grant (R/Dev/D/IoE/Incentive/2021- 22/32439) as well as funding from the SERB, New Delhi through Core Research Grant (CRG/2021/000907)
and IUCAA, Pune for associateship.
HZ acknowledges support from the Swedish Research Council (contract No 2020-03437).

\end{acknowledgements}

\clearpage
\onecolumn
\renewcommand{\arraystretch}{1.5}
\begin{longtable}{l l l l l l l}
\caption{The intensities (observed and corrected for extinction using foreground and intermingled formalisms) and column densities of the \molh lines in $v$=0 and $v$=1 states as observed in the five templates.} \label{tab:fluxes} \\
\hline \hline 
Line & Wavelength & Observed intensity & \multicolumn{2}{c}{Extinction corrected intensity} & \multicolumn{2}{c}{Column density}\\
 & ($\mu$m) &  
 \multicolumn{3}{c}{($\times 10^{-5}$ erg cm$^{-2}$s$^{-1}$sr$^{-1}$)} & \multicolumn{2}{c}{(cm$^{-2}$)}\\
 & & & Foreground & Intermingled & Foreground & Intermingled\\
      \hline 
\endfirsthead
\caption{continued.}\\
\hline \hline
Line & Wavelength & Observed intensity & \multicolumn{2}{c}{Extinction corrected intensity} & \multicolumn{2}{c}{Column density}\\
 & ($\mu$m) &  
 \multicolumn{3}{c}{($\times 10^{-5}$ erg cm$^{-2}$s$^{-1}$sr$^{-1}$)} & \multicolumn{2}{c}{(cm$^{-2}$)}\\
 & & & Foreground & Intermingled & Foreground & Intermingled \\
      \hline 
\endhead
\endlastfoot
\multicolumn{7}{c}{\textbf{\HII\ region}}\\
      0-0  S(8) & 5.053115155 & 2.50$_{2.41}^{2.59}$ & 2.69$_{2.60}^{2.79}$ & -- & 2.66 $_{2.57}^{2.75}$ & --  \\
0-0  S(9) & 4.694613923 & 5.72$_{5.43}^{6.01}$ & 6.20$_{5.88}^{6.51}$ & -- & 3.76$_{3.57}^{3.94}$ & --  \\
0-0  S(10) & 4.409790972 & 1.01$_{0.89}^{1.14}$ & 1.10$_{0.97}^{1.24}$ & -- & 0.44$_{0.39}^{0.49}$ & -- \\
0-0  S(11) & 4.181077199 & 3.34$_{3.14}^{3.53}$ & 3.66$_{3.44}^{3.87}$ & -- & 1.00$_{0.95}^{1.06}$ & -- \\
0-0  S(12) & 3.996146626 & 0.79$_{0.69}^{0.88}$ & 0.87$_{0.76}^{0.97}$ & -- & 0.17$_{0.15}^{0.19}$ & -- \\
0-0  S(13) & 3.846113193 & 2.14$_{2.06}^{2.21}$ & 2.37$_{2.28}^{2.45}$ & -- & 0.36$_{0.34}^{0.37}$ & --  \\
0-0  S(15) & 3.626166224 & 1.53$_{1.39}^{1.68}$ & 1.71$_{1.55}^{1.87}$ & -- & 0.16$_{0.15}^{0.18}$ & --  \\
1-0 Q(3) & 2.423729703 & 8.25$_{7.50}^{9.00}$ & 10.05$_{9.13}^{10.96}$ & -- & 5.45$_{4.96}^{5.95}$ & -- \\
1-0 Q(4) & 2.437489361 & 2.49$_{2.26}^{2.72}$ & 3.02$_{2.74}^{3.31}$ & -- & 1.71$_{1.56}^{1.87}$ & -- \\
1-0 Q(5) & 2.454751431 & 4.81$_{4.39}^{5.22}$ & 5.83$_{5.33}^{6.34}$ & -- & 3.41$_{3.12}^{3.71}$& -- \\
1-0 Q(7) & 2.499965526 & 2.72$_{2.35}^{3.09}$ & 3.28$_{2.83}^{3.72}$ & -- & 2.04$_{1.76}^{2.32}$ & --\\
1-0 S(0) & 2.223290181 & 3.94$_{3.63}^{4.25}$ & 4.92$_{4.54}^{5.31}$ & --  & 2.74$_{2.53}^{2.96}$ & -- \\
1-0 S(1) & 2.121833725 & 10.69$_{9.58}^{11.80}$ & 13.59$_{12.18}^{15.00}$ & --  & 5.26$_{4.710}^{5.80}$ & -- \\
1-0 S(2) & 2.033757812 & 4.12$_{3.97}^{4.27}$ & 5.32$_{5.13}^{5.51}$ & --  & 1.72$_{1.66}^{1.78}$ & -- \\
1-0 S(3) & 1.957558983 & 7.88$_{7.21}^{8.55}$ & 10.32$_{9.45}^{11.20}$ & --  & 3.04$_{2.78}^{3.30}$ & -- \\
1-0 S(4) & 1.891935929 & 1.83$_{1.59}^{2.06}$ & 2.42$_{2.12}^{2.74}$ & --  & 0.69$_{0.60}^{0.78}$ & -- \\
1-0 S(5) & 1.835759686 & 3.26$_{2.81}^{3.71}$ & 4.39$_{3.79}^{5.00}$ & --  & 1.29$_{1.11}^{1.47}$ & -- \\
1-0 S(7) & 1.747955195 & 0.94$_{0.47}^{1.42}$ & 1.30$_{0.64}^{1.95}$ & --  & 0.48$_{0.24}^{0.72}$ & -- \\
1-0 O(3) & 2.802516418 & 12.51$_{11.73}^{13.29}$ & 14.66$_{13.74}^{15.57}$ & -- & 6.16$_{5.77}^{6.54}$& -- \\
1-0 O(4) & 3.00386809 & 3.55$_{3.31}^{3.79}$ & 4.09$_{3.82}^{4.37}$ & -- & 2.69$_{2.51}^{2.87}$ & -- \\
1-0 O(5) & 3.23498763 & 4.69$_{4.52}^{4.86}$ & 5.33$_{5.14}^{5.52}$ & -- & 5.24$_{5.05}^{5.43}$ & -- \\
1-0 O(7) & 3.807418799 & 1.45$_{1.36}^{1.54}$ & 1.61$_{1.51}^{1.70}$ & -- & 3.65 $_{3.43 }^{3.86}$ & -- \\
1-1 S(9) & 4.954095222 & 0.84$_{0.75}^{0.92}$ & 0.90$_{0.81}^{0.99}$ & -- & 0.65$_{0.58}^{0.71}$ & -- \\
1-1 S(11) & 4.41661058 & 0.98$_{0.92}^{1.04}$ & 1.07$_{1.001}^{1.13}$ & -- & 0.35$_{0.33}^{0.38}$ & -- \\
1-1 S(12) & 4.223666262 & 0.21$_{0.15}^{0.26}$ & 0.23$_{0.17}^{0.29}$ & -- & 0.06$_{0.04}^{0.07}$ & -- \\
1-1 S(15) & 3.840506053 & 0.31$_{0.24}^{0.38}$ & 0.34$_{0.27}^{0.43}$ & -- & 0.04$_{0.03}^{0.05}$ & -- \\
1-1 S(19) & 3.619804857 & 0.46$_{0.40}^{0.52}$ & 0.52$_{0.45}^{0.59}$ & -- & 0.04$_{0.03}^{0.04}$ & -- \\
\hline\\[-5pt]
\multicolumn{7}{c}{\textbf{Atomic PDR}}\\
0-0 S(8) & 5.053115155 & 2.19$_{2.03}^{2.34}$ & 2.90$_{2.68}^{3.11}$ & 2.91$_{2.70}^{3.13}$ & 2.86$_{2.65}^{3.07}$ & 2.88$_{2.67}^{3.09}$\\
0-0 S(9) & 4.694613923 & 5.60$_{5.33}^{5.88}$ & 7.58 $_{7.20}^{7.95}$ & 7.62$_{7.25}^{8.00}$ & 4.59$_{4.37 }^{4.82}$ & 4.62$_{4.39}^{4.85}$ \\
0-0 S(10) & 4.409790972 & 1.26$_{1.19}^{1.33}$ & 1.73 $_{1.64}^{1.83}$ & 1.74$_{1.65}^{1.84}$ & 0.69$_{0.65}^{0.73}$ & 0.69$_{0.65}^{0.73}$ \\
0-0 S(11) & 4.181077199 & 3.42$_{3.24}^{3.60}$ & 4.81 $_{4.56}^{5.06}$ & 4.84$_{4.59}^{5.09}$ & 1.32$_{1.25 }^{1.39}$ & 1.33$_{1.26}^{1.40}$ \\
0-0 S(12) & 3.996146626 & 0.62$_{0.45}^{0.79}$ & 0.89$_{0.64}^{1.14}$ & 0.89$_{0.64}^{1.14}$ & 0.18$_{0.13}^{0.23}$ & 0.18$_{0.13}^{0.23}$ \\
0-0 S(13) & 3.846113193 & 2.21$_{2.13}^{2.29}$ & 3.23 $_{3.11}^{3.35}$ & 3.24$_{3.12}^{3.36}$ & 0.49$_{0.47}^{0.50}$ & 0.49$_{0.47}^{0.51}$ \\
0-0 S(14) & 3.724425896 & 0.97$_{0.84}^{1.10}$ & 1.44$_{1.24}^{1.64}$ & 1.45$_{1.25}^{1.65}$ & 0.17$_{0.15}^{0.19}$ & 0.17$_{0.15}^{0.19}$ \\
0-0 S(15) & 3.626166224 & 2.02$_{1.90}^{2.13}$ & 3.04$_{2.87}^{3.22}$ & 3.05$_{2.88}^{3.23}$ & 0.29$_{0.27}^{0.31}$ & 0.29$_{0.27}^{0.31}$ \\
1-0 Q(1) & 2.406591889 & 12.65$_{12.27}^{13.03}$ & 26.67$_{25.87}^{27.47}$ & 26.11$_{25.33}^{26.90}$ & 9.46$_{9.17}^{9.74}$ & 9.26$_{8.98}^{9.54}$ \\
1-0 Q(2) & 2.413438823 & 3.88$_{3.51}^{4.24}$ & 8.15 $_{7.39}^{8.92}$ & 7.99$_{7.23}^{8.74}$ & 4.08$_{3.70 }^{4.46}$ & 4.00$_{3.62}^{4.37}$ \\
1-0 Q(3) & 2.423729703 & 7.56$_{7.22}^{7.90}$ & 15.82$_{15.11}^{16.53}$ & 15.50$_{14.81}^{16.20}$ & 8.59$_{8.21}^{8.98}$ & 8.42$_{8.04}^{8.80}$ \\
1-0 Q(4) & 2.437489361 & 2.01$_{1.85}^{2.18}$ & 4.19 $_{3.84}^{4.54}$ & 4.11$_{3.76}^{4.45}$ & 2.37$_{2.18 }^{2.57}$ & 2.33$_{2.13}^{2.52}$ \\
1-0 Q(5) & 2.454751431 & 4.20$_{3.78}^{4.63}$ & 8.67 $_{7.80}^{9.54}$ & 8.51$_{7.65}^{9.37}$ & 5.08$_{4.57 }^{5.59}$ & 4.98$_{4.48}^{5.48}$ \\
1-0 Q(6) & 2.475558987 & 1.04$_{0.73}^{1.34}$ & 2.12 $_{1.50}^{2.74}$ & 2.08$_{1.48}^{2.69}$ & 1.28 $_{0.91}^{1.66}$ & 1.26$_{0.89}^{1.63}$ \\
1-0 Q(7) & 2.499965526 & 2.48$_{2.20}^{2.76}$ & 5.02 $_{4.45}^{5.59}$ & 4.94$_{4.38}^{5.49}$ & 3.12$_{2.77 }^{3.48}$ & 3.07$_{2.73}^{3.42}$ \\
1-0 Q(9) & 2.559850543 & 1.25$_{0.97}^{1.53}$ & 2.47 $_{1.91}^{3.03}$ & 2.43$_{1.88}^{2.98}$ & 1.64$_{1.27 }^{2.01}$ & 1.62$_{1.25}^{1.98}$ \\
1-0 S(0) & 2.223290181 & 3.04$_{2.53}^{3.55}$ & 7.04 $_{5.86}^{8.22}$ & 6.82$_{5.67}^{7.96}$ & 3.92$_{3.26 }^{4.58}$ & 3.80$_{3.16}^{4.43}$ \\
1-0 S(1) & 2.121833725 & 8.16 $_{7.32}^{9.00}$ & 20.07$_{18.00}^{22.14}$ & 19.27$_{17.29}^{21.26}$ & 7.76 $_{6.96}^{8.56}$ & 7.46$_{6.69}^{8.23}$ \\
1-0 S(2) & 2.033757812 & 2.90$_{2.73}^{3.07}$ & 7.57 $_{7.13}^{8.01}$ & 7.21 $_{6.79}^{7.63}$ & 2.45 $_{2.30}^{2.59}$ & 2.33$_{2.19}^{2.46}$ \\
1-0 S(3) & 1.957558983 & 5.87$_{5.51}^{6.24}$ & 16.19$_{15.19}^{17.20}$ & 15.29$_{14.34}^{16.24}$ & 4.77 $_{4.47}^{5.06}$ & 4.50$_{4.22}^{4.78}$ \\
1-0 S(4) & 1.891935929 & 1.16 $_{1.00}^{1.32}$ & 3.36 $_{2.90}^{3.82}$ & 3.15$_{2.71}^{3.58}$ & 0.96$_{0.83}^{1.09}$ & 0.90$_{0.77}^{1.02}$ \\
1-0 S(5) & 1.835759686 & 2.20 $_{1.91}^{2.48}$ & 6.69 $_{5.82}^{7.56}$ & 6.21$_{5.41}^{7.01}$ & 1.97$_{1.72 }^{2.22}$ & 1.82$_{1.59}^{2.06}$ \\
1-0 O(3) & 2.802516418 & 11.79$_{11.15}^{12.43}$ & 21.35$_{20.20}^{22.51}$ & 21.20$_{20.06}^{22.35}$ & 8.97 $_{8.48}^{9.45}$ & 8.90$_{8.42}^{9.39}$ \\
1-0 O(4) & 3.00386809 & 3.37$_{3.12}^{3.61}$ & 5.75 $_{5.33}^{6.18}$ & 5.73$_{5.31}^{6.16}$ & 3.78$_{3.50}^{4.05}$ & 3.76$_{3.49}^{4.04}$ \\
1-0 O(5) & 3.23498763 & 4.69$_{4.56}^{4.82}$ & 7.59 $_{7.38}^{7.79}$ & 7.59$_{7.39}^{7.79}$ & 7.46$_{7.26 }^{7.66}$ & 7.46$_{7.26}^{7.66}$ \\
1-0 O(7) & 3.807418799 & 1.46$_{1.38}^{1.53}$ & 2.14 $_{2.03}^{2.25}$ & 2.15$_{2.04}^{2.26}$ & 4.86$_{4.61 }^{5.11}$ & 4.88$_{4.63}^{5.13}$\\
1-1 S(9) & 4.954095222 & 1.07$_{0.95}^{1.20}$ & 1.43 $_{1.26}^{1.60}$ & 1.44$_{1.27}^{1.61}$ & 1.03 $_{0.90}^{1.15 }$ & 1.03$_{0.91}^{1.16}$\\
1-1 S(11) & 4.41661058 & 0.99$_{0.94}^{1.03}$ & 1.36 $_{1.30}^{1.42}$ & 1.37$_{1.31}^{1.43}$ & 0.45$_{0.43}^{0.47}$ & 0.45$_{0.43}^{0.48}$ \\
1-1 S(14) & 3.941609087 & 0.18$_{0.13}^{0.22}$ & 0.26$_{0.20}^{0.31}$ & 0.26$_{0.20}^{0.31}$ & 0.04$_{0.029}^{0.046}$ & 0.038$_{0.029}^{0.046}$\\
1-1 S(15) & 3.840506053 & 0.56$_{0.49}^{0.62}$ & 0.81$_{0.72}^{0.91}$ & 0.82$_{0.73}^{0.91}$ & 0.099$_{0.088}^{0.11}$ & 0.10$_{0.09}^{0.11}$ \\
1-1 S(19) & 3.619804857 & 0.48$_{0.43}^{0.52}$ & 0.72$_{0.66}^{0.79}$ & 0.73$_{0.66}^{0.79}$ & 0.05$_{0.05}^{0.06}$ & 0.05$_{0.04}^{0.06}$ \\
\hline\\[-5pt]
\multicolumn{7}{c}{\textbf{DF~1}}\\
0-0 S(8) & 5.053115155 & 5.72 $_{5.41}^{6.02}$ & 9.65 $_{9.13}^{10.16}$ & 14.91$_{14.11}^{15.70}$ & 9.53 $_{9.02}^{10.03}$ & 14.73$_{13.94}^{15.51}$ \\
0-0 S(9) & 4.694613923 & 14.30$_{13.92}^{14.69}$ & 25.06$_{24.38}^{25.73}$ & 39.43$_{38.36}^{40.49}$ & 15.19$_{14.78}^{15.60}$ & 23.90$_{23.25}^{24.54}$ \\
0-0 S(10) & 4.409790972 & 2.99$_{2.79}^{3.20}$ & 5.45 $_{5.08}^{5.82}$ & 8.72$_{8.13}^{9.31}$ & 2.16 $_{2.02 }^{2.31}$ & 3.46$_{3.23}^{3.70}$ \\
0-0 S(11) & 4.181077199 & 8.56$_{7.95}^{9.16}$ & 16.17$_{15.02}^{17.32}$ & 26.25$_{24.38}^{28.11}$ & 4.44$_{4.13}^{4.76}$ & 7.21$_{6.70}^{7.72}$ \\
0-0 S(12) & 3.996146626 & 1.97$_{1.79}^{2.16}$ & 3.87 $_{3.51}^{4.23}$ & 6.36$_{5.76}^{6.95}$ & 0.77$_{0.70}^{0.84}$ & 1.27$_{1.15}^{1.39}$ \\
0-0 S(13) & 3.846113193 & 5.48$_{5.34}^{5.62}$ & 11.11$_{10.83}^{11.39}$ & 18.45$_{17.98}^{18.91}$ & 1.67$_{1.63}^{1.71}$ & 2.78$_{2.71}^{2.85}$ \\
0-0 S(15) & 3.626166224 & 3.45$_{3.27}^{3.63}$ & 7.40 $_{7.01}^{7.79}$ & 12.48$_{11.82}^{13.14}$ & 0.71$_{0.67}^{0.74}$ & 1.19$_{1.13}^{1.25}$ \\
0-0 S(17) & 3.485802957 & 0.94$_{0.77}^{1.12}$ & 2.11 $_{1.72}^{2.51}$ & 3.60$_{2.93}^{4.26}$ & 0.14$_{0.12}^{0.17}$ & 0.24$_{0.20}^{0.29}$ \\
1-0 Q(1) & 2.406591889 & 32.22$_{30.93}^{33.52}$ & 129.0$_{123.8}^{134.2}$ & 218.7$_{209.9}^{227.5}$ & 45.73$_{43.90}^{47.57}$ & 77.53$_{74.43}^{80.64}$ \\
1-0 Q(2) & 2.413438823 & 8.81$_{8.60}^{9.02}$ & 35.05$_{34.23}^{35.87}$ & 59.49$_{58.09}^{60.88}$ & 17.54$_{17.13}^{17.95}$ & 29.77$_{29.07}^{30.47}$ \\
1-0 Q(3) & 2.423729703 & 17.91$_{17.43}^{18.40}$ & 70.67$_{68.74}^{72.59}$ & 120.1$_{116.8}^{123.4}$ & 38.37$_{37.32}^{39.41}$ & 65.22$_{63.44}^{66.99}$ \\
1-0 Q(4) & 2.437489361 & 4.34$_{3.94}^{4.74}$ & 16.92$_{15.36}^{18.48}$ & 28.82$_{26.16}^{31.47}$ & 9.589 $_{8.706 }^{10.47}$ & 16.33$_{14.83}^{17.83}$ \\
1-0 Q(5) & 2.454751431 & 9.06$_{8.31}^{9.82}$ & 34.84$_{31.93}^{37.74}$ & 59.46$_{54.51}^{64.42}$ & 20.40$_{18.70}^{22.10}$ & 34.82$_{31.92}^{37.72}$ \\
1-0 Q(6) & 2.475558987 & 1.52$_{1.38}^{1.66}$ & 5.74 $_{5.20}^{6.29}$ & 9.83$_{8.90}^{10.76}$ & 3.47 $_{3.14}^{3.80}$ & 5.94 $_{5.38}^{6.50}$ \\
1-0 Q(7) & 2.499965526 & 4.51 $_{4.01}^{5.01}$ & 16.71$_{14.84}^{18.57}$ & 28.68$_{25.48}^{31.88}$ & 10.41$_{9.246 }^{11.57}$ & 17.86$_{15.87}^{19.86}$ \\
1-0 Q(9) & 2.559850543 & 2.38$_{1.90}^{2.86}$ & 8.44 $_{6.74}^{10.14}$ & 14.57$_{11.64}^{17.51}$ & 5.61 $_{4.48}^{6.74}$ & 9.69 $_{7.74 }^{11.64}$ \\
1-0 S(0) & 2.223290181 & 7.17 $_{6.78}^{7.56 }$ & 34.17$_{32.32}^{36.02}$ & 55.78$_{52.77}^{58.80}$ & 19.02$_{18.00}^{20.05}$ & 31.06$_{29.38}^{32.74}$\\
1-0 S(1) & 2.121833725 & 18.00$_{16.11}^{19.89}$ & 95.98$_{85.89}^{106.1}$ & 152.0$_{136.0}^{168.0}$ & 37.13$_{33.22}^{41.03}$ & 58.79$_{52.61}^{64.97}$ \\
1-0 S(2) & 2.033757812 & 4.98$_{4.74}^{5.21}$ & 29.57$_{28.18}^{30.96}$ & 45.28$_{43.15}^{47.41}$ & 9.558 $_{9.109 }^{10.01}$ & 14.63$_{13.95}^{15.32}$ \\
1-0 S(3) & 1.957558983 & 9.94$_{9.13}^{10.75}$ & 65.49$_{60.14}^{70.84}$ & 96.79$_{88.88}^{104.7}$ & 19.28$_{17.71}^{20.86}$ & 28.50$_{26.17}^{30.82}$ \\
1-0 S(4) & 1.891935929 & 2.08$_{1.79}^{2.37}$ & 15.09$_{13.02}^{17.15}$ & 21.51$_{18.56}^{24.46}$ & 4.31$_{3.72}^{4.90}$ & 6.15$_{5.31}^{6.99}$ \\
1-0 S(5) & 1.835759686 & 3.29$_{2.93}^{3.66}$ & 26.10$_{23.22}^{28.99}$ & 35.94$_{31.96}^{39.91}$ & 7.667 $_{6.819 }^{8.515 }$ & 10.56$_{9.388 }^{11.72}$ \\
1-0 S(7) & 1.747955195 & 0.91$_{0.65}^{1.18}$ & 8.43 $_{5.99}^{10.88}$ & 10.88$_{7.72}^{14.03}$ & 3.13 $_{2.22}^{4.03}$ & 4.03 $_{2.86 }^{5.20 }$ \\
1-0 O(3) & 2.802516418 & 37.32$_{34.73}^{39.91}$ & 112.6$_{104.8}^{120.4}$ & 196.6$_{182.9}^{210.3}$ & 47.27$_{43.99}^{50.56}$ & 82.55$_{76.82}^{88.29}$ \\
1-0 O(4) & 3.00386809 & 9.58 $_{9.08 }^{10.08}$ & 25.94$_{24.58}^{27.30}$ & 45.23$_{42.86}^{47.60}$ & 17.03$_{16.13}^{17.92}$ & 29.69$_{28.14}^{31.25}$ \\
1-0 O(5) & 3.23498763 & 14.15$_{13.69}^{14.62}$ & 34.62$_{33.48}^{35.76}$ & 59.83$_{57.86}^{61.80}$ & 34.03$_{32.91}^{35.15}$ & 58.81$_{56.88}^{60.75}$ \\
1-0 O(7) & 3.807418799 & 3.88$_{3.66}^{4.10}$ & 7.94 $_{7.49}^{8.39}$ & 13.22$_{12.47}^{13.97}$ & 18.02$_{17.00}^{19.04}$ & 30.01$_{28.30}^{31.71}$ \\
1-0 O(8) & 4.162424968 & 0.74$_{0.63}^{0.85}$ & 1.40 $_{1.19}^{1.61}$ & 2.28$_{1.93}^{2.62}$ & 5.01 $_{4.26}^{5.77}$ & 8.15 $_{6.92}^{9.37}$ \\
1-0 O(9) & 4.575480546 & 0.87$_{0.79}^{0.95}$ & 1.54 $_{1.40}^{1.69}$ & 2.45$_{2.22}^{2.68}$ & 9.00$_{8.16 }^{9.85}$ & 14.26$_{12.92}^{15.59}$ \\
1-1 S(9) & 4.954095222 & 2.04$_{1.90}^{2.19}$ & 3.48 $_{3.24}^{3.72}$ & 5.40$_{5.03}^{5.78}$ & 2.49 $_{2.32 }^{2.67}$ & 3.87 $_{3.60 }^{4.14 }$ \\
1-1 S(11) & 4.41661058 & 1.80 $_{1.75 }^{1.85 }$ & 3.27 $_{3.18 }^{3.36 }$ & 5.23 $_{5.09 }^{5.37 }$ & 1.09 $_{1.06 }^{1.12 }$ & 1.74 $_{1.69 }^{1.78 }$ \\
1-1 S(12) & 4.223666262 & 0.60$_{0.55}^{0.65}$ & 1.12 $_{1.03}^{1.21}$ & 1.81$_{1.67}^{1.96}$ & 0.27$_{0.25}^{0.30}$ & 0.44$_{0.41}^{0.48}$ \\
1-1 S(13) & 4.067618249 & 1.68$_{1.53}^{1.83}$ & 3.25 $_{2.95}^{3.54}$ & 5.31$_{4.83}^{5.80}$ & 0.61$_{0.55}^{0.66}$ & 0.99$_{0.90}^{1.08}$ \\
1-1 S(14) & 3.941609087 & 0.29$_{0.23}^{0.36}$ & 0.58$_{0.45}^{0.70}$ & 0.96$_{0.75}^{1.16}$ & 0.09$_{0.07}^{0.10}$ & 0.14$_{0.11}^{0.17}$ \\
1-1 S(15) & 3.840506053 & 0.55$_{0.41}^{0.70}$ & 1.12 $_{0.82}^{1.42}$ & 1.86$_{1.37}^{2.35}$ & 0.14$_{0.10}^{0.17}$ & 0.23$_{0.17}^{0.29}$ \\
1-1 S(16) & 3.760417528 & 0.17$_{0.11}^{0.24}$ & 0.36$_{0.22}^{0.50}$ & 0.60$_{0.37}^{0.83}$ & 0.04$_{0.02}^{0.05}$ & 0.06$_{0.04}^{0.09}$ \\
1-1 S(17) & 3.698368298 & 0.35$_{0.27}^{0.42}$ & 0.73$_{0.57}^{0.89}$ & 1.23$_{0.95}^{1.50}$ & 0.06$_{0.05}^{0.08}$ & 0.11$_{0.08}^{0.13}$ \\
1-1 S(19) & 3.619804857 & 0.60$_{0.52}^{0.68}$ & 1.29 $_{1.13}^{1.46}$ & 2.18$_{1.90}^{2.46}$ & 0.09$_{0.08}^{0.10}$ & 0.16$_{0.14}^{0.18}$ \\
\hline\\[-5pt]
\multicolumn{7}{c}{\textbf{DF~2}}\\
0-0 S(8) & 5.053115155 & 8.27$_{7.74}^{8.80}$ & 11.08$_{10.36}^{11.79}$ & 11.20$_{10.48}^{11.92}$ & 10.94$_{10.24}^{11.64}$ & 11.06$_{10.35}^{11.77}$ \\
0-0 S(9) & 4.694613923 & 20.82$_{20.17}^{21.47}$ & 28.47$_{27.58}^{29.36}$ & 28.79$_{27.88}^{29.69}$ & 17.25$_{16.71}^{17.79}$ & 17.45$_{16.90}^{17.99}$ \\
0-0 S(10) & 4.409790972 & 4.49$_{4.17}^{4.81}$ & 6.27 $_{5.83}^{6.72}$ & 6.34$_{5.89}^{6.79}$ & 2.49$_{2.31 }^{2.67}$ & 2.52$_{2.34}^{2.70}$ \\
0-0 S(11) & 4.181077199 & 13.11$_{12.37}^{13.84}$ & 18.70$_{17.65}^{19.75}$ & 18.90$_{17.84}^{19.96}$ & 5.13$_{4.85}^{5.42}$ & 5.19$_{4.90}^{5.48}$ \\
0-0 S(12) & 3.996146626 & 3.10$_{2.81}^{3.40}$ & 4.52 $_{4.09}^{4.94}$ & 4.57$_{4.14}^{4.99}$ & 0.90$_{0.82}^{0.98}$ & 0.91$_{0.82}^{0.99}$ \\
0-0 S(13) & 3.846113193 & 8.74$_{8.50}^{8.99}$ & 12.96$_{12.60}^{13.32}$ & 13.09$_{12.73}^{13.46}$ & 1.95$_{1.90}^{2.00}$ & 1.97$_{1.92}^{2.03}$ \\
0-0 S(15) & 3.626166224 & 5.29$_{4.92}^{5.65}$ & 8.09 $_{7.53}^{8.65}$ & 8.16$_{7.60}^{8.73}$ & 0.77$_{0.72}^{0.82}$ & 0.78$_{0.72}^{0.83}$ \\
0-0 S(17) & 3.485802957 & 1.99$_{1.73}^{2.25}$ & 3.12 $_{2.71}^{3.53}$ & 3.15$_{2.73}^{3.56}$ & 0.21$_{0.18}^{0.24}$ & 0.21$_{0.19}^{0.24}$ \\
1-0 Q(1) & 2.406591889 & 64.38$_{63.12}^{65.63}$ & 139.6$_{136.8}^{142.3}$ & 137.3$_{134.6}^{140.0}$ & 49.48$_{48.52}^{50.44}$ & 48.68$_{47.73}^{49.63}$\\
1-0 Q(2) & 2.413438823 & 10.94$_{10.42}^{11.47}$ & 23.64$_{22.50}^{24.78}$ & 23.27$_{22.15}^{24.39}$ & 11.83$_{11.26}^{12.40}$ & 11.64$_{11.08}^{12.21}$ \\
1-0 Q(3) & 2.423729703 & 20.94$_{19.93}^{21.95}$ & 45.02$_{42.86}^{47.19}$ & 44.34$_{42.21}^{46.47}$ & 24.45$_{23.27}^{25.62}$ & 24.07$_{22.92}^{25.23}$ \\
1-0 Q(4) & 2.437489361 & 7.192 $_{6.718 }^{7.665 }$ & 15.36$_{14.35}^{16.38}$ & 15.14$_{14.14}^{16.14}$ & 8.71$_{8.13}^{9.28}$ & 8.58$_{8.01}^{9.15}$ \\
1-0 Q(5) & 2.454751431 & 17.90$_{16.42}^{19.39}$ & 37.94$_{34.80}^{41.09}$ & 37.42$_{34.32}^{40.53}$ & 22.22$_{20.38}^{24.06}$ & 21.91$_{20.10}^{23.73}$ \\
1-0 Q(6) & 2.475558987 & 3.93$_{3.76}^{4.10}$ & 8.24 $_{7.88}^{8.60}$ & 8.14$_{7.78}^{8.49}$ & 4.98$_{4.76 }^{5.19}$ & 4.92$_{4.70}^{5.13 }$ \\
1-0 Q(7) & 2.499965526 & 9.16 $_{8.26}^{10.06}$ & 19.03$_{17.16}^{20.90}$ & 18.81$_{16.96}^{20.66}$ & 11.85$_{10.69}^{13.02}$ & 11.72$_{10.56}^{12.87}$ \\
1-0 Q(9) & 2.559850543 & 4.55$_{3.56}^{5.54}$ & 9.21 $_{7.21}^{11.22}$ & 9.13$_{7.14}^{11.12}$ & 6.13 $_{4.79}^{7.46}$ & 6.07$_{4.75}^{7.40}$ \\
1-0 S(0) & 2.223290181 & 14.91$_{14.25}^{15.57}$ & 35.62$_{34.05}^{37.20}$ & 34.62$_{33.09}^{36.15}$ & 19.83$_{18.96}^{20.71}$ & 19.28$_{18.43}^{20.13}$ \\
1-0 S(1) & 2.121833725 & 40.43$_{36.68}^{44.17}$ & 102.8$_{93.32}^{112.4}$ & 99.05$_{89.88}^{108.2}$ & 39.78$_{36.10}^{43.46}$ & 38.32$_{34.77}^{41.86}$\\
1-0 S(2) & 2.033757812 & 11.52$_{11.12}^{11.92}$ & 31.13$_{30.05}^{32.21}$ & 29.70$_{28.67}^{30.73}$ & 10.06$_{9.71}^{10.41}$ & 9.60$_{9.27}^{9.93}$ \\
1-0 S(3) & 1.957558983 & 23.04$_{21.47}^{24.62}$ & 65.96$_{61.45}^{70.46}$ & 62.32$_{58.06}^{66.58}$ & 19.42$_{18.09}^{20.75}$ & 18.35$_{17.09}^{19.60}$ \\
1-0 S(4) & 1.891935929 & 5.49$_{5.27}^{5.71}$ & 16.59$_{15.92}^{17.26}$ & 15.52$_{14.89}^{16.15}$ & 4.74$_{4.55}^{4.93}$ & 4.43$_{4.26}^{4.61}$ \\
1-0 S(5) & 1.835759686 & 7.51$_{6.61}^{8.40}$ & 23.83$_{20.99}^{26.66}$ & 22.08$_{19.45}^{24.71}$ & 7.00 $_{6.17}^{7.83}$ & 6.49$_{5.71}^{7.26 }$ \\
1-0 S(7) & 1.747955195 & 2.80$_{2.25 }^{3.34 }$ & 9.67 $_{7.78 }^{11.57}$ & 8.81 $_{7.08}^{10.53}$ & 3.59 $_{2.89}^{4.29}$ & 3.27$_{2.63}^{3.91}$ \\
1-0 O(3) & 2.802516418 & 69.90$_{65.90}^{73.90}$ & 129.4$_{122.0}^{136.8}$ & 129.3$_{121.9}^{136.7}$ & 54.34$_{51.23}^{57.45}$ & 54.28$_{51.17}^{57.39}$ \\
1-0 O(4) & 3.00386809 & 16.67$_{15.55}^{17.79}$ & 29.06$_{27.11}^{31.01}$ & 29.15$_{27.19}^{31.10}$ & 19.07$_{17.79}^{20.35}$ & 19.13$_{17.85}^{20.42}$ \\
1-0 O(5) & 3.23498763 & 23.53$_{22.87}^{24.19}$ & 38.75$_{37.66}^{39.85}$ & 39.00$_{37.90}^{40.10}$ & 38.09$_{37.02}^{39.17}$ & 38.33$_{37.25}^{39.41}$ \\
1-0 O(7) & 3.807418799 & 5.85$_{5.56}^{6.14}$ & 8.72 $_{8.28}^{9.15}$ & 8.80$_{8.37}^{9.24}$ & 19.79$_{18.80}^{20.77}$ & 19.99$_{18.99}^{20.98}$ \\
1-0 O(8) & 4.162424968 & 1.15$_{1.04}^{1.26}$ & 1.64 $_{1.48}^{1.79}$ & 1.65$_{1.50}^{1.81}$ & 5.86 $_{5.30 }^{6.42}$ & 5.92$_{5.35}^{6.49}$ \\
1-0 O(9) & 4.575480546 & 1.51 $_{1.38 }^{1.63 }$ & 2.08 $_{1.91 }^{2.24 }$ & 2.10 $_{1.93 }^{2.27 }$ & 12.10$_{11.12}^{13.07}$ & 12.23$_{11.24}^{13.22}$ \\
1-1 S(9) & 4.954095222 & 3.25 $_{3.01 }^{3.49 }$ & 4.38 $_{4.06 }^{4.70 }$ & 4.42 $_{4.10 }^{4.75 }$ & 3.14 $_{2.91 }^{3.37 }$ & 3.17 $_{2.94 }^{3.41 }$ \\
1-1 S(11) & 4.41661058 & 3.31 $_{3.19 }^{3.43 }$ & 4.62 $_{4.45}^{4.79}$ & 4.67$_{4.50}^{4.84}$ & 1.53 $_{1.48 }^{1.59}$ & 1.55$_{1.49}^{1.61}$ \\
1-1 S(12) & 4.223666262 & 1.04 $_{0.99}^{1.10 }$ & 1.48 $_{1.40 }^{1.56 }$ & 1.50 $_{1.42 }^{1.58 }$ & 0.36$_{0.34}^{0.38}$ & 0.36$_{0.35}^{0.38}$ \\
1-1 S(14) & 3.941609087 & 0.59$_{0.49}^{0.70}$ & 0.87$_{0.72}^{1.02}$ & 0.88$_{0.73}^{1.03}$ & 0.13$_{0.11}^{0.15}$ & 0.13$_{0.11}^{0.15}$ \\
1-1 S(15) & 3.840506053 & 1.12 $_{0.92}^{1.32}$ & 1.66$_{1.36}^{1.96}$ & 1.68$_{1.38}^{1.98}$ & 0.20$_{0.17}^{0.24}$ & 0.20$_{0.17}^{0.24}$ \\
1-1 S(16) & 3.760417528 & 0.32$_{0.26}^{0.38}$ & 0.48$_{0.38}^{0.57}$ & 0.48$_{0.39}^{0.58}$ & 0.05$_{0.04}^{0.06}$ & 0.05$_{0.04}^{0.06}$ \\
\hline\\[-5pt]
\multicolumn{7}{c}{\textbf{DF~3}}\\
0-0 S(8) & 5.053115155 & 10.10$_{9.38}^{10.83}$ & 11.94$_{11.09}^{12.80}$ & 11.86$_{11.02}^{12.71}$ & 11.80$_{10.96}^{12.64}$ & 11.72$_{10.88}^{12.56}$ \\
0-0 S(9) & 4.694613923 & 24.29$_{23.44}^{25.14}$ & 29.07$_{28.05}^{30.09}$ & 28.85$_{27.85}^{29.86}$ & 17.62$_{17.00}^{18.24}$ & 17.49$_{16.88}^{18.10}$ \\
0-0 S(10) & 4.409790972 & 5.69$_{5.38}^{6.01}$ & 6.89 $_{6.51}^{7.27}$ & 6.84$_{6.46}^{7.22}$ & 2.74 $_{2.58 }^{2.89 }$ & 2.71 $_{2.56 }^{2.86 }$ \\
0-0 S(11) & 4.181077199 & 14.49$_{13.76}^{15.21}$ & 17.76$_{16.87}^{18.65}$ & 17.61$_{16.73}^{18.49}$ & 4.88$_{4.63 }^{5.12 }$ & 4.83 $_{4.59}^{5.08 }$ \\ 
0-0 S(12) & 3.996146626 & 3.45 $_{3.16 }^{3.75 }$ & 4.28 $_{3.92 }^{4.65 }$ & 4.24 $_{3.88 }^{4.61 }$ & 0.85$_{0.78}^{0.93}$ & 0.85$_{0.77}^{0.92}$ \\
0-0 S(13) & 3.846113193 & 9.29 $_{9.05 }^{9.53 }$ & 11.64$_{11.34}^{11.95}$ & 11.53$_{11.23}^{11.83}$ & 1.75 $_{1.71 }^{1.80 }$ & 1.73 $_{1.69 }^{1.78 }$ \\
0-0 S(14) & 3.724425896 & 3.24 $_{2.99 }^{3.49 }$ & 4.11 $_{3.79 }^{4.42 }$ & 4.06 $_{3.75 }^{4.38 }$ & 0.48$_{0.45}^{0.52}$ & 0.48$_{0.44}^{0.52}$ \\
0-0 S(15) & 3.626166224 & 5.31 $_{4.96 }^{5.67 }$ & 6.78 $_{6.33 }^{7.23 }$ & 6.71 $_{6.26 }^{7.16 }$ & 0.65$_{0.60}^{0.69}$ & 0.64$_{0.60}^{0.68}$ \\
0-0 S(16) & 3.547587207 & 1.44 $_{1.21 }^{1.68 }$ & 1.86 $_{1.55 }^{2.17 }$ & 1.84 $_{1.53 }^{2.14 }$ & 0.15$_{0.12}^{0.17}$ & 0.15$_{0.12}^{0.17}$ \\
0-0 S(17) & 3.485802957 & 1.98 $_{1.79 }^{2.17 }$ & 2.56 $_{2.32 }^{2.81 }$ & 2.53 $_{2.29 }^{2.78 }$ & 0.17$_{0.16}^{0.19}$ & 0.17$_{0.16}^{0.19}$ \\
0-0 S(19) & 3.40416291 & 1.19 $_{1.00}^{1.38 }$ & 1.55 $_{1.30 }^{1.80 }$ & 1.53 $_{1.29 }^{1.78 }$ & 0.08$_{0.07}^{0.10}$ & 0.08$_{0.07}^{0.09}$ \\
1-0 Q(1) & 2.406591889 & 76.61$_{74.64}^{78.58}$ & 119.4$_{116.3}^{122.5}$ & 116.5$_{113.5}^{119.5}$ & 42.34$_{41.25}^{43.43}$ & 41.31$_{40.25}^{42.37}$ \\
1-0 Q(4) & 2.437489361 & 4.45 $_{4.10 }^{4.81 }$ & 6.88 $_{6.34 }^{7.43 }$ & 6.72 $_{6.19 }^{7.25 }$ & 3.90 $_{3.59 }^{4.21 }$ & 3.81 $_{3.51 }^{4.11 }$ \\
1-0 Q(5) & 2.454751431 & 16.38$_{14.83}^{17.94}$ & 25.21$_{22.81}^{27.60}$ & 24.62$_{22.28}^{26.96}$ & 14.76$_{13.36}^{16.16}$ & 14.41$_{13.05}^{15.78}$ \\
1-0 Q(6) & 2.475558987 & 3.75 $_{3.60 }^{3.89 }$ & 5.73 $_{5.51 }^{5.95 }$ & 5.60 $_{5.39 }^{5.81 }$ & 3.46 $_{3.33 }^{3.59 }$ & 3.38 $_{3.25 }^{3.51 }$\\
1-0 Q(7) & 2.499965526 & 8.94$_{8.05}^{9.82}$ & 13.59$_{12.25}^{14.93}$ & 13.29$_{11.97}^{14.60}$ & 8.47$_{7.63 }^{9.30 }$ & 8.28 $_{7.46 }^{9.09 }$ \\
1-0 Q(9) & 2.559850543 & 4.90 $_{3.88 }^{5.92 }$ & 7.34 $_{5.82 }^{8.87 }$ & 7.19 $_{5.69 }^{8.68 }$ & 4.88 $_{3.87 }^{5.90 }$ & 4.78 $_{3.79 }^{5.77 }$\\
1-0 S(0) & 2.223290181 & 15.63$_{14.96}^{16.30}$ & 25.77$_{24.66}^{26.87}$ & 25.02$_{23.95}^{26.10}$ & 14.35$_{13.73}^{14.96}$ & 13.93$_{13.34}^{14.53}$ \\
1-0 S(1) & 2.121833725 & 42.16$_{38.44}^{45.88}$ & 72.04$_{65.69}^{78.39}$ & 69.75$_{63.60}^{75.90}$ & 27.87$_{25.41}^{30.32}$ & 26.98$_{24.60}^{29.36}$ \\
1-0 S(2) & 2.033757812 & 11.66$_{11.18}^{12.13}$ & 20.62$_{19.78}^{21.46}$ & 19.90$_{19.09}^{20.72}$ & 6.664 $_{6.39}^{6.94 }$ & 6.43 $_{6.17 }^{6.70 }$ \\
1-0 S(3) & 1.957558983 & 24.40$_{22.80}^{26.00}$ & 44.60$_{41.68}^{47.53}$ & 42.92$_{40.10}^{45.73}$ & 13.13$_{12.27}^{13.99}$ & 12.64$_{11.81}^{13.46}$ \\
1-0 S(4) & 1.891935929 & 5.641 $_{5.418 }^{5.864 }$ & 10.64$_{10.21}^{11.06}$ & 10.20$_{9.80}^{10.61}$ & 3.039 $_{2.92 }^{3.16 }$ & 2.92 $_{2.80 }^{3.03 }$ \\
1-0 S(5) & 1.835759686 & 8.22 $_{7.26 }^{9.19 }$ & 15.95$_{14.08}^{17.83}$ & 15.26$_{13.47}^{17.05}$ & 4.69 $_{4.14 }^{5.24 }$ & 4.48 $_{3.96 }^{5.10 }$ \\
1-0 O(3) & 2.802516418 & 76.13$_{72.13}^{80.14}$ & 108.4$_{102.7}^{114.1}$ & 106.5$_{100.9}^{112.1}$ & 45.52$_{43.12}^{47.91}$ & 44.71$_{42.36}^{47.06}$ \\
1-0 O(4) & 3.00386809 & 14.85$_{13.77}^{15.94}$ & 20.43$_{18.94}^{21.93}$ & 20.12$_{18.65}^{21.59}$ & 13.41$_{12.43}^{14.39}$ & 13.20$_{12.24}^{14.17}$ \\
1-0 O(5) & 3.23498763 & 20.90$_{20.38}^{21.43}$ & 27.83$_{27.14}^{28.53}$ & 27.46$_{26.77}^{28.15}$ & 27.36$_{26.68}^{28.04}$ & 26.99$_{26.32}^{27.67}$\\
1-0 O(7) & 3.807418799 & 5.89$_{5.64 }^{6.14 }$ & 7.41 $_{7.09 }^{7.73 }$ & 7.33 $_{7.02 }^{7.65 }$ & 16.82$_{16.10}^{17.54}$ & 16.65$_{15.94}^{17.36}$ \\
1-0 O(8) & 4.162424968 & 1.18 $_{1.08 }^{1.27 }$ & 1.45 $_{1.33 }^{1.56 }$ & 1.43 $_{1.32 }^{1.55 }$ & 5.18 $_{4.76 }^{5.60 }$ & 5.13 $_{4.72 }^{5.55 }$ \\
1-0 O(9) & 4.575480546 & 1.65 $_{1.53 }^{1.76 }$ & 1.98 $_{1.84 }^{2.12 }$ & 1.96 $_{1.82}^{2.11 }$ & 11.53$_{10.70}^{12.36}$ & 11.44$_{10.62}^{12.27}$\\
1-1 S(9) & 4.954095222 & 3.65 $_{3.26 }^{4.04 }$ & 4.32 $_{3.86 }^{4.78 }$ & 4.29 $_{3.84 }^{4.75 }$ & 3.10 $_{2.77 }^{3.43 }$ & 3.08 $_{2.75 }^{3.40 }$ \\
1-1 S(11) & 4.41661058 & 3.42 $_{3.34 }^{3.50 }$ & 4.14 $_{4.04}^{4.24 }$ & 4.10 $_{4.01 }^{4.20 }$ & 1.37 $_{1.34 }^{1.41 }$ & 1.37 $_{1.33 }^{1.39 }$ \\
1-1 S(12) & 4.223666262 & 0.84$_{0.73}^{0.96}$ & 1.03 $_{0.89}^{1.18 }$ & 1.02 $_{0.88}^{1.17 }$ & 0.25$_{0.22}^{0.29}$ & 0.25$_{0.21}^{0.28}$ \\
1-1 S(13) & 4.067618249 & 1.59 $_{1.30}^{1.88}$ & 1.97 $_{1.61}^{2.32}$ & 1.95 $_{1.60}^{2.30 }$ & 0.37$_{0.30}^{0.43}$ & 0.36$_{0.30}^{0.43}$ \\
1-1 S(14) & 3.941609087 & 0.57$_{0.49}^{0.64}$ & 0.70$_{0.61}^{0.80}$ & 0.70$_{0.60}^{0.79}$ & 0.10$_{0.09}^{0.12}$ & 0.10$_{0.09}^{0.12}$ \\
1-1 S(15) & 3.840506053 & 0.91$_{0.69}^{1.13 }$ & 1.14 $_{0.86}^{1.42 }$ & 1.13 $_{0.85}^{1.41 }$ & 0.14$_{0.10}^{0.17}$ & 0.14$_{0.10}^{0.17}$ \\
1-1 S(16) & 3.760417528 & 0.41$_{0.36}^{0.46}$ & 0.52$_{0.46}^{0.57}$ & 0.51$_{0.45}^{0.57}$ & 0.05$_{0.05}^{0.06}$ & 0.05$_{0.05}^{0.06}$ \\
1-1 S(17) & 3.698368298 & 0.54$_{0.44}^{0.64}$ & 0.68$_{0.56}^{0.81}$ & 0.68$_{0.56}^{0.80}$ & 0.06$_{0.05}^{0.07}$ & 0.06$_{0.05}^{0.07}$ \\
\hline
\end{longtable}
\twocolumn

\begin{appendix}
\section{Data reduction}
\label{app:datareduction}

We list our estimate of the cross-calibration factor between NIRCam and NIRSpec in Table~\ref{tab:xcal} and details about the extraction apertures employed for the five template spectra in Table~\ref{tab:aper}.  The cut employed in the paper connects coordinates (5:35:20.0785, -5:24:57.885) and  (5:35:21.0801, -5:25:31.157) ($\alpha$, $\delta$ (ICRS, J2000)). 

\begin{table}
     \caption{NIRSpec/NIRCam cross-calibration measurements.} 
    \label{tab:xcal}
   \begin{center}
    \begin{tabular}{lr@{}c@{}lr@{}c@{}l}
\hline \hline \\[-5pt]
\multicolumn{1}{c}{Filter} & \multicolumn{3}{c}{$a$\, \tablefootmark{1}} & \multicolumn{3}{c}{$b$\, \tablefootmark{1}} \\ 
& & & & \multicolumn{3}{c}{MJy sr$^{-1}$}   \\
    \multicolumn{1}{c}{(1)} & \multicolumn{3}{c}{(2)} & \multicolumn{3}{c}{(3)} \\ [5pt]
\hline \\[-5pt]
\multicolumn{7}{c}{\textbf{g235h-f170lp}} \\ [5pt]
\hline\\[-5pt]
  F187N & $0.994$ & $\pm$ & $0.001$ & $-19.115$ & $\pm$ & $1.150$  \\
  F210M & $0.897$ & $\pm$ & $0.003$ & $5.252$ & $\pm$ & $0.292$  \\
  F212N & $0.804$ & $\pm$ & $0.007$ & $13.579$ & $\pm$ & $0.474$  \\
Average & $0.9797$ & $\pm$ & $0.0011$ & $6.3578 $ & $\pm$ &  $0.2432$  \\[5pt]
\hline\\[-5pt]
 \multicolumn{7}{c}{\textbf{g395h-f290lp}} \\[5pt]
\hline\\[-5pt]
  F335M & $0.8709$ & $\pm$ & $0.0009$ & $4.6709$ & $\pm$ & $0.4649$  \\
 F405N & $0.9213$&  $\pm$ & $0.0038$ & $1.8698$ & $\pm$ & $2.6086$  \\
 Average & $0.8992$&  $\pm$ & $0.0005$ & $3.0398$ & $\pm$&  $0.1748$  \\[5pt]
 \hline
   \end{tabular}
   \end{center}
   \tablefoot{
    \tablefootmark{1}{Cross-calibration is parameterised by $I_\nu^\mathrm{NIRCam} = a I_\nu^\mathrm{NIRSpec} + b$. We multiply the F100LP and F170LP cubes by the average value of $a$ for F170LP, and we multiply the F290LP cubes by the average value of $a$ for that grating/filter combination. We do not use $b$ in our analysis.}}
\end{table}

\begin{table*}[]
\small
    \caption{Extraction apertures used in this paper.}
    \label{tab:aper}
    \begin{center}
    \begin{tabular}{lllcllc}
    \hline
    \hline\\[-5pt]
        Template & \multicolumn{2}{c}{Center} & \multicolumn{1}{c}{Size} & \multicolumn{1}{c}{PA} & \multicolumn{2}{c}{Projected distance of center to \oric}\\
        & \multicolumn{1}{c}{$\alpha$ (J2000)} & \multicolumn{1}{c}{$\delta$ (J2000)} & \multicolumn{1}{c}{\arcsec$\times$\arcsec} & \multicolumn{1}{c}{\textdegree}&\multicolumn{1}{c}{pc} & \multicolumn{1}{c}{\arcsec}\\
        \hline\\[-5pt]
\Ta & 5:35:20.1545 & -5:24:59.646 & 1.26$\times$2.5 & 43.738 & 0.224 & 111.4 \\
\Tb & 5:35:20.2307 & -5:25:02.555 & 1.7$\times$2.5 & 43.738 & 0.230 & 114.5\\
\Tc & 5:35:20.5105 & -5:25:11.931  &  1.0$\times$1.6 & 50.000 & 0.250 & 124.7 \\
\Td & 5:35:20.6135 & -5:25:14.691 &   1.0$\times$1.6 & 38.000 & 0.257 & 127.8 \\
\Te & 5:35:20.7095 & -5:25:20.351 &   1.0$\times$2.654 & 38.000 & 0.268 & 133.5 \\[5pt]
\hline
\end{tabular}
\end{center}
\end{table*}

\section{Template spectra}
\label{app:template-spectra}

\begin{figure*}
    \centering
    \includegraphics[width=\linewidth]{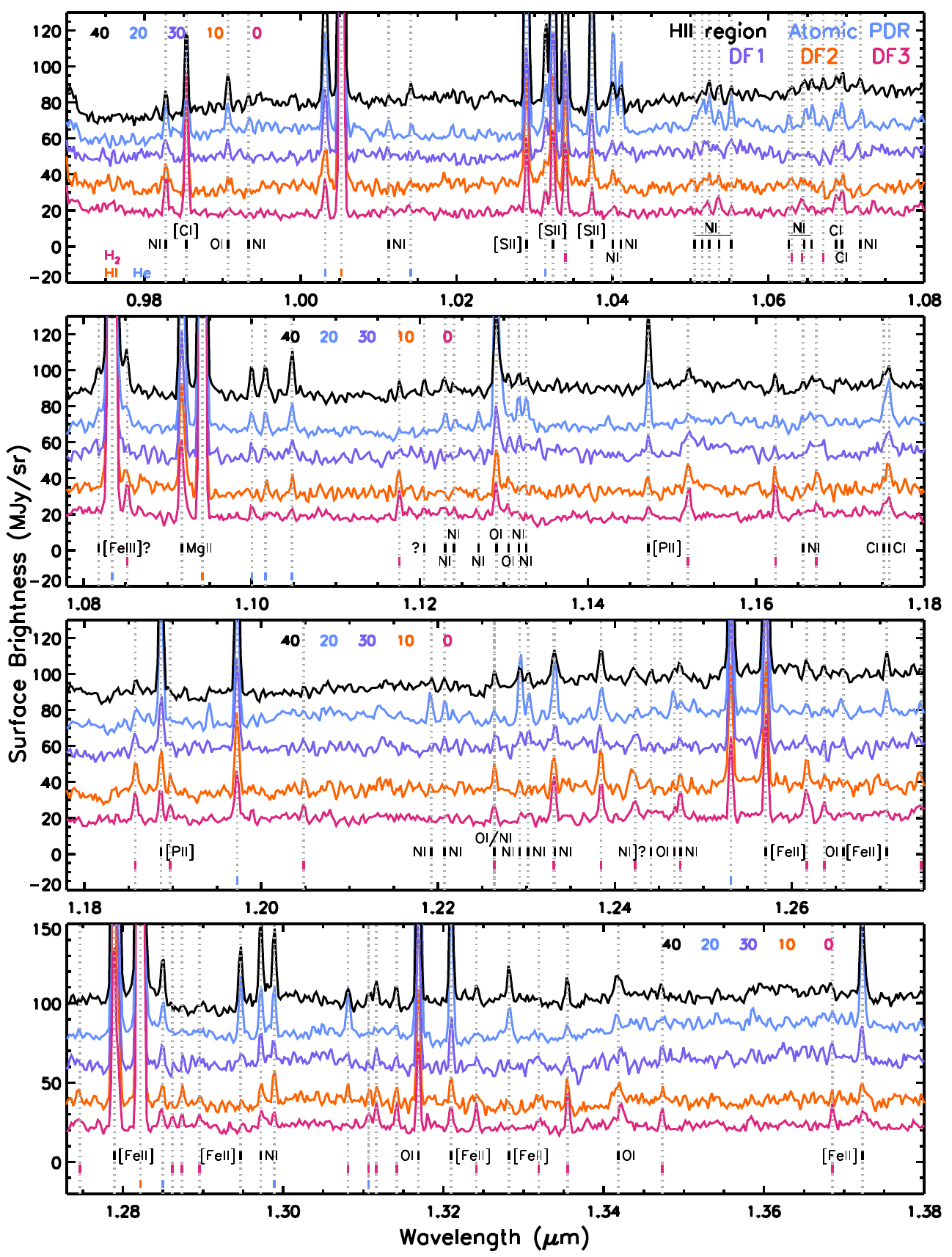}
    \caption{The spectral inventory of the five template spectra. Spectra are offset by the numbers given at the top of the panel. Areas susceptible to the wavelength gap are shown in light grey. The color coding is labeled in the top panel.  } 
    \label{fig:app:template1}
\end{figure*}

\begin{figure*}
    \centering
    \includegraphics[width=\linewidth]{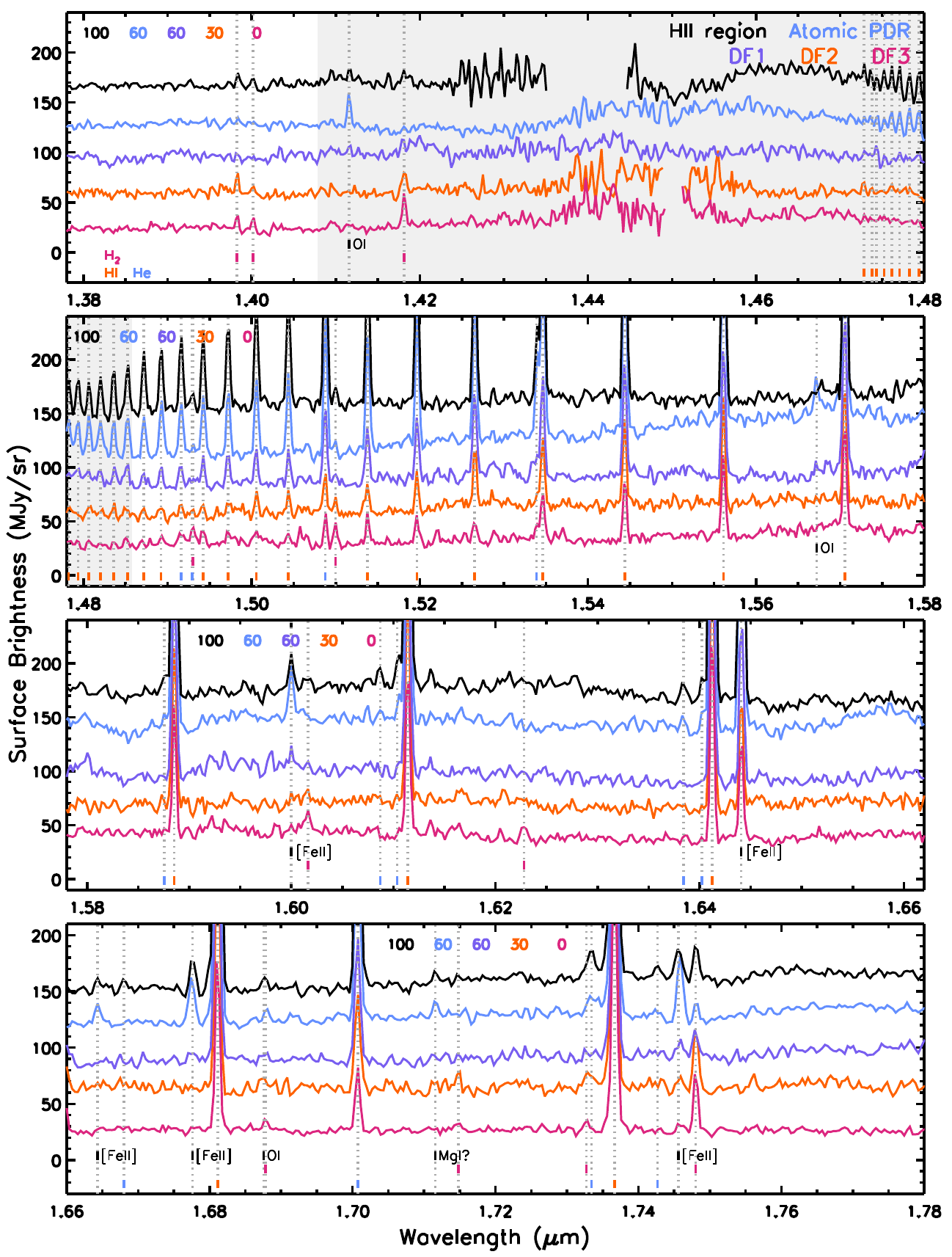}
    \caption{Fig.~\ref{fig:app:template1} continued. } 
    \label{fig:app:template2}
\end{figure*}

\begin{figure*}
    \centering
    \includegraphics[width=\linewidth]{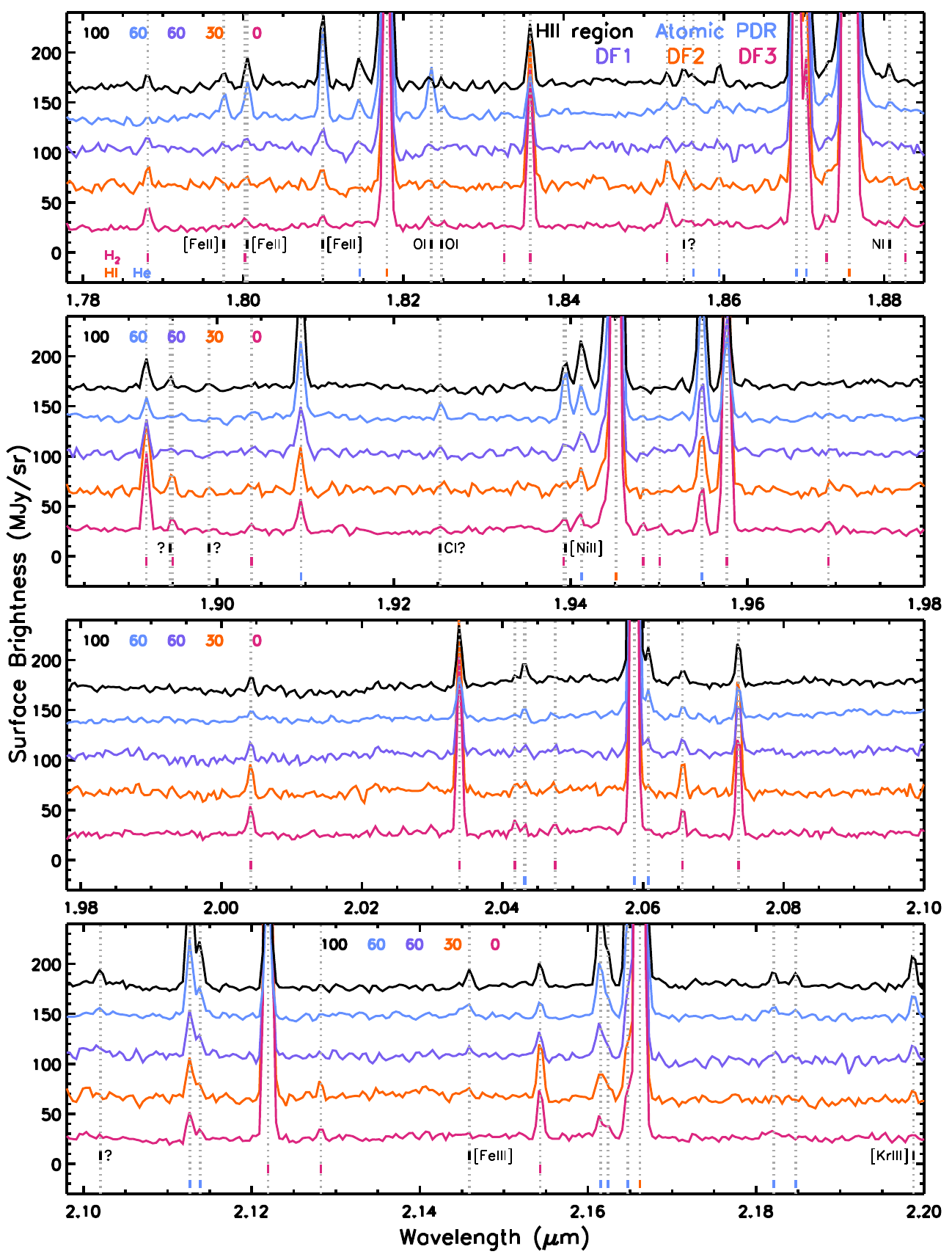}
    \caption{Fig.~\ref{fig:app:template1} continued. } 
    \label{fig:app:template3}
\end{figure*}

\begin{figure*}
    \centering
    \includegraphics[width=\linewidth]{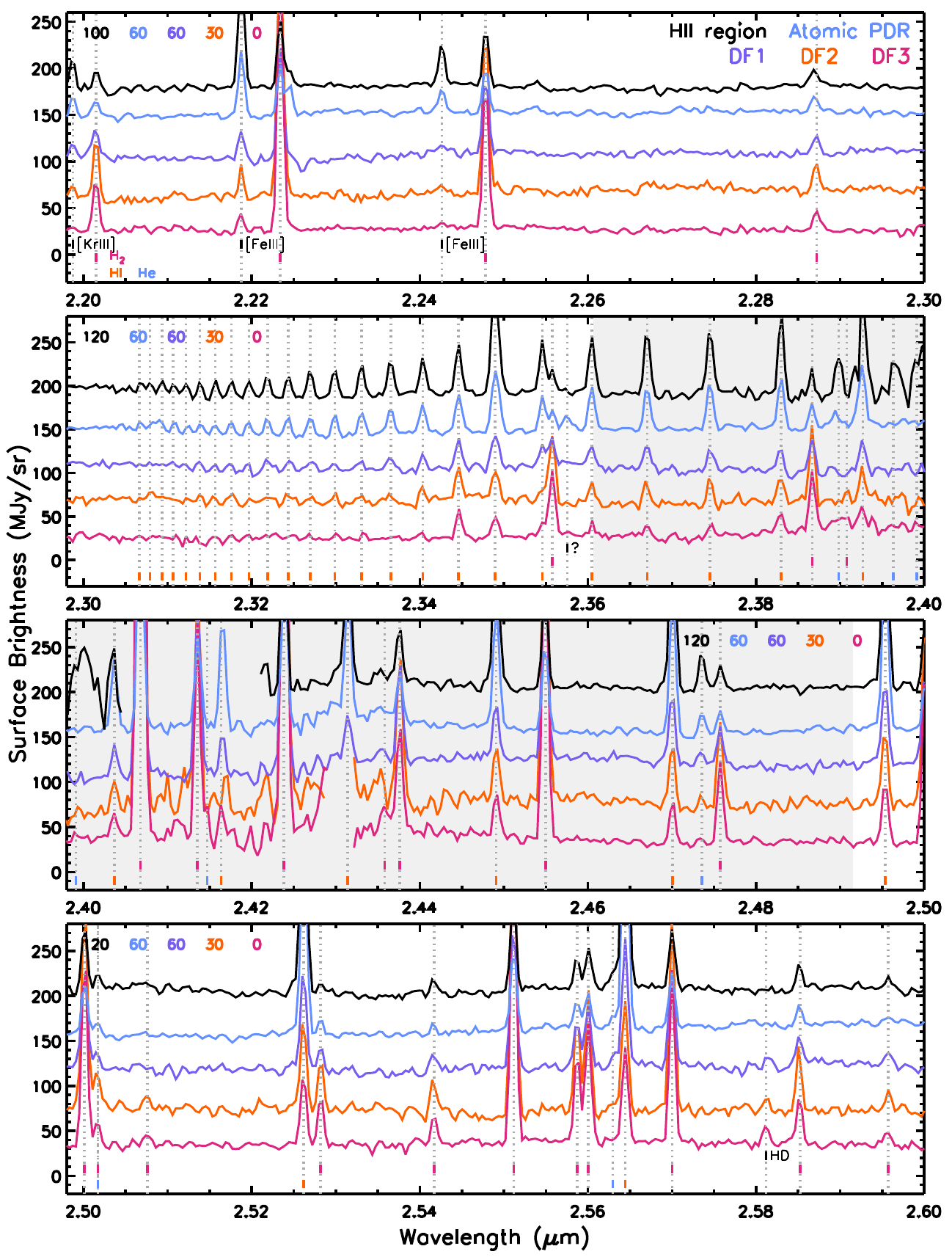}
    \caption{Fig.~\ref{fig:app:template1} continued. } 
    \label{fig:app:template4}
\end{figure*}

\begin{figure*}
    \centering
    \includegraphics[width=\linewidth]{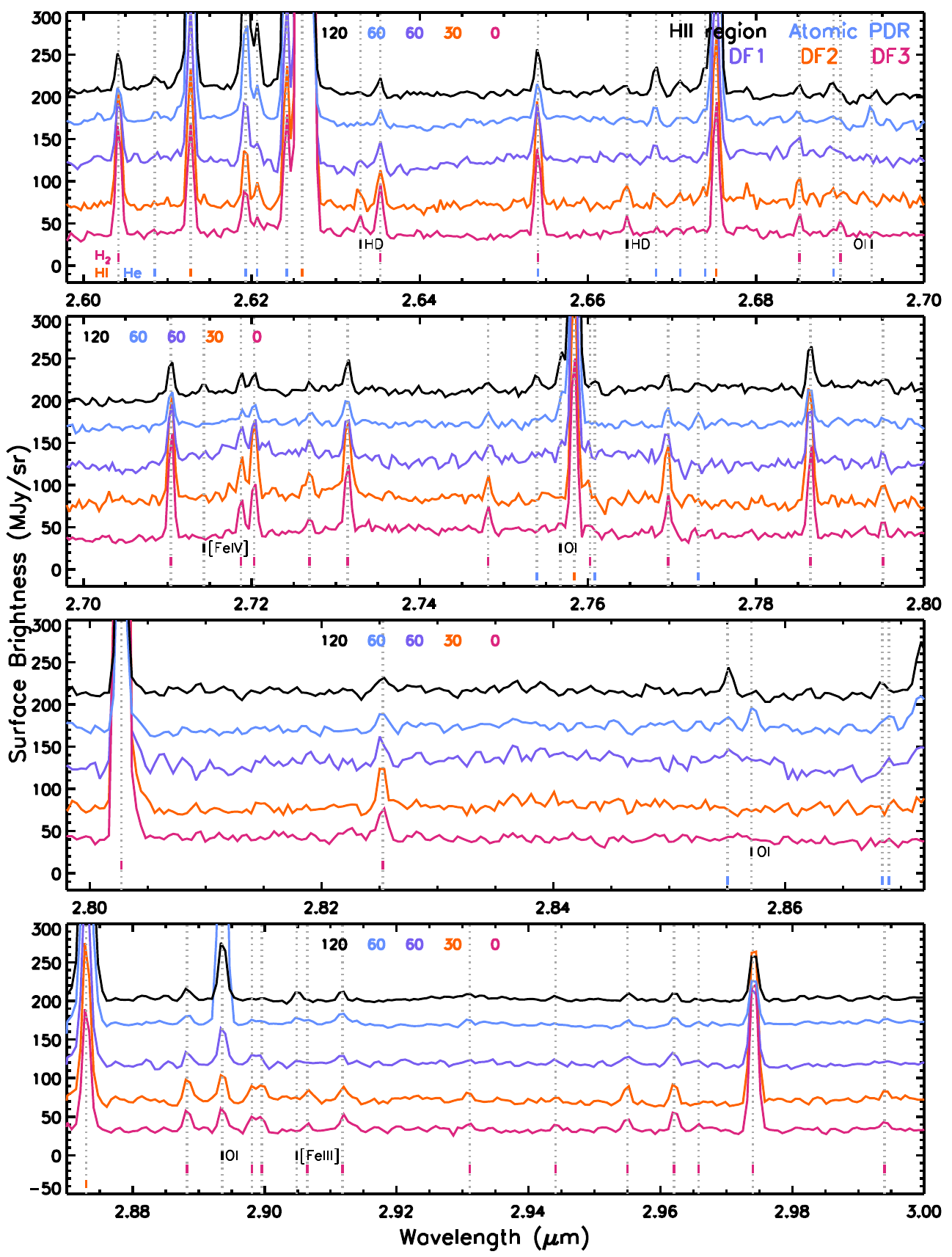}
    \caption{Fig.~\ref{fig:app:template1} continued. } 
    \label{fig:app:template5}
\end{figure*}

\begin{figure*}
    \centering
    \includegraphics[width=\linewidth]{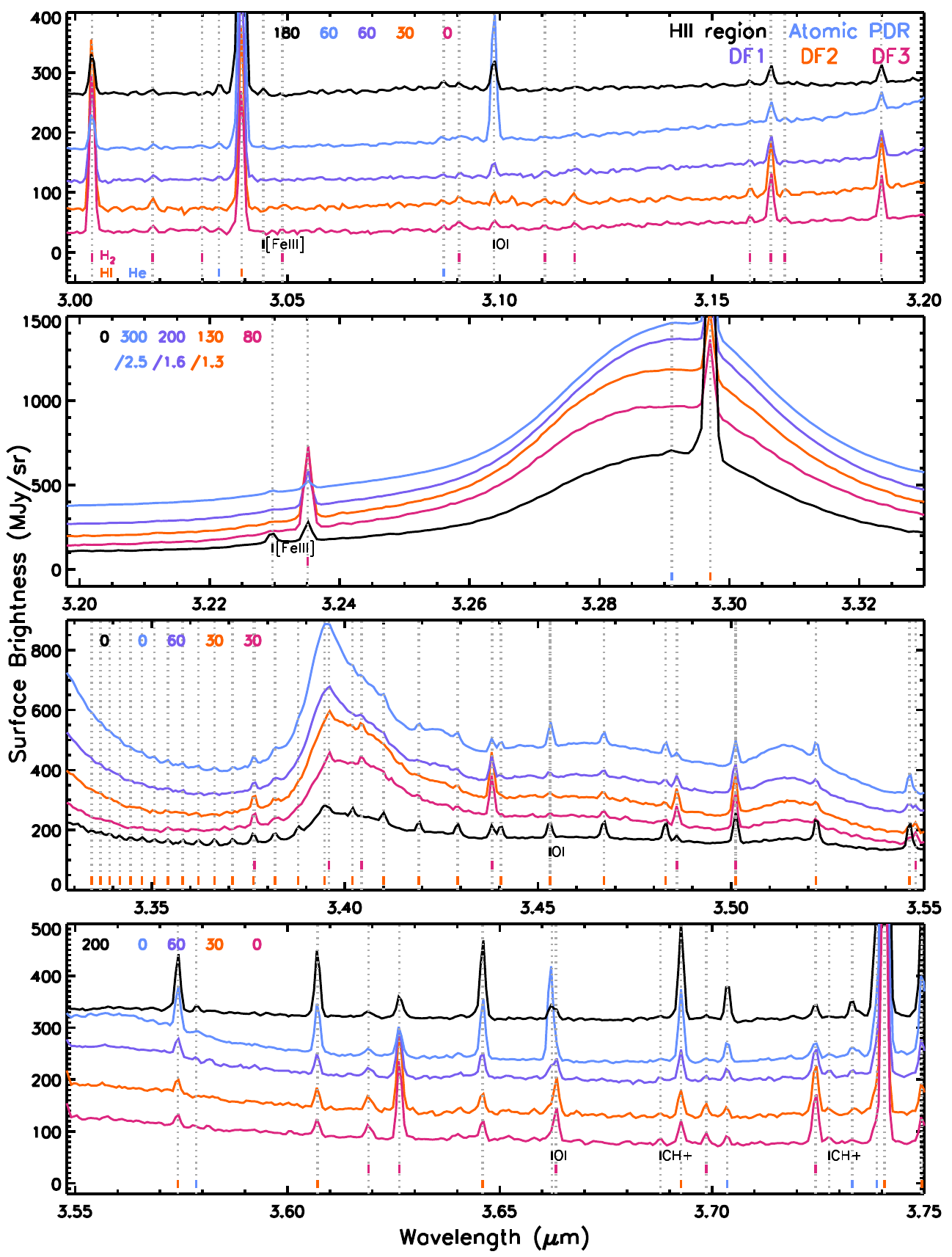}
    \caption{Fig.~\ref{fig:app:template1} continued. } 
    \label{fig:app:template6}
\end{figure*}

\begin{figure*}
    \centering
    \includegraphics[width=\linewidth]{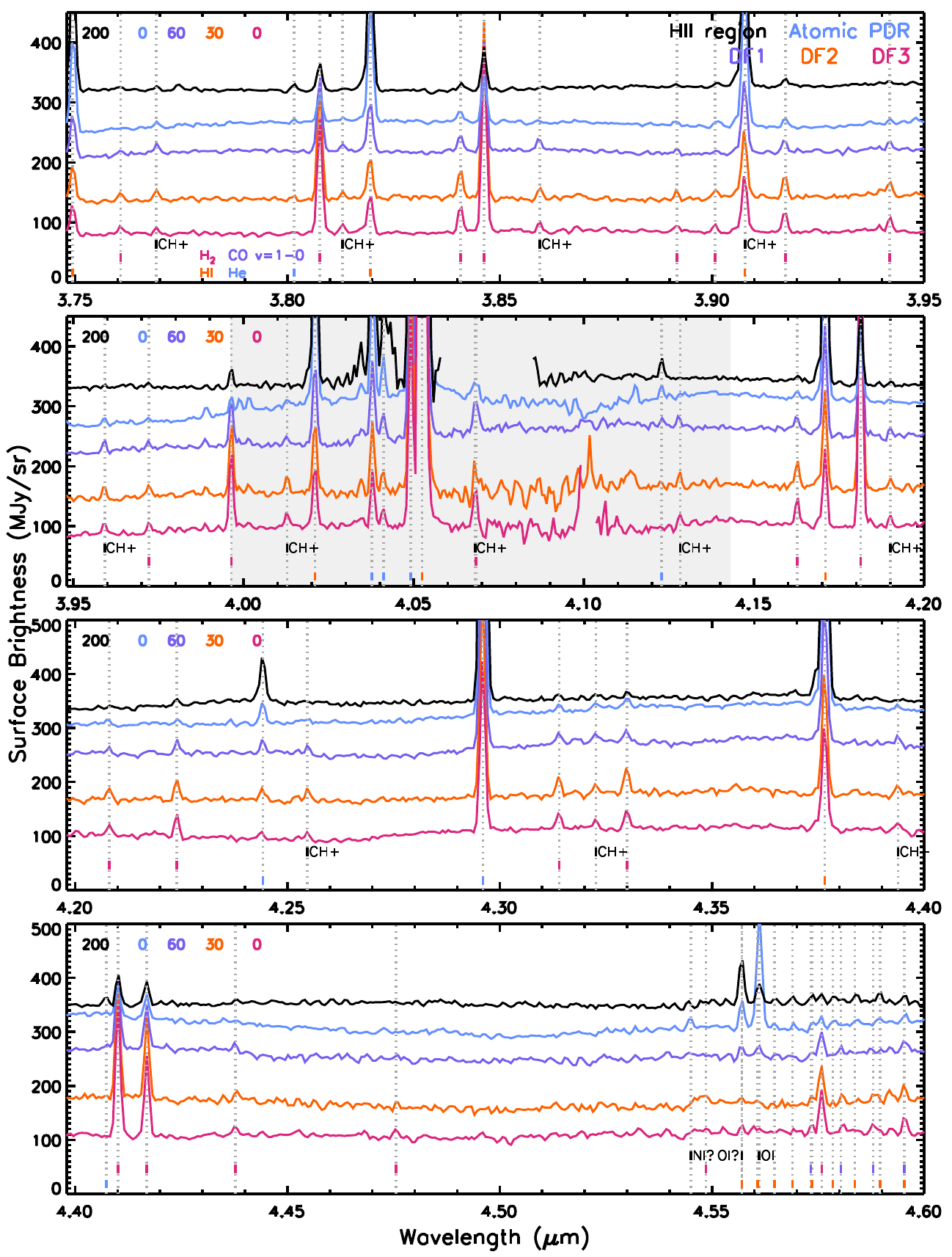}
    \caption{Fig.~\ref{fig:app:template1} continued. } 
    \label{fig:app:template7}
\end{figure*}

\begin{figure*}
    \centering
    \includegraphics[width=\linewidth]{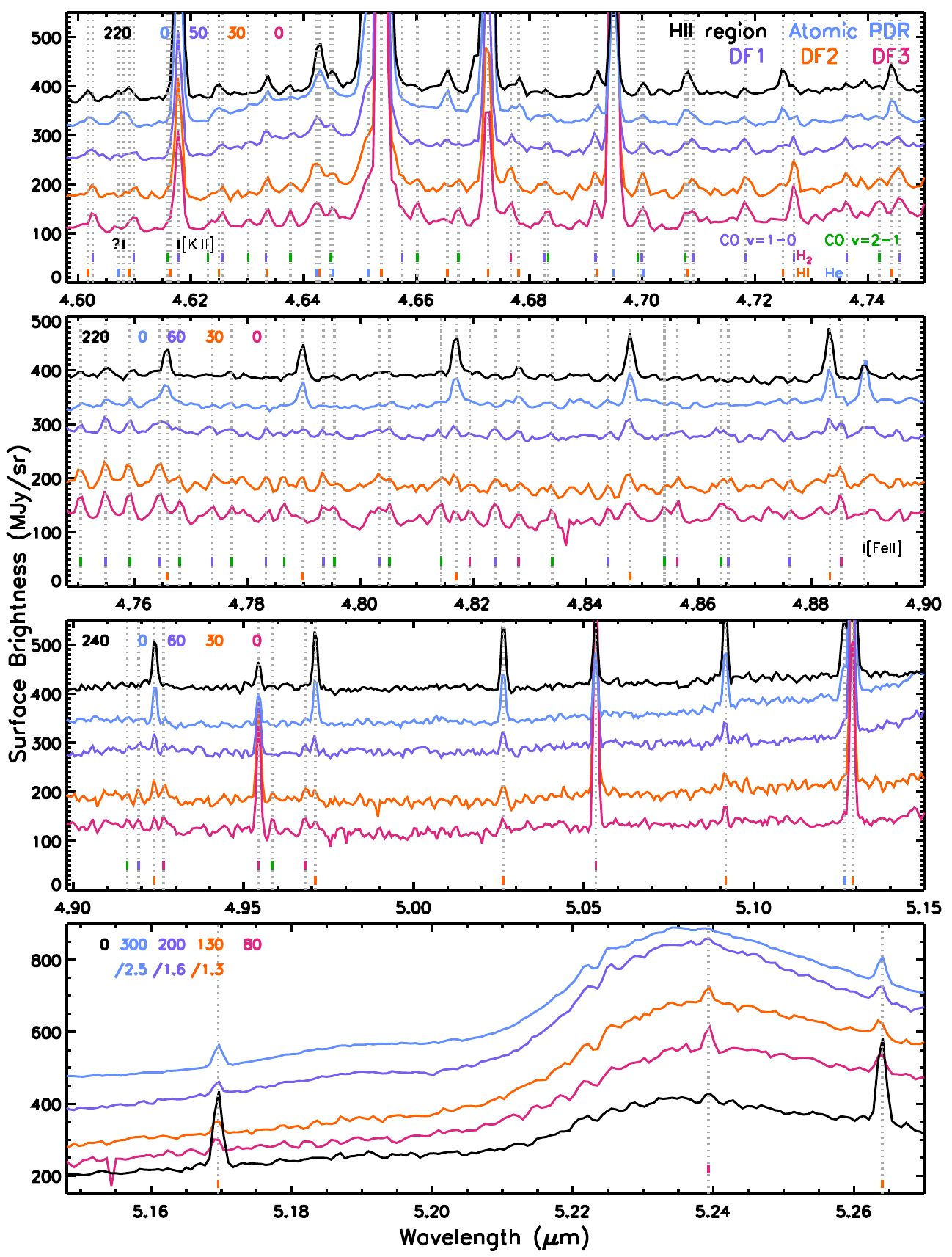}
    \caption{Fig.~\ref{fig:app:template1} continued. } 
    \label{fig:app:template8}
\end{figure*}

The spectral inventory of the five template spectra is shown in 
Figs.~\ref{fig:app:template1} to ~\ref{fig:app:template8}. Line intensities are given in Table~A.2, available at CDS. Column 1 lists the line identification, Column 2 the wavelength (\mum), Columns 3-7 the integrated line intensities in the five templates ($\times 10^{-5}$ erg cm$^{-2}$s$^{-1}$sr$^{-1}$). Intensities of the \molh lines in $v$=0 and $v$=1 states are given in Table~\ref{tab:fluxes}, intensities of selected \CI\ lines in Table~\ref{tab:CIfluxes}, and intensities of the AIB components in Table~\ref{tab:AIBflux}.

\section{HI recombination lines}
\label{app:HI}

The \HI\, recombination lines provide an estimate of the rms density in the ionised gas via:
\begin{equation}
\label{eq:EM}
    I_\lambda = \frac{hc}{\lambda}\frac{3.086\times 10^{18}}{4\pi}
    \alpha^{eff}_\lambda EM \,\,\,\,\,\,\,\,\,\,{\rm (erg\, cm^{-2}\,s^{-1}\,sr^{-1})},
\end{equation}
with $I_\lambda$ the intensity of the transition, $\alpha^{eff}_\lambda$ the effective recombination rate coefficient (cm$^3$\,s$^{-1}$) and EM the emission measure (cm$^{-6}$\,pc). We use $\alpha^{eff}_\lambda$ from case B recombination theory assuming an electron temperature of 10000~K and an electron density of $n_e = 1000\, {\rm cm^{-3}}$ \citep{Hummer:87}.

\section{\texorpdfstring{[He$^+$]/[H$^+$]}{He+/H+} abundance}
\label{app:he}

Based on the \HeI\, 1.70~\mum and the \HI\ 10-4 emission, \citet{Marconi98} estimated the [He$^+$]/[H$^+$] abundance from:
\begin{equation}
    \frac{F({\rm He\, {\textsc{i}}}\, 1.70~\mu{\rm m})}{F({\rm H\, {\textsc{i}}}\, 10-4)} = 3.61 \frac{[{\rm He}^+]}{[{\rm H}^+]},
\end{equation}
which is based on the model calculations of \citet{Smits:96} and assumes that the \HeI\, 1.70~\mum is only marginally affected by collisional excitations from the metastable 2$^3$S state as predicted by \citet{Osterbrock:92}.

\section{UV intensity}
\label{app:UVint}

We can estimate the UV continuum from UV pumped emission lines present in the NIRSpec wavelength range. The \OI\, $3{\rm d}\,^3{\rm D^o}-3{\rm p}\,^3{\rm P}$ 1.129~\mum and \OI\, $4{\rm s}\,^3{\rm S^o}-3{\rm p}\,^3{\rm P}$ 1.317~\mum emission result from UV pumping by photons of 1027 and 1040~${\mathring{\mathrm{A}}}$\, respectively. Hence, their UV intensity can be determined with: 
\begin{equation}
    I_\nu^{UV} = \frac{4\pi\sin(i)}{Af_b}\frac{\lambda_{IR}\lambda_{UV}}{cW_\lambda}I(IR) \hspace{1cm}{\rm (erg\, cm^{-2}\,s^{-1}\,Hz^{-1})},
\end{equation}
where $I(IR)$ is the observed intensity of the IR line in ${\rm erg\, cm^{-2}\,s^{-1}\,sr^{-1}}$, $i$ the inclination of the Bar with respect to the line-of-sight, $W_\lambda$ the equivalent width of the UV line, $W_\lambda/\lambda_{UV} = 3.6\times 10^{-5}$, $f_b$ the branching ratio or probability of IR emission following a UV photon absorption \citep[see Table 3 and 4 in][]{Walmsley00}, and A equals 3 for the \OI\, lines as their UV pumping lines are triplets with separation larger than $W_\lambda$ \citep{Marconi98, Walmsley00}.   
The \NI\, 1.2292~\mum emission is due to both the $3{\rm d}\,^4{\rm P}-3{\rm p}\, ^4{\rm S^o}$ and $4{\rm s}\,^4{\rm P}-3{\rm p}\, ^4{\rm P^o}$ transitions and occurs following absorption of UV photons of 953 and 965~${\mathring{\mathrm{A}}}$, respectively. The UV intensity can be estimated in a similar way as for the \OI\, lines, where A equals 1 because its UV pumping lines are a singlet. However, \NI\, has a more complex energy level diagram than \OI\, and, thus, this estimate is less straightforward. 

As this fluorescent emission originates from a narrow region in the ionisation front (see Sect.~\ref{sec:spatial-variation}), the calculated UV intensity represents the UV radiation emergent from the \HII\ region, where the PDR extinction is negligible. Hence, we only apply a foreground extinction, $exp(-\tau_{f, \lambda})$, with $\tau_{f, \lambda}$ the foreground optical depth as obtained in Sect.~\ref{subsec:HI}. We adopt an inclination $i$ of 4$^\circ$ \citep{Salgado16} and obtain the branching ratio from \citet[][their tables 3 and 4]{Walmsley00}. Assuming an interstellar radiation field of $1\, G_0 = 1.6\times 10^{-3}\, {\rm erg\, \rm cm^{-2}\,s^{-1}}$ between 6 and 13.6~eV, corresponding to $8.7\times 10^{-19}\, {\rm erg\, \rm cm^{-2}\,s^{-1}\,Hz^{-1}}$, the obtained UV line intensity can be converted to a normalised UV intensity, G$_0$.

\section{CI emission lines}
\label{app:CI}

The \CI\, emission lines provide the electron temperature, $T_e$, and gas density, $n_H$. 
We detect the forbidden fine-structure lines from 2p\,$^1$D$_2$ to 2p\,$^3$P$_1$ and 2p\,$^1$D$_2$ to 2p\,$^3$P$_2$ at respectively 0.9827 and 0.9854~\mum (Fig.~\ref{fig:app:template1}).  We do not detect the third fine-structure line from 2p\,$^1$D$_2$ to 2p\,$^3$P$_0$ that has a much smaller A value. In addition, we detect the multiplets 3s\,$^3$P$^0$ to 3p\,$^3$D at 1.0696~\mum and 3p\,$^3$D to 3d\,$^3$F$^0$ at 1.1759~\mum \citep[Fig.~\ref{fig:app:template1}; for wavelengths and transition probabilities, see][]{Walmsley00}.

The observed line intensities for the templates are given in Table~\ref{tab:CIfluxes}. The \CI\,  emission in the \Ta template likely originates from the background face-on PDR, whereas the \CI\, emission in the \Tb template originates in the edge-on PDR and in the \molh dissociation front templates it originates in the face-on PDR (see also the discussion in Sect.~\ref{sec:spatial-variation}). We adopt an internal extinction of $A_V=4$ and $A_V=10$ for, respectively, a face-on and edge-on PDR and apply the foreground extinction derived in Sect.~\ref{subsec:HI}. The resulting extinction corresponds to:
\begin{equation}
\label{eq:g}
    g(\tau_{p,\lambda}, \tau_{f,\lambda}) = exp(-\tau_{f,\lambda})\frac{1-exp(-\tau_{p,\lambda})}{\tau_{p,\lambda}},
\end{equation}
with $\tau_{p,\lambda}$ and $\tau_{f,\lambda}$ the PDR and foreground optical depths at the wavelength $\lambda$, respectively.

\begin{table}[!t]
    \centering
     \caption{\CI\ intensities for the five template spectra.}
    \label{tab:CIfluxes}
   \begin{tabular}{lccc}
   \hline \hline \\[-5pt]
      Wavelength & \multicolumn{1}{c}{I$_{Obs.}$} & \multicolumn{1}{c}{$g(\tau_{p,\lambda}, \tau_{f,\lambda})$} & \multicolumn{1}{c}{I$_{corr.}$}  \\
    \multicolumn{1}{c}{(1)} & \multicolumn{1}{c}{(2)} & \multicolumn{1}{c}{(3)} & \multicolumn{1}{c}{(4)}   \\
    \hline\\[-5pt]
    \multicolumn{4}{c}{\textbf{\Ta}} \\
    0.984 & 1.05$\pm$0.02 & 0.170 & 6.20$\pm$0.11 \\ 
1.0696 & 0.28$\pm$0.02 & 0.203 & 1.41$\pm$0.12 \\
1.1759 & 0.23$\pm$0.04 & 0.248 & 0.94$\pm$0.14 \\
1.0696/1.1759 & 1.22$\pm$0.21 & & 1.50$\pm$0.26 \\
0.984/1.0696 & 3.69$\pm$0.32 & & 4.41$\pm$0.38 \\ 
    \hline\\[-5pt]
    \multicolumn{4}{c}{\textbf{\Tb}} \\
0.984 & 1.32$\pm$0.02 & 0.083 & 15.81$\pm$0.24 \\
1.0696 & 0.37$\pm$0.03 & 0.101 & 3.68$\pm$0.28 \\
1.1759 & 0.41$\pm$0.02 & 0.128 & 3.23$\pm$0.18 \\
1.0696/1.1759 & 0.90$\pm$0.08 & & 1.14$\pm$0.11 \\
0.984/1.0696 & 3.53$\pm$0.27 & & 4.30$\pm$0.33 \\   
    \hline\\[-5pt]
    \multicolumn{4}{c}{\textbf{\Tc}} \\
    0.984 & 0.88$\pm$0.01 & 0.202 & 4.35$\pm$0.07 \\ 
1.0696 & 0.22$\pm$0.04 & 0.237 & 0.93$\pm$0.18 \\
1.1759 & 0.30$\pm$0.04 & 0.284 & 1.07$\pm$0.15 \\
1.0696/1.1759 & 0.73$\pm$0.18 & & 0.87$\pm$0.21 \\
0.984/1.0696 & 3.99$\pm$0.77 & & 4.67$\pm$0.90 \\    

    \hline\\[-5pt]
    \multicolumn{4}{c}{\textbf{\Td}} \\
0.984 & 1.13$\pm$0.02 & 0.201 & 5.63$\pm$0.09 \\ 
1.0696 & 0.20$\pm$0.03 & 0.236 & 0.86$\pm$0.12 \\
1.1759 & 0.26$\pm$0.04 & 0.283 & 0.92$\pm$0.15 \\
1.0696/1.1759 & 0.78$\pm$0.17 & & 0.93$\pm$0.20 \\
0.984/1.0696 & 5.60$\pm$0.80 & & 6.56$\pm$0.94 \\ 
    \hline\\[-5pt]
    \multicolumn{4}{c}{\textbf{\Te}} \\
0.984 & 1.75$\pm$0.02 & 0.184 & 9.53$\pm$0.10 \\ 
1.0696 & 0.34$\pm$0.02 & 0.217 & 1.56$\pm$0.09 \\
1.1759 & 0.21$\pm$0.05 & 0.264 & 0.79$\pm$0.17 \\
1.0696/1.1759 & 1.63$\pm$0.37 & & 1.98$\pm$0.45 \\
0.984/1.0696 & 5.15$\pm$0.31 & & 6.09$\pm$0.36 \\    
\hline
    \end{tabular}
\tablefoot{Columns: (1) wavelength (\mum). The 0.984 intensity is the sum of the 0.9827 and 0.9854~\mum line intensities, the 1.0696 intensity is the sum of the 1.0687 and 1.0695~\mum line intensities and the 1.1759 intensity is the sum of the 1.1752 and 1.1758~\mum line intensities.; (2) observed intensity ($10^{-4}$ \uint); (3) $g(\tau_{p,\lambda}, \tau_{f,\lambda})$ as defined in Eq.~\ref{eq:g}; (4) extinction corrected intensity ($10^{-4}$ \uint). }
\end{table}

\begin{figure}
\begin{center}
\resizebox{\hsize}{!}{%
\includegraphics{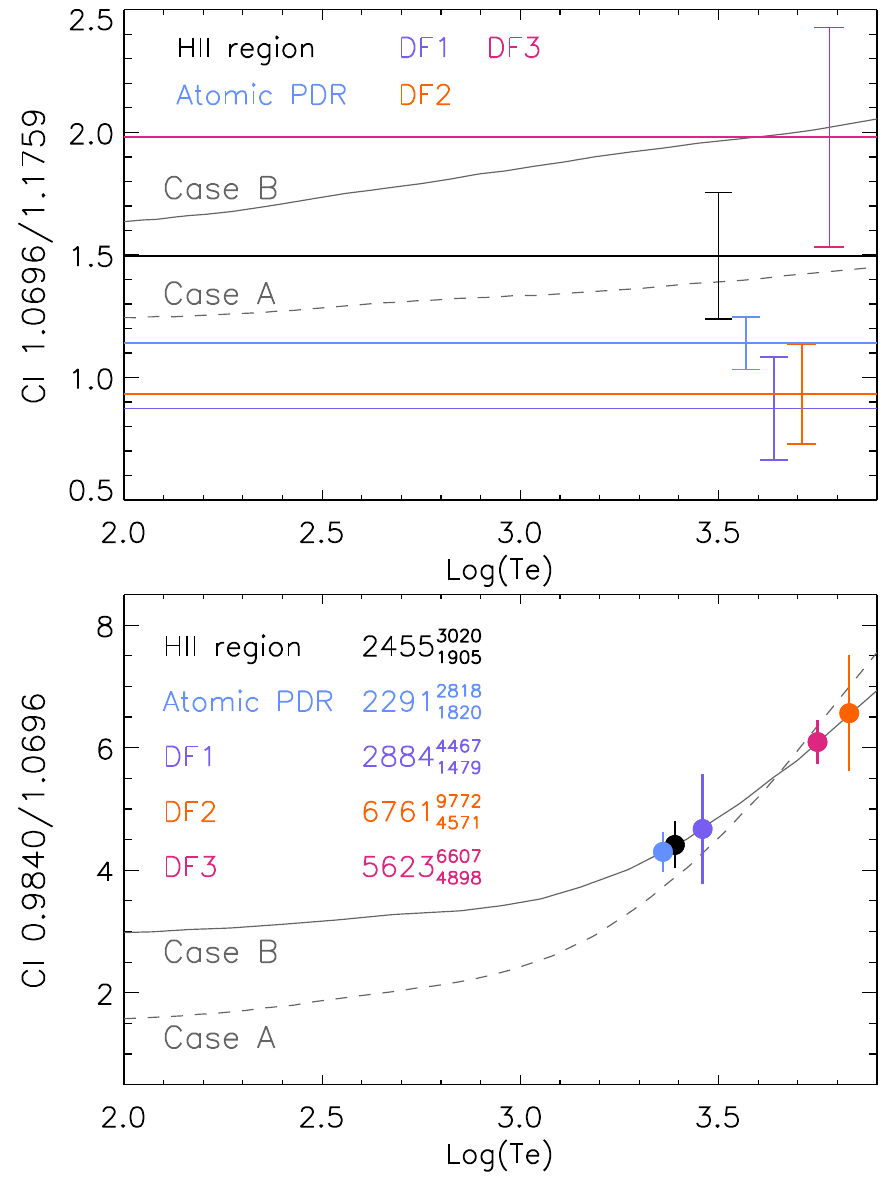}}
\caption{Comparison of extinction corrected \CI\ line ratios with computed ratios taken from \citet{Walmsley00} that are based on calculations of \citet{Escalante:90} for both case A and case B recombination theory. Uncertainties on the observed ratios are given by vertical lines that are placed at the derived electron temperature (lower panel) or in the $[3.5-3.8]$ x-range (top panel). }
\label{fig:CI_model}
\end{center}
\end{figure}

The 1.0696/1.1759 line ratio depends primarily on the optical depth of the UV pumping line \citep[][their Fig. 14]{Walmsley00}. We compare the observed intensities with their model calculations (Fig.~\ref{fig:CI_model}). 
We find that the extinction corrected 1.0696/1.1759 line ratio indicates case A conditions for the atomic PDR and case B conditions for \Te. The ratio for the \Ta can be consistent with either case A or B whereas the ratio falls below the theoretical curves for case A and B conditions for \Tc and \Td. \citet{Walmsley00} also reported case A conditions based on their observations (not extinction corrected). To investigate this further, we calculate the optical depth of a resonance line (for example 1261~\AA\, line) following \citet[][eq. 4.29]{Tielens:book2} assuming a typical line width of $\Delta v_{FWHM} = 3\,{\rm km/s}$ and a neutral carbon fraction of $10^{-5}$. The latter was determined using Eq.~9.6 in \citet{Tielens_book05} assuming a temperature of 1000~K, a UV radiation field G$_0$ of 10$^{4}$, a density of $3\times 10^6 {\rm cm^{-3}}$, and a visual extinction A$_V$ of 0. We evaluate the optical depth for two scenarios: one adopting a density $n_H = 10^{5}\,{\rm cm^{-3}}$ and a line-of-sight depth of $L=l^{los}_{PDR}=0.10\, {\rm pc}$ (Sect.~\ref{subsubsec:geometry}),  typical values for the Bar, and one adopting a density $n_H = 10^{7}\, {\rm cm^{-3}}$ and a line-of-sight depth of $L=10^{-3}\, {\rm pc}$ \citep[e.g.][]{Joblin18, goicoechea}, typical values for dense clumps reported in the Bar. In both cases we find that the resonance line is optically thick, suggesting case B conditions. The origin of the discrepancy with the result of the 1.0696/1.1759 line ratio will be investigated in a follow-up paper. For the remainder of this analysis we will assume case B conditions. 

The (0.983+0.985)/1.0696 line ratio depends primarily on the electron temperature \citep[][their Fig. 14]{Walmsley00}. Adopting case B conditions, the extinction corrected line ratio corresponds to electron temperatures of approximately 
2500, 2300, 2900, 6800, 5600 for the \Ta, the \Tb, \Tc, \Td, and \Te templates respectively (Fig.~\ref{fig:CI_model}). 
Given the uncertainty in the line ratio, the derived temperatures for the \Ta, the \Tb, and \Tc templates are consistent with each other. 
Likewise, the obtained temperature for the \Td and \Te templates are consistent within their uncertainty and are surprisingly high, similar to the electron temperature of around 4700~K obtained by \citet{Walmsley00} without correcting for extinction. No combination of foreground extinction (0-2 magn.) and internal extinction (0-15 magn.) results in an electron temperature below 1000~K for the \Te template (with the lowest obtained T being $\sim$3800~K for case B).

Lastly, we can obtain an estimate of the density from the [\CI]~0.984~\mum intensity that is the sum of the [\CI]~0.982 and 0.985~\mum intensities \citep[][Eq.~2]{Walmsley00}:
\begin{equation}
\label{eq:C:EM}
    I(0.984) = I_0\, T_3^{-0.6}\, EM\, g(\tau_{p,\lambda}, \tau_{f,\lambda}) \hspace{.5cm}{\rm (erg\, cm^{-2}\,s^{-1}\, sr^{-1})},
\end{equation}
with $I_0 = 6\times 10^{-7}$ for case B ($2.7\times 10^{-7}$ for case A), $T_3 = T/1000$~K, EM the carbon emission measure in ${\rm pc\, cm^{-6}}$ with ${\rm EM} = \int n_en_{C^+}ds$.  
As the \CI\ emission arises from a very thin layer of a few thousand degree gas (Sect.~\ref{subsec:Clines}), we adopt A$_V$ of 0.5 for this layer (i.e. $N = 1\times 10^{21} {\rm cm^{-2}}$). Assuming all C is ionised, a C gas phase abundance of $1.6\, 10^{-4}$, case B conditions, Eq.~\ref{eq:C:EM} can be written as:
\begin{multline}
\label{eq:C:nH}
    I(0.984) = 2.6 \times 10^{-5}\, \left( \frac{3000\, {\rm K}}{T} \right)^{0.6} \, \left( \frac{n}{1 \times 10^7 {\rm cm^{-3}}} \right) \\
    \hspace{.5cm}{\rm (erg\, cm^{-2}\,s^{-1}\, sr^{-1})},
\end{multline}

For the derived foreground extinction (Sect.~\ref{subsec:HI}), we obtain an extinction correction factor $g(\tau_{p,\lambda}, \tau_{f,\lambda})$ of 
0.170, 0.083, 0.202, 0.201, 0.184
for respectively the \Ta, \Tb, \Tc, \Td, and \Te templates. This is considerably smaller than the value of $\sim$0.3 used by \citet{Walmsley00} who adopt  $A_V=1.5$ for the PDR extinction. 
We obtain an emission measure ${\rm EM} = A\, T^{0.6}$, with A being 1034$\pm$18, 2635$\pm$40, 725$\pm$11, 938$\pm$15, and 1589$\pm$17 (cm$^{-6}$\,pc\, K$^{-0.6}$), respectively, for each of the five templates. Combined with the derived temperature, this results in an emission measure ${\rm EM}$ of 1.12$_{0.25}^{0.26}$, 2.73$_{0.56}^{0.63}$, 0.86$^{0.47}_{0.42}$, 1.86$^{0.83}_{0.60}$, and 2.82$^{0.49}_{0.37}$ 10$^{5}$cm$^{-6}$\,pc respectively.   
This results in a gas density, $n_H$, of $2.1$, $5.2$, $1.6$, $3.5$, and $5.3$ $\times\, 10^{8}\, {\rm cm^{-3}}$ for respectively, the \Ta, the \Tb, \Tc, \Td, and \Te templates.

\section{AIB emission}
\label{app:AIBs}

\subsection{AIB decomposition}
\label{app:AIBdecomposition}

\begin{table}[!h]
    \centering
     \caption{Fitting parameters used in the decomposition of the AIB emission.}
    \label{tab:AIBparam}
   \begin{tabular}{llllll}
   \hline \hline \\[-5pt]
      \multicolumn{3}{c}{PAHFIT} & \multicolumn{3}{c}{Gaussian decomposition} \\
    Band & \multicolumn{1}{c}{Position} & \multicolumn{1}{c}{FWHM} & Band &\multicolumn{1}{c}{Position} & \multicolumn{1}{c}{FWHM}\\
    \multicolumn{1}{c}{(1)} & \multicolumn{1}{c}{(2)} & \multicolumn{1}{c}{(3)} & \multicolumn{1}{c}{(4)} & \multicolumn{1}{c}{(5)} & \multicolumn{1}{c}{(6)}  \\
    \hline\\[-5pt]
     3.23 & 3.23  & 0.026  & 3.25 & 3.2465  & 0.0375 \\
   3.29 & 3.291 & 0.03762& 3.29 & 3.29027 & 0.0387 \\
        &       &        & 3.33 & 3.32821 & 0.0264 \\
   3.39 & 3.395 & 0.00995& 3.39 &  3.3944 & 0.0076 \\
   3.40 & 3.405 & 0.02691& 3.40 &  3.4031 & 0.0216 \\
   3.42 & 3.4253& 0.015  & 3.42 & 3.4242  & 0.0139 \\
   3.46 & 3.464 & 0.07012& 3.46 & 3.4649  & 0.0500 \\
   3.51 & 3.516 & 0.0271 & 3.51 & 3.5164  & 0.0224 \\
   3.56 & 3.561 & 0.02   & 3.56 & 3.5609  & 0.0352 \\
   & & & plateau & 3.4013 & 0.2438 \\
        \hline
    \end{tabular}
\tablefoot{Columns: (1)-(3) PAHFIT decomposition; (4)-(6) Gaussian decomposition; (1) AIB name; (2) peak position (\mum); (3) FWHM (\mum); (4) AIB name; (5) peak position (\mum); (6) FWHM (\mum).}
\end{table}

\begin{table}[!h]
    \centering
     \caption{Integrated intensities of the AIB components in the five template spectra ($10^{-3}$ \uint). }
    \label{tab:AIBflux}
   \begin{tabular}{crrrrr}
   \hline \hline \\[-5pt]
     Band & \multicolumn{1}{c}{\HII} & \multicolumn{1}{c}{Atomic} & \multicolumn{1}{c}{\Tc} &\multicolumn{1}{c}{\Td} & \multicolumn{1}{c}{\Te}\\
     & \multicolumn{1}{c}{region} & \multicolumn{1}{c}{PDR}\\
    \hline\\[-5pt]
   \multicolumn{6}{c}{\textbf{PAHFIT}}  \\
   3.23 &  0 & 0 & 0 & 0 & 0\\
   3.29 & 10.58 & 45.27 & 30.43 & 24.09 & 15.50\\
   3.39 & 0.30 & 1.59 & 1.05 & 0.78 & 0.47\\
   3.40 & 1.01 & 3.21 & 2.86 & 3.35 & 2.73\\
   3.42 & 0.15 & 0.51  & 0.32 & 0.33 & 0.25\\
   3.46 & 1.61 & 6.94 & 5.16 & 4.79 & 3.71\\
   3.51 & 0.39 & 1.93 & 1.43 & 1.28 & 1.08\\
   3.56 & 0.14 & 0.64 & 0.53 & 0.45 & 0\\
    \hline\\[-5pt]
\multicolumn{6}{c}{\textbf{Gaussian decomposition}} \\
3.25  & 0.85 & 3.48 &  2.23&  1.71&  1.24\\  
3.29 &  6.46 &  28.63 &  18.84 &  14.50 &  9.31\\
3.33 & 0.32 &  1.44 & 1.00 & 0.60 & 0.37 \\
3.39  &  0.08 & 0.63 & 0.37 & 0.25 & 0.15\\
3.40 & 0.51 &  1.53 &  1.42 &  1.60 & 1.27 \\
3.42  &   0.09 & 0.25& 0.18 & 0.24 & 0.20\\
3.46 & 0.27 &  1.06 & 0.84 & 0.75& 0.60 \\
3.51 & 0.14 & 0.71 & 0.50 & 0.44 & 0.35\\ 
3.56 & 0.06 & 0.21 & 0.17 & 0.09 &  0.08 \\
plat &  3.89 &  16.60 &  11.46 &  11.00 & 8.30 \\
        \hline\\[-5pt]
        \multicolumn{6}{c}{\textbf{Deuterated PAHs}\tablefootmark{a}} \\
        4.64 & 0.10 & 0.35 & 0.22 & 0.09 & 0.05 \\
        4.75 & 0.04 & 0.05 & 0.09 & 0.08 & 0.06\\
        \hline
    \end{tabular}
\tablefoot{\tablefoottext{a}{See Appendix~\ref{app:PADs} for details on the flux estimates. }}
\end{table}

\begin{figure}
\begin{center}
\resizebox{\hsize}{!}{%
\includegraphics{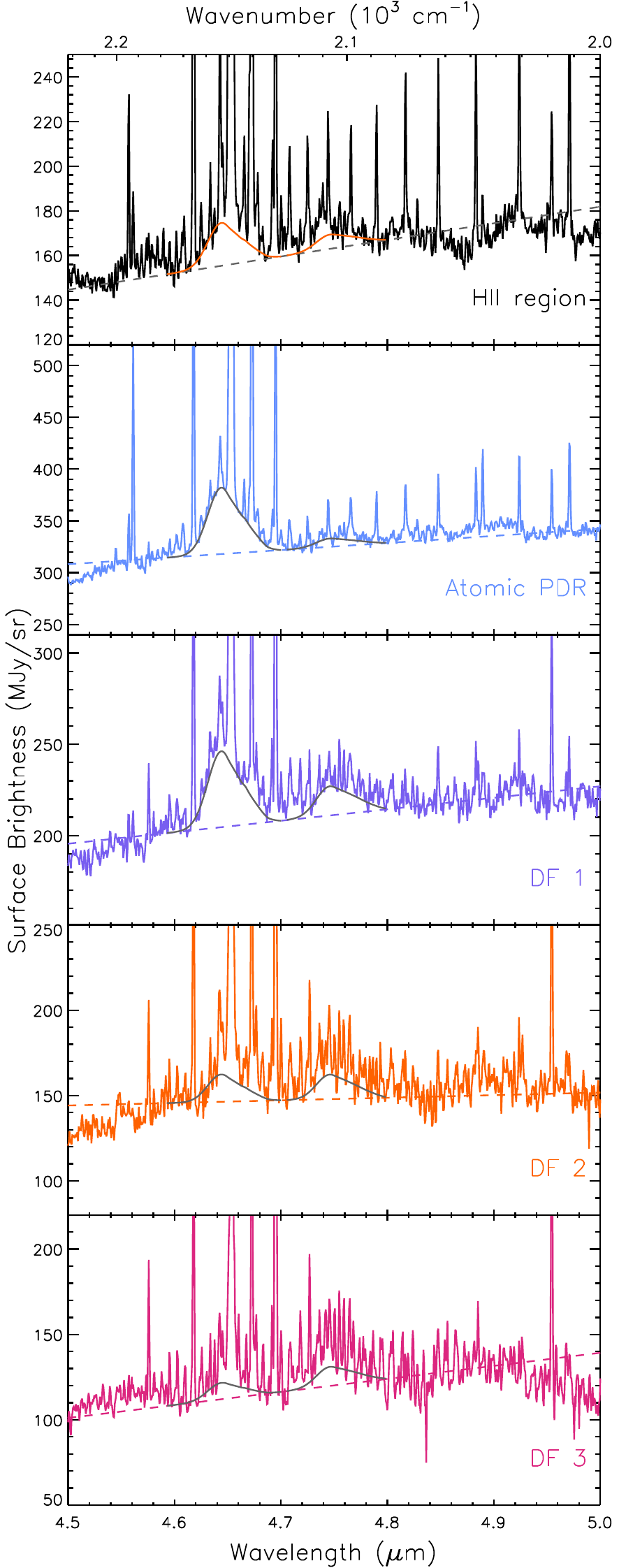}}
\caption{The tentative 4.644 and 4.746~\mum dust features attributed to deuterated PAHs (Sect.~\ref{sec:spectral-characteristics}). Local linear continua are shown by the dashed lines. The 4.646 and 4.746~\mum bands are plotted on top of the continuum as a solid line. The same band profile for each feature is used for all templates. See Appendix~\ref{app:PADs} for details. }
\label{fig:pad_app}
\end{center}
\end{figure}

We have performed two spectral decompositions of the AIB emission, which are applied to every pixel of the NIRSpec mosaic and to the five template spectra. First, we employ an updated version of PAHFIT \citep{PAHFIT2007}\footnote{available at \url{https://github.com/PAHFIT} We note that PAHFIT fit the spectrum expressed in \mum vs. MJy/sr.}. In PAHFIT-based models, the AIBs are represented using Drude profiles for simplicity\footnote{An isolated harmonic oscillator would give a Lorentz profile while an electron gas without restoring force would give a Drude profile.}. The continuum is fit to the entire F290LP range ($2.87 - 5.27$~\mum) using a superposition of fixed-temperature blackbody emission components and the emission lines (see Sect.~\ref{sec:spectral-characteristics}) are fit using Gaussian profiles with a FWHM that is determined by the resolution curve of F290LP. The AIB emission in the NIRSpec range is decomposed into seven components, their peak position and FWHM are listed in Table~\ref{tab:AIBparam}. The obtained fit reproduces the observations very well (Fig.~\ref{fig:AIBfit}). The component near 3.23~\mum cannot be fit because the blue wing of the 3.29~\mum feature is not reproduced well by a single Drude component (the PAHFIT model overestimates the AIB emission shortwards of 3.25~\mum). On the other hand, the Drude profiles can reproduce the overlap region between the 3.29 and 3.4~\mum AIBs without the requirement for an extra plateau-like component. The 3.4~\mum AIB consists of two components with different widths, referred to as the ``3.39'' and ``3.40'' components. The AIB emission at wavelengths longer than 3.4~\mum consists of a very broad band at 3.46~\mum, with two narrower bands at 3.42~\mum and 3.51~\mum. There is also a noticeable weaker and broad feature at 3.56~\mum, but just like the wing on the blue side of 3.29~\mum, the wing on the red side of the 3.51~\mum band is not fit as well. The width and power of 3.56~\mum AIB were therefore harder to determine. 
Second, we have employed a Gaussian decomposition of the AIB emission in the 3.2 to 3.7~\mum region after subtracting a linear dust continuum emission (determined in the [2.97, 3.03] and [3.65, 3.720]~\mum wavelength ranges)\footnote{We note that we fit the spectrum expressed in \mum vs. ${\rm W\, m^{-2}\,}$ \mum$^{-1}\,{\rm sr}^{-1}$.  }. Narrow emission lines were removed prior to fitting. We fitted the AIB emission with ten Gaussians that were highly constrained in peak position and FWHM ($\pm$0.0005 and $\pm$0.001, respectively; Table~\ref{tab:AIBparam}). The resulting fit reproduces the observations very well (Fig~\ref{fig:AIBfit}). We point out that in contrast to the PAHFIT method, one Gaussian represents the underlying plateau emission and one Gaussian represents the extended red wing of the 3.29~\mum AIB.  The remaining components are comparable between both decomposition methods. The integrated intensity of the AIB components in the five templates for both decomposition methods are given in Table~\ref{tab:AIBflux}.

\subsection{Deuterated PAHs}
\label{app:PADs}

While an emission band at 4.646~\mum is clearly visible in the \Ta, the \Tb, and \Tc templates (Fig.~\ref{fig:tentativebands}), we here investigate its potential presence in the \Td and \Te templates. This is severely hampered by the detection of gas-phase CO emission in the molecular PDR (Sect.\ref{sec:spectral-characteristics}) which coincides in wavelength with the potential 4.646~\mum band. We therefore fit the \Te template with an optically thin and optically thick LTE model of $^{12}$CO and $^{13}$CO to assess whether the $4.6-4.8$~\mum emission (in addition to the continuum emission) can be solely due to CO emission. The optically thin model does not provide a good fit to the data whereas the optically thick model provides a better fit to the data in terms of both the relative intensities and the density of lines. While the CO emission clearly requires more advanced modelling, this simple exercise suggests that the $4.62-4.68$~\mum and $4.71-4.79$~\mum range has additional broad band emission that is not reproduced by the CO models. Next, we extract the (asymmetric) 4.646~\mum band profile from the \Tb template where the band is strongest. We then scale this 4.646~\mum band profile to match the emission in the other templates (Fig.~\ref{fig:pad_app}). Given the presence of numerous emission lines in this wavelength range and the uncertainty on the continuum determination, this provides an approximate estimate of its intensity in the five templates which is an upper limit for those templates with strong CO emission. The derived intensities are 
given in Table~\ref{tab:AIBflux}. We note that within the uncertainties, the band profile does not vary between the templates. 

As the CO model also indicated additional emission in the $4.71-4.79$~\mum range, we applied the same method here to derive rough estimates of this broad band's intensity. In this case, we extract the band profile in the \Tc template where it is strongest. Similar as for the 4.646~\mum band, this 4.746~\mum band profile is also asymmetric and, when scaled, matches the observations in all templates, albeit it is relatively very weak in the \Ta and \Tb templates. For completeness, we give the approximate intensities of this band in Table~\ref{tab:AIBflux}.

\end{appendix}


\begin{thebibliography}{181}
\expandafter\ifx\csname natexlab\endcsname\relax\def\natexlab#1{#1}\fi

\bibitem[{{Abgrall} {et~al.}(1992){Abgrall}, {Le Bourlot}, {Pineau Des Forets},
  {Roueff}, {Flower}, \& {Heck}}]{abgrall}
{Abgrall}, H., {Le Bourlot}, J., {Pineau Des Forets}, G., {et~al.} 1992, \aap,
  253, 525

\bibitem[{{Ag{\'u}ndez} {et~al.}(2010){Ag{\'u}ndez}, {Goicoechea},
  {Cernicharo}, {Faure}, \& {Roueff}}]{Agundez10}
{Ag{\'u}ndez}, M., {Goicoechea}, J.~R., {Cernicharo}, J., {Faure}, A., \&
  {Roueff}, E. 2010, \apj, 713, 662

\bibitem[{{Allamandola} {et~al.}(2021){Allamandola}, {Boersma}, {Lee},
  {Bregman}, \& {Temi}}]{Allamandola:21}
{Allamandola}, L.~J., {Boersma}, C., {Lee}, T.~J., {Bregman}, J.~D., \& {Temi},
  P. 2021, \apjl, 917, L35

\bibitem[{{Allamandola} {et~al.}(1989){Allamandola}, {Tielens}, \&
  {Barker}}]{ATB}
{Allamandola}, L.~J., {Tielens}, A.~G.~G.~M., \& {Barker}, J.~R. 1989, ApJS,
  71, 733

\bibitem[{{Allers} {et~al.}(2005){Allers}, {Jaffe}, {Lacy}, {Draine}, \&
  {Richter}}]{Allers05}
{Allers}, K.~N., {Jaffe}, D.~T., {Lacy}, J.~H., {Draine}, B.~T., \& {Richter},
  M.~J. 2005, \apj, 630, 368

\bibitem[{{Andree-Labsch} {et~al.}(2017){Andree-Labsch}, {Ossenkopf-Okada}, \&
  {R{\"o}llig}}]{Andree17}
{Andree-Labsch}, S., {Ossenkopf-Okada}, V., \& {R{\"o}llig}, M. 2017, \aap,
  598, A2

\bibitem[{{Arab} {et~al.}(2012){Arab}, {Abergel}, {Habart}, {Bernard-Salas},
  {Ayasso}, {Dassas}, {Martin}, \& {White}}]{Arab12}
{Arab}, H., {Abergel}, A., {Habart}, E., {et~al.} 2012, \aap, 541, A19

\bibitem[{{Baldwin} {et~al.}(2000){Baldwin}, {Verner}, {Verner}, {Ferland},
  {Martin}, {Korista}, \& {Rubin}}]{Baldwin:00}
{Baldwin}, J.~A., {Verner}, E.~M., {Verner}, D.~A., {et~al.} 2000, \apjs, 129,
  229

\bibitem[{{Bally}(2008)}]{Bally:08}
{Bally}, J. 2008, {Overview of the Orion Complex}, ed. B.~{Reipurth}
  (Astronomical Society of the Pacific), 459

\bibitem[{{Bally} {et~al.}(2000){Bally}, {O'Dell}, \& {McCaughrean}}]{Bally00}
{Bally}, J., {O'Dell}, C.~R., \& {McCaughrean}, M.~J. 2000, \aj, 119, 2919

\bibitem[{{Bernard-Salas} {et~al.}(2012){Bernard-Salas}, {Habart}, {Arab},
  {Abergel}, {Dartois}, {Martin}, {Bontemps}, {Joblin}, {White}, {Bernard}, \&
  {Naylor}}]{Bernard-Salas12}
{Bernard-Salas}, J., {Habart}, E., {Arab}, H., {et~al.} 2012, \aap, 538, A37

\bibitem[{{Bernard-Salas} \& {Tielens}(2005)}]{bernard2005physical}
{Bernard-Salas}, J. \& {Tielens}, A.~G.~G.~M. 2005, \aap, 431, 523

\bibitem[{{Bern{\'e}} {et~al.}(2022){Bern{\'e}}, {Habart}, {Peeters},
  {Abergel}, {Bergin}, {Bernard-Salas}, {Bron}, {Cami}, {Dartois}, {Fuente},
  {Goicoechea}, {Gordon}, {Okada}, {Onaka}, {Robberto}, {R{\"o}llig},
  {Tielens}, {Vicente}, {Wolfire}, {Alarc{\'o}n}, {Boersma}, {Canin}, {Chown},
  {Dicken}, {Languignon}, {Le Gal}, {Pound}, {Trahin}, {Simmer}, {Sidhu}, {Van
  De Putte}, {Cuadrado}, {Guilloteau}, {Maragkoudakis}, {Schefter}, {Schirmer},
  {Cazaux}, {Aleman}, {Allamandola}, {Auchettl}, {Antonio Baratta}, {Bejaoui},
  {Bera}, {Bilalbegovi{\'c}}, {Black}, {Boulanger}, {Bouwman}, {Brandl},
  {Brechignac}, {Br{\"u}nken}, {Burkhardt}, {Candian}, {Cernicharo}, {Chabot},
  {Chakraborty}, {Champion}, {Colgan}, {Cooke}, {Coutens}, {Cox}, {Demyk},
  {Donovan Meyer}, {Engrand}, {Foschino}, {Garc{\'\i}a-Lario}, {Gavilan},
  {Gerin}, {Godard}, {Gottlieb}, {Guillard}, {Gusdorf}, {Hartigan}, {He},
  {Herbst}, {Hornekaer}, {J{\"a}ger}, {Janot-Pacheco}, {Joblin}, {Kaufman},
  {Kemper}, {Kendrew}, {Kirsanova}, {Klaassen}, {Knight}, {Kwok}, {Labiano},
  {Lai}, {Lee}, {Lefloch}, {Le Petit}, {Li}, {Linz}, {Mackie}, {Madden},
  {Mascetti}, {McGuire}, {Merino}, {Micelotta}, {Misselt}, {Morse}, {Mulas},
  {Neelamkodan}, {Ohsawa}, {Omont}, {Paladini}, {Elisabetta Palumbo}, {Pathak},
  {Pendleton}, {Petrignani}, {Pino}, {Puga}, {Rangwala}, {Rapacioli}, {Ricca},
  {Roman-Duval}, {Roser}, {Roueff}, {Rouill{\'e}}, {Salama}, {Sales},
  {Sandstrom}, {Sarre}, {Sciamma-O'Brien}, {Sellgren}, {Shannon}, {Shenoy},
  {Teyssier}, {Thomas}, {Togi}, {Verstraete}, {Witt}, {Wootten}, {Ysard},
  {Zettergren}, {Zhang}, {Zhang}, \& {Zhen}}]{pdrs4all}
{Bern{\'e}}, O., {Habart}, {\'E}., {Peeters}, E., {et~al.} 2022, \pasp, 134,
  054301

\bibitem[{{Bern{\'e}} {et~al.}(2023{\natexlab{a}}){Bern{\'e}}, {Habart},
  {Peeters}, \& {et al.}}]{Berne:proplyd}
{Bern{\'e}}, O., {Habart}, {\'E}., {Peeters}, E., \& {et al.}
  2023{\natexlab{a}}, Science, 000, 000

\bibitem[{{Bern{\'e}} {et~al.}(2023{\natexlab{b}}){Bern{\'e}}, {Martin-Drumel},
  {Schroetter}, \& {et al.}}]{Berne:proplyd2}
{Bern{\'e}}, O., {Martin-Drumel}, M.-A., {Schroetter}, I., \& {et al.}
  2023{\natexlab{b}}, Nature, 000, 000

\bibitem[{{Black} \& {Dalgarno}(1976)}]{Black76}
{Black}, J.~H. \& {Dalgarno}, A. 1976, \apj, 203, 132

\bibitem[{{Blagrave} {et~al.}(2007){Blagrave}, {Martin}, {Rubin}, {Dufour},
  {Baldwin}, {Hester}, \& {Walter}}]{Blagrave:07}
{Blagrave}, K.~P.~M., {Martin}, P.~G., {Rubin}, R.~H., {et~al.} 2007, \apj,
  655, 299

\bibitem[{{Boersma} {et~al.}(2012){Boersma}, {Rubin}, \&
  {Allamandola}}]{Boersma2012}
{Boersma}, C., {Rubin}, R.~H., \& {Allamandola}, L.~J. 2012, \apj, 753, 168

\bibitem[{{B{\"o}ker} {et~al.}(2022{\natexlab{a}}){B{\"o}ker}, {Arribas},
  {L{\"u}tzgendorf}, {Alves de Oliveira}, {Beck}, {Birkmann}, {Bunker},
  {Charlot}, {de Marchi}, {Ferruit}, {Giardino}, {Jakobsen}, {Kumari},
  {L{\'o}pez-Caniego}, {Maiolino}, {Manjavacas}, {Marston}, {Moseley},
  {Muzerolle}, {Ogle}, {Pirzkal}, {Rauscher}, {Rawle}, {Rix}, {Sabbi},
  {Sargent}, {Sirianni}, {te Plate}, {Valenti}, {Willott}, \&
  {Zeidler}}]{IFUnirspec}
{B{\"o}ker}, T., {Arribas}, S., {L{\"u}tzgendorf}, N., {et~al.}
  2022{\natexlab{a}}, \aap, 661, A82

\bibitem[{{B{\"o}ker} {et~al.}(2022{\natexlab{b}}){B{\"o}ker}, {Arribas},
  {L{\"u}tzgendorf}, {Alves de Oliveira}, {Beck}, {Birkmann}, {Bunker},
  {Charlot}, {de Marchi}, {Ferruit}, {Giardino}, {Jakobsen}, {Kumari},
  {L{\'o}pez-Caniego}, {Maiolino}, {Manjavacas}, {Marston}, {Moseley},
  {Muzerolle}, {Ogle}, {Pirzkal}, {Rauscher}, {Rawle}, {Rix}, {Sabbi},
  {Sargent}, {Sirianni}, {te Plate}, {Valenti}, {Willott}, \&
  {Zeidler}}]{boker2022}
{B{\"o}ker}, T., {Arribas}, S., {L{\"u}tzgendorf}, N., {et~al.}
  2022{\natexlab{b}}, aa, 661, A82

\bibitem[{{Boogert} {et~al.}(2015){Boogert}, {Gerakines}, \&
  {Whittet}}]{boogert2015}
{Boogert}, A.~C.~A., {Gerakines}, P.~A., \& {Whittet}, D. C.~B. 2015, \araa,
  53, 541

\bibitem[{{Bregman} {et~al.}(1989){Bregman}, {Allamandola}, {Tielens},
  {Geballe}, \& {Witteborn}}]{Bregman89}
{Bregman}, J.~D., {Allamandola}, L.~J., {Tielens}, A.~G.~G.~M., {Geballe},
  T.~R., \& {Witteborn}, F.~C. 1989, \apj, 344, 791

\bibitem[{{Brieva} {et~al.}(2016){Brieva}, {Gredel}, {J{\"a}ger}, {Huisken}, \&
  {Henning}}]{brieva2016}
{Brieva}, A.~C., {Gredel}, R., {J{\"a}ger}, C., {Huisken}, F., \& {Henning}, T.
  2016, \apj, 826, 122

\bibitem[{{Brooke} {et~al.}(2016){Brooke}, {Bernath}, {Western}, {Sneden},
  {Af{\c{s}}ar}, {Li}, \& {Gordon}}]{brooke2016}
{Brooke}, J. S.~A., {Bernath}, P.~F., {Western}, C.~M., {et~al.} 2016, \jqsrt,
  168, 142

\bibitem[{{Buragohain} {et~al.}(2015){Buragohain}, {Pathak}, {Sarre}, {Onaka},
  \& {Sakon}}]{Buragohain:15}
{Buragohain}, M., {Pathak}, A., {Sarre}, P., {Onaka}, T., \& {Sakon}, I. 2015,
  \mnras, 454, 193

\bibitem[{{Burton} {et~al.}(1990){Burton}, {Hollenbach}, \&
  {Tielens}}]{Burton90}
{Burton}, M.~G., {Hollenbach}, D.~J., \& {Tielens}, A.~G.~G.~M. 1990, \apj,
  365, 620

\bibitem[{{Cardelli} {et~al.}(1989){Cardelli}, {Clayton}, \&
  {Mathis}}]{Cardelli89}
{Cardelli}, J.~A., {Clayton}, G.~C., \& {Mathis}, J.~S. 1989, \apj, 345, 245

\bibitem[{{Champion} {et~al.}(2017){Champion}, {Bern{\'e}}, {Vicente}, {Kamp},
  {Le Petit}, {Gusdorf}, {Joblin}, \& {Goicoechea}}]{Champion17}
{Champion}, J., {Bern{\'e}}, O., {Vicente}, S., {et~al.} 2017, \aap, 604, A69

\bibitem[{{Changala} {et~al.}(2021){Changala}, {Neufeld}, \&
  {Godard}}]{Changala21}
{Changala}, P.~B., {Neufeld}, D.~A., \& {Godard}, B. 2021, \apj, 917, 16

\bibitem[{{Chown} {et~al.}(2023){Chown}, {Sidhu}, {Peeters}, {Tielens}, {Cami},
  {Habart}, {Bern{\'e}}, \& {the PDRs4All team}}]{Chown:23}
{Chown}, R., {Sidhu}, A., {Peeters}, E., {et~al.} 2023, \aap, 000, 000

\bibitem[{{Code}(1973)}]{Code:73}
{Code}, A.~D. 1973, in Interstellar Dust and Related Topics, ed. J.~M.
  {Greenberg} \& H.~C. {van de Hulst}, Vol.~52, 505

\bibitem[{{Cuadrado} {et~al.}(2017){Cuadrado}, {Goicoechea}, {Cernicharo},
  {Fuente}, {Pety}, \& {Tercero}}]{Cuadrado17}
{Cuadrado}, S., {Goicoechea}, J.~R., {Cernicharo}, J., {et~al.} 2017, \aap,
  603, A124

\bibitem[{{Cuadrado} {et~al.}(2015){Cuadrado}, {Goicoechea}, {Pilleri},
  {Cernicharo}, {Fuente}, \& {Joblin}}]{Cuadrado15}
{Cuadrado}, S., {Goicoechea}, J.~R., {Pilleri}, P., {et~al.} 2015, \aap, 575,
  A82

\bibitem[{{Cuadrado} {et~al.}(2019){Cuadrado}, {Salas}, {Goicoechea},
  {Cernicharo}, {Tielens}, \& {B{\'a}ez-Rubio}}]{Cuadrado19}
{Cuadrado}, S., {Salas}, P., {Goicoechea}, J.~R., {et~al.} 2019, \aap, 625, L3

\bibitem[{{Del Zanna} \& {Storey}(2022)}]{DelZanna:22}
{Del Zanna}, G. \& {Storey}, P.~J. 2022, \mnras, 513, 1198

\bibitem[{{Doney} {et~al.}(2016){Doney}, {Candian}, {Mori}, {Onaka}, \&
  {Tielens}}]{Doney:16}
{Doney}, K.~D., {Candian}, A., {Mori}, T., {Onaka}, T., \& {Tielens},
  A.~G.~G.~M. 2016, \aap, 586, A65

\bibitem[{{Egorov} {et~al.}(2017){Egorov}, {Lozinskaya}, {Moiseev}, \&
  {Shchekinov}}]{Egorov:17}
{Egorov}, O.~V., {Lozinskaya}, T.~A., {Moiseev}, A.~V., \& {Shchekinov}, Y.~A.
  2017, \mnras, 464, 1833

\bibitem[{{Egorov} {et~al.}(2014){Egorov}, {Lozinskaya}, {Moiseev}, \&
  {Smirnov-Pinchukov}}]{Egorov:14}
{Egorov}, O.~V., {Lozinskaya}, T.~A., {Moiseev}, A.~V., \& {Smirnov-Pinchukov},
  G.~V. 2014, \mnras, 444, 376

\bibitem[{{Elliott} \& {Meaburn}(1974)}]{Elliott1974}
{Elliott}, K.~H. \& {Meaburn}, J. 1974, \apss, 28, 351

\bibitem[{{Elmegreen} \& {Lada}(1977)}]{Elmegreen:77}
{Elmegreen}, B.~G. \& {Lada}, C.~J. 1977, \apj, 214, 725

\bibitem[{{Escalante} \& {Victor}(1990)}]{Escalante:90}
{Escalante}, V. \& {Victor}, G.~A. 1990, \apjs, 73, 513

\bibitem[{{Esteban} {et~al.}(1998){Esteban}, {Peimbert}, {Torres-Peimbert}, \&
  {Escalante}}]{Esteban:98}
{Esteban}, C., {Peimbert}, M., {Torres-Peimbert}, S., \& {Escalante}, V. 1998,
  \mnras, 295, 401

\bibitem[{{Esteban} {et~al.}(1994){Esteban}, {Vilchez}, \&
  {Smith}}]{Esteban:94}
{Esteban}, C., {Vilchez}, J.~M., \& {Smith}, L.~J. 1994, \aj, 107, 1041

\bibitem[{{Ferland} {et~al.}(2017){Ferland}, {Chatzikos}, {Guzm{\'a}n},
  {Lykins}, {van Hoof}, {Williams}, {Abel}, {Badnell}, {Keenan}, {Porter}, \&
  {Stancil}}]{cloudy}
{Ferland}, G.~J., {Chatzikos}, M., {Guzm{\'a}n}, F., {et~al.} 2017, \rmxaa, 53,
  385

\bibitem[{{Fuente} {et~al.}(2003){Fuente}, {Rodr{\i}guez-Franco},
  {Garc{\i}a-Burillo}, {Mart{\i}n-Pintado}, \& {Black}}]{Fuente03}
{Fuente}, A., {Rodr{\i}guez-Franco}, A., {Garc{\i}a-Burillo}, S.,
  {Mart{\i}n-Pintado}, J., \& {Black}, J.~H. 2003, \aap, 406, 899

\bibitem[{Gardner {et~al.}(2006)Gardner, Mather, Clampin, Doyon, Greenhouse,
  Hammel, Hutchings, Jakobsen, Lilly, Long, Lunine, Mccaughrean, Mountain,
  Nella, Rieke, Rieke, Rix, Smith, Sonneborn, Stiavelli, Stockman, Windhorst,
  \& Wright}]{gardner_JWST2006}
Gardner, J.~P., Mather, J.~C., Clampin, M., {et~al.} 2006, Space Science
  Reviews, 123, 485

\bibitem[{{Geballe} {et~al.}(1989){Geballe}, {Tielens}, {Allamandola},
  {Moorhouse}, \& {Brand}}]{geballe}
{Geballe}, T.~R., {Tielens}, A.~G.~G.~M., {Allamandola}, L.~J., {Moorhouse},
  A., \& {Brand}, P.~W.~J.~L. 1989, ApJ, 341, 278

\bibitem[{{Genzel} \& {Stutzki}(1989)}]{Genzel89}
{Genzel}, R. \& {Stutzki}, J. 1989, ARA\&A, 27, 41

\bibitem[{{Gerin} {et~al.}(2016){Gerin}, {Neufeld}, \& {Goicoechea}}]{Gerin16}
{Gerin}, M., {Neufeld}, D.~A., \& {Goicoechea}, J.~R. 2016, \araa, 54, 181

\bibitem[{{Giard} {et~al.}(1994){Giard}, {Bernard}, {Lacombe}, {Normand}, \&
  {Rouan}}]{Giard94}
{Giard}, M., {Bernard}, J.~P., {Lacombe}, F., {Normand}, P., \& {Rouan}, D.
  1994, \aap, 291, 239

\bibitem[{{Gibb} {et~al.}(2004){Gibb}, {Whittet}, {Boogert}, \&
  {Tielens}}]{gibb2004}
{Gibb}, E.~L., {Whittet}, D.~C.~B., {Boogert}, A.~C.~A., \& {Tielens},
  A.~G.~G.~M. 2004, \apjs, 151, 35

\bibitem[{{Godard} \& {Cernicharo}(2013)}]{Godard13}
{Godard}, B. \& {Cernicharo}, J. 2013, \aap, 550, A8

\bibitem[{{Goicoechea} {et~al.}(2017){Goicoechea}, {Cuadrado}, {Pety}, {Bron},
  {Black}, {Cernicharo}, {Chapillon}, {Fuente}, \& {Gerin}}]{Goicoechea17}
{Goicoechea}, J.~R., {Cuadrado}, S., {Pety}, J., {et~al.} 2017, \aap, 601, L9

\bibitem[{{Goicoechea} {et~al.}(2016){Goicoechea}, {Pety}, {Cuadrado},
  {Cernicharo}, {Chapillon}, {Fuente}, {Gerin}, {Joblin}, {Marcelino}, \&
  {Pilleri}}]{goicoechea}
{Goicoechea}, J.~R., {Pety}, J., {Cuadrado}, S., {et~al.} 2016, Nature, 537,
  207

\bibitem[{{Goicoechea} \& {Roncero}(2022)}]{Goicoechea22}
{Goicoechea}, J.~R. \& {Roncero}, O. 2022, \aap, 664, A190

\bibitem[{{Goicoechea} {et~al.}(2015){Goicoechea}, {Teyssier}, {Etxaluze},
  {Goldsmith}, {Ossenkopf}, {Gerin}, {Bergin}, {Black}, {Cernicharo},
  {Cuadrado}, {Encrenaz}, {Falgarone}, {Fuente}, {Hacar}, {Lis}, {Marcelino},
  {Melnick}, {M{\"u}ller}, {Persson}, {Pety}, {R{\"o}llig}, {Schilke}, {Simon},
  {Snell}, \& {Stutzki}}]{Goicoechea15}
{Goicoechea}, J.~R., {Teyssier}, D., {Etxaluze}, M., {et~al.} 2015, \apj, 812,
  75

\bibitem[{{Gordon} {et~al.}(2022){Gordon}, {Rothman}, {Hargreaves}, {Hashemi},
  {Karlovets}, {Skinner}, {Conway}, {Hill}, {Kochanov}, {Tan}, {Wcis{\l}o},
  {Finenko}, {Nelson}, {Bernath}, {Birk}, {Boudon}, {Campargue}, {Chance},
  {Coustenis}, {Drouin}, {Flaud}, {Gamache}, {Hodges}, {Jacquemart}, {Mlawer},
  {Nikitin}, {Perevalov}, {Rotger}, {Tennyson}, {Toon}, {Tran}, {Tyuterev},
  {Adkins}, {Baker}, {Barbe}, {Can{\`e}}, {Cs{\'a}sz{\'a}r}, {Dudaryonok},
  {Egorov}, {Fleisher}, {Fleurbaey}, {Foltynowicz}, {Furtenbacher}, {Harrison},
  {Hartmann}, {Horneman}, {Huang}, {Karman}, {Karns}, {Kassi}, {Kleiner},
  {Kofman}, {Kwabia-Tchana}, {Lavrentieva}, {Lee}, {Long}, {Lukashevskaya},
  {Lyulin}, {Makhnev}, {Matt}, {Massie}, {Melosso}, {Mikhailenko}, {Mondelain},
  {M{\"u}ller}, {Naumenko}, {Perrin}, {Polyansky}, {Raddaoui}, {Raston},
  {Reed}, {Rey}, {Richard}, {T{\'o}bi{\'a}s}, {Sadiek}, {Schwenke},
  {Starikova}, {Sung}, {Tamassia}, {Tashkun}, {Vander Auwera}, {Vasilenko},
  {Vigasin}, {Villanueva}, {Vispoel}, {Wagner}, {Yachmenev}, \&
  {Yurchenko}}]{Hitran:22}
{Gordon}, I.~E., {Rothman}, L.~S., {Hargreaves}, R.~J., {et~al.} 2022, \jqsrt,
  277, 107949

\bibitem[{{Gordon} {et~al.}(2023){Gordon}, {Clayton}, {Decleir}, {Fitzpatrick},
  {Massa}, {Misselt}, \& {Tollerud}}]{gordon2023}
{Gordon}, K.~D., {Clayton}, G.~C., {Decleir}, M., {et~al.} 2023, arXiv
  e-prints, arXiv:2304.01991

\bibitem[{{Gorti} \& {Hollenbach}(2002)}]{Gorti02}
{Gorti}, U. \& {Hollenbach}, D. 2002, \apj, 573, 215

\bibitem[{{Gro{\ss}schedl} {et~al.}(2018){Gro{\ss}schedl}, {Alves}, {Meingast},
  {Ackerl}, {Ascenso}, {Bouy}, {Burkert}, {Forbrich}, {F{\"u}rnkranz},
  {Goodman}, {Hacar}, {Herbst-Kiss}, {Lada}, {Larreina}, {Leschinski},
  {Lombardi}, {Moitinho}, {Mortimer}, \& {Zari}}]{Gross:18}
{Gro{\ss}schedl}, J.~E., {Alves}, J., {Meingast}, S., {et~al.} 2018, \aap, 619,
  A106

\bibitem[{{Guan} {et~al.}(2021){Guan}, {Jiang}, {Zhang}, {Yin}, {Cheng}, \&
  {Gao}}]{Guan:21}
{Guan}, L., {Jiang}, P., {Zhang}, G., {et~al.} 2021, \aap, 647, A127

\bibitem[{{G{\"u}del} {et~al.}(2008){G{\"u}del}, {Briggs}, {Montmerle},
  {Audard}, {Rebull}, \& {Skinner}}]{Gudel08}
{G{\"u}del}, M., {Briggs}, K.~R., {Montmerle}, T., {et~al.} 2008, Science, 319,
  309

\bibitem[{{Habart} {et~al.}(2010){Habart}, {Dartois}, {Abergel}, {Baluteau},
  {Naylor}, {Polehampton}, {Joblin}, {Ade}, {Anderson}, {Andr{\'e}}, {Arab},
  {Bernard}, {Blagrave}, {Bontemps}, {Boulanger}, {Cohen}, {Compiegne}, {Cox},
  {Davis}, {Emery}, {Fulton}, {Gry}, {Huang}, {Jones}, {Kirk}, {Lagache},
  {Lim}, {Madden}, {Makiwa}, {Martin}, {Miville-Desch{\^e}nes}, {Molinari},
  {Moseley}, {Motte}, {Okumura}, {Pinheiro Gon{\c c}alves}, {Rodon}, {Russeil},
  {Saraceno}, {Sidher}, {Spencer}, {Swinyard}, {Ward-Thompson}, {White}, \&
  {Zavagno}}]{Habart10}
{Habart}, E., {Dartois}, E., {Abergel}, A., {et~al.} 2010, \aap, 518, L116

\bibitem[{Habart {et~al.}(2023)Habart, Le~Gal, Alvarez, Olivier~Berne, Peeters,
  Wolfire, Goicoechea, \& Wolfire}]{Habart2022}
Habart, E., Le~Gal, R., Alvarez, C., {et~al.} 2023, \aap

\bibitem[{{Habart} {et~al.}(2023){Habart}, {Peeters}, \&
  {Bern{\'e}}}]{Habart:im}
{Habart}, {\'E}., {Peeters}, E., \& {Bern{\'e}}, O. 2023, \aap, 000, 000

\bibitem[{{Henney}(2021)}]{Henney2021}
{Henney}, W.~J. 2021, \mnras, 502, 4597

\bibitem[{{Herrmann} {et~al.}(1997){Herrmann}, {Madden}, {Nikola}, {Poglitsch},
  {Timmermann}, {Geis}, {Townes}, \& {Stacey}}]{Herrmann97}
{Herrmann}, F., {Madden}, S.~C., {Nikola}, T., {et~al.} 1997, \apj, 481, 343

\bibitem[{{Hogerheijde} {et~al.}(1995){Hogerheijde}, {Jansen}, \& {van
  Dishoeck}}]{Hoger95}
{Hogerheijde}, M.~R., {Jansen}, D.~J., \& {van Dishoeck}, E.~F. 1995, \aap,
  294, 792

\bibitem[{{Hopkins} {et~al.}(2012){Hopkins}, {Quataert}, \&
  {Murray}}]{hopkins2012}
{Hopkins}, P.~F., {Quataert}, E., \& {Murray}, N. 2012, \mnras, 421, 3522

\bibitem[{{Hudgins} {et~al.}(2004){Hudgins}, {Bauschlicher}, \&
  {Sandford}}]{Hudgins:04}
{Hudgins}, D.~M., {Bauschlicher}, C.~W., J., \& {Sandford}, S.~A. 2004, \apj,
  614, 770

\bibitem[{{Hummer} \& {Storey}(1987)}]{Hummer:87}
{Hummer}, D.~G. \& {Storey}, P.~J. 1987, \mnras, 224, 801

\bibitem[{{Jakobsen} {et~al.}(2022){Jakobsen}, {Ferruit}, {Alves de Oliveira},
  {Arribas}, {Bagnasco}, {Barho}, {Beck}, {Birkmann}, {B{\"o}ker}, {Bunker},
  {Charlot}, {de Jong}, {de Marchi}, {Ehrenwinkler}, {Falcolini}, {Fels},
  {Franx}, {Franz}, {Funke}, {Giardino}, {Gnata}, {Holota}, {Honnen}, {Jensen},
  {Jentsch}, {Johnson}, {Jollet}, {Karl}, {Kling}, {K{\"o}hler}, {Kolm},
  {Kumari}, {Lander}, {Lemke}, {L{\'o}pez-Caniego}, {L{\"u}tzgendorf},
  {Maiolino}, {Manjavacas}, {Marston}, {Maschmann}, {Maurer}, {Messerschmidt},
  {Moseley}, {Mosner}, {Mott}, {Muzerolle}, {Pirzkal}, {Pittet}, {Plitzke},
  {Posselt}, {Rapp}, {Rauscher}, {Rawle}, {Rix}, {R{\"o}del}, {Rumler},
  {Sabbi}, {Salvignol}, {Schmid}, {Sirianni}, {Smith}, {Strada}, {te Plate},
  {Valenti}, {Wettemann}, {Wiehe}, {Wiesmayer}, {Willott}, {Wright}, {Zeidler},
  \& {Zincke}}]{nirspec}
{Jakobsen}, P., {Ferruit}, P., {Alves de Oliveira}, C., {et~al.} 2022, \aap,
  661, A80

\bibitem[{Jansen {et~al.}(1995)Jansen, Spaans, Hogerheijde, \&
  Van~Dishoeck}]{jansen1995millimeter}
Jansen, D.~J., Spaans, M., Hogerheijde, M.~R., \& Van~Dishoeck, E.~F. 1995,
  \aap

\bibitem[{{Joblin} {et~al.}(2018){Joblin}, {Bron}, {Pinto}, {Pilleri}, {Le
  Petit}, {Gerin}, {Le Bourlot}, {Fuente}, {Bern{\'e}}, {Goicoechea}, {Habart},
  {K{\"o}hler}, {Teyssier}, {Nagy}, {Montillaud}, {Vastel}, {Cernicharo},
  {R{\"o}llig}, {Ossenkopf-Okada}, \& {Bergin}}]{Joblin18}
{Joblin}, C., {Bron}, E., {Pinto}, C., {et~al.} 2018, \aap, 615, A129

\bibitem[{{Joblin} {et~al.}(1996){Joblin}, {Tielens}, {Allamandola}, \&
  {Geballe}}]{Joblin:3umvsmethyl:96}
{Joblin}, C., {Tielens}, A.~G.~G.~M., {Allamandola}, L.~J., \& {Geballe}, T.~R.
  1996, \apj, 458, 610

\bibitem[{{Kaplan} {et~al.}(2021){Kaplan}, {Dinerstein}, {Kim}, \&
  {Jaffe}}]{Kaplan21}
{Kaplan}, K.~F., {Dinerstein}, H.~L., {Kim}, H., \& {Jaffe}, D.~T. 2021, \apj,
  919, 27

\bibitem[{{Kaplan} {et~al.}(2017){Kaplan}, {Dinerstein}, {Oh}, {Mace}, {Kim},
  {Sokal}, {Pavel}, {Lee}, {Pak}, {Park}, {Sok Oh}, \& {Jaffe}}]{Kaplan17}
{Kaplan}, K.~F., {Dinerstein}, H.~L., {Oh}, H., {et~al.} 2017, \apj, 838, 152

\bibitem[{{Kim} {et~al.}(2013){Kim}, {Ostriker}, \& {Kim}}]{kim2013}
{Kim}, C.-G., {Ostriker}, E.~C., \& {Kim}, W.-T. 2013, \apj, 776, 1

\bibitem[{{Kirsanova} {et~al.}(2008){Kirsanova}, {Sobolev}, {Thomasson},
  {Wiebe}, {Johansson}, \& {Seleznev}}]{Kirsanova:08}
{Kirsanova}, M.~S., {Sobolev}, A.~M., {Thomasson}, M., {et~al.} 2008, \mnras,
  388, 729

\bibitem[{{Knight} {et~al.}(2021{\natexlab{a}}){Knight}, {Peeters}, {Stock},
  {Vacca}, \& {Tielens}}]{Knight21}
{Knight}, C., {Peeters}, E., {Stock}, D.~J., {Vacca}, W.~D., \& {Tielens},
  A.~G.~G.~M. 2021{\natexlab{a}}, \apj, 918, 8

\bibitem[{{Knight} {et~al.}(2021{\natexlab{b}}){Knight}, {Peeters}, {Tielens},
  \& {Vacca}}]{Knight21:Orion}
{Knight}, C., {Peeters}, E., {Tielens}, A.~G.~G.~M., \& {Vacca}, W.~D.
  2021{\natexlab{b}}, \mnras

\bibitem[{{Koenig} {et~al.}(2008){Koenig}, {Allen}, {Gutermuth}, {Hora},
  {Brunt}, \& {Muzerolle}}]{Koenig:08}
{Koenig}, X.~P., {Allen}, L.~E., {Gutermuth}, R.~A., {et~al.} 2008, \apj, 688,
  1142

\bibitem[{{Kounkel} {et~al.}(2018){Kounkel}, {Covey}, {Su{\'a}rez},
  {Rom{\'a}n-Z{\'u}{\~n}iga}, {Hernandez}, {Stassun}, {Jaehnig}, {Feigelson},
  {Pe{\~n}a Ram{\'\i}rez}, {Roman-Lopes}, {Da Rio}, {Stringfellow}, {Kim},
  {Borissova}, {Fern{\'a}ndez-Trincado}, {Burgasser},
  {Garc{\'\i}a-Hern{\'a}ndez}, {Zamora}, {Pan}, \& {Nitschelm}}]{Kounkel2018}
{Kounkel}, M., {Covey}, K., {Su{\'a}rez}, G., {et~al.} 2018, \aj, 156, 84

\bibitem[{{Kraemer} {et~al.}(2022){Kraemer}, {Engelke}, {Renger}, \&
  {Sloan}}]{kraemer2022}
{Kraemer}, K.~E., {Engelke}, C.~W., {Renger}, B.~A., \& {Sloan}, G.~C. 2022,
  \aj, 164, 161

\bibitem[{{Krasnobaev} \& {Tagirova}(2017)}]{Krasnobaev:17}
{Krasnobaev}, K.~V. \& {Tagirova}, R.~R. 2017, \mnras, 469, 1403

\bibitem[{{Krasnobaev} {et~al.}(2016){Krasnobaev}, {Tagirova}, {Arafailov}, \&
  {Kotova}}]{Krasnobaev:16}
{Krasnobaev}, K.~V., {Tagirova}, R.~R., {Arafailov}, S.~I., \& {Kotova}, G.~Y.
  2016, Astronomy Letters, 42, 460

\bibitem[{{Lai} {et~al.}(2020){Lai}, {Smith}, {Baba}, {Spoon}, \&
  {Imanishi}}]{Lai:20}
{Lai}, T. S.~Y., {Smith}, J.~D.~T., {Baba}, S., {Spoon}, H. W.~W., \&
  {Imanishi}, M. 2020, \apj, 905, 55

\bibitem[{{Le Bourlot} {et~al.}(1993){Le Bourlot}, {Pineau Des Forets},
  {Roueff}, \& {Flower}}]{lebourlot}
{Le Bourlot}, J., {Pineau Des Forets}, G., {Roueff}, E., \& {Flower}, D.~R.
  1993, A\&A, 267, 233

\bibitem[{{Le Petit} {et~al.}(2006){Le Petit}, {Nehm{\'e}}, {Le Bourlot}, \&
  {Roueff}}]{lepetit}
{Le Petit}, F., {Nehm{\'e}}, C., {Le Bourlot}, J., \& {Roueff}, E. 2006, ApJS,
  164, 506

\bibitem[{{Leurini} {et~al.}(2006){Leurini}, {Rolffs}, {Thorwirth}, {Parise},
  {Schilke}, {Comito}, {Wyrowski}, {G{\"u}sten}, {Bergman}, {Menten}, \&
  {Nyman}}]{Leurini06}
{Leurini}, S., {Rolffs}, R., {Thorwirth}, S., {et~al.} 2006, \aap, 454, L47

\bibitem[{{Lis} \& {Schilke}(2003)}]{Lis03}
{Lis}, D.~C. \& {Schilke}, P. 2003, \apjl, 597, L145

\bibitem[{{Luhman} {et~al.}(1994){Luhman}, {Jaffe}, {Keller}, \&
  {Pak}}]{Luhman94}
{Luhman}, M.~L., {Jaffe}, D.~T., {Keller}, L.~D., \& {Pak}, S. 1994, \apjl,
  436, L185

\bibitem[{{Mackie} {et~al.}(2015){Mackie}, {Candian}, {Huang}, {Maltseva},
  {Petrignani}, {Oomens}, {Buma}, {Lee}, \& {Tielens}}]{Mackie:15}
{Mackie}, C.~J., {Candian}, A., {Huang}, X., {et~al.} 2015, J. Chem. Phys.,
  143, 224314

\bibitem[{{Mackie} {et~al.}(2016){Mackie}, {Candian}, {Huang}, {Maltseva},
  {Petrignani}, {Oomens}, {Mattioda}, {Buma}, {Lee}, \& {Tielens}}]{Mackie:16}
{Mackie}, C.~J., {Candian}, A., {Huang}, X., {et~al.} 2016, Journal of Chemical
  Physics, 145, 084313

\bibitem[{Mackie {et~al.}(2022)Mackie, Candian, Lee, \& Tielens}]{mackie2022}
Mackie, C.~J., Candian, A., Lee, T.~J., \& Tielens, A. G. G.~M. 2022, The
  Journal of Physical Chemistry A, 126, 3198, pMID: 35544706

\bibitem[{Maltseva {et~al.}(2015)Maltseva, Petrignani, Candian, Mackie, Huang,
  Lee, Tielens, Oomens, \& Buma}]{Maltseva:2015}
Maltseva, E., Petrignani, A., Candian, A., {et~al.} 2015, The Astrophysical
  Journal, 814, 23

\bibitem[{Maltseva {et~al.}(2016)Maltseva, Petrignani, Candian, Mackie, Huang,
  Lee, Tielens, Oomens, \& Buma}]{Maltseva:2016}
Maltseva, E., Petrignani, A., Candian, A., {et~al.} 2016, The Astrophysical
  Journal, 831, 58

\bibitem[{{Marconi} {et~al.}(1998){Marconi}, {Testi}, {Natta}, \&
  {Walmsley}}]{Marconi98}
{Marconi}, A., {Testi}, L., {Natta}, A., \& {Walmsley}, C.~M. 1998, \aap, 330,
  696

\bibitem[{{Mart{\'\i}n-Hern{\'a}ndez}
  {et~al.}(2003){Mart{\'\i}n-Hern{\'a}ndez}, {Bik}, {Kaper}, {Tielens}, \&
  {Hanson}}]{MartinHernandez:03}
{Mart{\'\i}n-Hern{\'a}ndez}, N.~L., {Bik}, A., {Kaper}, L., {Tielens},
  A.~G.~G.~M., \& {Hanson}, M.~M. 2003, \aap, 405, 175

\bibitem[{{McLeod} {et~al.}(2015){McLeod}, {Dale}, {Ginsburg}, {Ercolano},
  {Gritschneder}, {Ramsay}, \& {Testi}}]{McLeod}
{McLeod}, A.~F., {Dale}, J.~E., {Ginsburg}, A., {et~al.} 2015, MNRAS, 450, 1057

\bibitem[{{Menten} {et~al.}(2007{\natexlab{a}}){Menten}, {Reid}, {Forbrich}, \&
  {Brunthaler}}]{Menten:07}
{Menten}, K.~M., {Reid}, M.~J., {Forbrich}, J., \& {Brunthaler}, A.
  2007{\natexlab{a}}, \aap, 474, 515

\bibitem[{{Menten} {et~al.}(2007{\natexlab{b}}){Menten}, {Reid}, {Forbrich}, \&
  {Brunthaler}}]{Menten07}
{Menten}, K.~M., {Reid}, M.~J., {Forbrich}, J., \& {Brunthaler}, A.
  2007{\natexlab{b}}, \aap, 474, 515

\bibitem[{{Mori} {et~al.}(2014){Mori}, {Onaka}, {Sakon}, {Ishihara},
  {Shimonishi}, {Ohsawa}, \& {Bell}}]{Mori:14}
{Mori}, T.~I., {Onaka}, T., {Sakon}, I., {et~al.} 2014, \apj, 784, 53

\bibitem[{{Nagy} {et~al.}(2013){Nagy}, {Van der Tak}, {Ossenkopf}, {Gerin}, {Le
  Petit}, {Le Bourlot}, {Black}, {Goicoechea}, {Joblin}, {R{\"o}llig}, \&
  {Bergin}}]{Nagy13}
{Nagy}, Z., {Van der Tak}, F.~F.~S., {Ossenkopf}, V., {et~al.} 2013, \aap, 550,
  A96

\bibitem[{{Neufeld} {et~al.}(2021){Neufeld}, {Godard}, {Bryan Changala},
  {Faure}, {Geballe}, {G{\"u}sten}, {Menten}, \& {Wiesemeyer}}]{Neufeld21}
{Neufeld}, D.~A., {Godard}, B., {Bryan Changala}, P., {et~al.} 2021, \apj, 917,
  15

\bibitem[{{O'Dell}(2001)}]{Odell01}
{O'Dell}, C.~R. 2001, \araa, 39, 99

\bibitem[{{O'Dell} {et~al.}(2020){O'Dell}, {Abel}, \& {Ferland}}]{ODell:20}
{O'Dell}, C.~R., {Abel}, N.~P., \& {Ferland}, G.~J. 2020, \apj, 891, 46

\bibitem[{{O'Dell} {et~al.}(2017){O'Dell}, {Kollatschny}, \&
  {Ferland}}]{ODell17}
{O'Dell}, C.~R., {Kollatschny}, W., \& {Ferland}, G.~J. 2017, \apj, 837, 151

\bibitem[{{Ojha} {et~al.}(2011){Ojha}, {Samal}, {Pandey}, {Bhatt}, {Ghosh},
  {Sharma}, {Tamura}, {Mohan}, \& {Zinchenko}}]{Ojha:11}
{Ojha}, D.~K., {Samal}, M.~R., {Pandey}, A.~K., {et~al.} 2011, \apj, 738, 156

\bibitem[{{Onaka} {et~al.}(2014){Onaka}, {Mori}, {Sakon}, {Ohsawa}, {Kaneda},
  {Okada}, \& {Tanaka}}]{Onaka:14}
{Onaka}, T., {Mori}, T.~I., {Sakon}, I., {et~al.} 2014, \apj, 780, 114

\bibitem[{{Onaka} {et~al.}(2022){Onaka}, {Sakon}, \& {Shimonishi}}]{Onaka:22}
{Onaka}, T., {Sakon}, I., \& {Shimonishi}, T. 2022, \apj, 941, 190

\bibitem[{{Ossenkopf} {et~al.}(2013){Ossenkopf}, {R{\"o}llig}, {Neufeld},
  {Pilleri}, {Lis}, {Fuente}, {van der Tak}, \& {Bergin}}]{Ossenkopf13}
{Ossenkopf}, V., {R{\"o}llig}, M., {Neufeld}, D.~A., {et~al.} 2013, \aap, 550,
  A57

\bibitem[{{Osterbrock} {et~al.}(1992){Osterbrock}, {Tran}, \&
  {Veilleux}}]{Osterbrock:92}
{Osterbrock}, D.~E., {Tran}, H.~D., \& {Veilleux}, S. 1992, \apj, 389, 305

\bibitem[{{Pabst} {et~al.}(2019){Pabst}, {Higgins}, {Goicoechea}, {Teyssier},
  {Bern{\'e}}, {Chambers}, {Wolfire}, {Suri}, {Guesten}, {Stutzki}, {Graf},
  {Risacher}, \& {Tielens}}]{Pabst19}
{Pabst}, C., {Higgins}, R., {Goicoechea}, J.~R., {et~al.} 2019, \nat, 565, 618

\bibitem[{{Pabst} {et~al.}(2020){Pabst}, {Goicoechea}, {Teyssier}, {Bern{\'e}},
  {Higgins}, {Chambers}, {Kabanovic}, {G{\"u}sten}, {Stutzki}, \&
  {Tielens}}]{Pabst20}
{Pabst}, C.~H.~M., {Goicoechea}, J.~R., {Teyssier}, D., {et~al.} 2020, \aap,
  639, A2

\bibitem[{{Parikka} {et~al.}(2018){Parikka}, {Habart}, {Bernard-Salas},
  {K{\"o}hler}, \& {Abergel}}]{Parikka:2018}
{Parikka}, A., {Habart}, E., {Bernard-Salas}, J., {K{\"o}hler}, M., \&
  {Abergel}, A. 2018, \aap, 617, A77

\bibitem[{{Parmar} {et~al.}(1991){Parmar}, {Lacy}, \& {Achtermann}}]{Parmar91}
{Parmar}, P.~S., {Lacy}, J.~H., \& {Achtermann}, J.~M. 1991, \apjl, 372, L25

\bibitem[{{Peeters} {et~al.}(2004){Peeters}, {Allamandola}, {Bauschlicher},
  {Hudgins}, {Sandford}, \& {Tielens}}]{Peeters:pads:04}
{Peeters}, E., {Allamandola}, L.~J., {Bauschlicher}, C.~W., J., {et~al.} 2004,
  \apj, 604, 252

\bibitem[{{Peeters} {et~al.}(2017){Peeters}, {Bauschlicher}, {Allamandola},
  {Tielens}, {Ricca}, \& {Wolfire}}]{peeters17}
{Peeters}, E., {Bauschlicher}, Jr., C.~W., {Allamandola}, L.~J., {et~al.} 2017,
  ApJ, 836, 198

\bibitem[{{Peeters} {et~al.}(2012){Peeters}, {Tielens}, {Allamandola}, \&
  {Wolfire}}]{Peeters:12}
{Peeters}, E., {Tielens}, A.~G.~G.~M., {Allamandola}, L.~J., \& {Wolfire},
  M.~G. 2012, \apj, 747, 44

\bibitem[{{Pellegrini} {et~al.}(2009){Pellegrini}, {Baldwin}, {Ferland},
  {Shaw}, \& {Heathcote}}]{Pellegrini09}
{Pellegrini}, E.~W., {Baldwin}, J.~A., {Ferland}, G.~J., {Shaw}, G., \&
  {Heathcote}, S. 2009, \apj, 693, 285

\bibitem[{{Pilleri} {et~al.}(2015){Pilleri}, {Joblin}, {Boulanger}, \&
  {Onaka}}]{pilleri2015}
{Pilleri}, P., {Joblin}, C., {Boulanger}, F., \& {Onaka}, T. 2015, \aap, 577,
  A16

\bibitem[{{Pilleri} {et~al.}(2012){Pilleri}, {Montillaud}, {Bern{\'e}}, \&
  {Joblin}}]{Pilleri12}
{Pilleri}, P., {Montillaud}, J., {Bern{\'e}}, O., \& {Joblin}, C. 2012, A\&A,
  542, A69

\bibitem[{{Pound} \& {Wolfire}(2023)}]{2023AJ....165...25P}
{Pound}, M.~W. \& {Wolfire}, M.~G. 2023, \aj, 165, 25

\bibitem[{{Preibisch} \& {Zinnecker}(1999)}]{Preibisch:99}
{Preibisch}, T. \& {Zinnecker}, H. 1999, \aj, 117, 2381

\bibitem[{{Prozesky} \& {Smits}(2018)}]{Prozesky:18}
{Prozesky}, A. \& {Smits}, D.~P. 2018, \mnras, 478, 2766

\bibitem[{{Reiter} {et~al.}(2019){Reiter}, {McLeod}, {Klaassen}, {Guzm{\'a}n},
  {Dale}, {Mottram}, \& {Garay}}]{Reiter:19}
{Reiter}, M., {McLeod}, A.~F., {Klaassen}, P.~D., {et~al.} 2019, \mnras, 490,
  2056

\bibitem[{{Riashchikov} {et~al.}(2022){Riashchikov}, {Pomelnikov}, \&
  {Molevich}}]{Riashchikov:22}
{Riashchikov}, D.~S., {Pomelnikov}, I.~A., \& {Molevich}, N.~E. 2022, Bulletin
  of the Lebedev Physics Institute, 49, 307

\bibitem[{{R{\"o}llig} {et~al.}(2007){R{\"o}llig}, {Abel}, {Bell}, {Bensch},
  {Black}, {Ferland}, {Jonkheid}, {Kamp}, {Kaufman}, {Le Bourlot}, {Le Petit},
  {Meijerink}, {Morata}, {Ossenkopf}, {Roueff}, {Shaw}, {Spaans}, {Sternberg},
  {Stutzki}, {Thi}, {van Dishoeck}, {van Hoof}, {Viti}, \&
  {Wolfire}}]{Rollig2007}
{R{\"o}llig}, M., {Abel}, N.~P., {Bell}, T., {et~al.} 2007, \aap, 467, 187

\bibitem[{{Roueff} {et~al.}(2019){Roueff}, {Abgrall}, {Czachorowski},
  {Pachucki}, {Puchalski}, \& {Komasa}}]{Roueff19}
{Roueff}, E., {Abgrall}, H., {Czachorowski}, P., {et~al.} 2019, \aap, 630, A58

\bibitem[{{Rubin} {et~al.}(2011){Rubin}, {Simpson}, {O'Dell}, {McNabb},
  {Colgan}, {Zhuge}, {Ferland}, \& {Hidalgo}}]{Rubin2011}
{Rubin}, R.~H., {Simpson}, J.~P., {O'Dell}, C.~R., {et~al.} 2011, \mnras, 1526

\bibitem[{{Salas} {et~al.}(2019){Salas}, {Oonk}, {Emig}, {Pabst}, {Toribio},
  {R{\"o}ttgering}, \& {Tielens}}]{Salas19}
{Salas}, P., {Oonk}, J.~B.~R., {Emig}, K.~L., {et~al.} 2019, \aap, 626, A70

\bibitem[{{Salgado} {et~al.}(2016){Salgado}, {Bern{\'e}}, {Adams}, {Herter},
  {Keller}, \& {Tielens}}]{Salgado16}
{Salgado}, F., {Bern{\'e}}, O., {Adams}, J.~D., {et~al.} 2016, \apj, 830, 118

\bibitem[{{Schirmer} {et~al.}(2022){Schirmer}, {Ysard}, {Habart}, {Jones},
  {Abergel}, \& {Verstraete}}]{Schirmer2022}
{Schirmer}, T., {Ysard}, N., {Habart}, E., {et~al.} 2022, \aap, 666, A49

\bibitem[{{Sellgren} {et~al.}(1990){Sellgren}, {Tokunaga}, \&
  {Nakada}}]{Sellgren90}
{Sellgren}, K., {Tokunaga}, A.~T., \& {Nakada}, Y. 1990, \apj, 349, 120

\bibitem[{{Shaw} {et~al.}(2009){Shaw}, {Ferland}, {Henney}, {Stancil}, {Abel},
  {Pellegrini}, {Baldwin}, \& {van Hoof}}]{Shaw09}
{Shaw}, G., {Ferland}, G.~J., {Henney}, W.~J., {et~al.} 2009, \apj, 701, 677

\bibitem[{Sheffer {et~al.}(2011)Sheffer, Wolfire, Hollenbach, Kaufman, \&
  Cordier}]{Sheffer11}
Sheffer, Y., Wolfire, M.~G., Hollenbach, D.~J., Kaufman, M.~J., \& Cordier, M.
  2011, ApJ, 741, 45

\bibitem[{{Simon} {et~al.}(1997){Simon}, {Stutzki}, {Sternberg}, \&
  {Winnewisser}}]{Simon97}
{Simon}, R., {Stutzki}, J., {Sternberg}, A., \& {Winnewisser}, G. 1997, \aap,
  327, L9

\bibitem[{{Sloan} {et~al.}(1997){Sloan}, {Bregman}, {Geballe}, {Allamandola},
  \& {Woodward}}]{Sloan:97}
{Sloan}, G.~C., {Bregman}, J.~D., {Geballe}, T.~R., {Allamandola}, L.~J., \&
  {Woodward}, E. 1997, \apj, 474, 735

\bibitem[{{Smith} {et~al.}(2007){Smith}, {Draine}, {Dale}, {Moustakas},
  {Kennicutt}, {Helou}, {Armus}, {Roussel}, {Sheth}, {Bendo}, {Buckalew},
  {Calzetti}, {Engelbracht}, {Gordon}, {Hollenbach}, {Li}, {Malhotra},
  {Murphy}, \& {Walter}}]{PAHFIT2007}
{Smith}, J.~D.~T., {Draine}, B.~T., {Dale}, D.~A., {et~al.} 2007, \apj, 656,
  770

\bibitem[{{Smits}(1996)}]{Smits:96}
{Smits}, D.~P. 1996, \mnras, 278, 683

\bibitem[{{Sota} {et~al.}(2011){Sota}, {Ma{\'\i}z Apell{\'a}niz}, {Walborn},
  {Alfaro}, {Barb{\'a}}, {Morrell}, {Gamen}, \& {Arias}}]{Sota:11}
{Sota}, A., {Ma{\'\i}z Apell{\'a}niz}, J., {Walborn}, N.~R., {et~al.} 2011,
  \apjs, 193, 24

\bibitem[{{Stacey} {et~al.}(1993){Stacey}, {Jaffe}, {Geis}, {Grenzel},
  {Harris}, {Poglitsch}, {Stutzki}, \& {Townes}}]{Stacey93}
{Stacey}, G.~J., {Jaffe}, D.~T., {Geis}, N., {et~al.} 1993, \apj, 404, 219

\bibitem[{{Sternberg} \& {Dalgarno}(1989)}]{Sternberg89}
{Sternberg}, A. \& {Dalgarno}, A. 1989, \apj, 338, 197

\bibitem[{{Sternberg} \& {Dalgarno}(1995)}]{Sternberg95}
{Sternberg}, A. \& {Dalgarno}, A. 1995, \apjs, 99, 565

\bibitem[{{Stock} \& {Peeters}(2017)}]{Stock2017}
{Stock}, D.~J. \& {Peeters}, E. 2017, \apj, 837, 129

\bibitem[{{Stoerzer} {et~al.}(1995){Stoerzer}, {Stutzki}, \&
  {Sternberg}}]{Stoerzer95}
{Stoerzer}, H., {Stutzki}, J., \& {Sternberg}, A. 1995, \aap, 296, L9

\bibitem[{{Tabone} {et~al.}(2021){Tabone}, {van Hemert}, {van Dishoeck}, \&
  {Black}}]{Tabone2021}
{Tabone}, B., {van Hemert}, M.~C., {van Dishoeck}, E.~F., \& {Black}, J.~H.
  2021, \aap, 650, A192

\bibitem[{{Tauber} {et~al.}(1995){Tauber}, {Lis}, {Keene}, {Schilke}, \&
  {Buettgenbach}}]{Tauber95}
{Tauber}, J.~A., {Lis}, D.~C., {Keene}, J., {Schilke}, P., \& {Buettgenbach},
  T.~H. 1995, \aap, 297, 567

\bibitem[{{Tauber} {et~al.}(1994){Tauber}, {Tielens}, {Meixner}, \&
  {Goldsmith}}]{Tauber94}
{Tauber}, J.~A., {Tielens}, A.~G.~G.~M., {Meixner}, M., \& {Goldsmith}, P.~F.
  1994, \apj, 422, 136

\bibitem[{Tielens(2005)}]{Tielens_book05}
Tielens, A. 2005, The Physics and Chemistry of the Interstellar Medium
  (Springer Netherlands)

\bibitem[{Tielens(2021)}]{Tielens:book2}
Tielens, A.~G. 2021, Molecular astrophysics (Cambridge University Press)

\bibitem[{{Tielens} \& {Hollenbach}(1985{\natexlab{a}})}]{tielens:85}
{Tielens}, A.~G.~G.~M. \& {Hollenbach}, D. 1985{\natexlab{a}}, \apj, 291, 722

\bibitem[{{Tielens} \& {Hollenbach}(1985{\natexlab{b}})}]{tielens:85b}
{Tielens}, A.~G.~G.~M. \& {Hollenbach}, D. 1985{\natexlab{b}}, \apj, 291, 747

\bibitem[{{Tielens} {et~al.}(1993){Tielens}, {Meixner}, {van der Werf},
  {Bregman}, {Tauber}, {Stutzki}, \& {Rank}}]{Tielens93}
{Tielens}, A.~G.~G.~M., {Meixner}, M.~M., {van der Werf}, P.~P., {et~al.} 1993,
  Science, 262, 86

\bibitem[{{Van De Putte} {et~al.}(2023){Van De Putte}, {Trahin}, {Habart},
  {Peeters}, \& {Bern{\'e}}}]{vandeputte23}
{Van De Putte}, D., {Trahin}, B., {Habart}, {\'E}., {Peeters}, E., \&
  {Bern{\'e}}, O. 2023, \aap, 000, 000

\bibitem[{{van der Tak} {et~al.}(2013){van der Tak}, {Nagy}, {Ossenkopf},
  {Makai}, {Black}, {Faure}, {Gerin}, \& {Bergin}}]{Tak13}
{van der Tak}, F.~F.~S., {Nagy}, Z., {Ossenkopf}, V., {et~al.} 2013, \aap, 560,
  A95

\bibitem[{{van der Werf} {et~al.}(2013){van der Werf}, {Goss}, \&
  {O'Dell}}]{vanderWerf13}
{van der Werf}, P.~P., {Goss}, W.~M., \& {O'Dell}, C.~R. 2013, \apj, 762, 101

\bibitem[{{van der Werf} {et~al.}(1996){van der Werf}, {Stutzki}, {Sternberg},
  \& {Krabbe}}]{vanderWerf96}
{van der Werf}, P.~P., {Stutzki}, J., {Sternberg}, A., \& {Krabbe}, A. 1996,
  \aap, 313, 633

\bibitem[{{van der Wiel} {et~al.}(2009){van der Wiel}, {van der Tak},
  {Ossenkopf}, {Spaans}, {Roberts}, {Fuller}, \& {Plume}}]{Wiel09}
{van der Wiel}, M.~H.~D., {van der Tak}, F.~F.~S., {Ossenkopf}, V., {et~al.}
  2009, \aap, 498, 161

\bibitem[{{van Diedenhoven} {et~al.}(2004){van Diedenhoven}, {Peeters}, {Van
  Kerckhoven}, {Hony}, {Hudgins}, {Allamandola}, \&
  {Tielens}}]{vandiedenhoven2004}
{van Diedenhoven}, B., {Peeters}, E., {Van Kerckhoven}, C., {et~al.} 2004,
  \apj, 611, 928

\bibitem[{{van Dishoeck} \& {Black}(1988)}]{vanDishoeck:88}
{van Dishoeck}, E.~F. \& {Black}, J.~H. 1988, \apj, 334, 771

\bibitem[{{van Hoof}(2018)}]{vanhoof2018}
{van Hoof}, P. A.~M. 2018, Galaxies, 6, 63

\bibitem[{Vicente {et~al.}(2013)Vicente, Bern{\'e}, Tielens, Hu{\'e}lamo,
  Pantin, Kamp, \& Carmona}]{vicente2013polycyclic}
Vicente, S., Bern{\'e}, O., Tielens, A., {et~al.} 2013, The Astrophysical
  Journal Letters, 765, L38

\bibitem[{{Visser} {et~al.}(2009){Visser}, {van Dishoeck}, \&
  {Black}}]{Visser:09}
{Visser}, R., {van Dishoeck}, E.~F., \& {Black}, J.~H. 2009, \aap, 503, 323

\bibitem[{{Walmsley} {et~al.}(2000){Walmsley}, {Natta}, {Oliva}, \&
  {Testi}}]{Walmsley00}
{Walmsley}, C.~M., {Natta}, A., {Oliva}, E., \& {Testi}, L. 2000, \aap, 364,
  301

\bibitem[{{Weilbacher} {et~al.}(2015){Weilbacher}, {Monreal-Ibero},
  {Kollatschny}, {Ginsburg}, {McLeod}, {Kamann}, {Sandin}, {Palsa}, {Wisotzki},
  {Bacon}, {Selman}, {Brinchmann}, {Caruana}, {Kelz}, {Martinsson},
  {P{\'e}contal-Rousset}, {Richard}, \& {Wendt}}]{Weilbacher15}
{Weilbacher}, P.~M., {Monreal-Ibero}, A., {Kollatschny}, W., {et~al.} 2015,
  \aap, 582, A114

\bibitem[{Wen \& O'dell(1995)}]{wen1995three}
Wen, Z. \& O'dell, C. 1995, The Astrophysical Journal, 438, 784

\bibitem[{{Wiersma} {et~al.}(2020){Wiersma}, {Candian}, {Bakker}, {Martens},
  {Berden}, {Oomens}, {Buma}, \& {Petrignani}}]{Wiersma:20}
{Wiersma}, S.~D., {Candian}, A., {Bakker}, J.~M., {et~al.} 2020, \aap, 635, A9

\bibitem[{{Williams} \& {McKee}(1997)}]{williams1997}
{Williams}, J.~P. \& {McKee}, C.~F. 1997, \apj, 476, 166

\bibitem[{{Wolfire} {et~al.}(2010){Wolfire}, {Hollenbach}, \&
  {McKee}}]{Wolfire:10}
{Wolfire}, M.~G., {Hollenbach}, D., \& {McKee}, C.~F. 2010, \apj, 716, 1191

\bibitem[{Wolfire {et~al.}(2003)Wolfire, McKee, Hollenbach, \&
  Tielens}]{wolfire_neutral_2003}
Wolfire, M.~G., McKee, C.~F., Hollenbach, D., \& Tielens, A. G. G.~M. 2003, The
  Astrophysical Journal, 587, 278

\bibitem[{{Wolfire} {et~al.}(2022){Wolfire}, {Vallini}, \&
  {Chevance}}]{wolfire2022}
{Wolfire}, M.~G., {Vallini}, L., \& {Chevance}, M. 2022, \araa, 60, 247

\bibitem[{{Wyrowski} {et~al.}(1997){Wyrowski}, {Schilke}, {Hofner}, \&
  {Walmsley}}]{Wyrowski97}
{Wyrowski}, F., {Schilke}, P., {Hofner}, P., \& {Walmsley}, C.~M. 1997, \apjl,
  487, L171

\bibitem[{{Yang} {et~al.}(2020){Yang}, {Li}, \& {Glaser}}]{Yang:20}
{Yang}, X.~J., {Li}, A., \& {Glaser}, R. 2020, \apjs, 251, 12

\bibitem[{{Yang} {et~al.}(2021){Yang}, {Li}, {He}, \& {Glaser}}]{Yang:21}
{Yang}, X.~J., {Li}, A., {He}, C.~Y., \& {Glaser}, R. 2021, \apjs, 255, 23

\bibitem[{{Young Owl} {et~al.}(2000){Young Owl}, {Meixner}, {Wolfire},
  {Tielens}, \& {Tauber}}]{YoungOwl00}
{Young Owl}, R.~C., {Meixner}, M.~M., {Wolfire}, M., {Tielens}, A.~G.~G.~M., \&
  {Tauber}, J. 2000, \apj, 540, 886

\bibitem[{{Yousefi} {et~al.}(2018){Yousefi}, {Bernath}, {Hodges}, \&
  {Masseron}}]{yousefi2018}
{Yousefi}, M., {Bernath}, P.~F., {Hodges}, J., \& {Masseron}, T. 2018, \jqsrt,
  217, 416

\bibitem[{{Zannese} {et~al.}(2023){Zannese}, {Tabone}, {Habart}, {Le Petit},
  {van Dishoeck}, \& {Bron}}]{zannese2023}
{Zannese}, M., {Tabone}, B., {Habart}, E., {et~al.} 2023, \aap, 671, A41

\bibitem[{{Zatsarinny} \& {Bartschat}(2013)}]{Zatsarinny:13}
{Zatsarinny}, O. \& {Bartschat}, K. 2013, Journal of Physics B Atomic Molecular
  Physics, 46, 112001

\bibitem[{{Zatsarinny} {et~al.}(2005){Zatsarinny}, {Bartschat}, {Bandurina}, \&
  {Gedeon}}]{Zatsarinny:05}
{Zatsarinny}, O., {Bartschat}, K., {Bandurina}, L., \& {Gedeon}, V. 2005, \pra,
  71, 042702

\end{thebibliography}
\end{document}